\documentclass[12pt]{article}

\usepackage{latexsym}
\usepackage{amssymb,amsfonts,amsmath}
\usepackage{graphicx} 
\usepackage{indentfirst}
\usepackage{bbm}
\usepackage{amssymb}
\usepackage{verbatim}
\usepackage{amsmath, amsthm,amssymb}
\usepackage{mathrsfs}
\usepackage{hyperref}
\usepackage{amsfonts}
\usepackage{dsfont}
\usepackage{cite}
\usepackage{xcolor}
\usepackage{slashed}
\usepackage{booktabs}
\usepackage{changepage}
\usepackage{setspace}

\PassOptionsToPackage{linecolor=blue,backgroundcolor=blue!25,bordercolor=blue,textsize=scriptsize}{todonotes}
\usepackage{todonotes}

\parskip=\medskipamount

\usepackage[mathscr]{euscript}

\newcommand{\de}{{\nabla}}             
\newcommand{\deb}{{\bar{\nabla}}}

\numberwithin{equation}{section}

\newcommand {\cC}{{\cal C}}
\newcommand {\cD}{{\cal D}}
\newcommand {\cE}{{\cal E}}
\newcommand {\cF}{{\cal F}}
\newcommand {\cG}{{\cal G}}
\newcommand {\cH}{{\cal H}}

\newcommand {\cK}{{\cal K}}
\newcommand {\cL}{{\cal L}}

\newcommand {\cN}{{\cal N}}

\newcommand {\cU}{{\cal U}}

\newcommand {\cW}{{\cal W}}

\newcommand{\bF}{{\bf F}}

\newcommand{\bM}{{\bf M}}

\newcommand{\bW}{{\bf W}}
\newcommand{\bX}{{\bf X}}

\newcommand{\bl}{{\boldsymbol{\l}}}
\newcommand{\Wb}{{\bar{W}}}

\def\a{\alpha}
\def\b{\beta}
\def\c{\chi}
\def\d{\delta}
\def\e{\epsilon}

\def\g{\gamma}

\def\l{\lambda}

\def\o{\omega}

\def\q{\theta}

\def\s{\sigma}

\def\z{\zeta}
\def\D{\Delta}

\def\L{\Lambda}
\def\O{\Omega}

\def\S{\Sigma}
\def\U{\Upsilon}

\def\ri{{\rm i}}
\def\re{{\rm e}}

\newcommand{\gd}{{\dot\g}}
\newcommand{\dd}{{\dot\d}}
\newcommand{\ad}{{\dot{\alpha}}}
\newcommand{\bd}{{\dot{\beta}}}
\newcommand{\ld}{{\dot{\lambda}}}

\newcommand{\hm}{{\hat{m}}}

\newcommand{\hA}{{\hat{A}}}

\newcommand{\ve}{\varepsilon}

\newcommand{\pa}{\partial}
\newcommand{\hf}{\frac12}

\newcommand{\psib}{\bar{\psi}}

\newcommand{\be}{\begin{equation}}
\newcommand{\ee}{\end{equation}}
\newcommand{\bea}{\begin{eqnarray}}
\newcommand{\eea}{\end{eqnarray}}
\newcommand{\non}{\nonumber}
\newcommand{\ba}{\begin{array}}
\newcommand{\ea}{\end{array}}

\newcommand{\bm}[1]{\mbox{\boldmath$#1$}}

\def\double #1{#1{\hbox{\kern-2pt $#1$}}}

\newcommand{\ts}{{\tilde{\s}}}

\newcommand{\bsubeq}{\begin{subequations}}
\newcommand{\esubeq}{\end{subequations}}

\newcommand{\rd}{\mathrm d}
\newcommand{\veps}{\varepsilon}

\newcommand{\bphi}{{\bar\phi}}
\newcommand{\bpsi}{{\bar\psi}}

\newcommand{\eps}{{\epsilon}}

\newcommand{\eol}{\notag \\}
\newcommand{\lc}{{\vert}}

\def\tr{{\rm tr}}

\expandafter\ifx\csname url\endcsname\relax
  \def\url#1{\texttt{#1}}\fi
\expandafter\ifx\csname urlprefix\endcsname\relax\fi
\providecommand{\eprint}[2][]{\url{#2}}

\renewcommand{\eps}{\ve}

\newcommand{\tsigma}{{\tilde{\sigma}}}

\newcommand{\loco}{\vert}
\newcommand{\doubar}{{{\loco}\!{\loco}}}

\begin{document}

\begin{titlepage}
\begin{flushright}
September, 2024
\end{flushright}
\vspace{5mm}

\begin{center}
{\Large \bf 
On 4D, ${\mathcal{N}=2}$ deformed vector multiplets and partial supersymmetry breaking in off-shell supergravity
}
\end{center}

\begin{center}

{\bf
Gregory Gold,
Saurish Khandelwal,
and Gabriele Tartaglino-Mazzucchelli
} \\
\vspace{5mm}

\footnotesize{
{\it 
School of Mathematics and Physics, University of Queensland,
\\
 St Lucia, Brisbane, Queensland 4072, Australia}
}
\vspace{2mm}
~\\
\texttt{g.gold@uq.edu.au; 
s.khandelwal@uq.edu.au; g.tartaglino-mazzucchelli@uq.edu.au}\\
\vspace{2mm}

\end{center}

\begin{abstract}
\baselineskip=14pt

Electric and magnetic Fayet-Ilioupulous (FI) terms are used to engineer partial breaking of ${\mathcal{N}=2}$ global supersymmetry for systems of vector multiplets. The magnetic FI term induces a deformation of the off-shell field transformations associated with an imaginary constant shift of the triplet of auxiliary fields of the vector multiplet. In this paper, we elaborate on the deformation of off-shell vector multiplets in supergravity, both in components and superspace. In a superconformal framework, the deformations are associated with (composite) linear multiplets. We engineer an off-shell model that exhibits partial local supersymmetry breaking with a zero cosmological constant. This is based on the hyper-dilaton Weyl multiplet introduced in arXiv:2203.12203, coupled to the SU(1,1)/U(1) special-K\"ahler sigma model in a symplectic frame admitting a holomorphic prepotential, with one compensating and one physical vector multiplet, the latter magnetically deformed.

\end{abstract}

\vspace{0.5cm}
\begin{flushright}
{\it Dedicated to the memory of Luciano Girardello}
\end{flushright}
\vspace{5mm}
\vfill
\end{titlepage}


\newpage
\renewcommand{\thefootnote}{\arabic{footnote}}
\setcounter{footnote}{0}

\tableofcontents{}
\vspace{1cm}
\bigskip\hrule

\allowdisplaybreaks
\section{Introduction}

Partial supersymmetry (SUSY) breaking is a fascinating subject that has been studied for several decades. 
It has witnessed no-go theorems \cite{Witten:1981nf,Cecotti:1984rk} and their eventual disproof
\cite{Hughes:1986dn,Hughes:1986fa,Cecotti:1985sf,Antoniadis:1995vb,Ferrara:1995gu}, and to date, still remains a subject with various interesting open questions. 
In the case of four space-time dimensions (4D), which is the subject of this work, partial breaking of global $\cN=2\to\cN=1$ supersymmetry requires the deformation of supersymmetry transformations \cite{Antoniadis:1995vb,ADtM,IZ1,IZ2,Kuzenko:2009ym,Antoniadis:2017jsk,Antoniadis:2019gbd,Kuzenko:2017gsc}. This can be engineered with off-shell supersymmetry, meaning that the algebra of supersymmetry transformations closes without the aid of any equations of motion. In particular, possibly the simplest model exhibiting spontaneous global supersymmetry breaking is the one introduced by Antoniadis-Partouche-Taylor (APT) in 1995 \cite{Antoniadis:1995vb} where a single $\cN=2$ vector multiplet deformed by both electric and a magnetic Fayet-Iliopoulos (FI) terms suffices to engineer the $\cN=2\to\cN=1$ supersymmetry breaking. 
A motivation for our work is to look for APT-type models in supergravity engineered with manifest off-shell local supersymmetry.

An alternative way to construct partial supersymmetry breaking models employs non-linear realization techniques, including nilpotent Goldstone multiplet analyses, where only one supersymmetry is manifestly preserved and linearly realized
\cite{Antoniadis:2017jsk,Antoniadis:2019gbd,Bagger:1996wp,Bagger:1997pi,Bagger:1997me,Rocek:1997hi,GonzalezRey:1998kh,Ambrosetti:2009za,Kuzenko:2015rfx,Kuzenko:2017gsc,Ferrara:2014oka,Dudas:2017sbi}. 
The non-linear realization of partial supersymmetry breaking also naturally takes place in theories with supersymmetric extended objects (like membranes), which lead to supersymmetric Dirac-Born-Infeld (DBI) type actions \cite{DP,Cecotti:1986gb,Rocek:1997hi,GonzalezRey:1998kh}. 
Moreover, understanding the mechanisms of partial supersymmetry breaking is also motivated by phenomenology as it would be welcome to have feasible mechanisms to control the breaking of extended supersymmetries in some high-energy scale while allowing a single ${\cal N}=1$ supersymmetry at low energy, see, e.g., \cite{Antoniadis:2012cg} and references therein.

Returning to the APT-type model, it is worth mentioning that electric and magnetic FI terms are related to each other under the electric-magnetic duality of the associated vector multiplets \cite{Antoniadis:1995vb,IZ1,IZ2}. However, the two possess different features in the context of supersymmetry. The electric FI term is a supersymmetric deformation of an $\cN=2$ Lagrangian that on shell can induce a vev for the triplet of real auxiliary fields of a vector multiplet. The magnetic FI term is a deformation of the supersymmetry algebra that results from a constant imaginary shift of the same vector multiplet, resulting in an inherited deformation of the supersymmetric constraint off shell. The possibility of turning on both types of FI terms in an off-shell setting and tuning them appropriately allows for the simple engineering of general matter systems with global spontaneous supersymmetry breaking. To the best of our knowledge, one of the remaining open questions on the subject is to explicitly engineer fully off-shell models that lead to local partial supersymmetry breaking in the supergravity context. In our paper, we revisit this question and propose a solution to this problem.

Local partial supersymmetry breaking in supergravity is a subject that has obtained substantial attention with a non-geodesic history. A limited set of references on the subject can be found here
\cite{Cecotti:1984rk,Cecotti:1984fn,Cecotti:1984wn,Cecotti:1985sf,Ferrara:1995gu,Ferrara:1995xi,Fre:1996js,Louis:2012ux,Louis1,Louis2,Andrianopoli:2015wqa,Antoniadis:2018blk,Abe:2019svc,Abe:2019vzi,Lauria:2020rhc}. 
In line with the early result of \cite{Witten:1981nf}, in 1984 Cecotti-Girardello-Porrati did prove a no-go theorem for local partial supersymmetry breaking in supergravity \cite{Cecotti:1984rk}. These were based on a clever analysis of the general aspects of local supersymmetry algebras with different field content and the employment of superconformal techniques \cite{deWit:1983xhu,deWit:1984wbb,deWit:1984rvr,Cremmer:1984hj,Freedman:2012zz,Lauria:2020rhc}. The next year, it was realised by the same authors that local partial supersymmetry breaking with a zero cosmological constant can be realised if one lifts the technical assumption of the existence of a holomorphic prepotential for the special-K\"ahler geometry of the vector multiplets \cite{Cecotti:1985sf}. The resulting model has one physical vector multiplet parametrising a ${\rm SU(1,1)/U(1)}$ special-K\"ahler sigma model together with a hypermultiplet parametrising a ${\rm SO(4,1)/SO(4)}$ quaternion-K\"ahler
manifold and a set of (electric) gaugings. This set-up has seen several generalisations, see, e.g., \cite{Ferrara:1995xi,Fre:1996js,Louis1,Louis2,Andrianopoli:2015wqa,Antoniadis:2018blk,Abe:2019svc,Abe:2019vzi,Lauria:2020rhc}, however two common features are: (i) these local partial supersymmetry breaking models include physical vector multiplets and at least one physical hypermultiplet; (ii) due to the presence of the hypermultiplet, which in the component superconformal tensor calculus with a finite number of auxiliary fields is on shell \cite{Lauria:2020rhc},\footnote{One way to overcome this difficulty is to employ multiplets with gauged central charges, see for example \cite{deWit:1984rvr}, but, to the best of our knowledge, it remains an open question whether most general supergravity-matter couplings can be engineered this way, see also the discussion in \cite{Lauria:2020rhc}.} the resulting models have local supersymmetry that only closes on shell. We will see in our work that the second restriction can be lifted for an off-shell model of supergravity that has partial supersymmetry breaking and, in fact, is closely related to the original set-up of \cite{Cecotti:1985sf}. However, for the model in our paper, we will employ an off-shell magnetically deformed vector multiplet, together with a compensating vector multiplet in a hyper-dilaton Weyl multiplet background (which we will comment about shortly) with no other matter multiplets. The structure resembles the APT model with both electric and magnetic FI terms, and the resulting construction is fully off shell, due to the modification of one of the building blocks in the superconformal tensor calculus approach that leads to a new spectrum of fields.

Other than the examples mentioned above, it is worth reminding the reader that conformal supergravity has played an important role in several research 
avenues in the last five decades --- we refer the reader to a few books
and reviews for a more detailed discussion and list 
of references 
\cite{Freedman:2012zz,Lauria:2020rhc,SUPERSPACE,Buchbinder-Kuzenko,Kuzenko:2022skv,Kuzenko:2022ajd}. 
Similar to superspace approaches
(see \cite{SUPERSPACE,Buchbinder-Kuzenko,Kuzenko:2022skv,Kuzenko:2022ajd} for introductory reviews
and, e.g., 
\cite{Gold:2022bdk,Howe1,Howe:1981gz,Galperin:1984av,Galperin:1987ek,Galperin:1987em,Galperin:2001seg,KLR,LR3,LR:SYM,Kuzenko:SPH,Kuzenko:2008ep,Kuzenko:2008qz,Kuzenko:2009zu,Butter:2011sr,Butter:2012xg,Butter:2014gha,Butter:2014xua,Butter:2015nza}
and references therein, 
for the 4D,  $\cN=2$ case)
a main advantage of the superconformal tensor calculus 
is to provide an off-shell
description of potentially general supergravity-matter couplings.
This allows one to formulate models where local supersymmetry is
engineered in a completely model-independent way. 
The approach has been very successful in helping to decipher
many of the intricate geometrical structures associated to 
(two-derivative) sigma models in supergravity-matter 
systems with eight real supercharges, see, e.g.,
\cite{deWit:1984wbb,Cremmer:1984hj,deWit:1999fp,deWit:2001brd,deWit:2001bk,Lauria:2020rhc}. 
The off-shell nature of the formalism has been a central 
ingredient in its employment in the study of supersymmetric 
localisation and supersymmetric quantum field theories 
on curved space-times --- see \cite{Pestun:2016zxk} 
for a recent extensive review.
Moreover, off-shell supersymmetry has also been a crucial 
ingredient when using superconformal tensor calculus
to construct higher-derivative supergravity invariants
\cite{BSS1,LopesCardoso:1998tkj,Mohaupt:2000mj,Hanaki:2006pj,Butter:2010jm,CVanP,Bergshoeff:2012ax,Butter:2013rba,Butter:2013lta,Kuzenko:2013vha,OP131,OP132,OzkanThesis,Butter:2014xxa,Kuzenko:2015jxa,BKNT16,Butter:2016mtk,BNT-M17,NOPT-M17,Butter:2018wss,Butter:2019edc,Hegde:2019ioy,Mishra:2020jlc,Gold:2023dfe,Gold:2023ymc,Gold:2023ykx,Ozkan:2024euj,Casarin:2024qdn,Gold:2024nbw}.
These play an important role, e.g., in the study 
of black-hole entropy and other applications in next to leading order 
AdS/CFT
--- see the recent works
\cite{Bobev:2020egg,Bobev:2021oku,Bobev:2021qxx,Liu:2022sew,Hristov:2022lcw,Cassani:2022lrk,Gold:2023ymc,Cassani:2024tvk,Ma:2024ynp,Saskowski:2024otc,Hristov:2024cgj} 
and references therein.

Within the superconformal tensor calculus, general 
supergravity-matter couplings are engineered by a few 
ingredients. 
First, one needs a multiplet of conformal supergravity 
--- named the \emph{Weyl} multiplet ---
which forms an off-shell 
representation of the local superconformal algebra
and contains the vielbein as one of its independent fields.
This multiplet defines the geometry 
(soft algebra) associated with the gauging of the superconformal
space-time symmetry.
Next, one identifies off-shell matter multiplets with 
local superconformal transformation rules in a Weyl multiplet
background. These two ingredients provide the kinematic 
data of a specific supergravity-matter system. 
Finally, one engineers locally superconformal invariant action
principles
constructed out of these multiplets to obtain well-defined
supergravity theories.\footnote{These tasks can be
simplified  by manifestly gauging the superconformal
algebra in superspace through so-called 
\emph{conformal superspace}.
Conformal superspace was first introduced for 4D, $\cN=1,2$ supergravity in \cite{Butter:2009cp,Butter:2011sr}
(see also the seminal work  \cite{Kugo:1983mv})
and it was then developed for 
3D, $\cN$-extended supergravity \cite{Butter:2013goa},
5D, $\cN=1$ supergravity \cite{Butter:2014xxa},
6D, $\cN=(1,0)$ supergravity \cite{BKNT16,BNT-M17}, and recently 4D $\cN=3$ supergravity \cite{Kuzenko:2023qkg} --- see \cite{Kuzenko:2022skv,Kuzenko:2022ajd} for recent reviews.}

Assuming the matter multiplets contain
enough ``compensating'' degrees of freedom, one can  
suitably gauge fix part of the superconformal group, specifically dilatations, 
special conformal transformations, 
$S$-supersymmetry, and  $R$-symmetry, 
to obtain
supergravity models where only the 
super-Poincar\'e symmetry survives and is gauged.
For instance, pure 4D,  $\cN = 2$ Poincar\'e supergravity
can arise by the coupling of the 
\emph{standard Weyl multiplet}
\cite{deWit:1979dzm,deWit:1980gt,deWit:1980lyi,deWit:1983xhu,deWit:1984rvr}
to two compensating multiplets.
 There is significant freedom in doing so. 
 Typically, one uses 
 a vector multiplet and a hypermultiplet
 (the hyper can take several forms, e.g., a linear, non-linear, or hypermultiplet with or without a central charge)  as compensators
 --- see \cite{Lauria:2020rhc,Kuzenko:2022ajd} for recent reviews.
Note that, in this approach, historically the first step has predominantly been the same (standard Weyl multiplet), 
while most of the freedom that has been used concerned
the matter (compensators) side of this story.
However, it is known that variant Weyl multiplets exist and can be used to engineer theories of Poincar\'e supergravity. These go by the name of dilaton Weyl multiplets.

The first example of a dilaton Weyl multiplet was introduced for 6D,  $\cN=(1,0)$ supergravity in 1986 \cite{Bergshoeff:1985mz}, and similar ideas were then employed to construct a variant dilaton Weyl multiplet for 5D, $\cN=1$ conformal supergravity \cite{Bergshoeff:2001hc,Coomans:2012cf}.
For the 4D, $\cN=2$ case, the existence of a variant representation of the Weyl multiplet of conformal supergravity was argued in \cite{Siegel:1995px} and was explicitly constructed only recently in \cite{Butter:2017pbp} by coupling an on-shell vector multiplet to a standard Weyl multiplet --- for this reason, we sometimes refer to this as the \emph{vector-dilaton Weyl} multiplet. Two years ago, we did show the existence of a so-called \emph{hyper-dilaton Weyl} multiplet for 4D, $\cN=2$ conformal supergravity engineered by coupling an on-shell hypermultiplet to the standard Weyl multiplet and by reinterpreting the resulting system as a variant off-shell Weyl multiplet \cite{Gold:2022bdk}. This is the Weyl multiplet that we will use to engineer an off-shell model for local partial supersymmetry breaking. A similar analysis was then performed to define hyper-dilaton Weyl multiplets also in five and six dimensions \cite{Hutomo:2022hdi}.
It is also worth mentioning that new dilaton Weyl multiplets were recently engineered for maximal conformal supergravity in four and five space-time dimensions \cite{Adhikari:2023tzi,Ciceri:2024xxf}.

Considering the role played by the hyper-dilaton Weyl multiplet in our paper, let us now review some of its key features.
The off-shell standard Weyl multiplet of 4D, $\cN=2$ conformal supergravity comprises $24+24$ independent fields. Besides the vielbein, gravitini, ${\rm U(1)_R\times SU(2)_R}$, and dilatation symmetry connections, the multiplet comprises a set of covariant matter (auxiliary) fields: a real antisymmetric tensors, $W_{ab}$,
the real scalar field, $D$, and the spinor fields that we denote by $(\S^{\a i},\bar{\S}_{\ad i})$. The presence of the matter fields is key to obtaining a set of local superconformal field transformations that close off shell.
To define the hyper-dilaton Weyl multiplet one starts with an on-shell hypermultiplet \cite{Fayet:1975yi,FS2} in a standard Weyl multiplet background \cite{deWit:1980gt,deWit:1980lyi,deWit:1983xhu,deWit:1984rvr,deWit:1999fp}. The constraints that arise by requiring the algebra of local superconformal transformations to close on the fields of the hypermultiplet can then be interpreted as algebraic equations for some of the fields of the standard Weyl multiplet. More precisely, 
the standard Weyl multiplet's matter fields $(\S^{\a i},\bar{\S}_{\ad i})$ and $D$, together with the ${\rm SU(2)_R}$ symmetry connection $\phi_m{}^{ij}$ become composite fields. On the other hand, the four bosonic $q^{i\underline{i}}$ and four fermionic $(\rho_\a^{\underline{i}},\bar{\rho}^\ad_{\underline{i}})$ fields of the hypermultiplet, together with an emerging triplet of real gauge two-forms $b_{mn}{}^{\underline{i}\underline{j}}=b_{mn}{}^{\underline{j}\underline{i}}=-b_{nm}{}^{\underline{i}\underline{j}}$, are independent and not subject to any equations of motion. In turn, the new set of independent fields describes another $24+24$ representation of the local superconformal algebra that closes off shell.
An interesting feature of the hyper-dilaton Weyl multiplet is that not only dilatation but also ${\rm SU(2)_R}$ becomes pure gauge, while a triplet of one-form symmetry takes place.

To construct a multiplet of $\cN=2$ Poincar\'e supergravity, where only local Lorentz symmetry and local $Q$-supersymmetry are unbroken, it then suffices to couple the hyper-dilaton Weyl multiplet to a single compensating vector multiplet.
The result is an off-shell $32+32$ hyper-dilaton Poincar\'e supergravity multiplet, which was originally constructed by M\"uller in \cite{Muller_hyper:1986ts} with a different approach. Even though the off-shell field content is minimal, the on-shell theory is non-minimal and comprises the $\cN=2$ Poincar\'e supergravity multiplet with a vielbein, gravitini, and a graviphoton together with an on-shell hypermultiplet where three of the real scalar field are dualised to a triplet of gauge two-forms and one of the scalars plays the role of a dilaton. This is precisely the Poincar\'e supergravity that we will use to engineer an off-shell model that comprises an extra physical vector multiplet (similar to the one of the APT model) where local partial supersymmetry breaking easily takes place.

An interesting feature of M\"uller's supergravity, and our implementation in terms of the hyper-dilaton Weyl multiplet, is the alternative way with which it is possible to generate scalar field potentials with a mechanism different than that of gauging the $R$-symmetry. It is well known that in the standard engineering of general supergravity-matter couplings in 4D, extended ($\cN=2$) supergravity, scalar potentials are associated with moment maps of the embedding of the scalar fields sigma model gauged isometries in the ${\rm SU(2)_R}$ group. In a superconformal setting based on the $\cN=2$ standard Weyl multiplet, this emerges by integrating out the ${\rm SU(2)_R}$ gauge connection, which in a two-derivative theory is an auxiliary field --- see \cite{Lauria:2020rhc} for review. In the hyper-dilaton Weyl and M\"uller multiplets, ${\rm SU(2)_R}$ can be fixed without the aid of a compensating multiplet as, in fact, its gauge field is a composite field that turns into the Hodge dual of the field strength of a triplet of gauge two-forms $b_{mn}{}^{\underline{i}\underline{j}}$. The result is a coupling of the supergravity multiplet to new two-form physical fields and not a mechanism that makes fields (as, for example, the gravitini) charged under the physical gauge group. This was explained in \cite{Gold:2022bdk}, and it generalises to generic couplings with vector multiplets. The result is an alternative, yet simple, off-shell engineering of non-trivial scalar potentials in 4D, $\cN=2$ supergravity.

In our work, we will focus on off-shell Poincar\'e supergravity based on the hyper-dilaton Weyl multiplet coupled to a system of off-shell Abelian vector multiplets, one of which is a conformal compensator. We will not add other matter fields in the system, in particular, no hypermultiplets other than the on-shell one which defines the hyper-dilaton Weyl multiplet. It is well known that, by using the standard Weyl multiplet, for pure systems of physical vector multiplets, non-trivial scalar potentials in 4D, $\cN=2$ supergravity are engineered through gauging by local FI terms. As in the rigid case, FI terms are either electric or magnetic.  To the best of our knowledge, in supergravity, the off-shell description of 4D, $\cN=2$ magnetic FI terms (and magnetic gaugings) has not been developed in full generality yet, though they are expected to play an important role in engineering scalar potentials in supergravity models possessing vacua with both positive and negative cosmological constants -- see, for instance the recent discussion of magnetic 4D, $\cN=1$ FI terms \cite{Antoniadis:2020qoj}. Part of our work is to extend this analysis to the 4D, $\cN=2$ case and elaborate on off-shell magnetically deformed vector multiplets.

The curved superspace constraints for off-shell magnetic FI-type terms were introduced in \cite{Kuzenko:2013gva,Kuzenko:2015rfx} and in depth supergravity analyses in components (though not fully off shell) were presented earlier in \cite{deWS,deVroome:2007unr,deWit:2011gk}.
By using a hyper-dilaton Weyl multiplet it is straightforward to engineer generic electric and magnetic FI-type terms by means of composite linear multiplets. The result is similar to the global case, where the two types of deformations induce a real or imaginary shift of the vector multiplets.
The supergravity extensions of electric FI terms, which we will parameterize with $\xi$, can be obtained by using the $BF$-coupling between a vector and a linear multiplet. In a hyper-dilaton Weyl background, one can construct composite linear multiplets by using a quadratic combination of the fields of the on-shell hypermultiplet. In the case of a $\xi$-deformation, the bottom component of such a composite linear multiplet is given by ${G}_{\xi}{}_{ij} = \xi_{\underline{i}\underline{j}} q_i{}^{\underline{i}} q_j{}^{\underline{j}}$.
We will see that off-shell magnetic FI-type deformations 
in a hyper-dilaton Weyl multiplet background can easily be engineered in terms of the same type of composite linear multiplet. 
This would, for example, appear as an imaginary deformation of
the $X^{ij}$-auxiliary real
field of a vector multiplet.
These deformations are parametrised by the composite field
$G_\z{}_{ij}=\zeta_{\underline{i}\underline{j}}q_i{}^{\underline{i}}q_j{}^{\underline{j}}$ with 
$\zeta_{\underline{i}\underline{j}}=\zeta_{\underline{j}\underline{i}}$, $(\zeta_{\underline{i}\underline{j}})^*=\zeta^{\underline{i}\underline{j}}$
constants that generalise the magnetic FI terms of global 
supersymmetry.
Given a system of $n+1$ vector multiplets 
with scalar fields $\phi^I$ (with $I=0,1,\cdots,n$)
coupled to the off-shell 
hyper-dilaton Weyl multiplet, 
it is then straightforward to introduce $3(n+1)$ 
off-shell deformations each associated to either
a $\xi_I^{\underline{i}\underline{j}}$ electric deformation
or a $\z^I_{\underline{i}\underline{j}}$ magnetic deformation.
These in general induce non-trivial scalar potentials and vacuum structures.

Remarkably, due to the fact that each of the $\xi$ and $\zeta$ deformations can take three SU(2) directions independently, there is enough freedom to obtain local partial supersymmetry breaking. We will prove this by considering a very simple model given by one physical and one compensating vector multiplets, the first magnetically deformed, the second having an electric deformation turned on. By taking a special-K\"ahler holomorphic prepotential of the form $\cF = c \phi \boldsymbol{\phi}$ (the reader can look at \cite{Freedman:2012zz, Lauria:2020rhc, Kuzenko:2022skv, Kuzenko:2022ajd} for reviews on structures of general Lagrangians for off-shell vector multiples), with $c$ a nonzero real constant while $\phi$ being the complex scalar field of the compensating vector multiplet and $\boldsymbol{\phi}$ the same for the physical vector multiplets, and by choosing the determinant of the matrix
\bea
\bM_{\underline{i}\underline{j}} 
  =- \frac{2}{c} \xi_{\underline{i} \underline{j}} 
     + 
    \ri \zeta_{\underline{i} \underline{j}}
    ~,
\eea
to be zero, $\det{\bM}=0$, we find local partial supersymmetry breaking in a Minkowski vacua. A zero determinant condition of a matrix given by a linear combination of an electric and a magnetic FI term, is precisely the one for partial breaking in the global APT model. In fact, in the local model that we consider, the mechanism is very similar, since all shift symmetry terms in the supersymmetry variation of the fermions, together with the fermionic mass matrices, are all parametrized by the matrix $\bM_{\underline{i}\underline{j}}$ given above. A difference, however, is the fact that here, one FI term belongs to the physical multiplet and the other to the compensator. Still, the simplicity of the construction is inspiring, also because at all steps supersymmetry is off shell, allowing one to potentially add other couplings to the model, including higher-derivative ones, in a fairly straightforward way.

Before moving to the technical part of our paper, we would like to comment on the form of the special-K\"ahler potential, $\cF = c \phi \boldsymbol{\phi}$, that we have chosen for the example that we discuss in detail in this paper. This is a natural choice. In fact, as mentioned, e.g., in footnote four of \cite{Ferrara:1995gu}, this holomorphic prepotential is precisely the one of the ${\rm SU(1,1)/U(1)}$ special-K\"ahler sigma model which was employed in the seminal work on local partial supersymmetry breaking \cite{Cecotti:1985sf} but in a symplectic frame, obtained after an electric-magnetic duality, where a holomorphic potential actually exists. From the point of view of the special-K\"ahler geometry, our exemplary model is inspired by the one of \cite{Cecotti:1985sf} after a duality transformation, where, however, the hypermultiplet sector in our case becomes part of the conformal supergravity multiplet with three scalar fields turned into gauge two-forms. The emergence of a magnetically deformed vector multiplet is then expected. However, the absence of gauging in our setup, as well as the new spectrum of the on-shell theory, are intriguing features of working with hyper-dilaton supergravity.

This paper is organised as follows. In Section \ref{sec:multiplets}, we introduce the relevant superconformal multiplets in superspace and components. This includes the standard Weyl multiplet, the abelian vector multiplet, the linear multiplet (often referred to tensor multiplet), and the hyper-dilaton Weyl multiplet (constructed by an on-shell hypermultiplet). The reader familiar with these results can skip this review section, but note that this section does review a wealth of results in our notations. In Section \ref{sec:deformedVec}, we introduce the deformed abelian vector multiplet with which we induce the previously mentioned magnetic deformation. In Section \ref{sec:actions}, we give the component actions used in this paper being the Abelian deformed vector multiplet action and the standard FI term by a linear multiplet action. In this section, the linear multiplet is considered as a general one (not necessarily composite), whereas in the successive sections of the paper, all deformations will be parametrised by linear multiplets that are composite of a hypermultiplet. The general off-shell action for deformed $\cN=2$ conformal supergravity in a hyper-dilaton Weyl multiplet background is then given in Section \ref{section:gauged} followed by the covariant equations of motion for this general model's auxiliary fields. In Section \ref{sec:SU11U1}, we then proceed with a specific choice being the SU(1,1)/U(1) special-K\"ahler sigma model and give the corresponding off-shell action. This is followed by a process of gauge fixing and integrating out auxiliary fields resulting in an on-shell supergravity model with partial supersymmetry breaking. Finally, in Section \ref{sec:conclusion}, we collect concluding comments and an outlook for our paper. We also present a few technical appendices. This first includes our notations and conventions and details on 4D, $\cN=2$ conformal superspace in Appendices \ref{NC} and \ref{Appendix-B}, respectively. For the reader's convenience, we then give the $S$-supersymmetry and local superconformal transformations of various multiplet fields seen throughout this paper in Appendices \ref{S-transformation} and \ref{local-superconformal-transformation}, respectively. Lastly, we accompany our paper with a supplementary file where we give the fermionic counterparts to various component actions, the bosonic parts of which are given in Sections \ref{section:gauged} and \ref{sec:SU11U1} of this paper.

\section{Superconformal multiplets in 4D, $\mathcal{N}=2$} \label{sec:multiplets}

We review here various superconformal multiplets which serve as the building blocks for the invariants of 4D, $\mathcal{N}=2$ conformal supergravity both in superspace and components.

\subsection{The standard Weyl multiplet}
\subsubsection{The standard Weyl multiplet in superspace}
The 4D, $\cN=2$ conformal superspace is parametrised by local bosonic $(x^{m})$ and fermionic $(\theta_i^{\mu}, \bar{\theta}^i_{\dot{\mu}})$ coordinates 
$z^{M} = (x^{\hm},\theta_i^{\mu}, \bar{\theta}^i_{\dot{\mu}})$, 
where $m = 0, 1, 2,3$, ${\mu},\dot{\mu} = 1, 2$, and $i = 1, 2$.
To gauge the superconformal algebra, one introduces the covariant derivatives $ {\nabla}_{A} = (\nabla_{a} , \nabla_{\a}^i, \bar{\nabla}^{\ad}_i)$ which have the form
\bea\label{eq:covD}
\de_{A} 
= E_{A} - \o_{A}{}^{\underline{b}}X_{\underline{b}} 
&=& E_{\hA} - \hf \O_{A}{}^{a b} M_{a b} - \ri \Phi_A Y - \Phi_{A}{}^{ij} J_{ij} - B_{A} \mathbb{D} - \mathfrak{F}_{A}{}^{B} K_{B}~.~~~~
\eea
Here $E_{A} = E_{A}{}^{M} \partial_{M}$ is the inverse super-vielbein, 
$M_{a b}$ are the Lorentz generators, $Y$ is the generator of the chiral rotation
group $\rm U(1)_R$, $J_{ij}$ are generators of the
$\rm SU(2)_R$ $R$-symmetry group,
$\mathbb D$ is the dilatation generator, and $K_{A} = (K_{a}, S^\a_i , \bar S_\ad^i)$ are the special superconformal
generators.
The super-vielbein one-form is $E^{A} =\rd z^{M} E_{M}{}^{A}$ with $E_{M}{}^{A} E_{A}{}^{N} =\d_{M}^{N}$ and
 $E_{A}{}^{M} E_{M}{}^{B}=\d_{A}^{B}$.
Associated with each generator $X_{\underline{a}} = ( M_{ab} , Y, J_{ij} ,{\mathbb D},K^a, S^\a_i , \bar S_\ad^i)$ is the connection super one-form 
$\omega^{\underline{a}} = (\O^{ab},\Phi,\Theta^{ij},B,\mathfrak{F}_{A})= (\O^{ab},\Phi,\Theta^{ij},B,\mathfrak{F}_{a},\mathfrak{F}_{\a}^i,\bar{\mathfrak{F}}^{\ad}_i)= \rd z^M \omega_M{}^{\underline{a}} = E^{A} \o_{A}{}^{\underline{a}}$.

The algebra of covariant derivatives is
\begin{align}
[ \nabla_A , \nabla_B \}
	&= -\mathscr{T}_{A B}{}^C \nabla_C
	- \frac{1}{2} {\mathscr{R}(M)}_{AB}{}^{cd} M_{cd}
	- {\mathscr{R}(J)}_{AB}{}^{kl} J_{kl} - {\mathscr{R}(\mathbb{D})}_{AB} \mathbb D 
	\non \\ & \quad
	- \ri {\mathscr{R}(Y)}_{AB} \mathbb Y
	- {\mathscr{R}(S)}_{AB}{}_{\a}^{i} S^{\a}_{i} - {\mathscr{R}(\bar{S})}_{AB}{}^{\ad}_{i} S_{\ad}^{i}
	- {\mathscr{R}(K)}_{AB}{}^c K_c~,
	\label{nablanabla}
\end{align}
and is constrained to be  expressed in terms of a single primary superfield, the super-Weyl tensor $(W_{\a \b}, \overline{W}^{\ad \bd})$,\footnote{Here and in what follows, an antisymmetric rank-2 tensor $T_{a b} = -T_{b a}$ can equivalently be written as: $T_{a b} = (\s_{a b})^{\a \b} T_{\a \b} - (\s_{a b})_{\ad \bd} \bar{T}^{\ad \bd}$ with $T_{\a \b} = \frac{1}{2} (\s^{a b})_{\a \b} T_{a b}$ and $\bar{T}_{\ad \bd} = - \frac{1}{2} (\tilde{\s}^{a b})_{\ad \bd} T_{a b}$.} which has the following properties
 \bsubeq
\bea
&& W_{\a \b} = W_{\b \a} \ , \quad
K_{A} W_{\a \b} = 0 \ ,
 \quad \mathbb D W_{\a \b} = W_{\a \b} \ , \quad Y W_{\a \b} = -2 W_{\a \b} ~, \\
 && \overline{W}^{\ad \bd} = \overline{W}^{\bd \ad} \ , \quad
K_{A} \overline{W}^{\ad \bd} = 0 \ ,
 \quad \mathbb D \overline{W}^{\ad \bd} = \overline{W}^{\ad \bd} \ , \quad Y \overline{W}^{\ad \bd} = 2 \overline{W}^{\ad \bd} ~.
\eea
\esubeq
The super-Weyl tensor also obeys  the additional constraints
\begin{align}
    \bar{\nabla}^\ad_i W_{\b\g} = 0~,\qquad
    \nabla_{\a\b} W^{\a\b} &= \bar{\nabla}^{\ad\bd} \overline{W}_{\ad\bd} ~,
\end{align}
where we have introduced the following notation,
\begin{align}
    \nabla_{\a\b} := \nabla_{(\a}^k \nabla_{\b) k} \ , \quad
    \bar{\nabla}^{\ad\bd} := \bar\nabla^{(\ad}_k \bar\nabla^{\bd) k} ~.
\end{align}

In eq.~\eqref{nablanabla}, $\mathscr{T}_{A B}{}^C$ is the torsion, while $ \mathscr{R}(M)_{A B}{}^{c d}$,
$ \mathscr{R}(J)_{A B}{}^{kl}$, ${\mathscr{R}(\mathbb{D})}_{AB}$, ${\mathscr{R}(Y)}_{AB}$, $ {\mathscr{R}(S)}_{AB}{}_{\a}^{i}$, ${\mathscr{R}(\bar{S})}_{AB}{}^{\ad}_{i} $, and ${\mathscr{R}(K)}_{AB}{}^c$
are the curvatures associated with Lorentz, $\rm SU(2)_R$,
dilatation, $\rm U(1)_R$, $S$-supersymmetry, and special conformal
boosts, respectively. The full algebra of covariant derivatives \eqref{nablanabla} (including the explicit expressions for the torsion and curvature tensors) are given in Appendix \ref{Appendix-B} of our paper.

Let us introduce the descendant superfields constructed by acting successive spinor covariant derivatives on $W_{\a \b}$. These are the dimension-3/2 descendant superfields
\bsubeq \label{descendantsW-5d}
\bea
    && W_{\a \b \g}{}^k := \nabla_{(\a}^k W_{\b \g )} \ ,  \quad \S^{\a i} := \frac{1}{3} \nabla_\b^i W^{\a\b} \ , 
    \\&&  
    \overline{W}^{\ad \bd \gd}{}_k := \bar{\nabla}^{(\ad}_k \overline{W}^{\bd \gd)} \ , \quad \bar{\S}_{\ad i} := - \frac{1}{3} \bar{\nabla}^\bd_i \overline{W}_{\ad\bd}  \ ,
\eea
other dimension-2 descendant superfields
\bea
&& W_{\a \b \g \d} := \nabla_{(\a}^k W_{\b \g \d) k} \ , \quad 
\S_{\a \b}{}^{ij} := \nabla_{(\a}^{(i} \S_{\b)}^{j)} 
\ , \quad 
\S_{\a \b} := \nabla_{(\a}^{i} \S_{\b) i} 
\ , \\
&& \overline{W}^{\ad \bd \gd \dd} := \bar{\nabla}^{(\ad}_k \overline{W}^{\bd \gd \dd) k} \ , \quad 
\bar{\S}^{\ad \bd}{}_{ij} := \bar{\nabla}^{(\ad}_{(i} \bar{\S}^{\bd)}_{j)} 
\ , \quad 
\bar{\S}^{\ad \bd} := \nabla^{(\ad}_{i} \bar{\S}^{\bd) i} 
\ , 
\\ && D := \frac{1}{12} \nabla^{\a\b} W_{\a\b} = \frac{1}{12} \bar{\nabla}_{\ad\bd} \overline{W}^{\ad\bd} =  \frac{1}{4} \nabla_{\a}^{k} \S^{\a}_{k}  = - \frac{1}{4} \bar{\nabla}^{\ad}_{k} \bar{\S}_{\ad}^{k}
 \ ,
\eea
and the dimension-5/2 descendant superfields
\bea
    \S_{\a \b \g}{}^k := \nabla_{(\a}^k \S_{\b \g )}  \ , \quad \bar{\S}^{\ad \bd \gd}{}_k := \bar{\nabla}^{(\ad}_k \S^{\bd \gd )} ~.
\eea
\esubeq
It can be checked that only the superfields \eqref{descendantsW-5d} and their vector derivatives appear upon taking successive spinor derivatives of $W_{\a \b}$. 
The independent descendant superfields of $(W_{\a\b}, W^{\ad \bd})$ are all annihilated by $K_{a}$. However, under $S$-supersymmetry, they transform non-trivially, as given in the Appendix \ref{S-transformation}.

The gauge group of conformal supergravity is denoted by $\cG$. It is generated by
{\it covariant general coordinate transformations},
$\delta_{\rm cgct}$, associated with a local superdiffeomorphism parameter $\xi^{A}$ and
{\it standard superconformal transformations},
$\delta_{\cH}$, associated with the following other local superfield parameters:
the dilatation $\s$, Lorentz $\L^{a b}=-\L^{b a}$,  $\rm SU(2)_R$ $\L^{ij}=\L^{ji}$, $\rm U(1)_R$ $\L$,
 and special conformal (bosonic and fermionic) transformations $\L^{A}=(\eta_{\a}^{ i}, \bar{\eta}^{\ad}_{i}, \L^{a}_{K})$.
The covariant derivatives transform as
\bsubeq
\label{sugra-group}
\bea
\d_\cG \nabla_{A} &=& [\cK , \nabla_{A}] \ ,
\label{TransCD}
\eea
with
\bea
\cK = \xi^{C}  {\nabla}_{C} + \hf  {\L}^{a b} M_{a b} +  {\L}^{ij} J_{ij} +  \s \mathbb D + \L Y +  {\L}^{A} K_{A} ~.
\eea
\esubeq
A covariant (or tensor) superfield $U$ transforms as
\be
\d_{\cG} U =
(\d_{\rm cgct}
+\d_{\cH}) U =
 \cK U
 ~. \label{trans-primary}
\ee
The superfield $U$ is said to be \emph{superconformal primary} of dimension $\D$ and $\rm U(1)_R$ charge $q_\textrm{R}$ if $K_{A} U = 0$ (it suffices to require that $S^\a_{i} U = \bar{S}_\ad^{i} U =0$), $\mathbb D U = \D U$, and $Y U = q_\textrm{R} U$.

\subsubsection{The standard Weyl multiplet in components}
\label{SWM}

The standard Weyl multiplet  of 4D, $\cN=2$ conformal supergravity
is associated with the local off-shell gauging in space-time of the
superconformal group SU$(2,2|2)$ \cite{deWit:1979dzm},
see also 
\cite{deWit:1980gt,deWit:1980lyi,deWit:1983xhu,deWit:1984rvr}
and \cite{Freedman:2012zz,Lauria:2020rhc} for reviews.
The multiplet is comprised of $24+24$ physical components described by a set of independent gauge fields:
the vielbein $e_m{}^a$ and a dilatation connection $b_m$;
the gravitino $(\psi_m{}^\alpha_i,\bar\psi_m{}_\ad^i)$, associated with the gauging of $Q$-supersymmetry;
a $\rm U(1)_R$ gauge field $A_m$;
and $\rm SU(2)_R$ gauge fields $\phi_m{}^{ij}=\phi_m{}^{ji}$.
The fields associated to the remaining generators of SU$(2,2|2)$ are composite and include the Lorentz connections $\o_m{}^{cd}$, 
$S$-supersymmetry connection $(\phi_m{}_\alpha^i,\bar\phi_m{}^\ad_i)$,
and the special conformal connection $\mathfrak{f}_m{}_a$.
The connections define the locally superconformal covariant derivatives\footnote{The notations of this paper are adopted from \cite{Butter:2012xg} as well as the generator algebra of SU$(2,2|2)$.
We, for example, use two-component 
SL$(2,\mathbb{C})$ spinor indices
according to the notation of \cite{Buchbinder-Kuzenko}.
In general, we closely adhere to the conventions given in
\cite{Butter:2012xg} with the exception of the overall signs in the definition of the connections, 
${\omega}_m{}^{cd}$,
${b}_m$,
$A_m$,
$\phi_m{}^{kl}$,
$({\phi}_m{}_{\a}^{ i},\bar{{\phi}}_m{}^{\ad}_{i})$,
and 
${\mathfrak f}_m{}_{a}$, 
as well as the overall signs for all the superconformal curvatures.}
\bea
{\nabla}_a
=e_a{}^m{\nabla}_m&=& 
e_a{}^m\Big(\pa_m
- \frac{1}{2} \psi_m{}^\a_i Q_{\a}^i 
- \frac{1}{2} \bar{\psi}_m{}_\ad^i \bar{Q}^{\ad}_i 
- \frac{1}{2}  {\omega}_m{}^{cd} M_{cd}
-\ri   A_m Y
-  \phi_m{}^{kl} J_{kl}
\non\\
&&~~~~~~
-  {b}_m \mathbb D
- \frac{1}{2}  {\phi}_m{}_{\a}^{ i} S^{\a}_{ i}
- \frac{1}{2}  \bar{{\phi}}_m{}^{\ad}_{i} \bar{S}_{\ad}^{ i}
-  {{\mathfrak f}}_m{}_{c} K^c
\Big)
~.
\label{conf-derivative-components}
\eea
In addition to the independent gauge connections, the standard Weyl 
multiplet is comprised of a set of covariant matter fields 
that are needed to close the local superconformal 
algebra off shell. These include
an anti-symmetric real tensor $W_{ab}=W^+_{ab}+W^-_{ab}$, 
which decomposes
into its imaginary-(anti-)self-dual components $W^{\pm}_{ab}$, 
a real scalar field $D$,
and fermions $(\Sigma^{\a i},\bar{\Sigma}_{\ad i})$.
The covariant derivatives satisfy the algebra 
\bea
{[}\de_a,\de_b{]}
&=&
-{R}(P)_{ab}{}^c\de_c
-{R}(Q)_{ab}{}^\a_i Q_\a^i
-{R}(\bar{Q})_{ab}{}_\ad^i \bar{Q}^\ad_i
- \frac{1}{2}  {R}(M)_{ab}{}^{cd} M_{cd}
-  {R}(\mathbb D)_{ab}\mathbb D
\non\\
&&
-\ri  R(Y)_{ab}Y
-  R(J)_{ab}{}^{kl} J_{kl}
- R(S)_{ab}{}_{\a}^{ i} S^{\a}_{ i}
- R(\bar{S})_{ab}{}^{\ad}_{i} \bar{S}_{\ad}^{ i}
- R(K)_{ab}{}_{c} K^c
~.
\label{soft-algebra-components}
\eea
It is also useful to list the non-trivial 
conjugation properties
\bsubeq\bea
&(\psi_m{}^\a_i)^*=\bar{\psi}_m{}^{\ad i}
~,~~~
(\phi_m{}_\a^i)^*=\bar{\phi}_m{}_{\ad i}
~,~~~
(\phi_m{}^{ij})^*=\phi_m{}_{ij}
~,~~~
(\S^{\a i})^*=\bar{\S}^\ad_i
~,~~~~~~\\
&(R(Q)_{ab}{}^\a_i)^*=R(\bar{Q})_{ab}{}^{\ad i}
~,~~~
(R(S)_{ab}{}_\a^i)^*=R(\bar{S})_{ab}{}_{\ad i}
~,~~~
(R(J)_{ab}{}^{kl})^*=R(J)_{ab}{}_{kl}
~,~~~~~~
\eea\esubeq
while all other fields and curvatures are real.

A set of conventional constraints express the superconformal curvatures in terms of connections and covariant matter fields and 
render the connections $\o_m{}^{cd}$, $(\phi_m{}_\alpha^i,\bar\phi_m{}^\ad_i)$,
and $\mathfrak{f}_m{}_a$ composite.
There is considerable freedom in the choice of these conventional constraints which is apparent from the fact that different papers often make different choices.
The conventional constraints from \cite{Butter:2012xg} are adopted in this paper but adapted to our conventions (see Appendix \ref{NC}). These constraints descend from those used in superspace, see, e.g., \cite{Butter:2011sr}.
It is, in fact, straightforward to obtain the component structures from the superspace ones by appropriately identifying component fields with the lowest components of superfields. The vielbein ($e_m{}^a$) and gravitini ($\psi_m{}^\a_i,\bar{\psi}_m{}_\ad^i$) appear as the $\q=0$ projections of the coefficients of 
$\rd x^m$ in the supervielbein $E^A$ one-form,
\begin{align}
e{}^a = \rd x^m e_m{}^a = E^a\doubar~,~~~
\psi^\a_i = \rd x^m \psi_m{}^\a_i =  2 E^\a_i \doubar
~,~~~
\bar{\psi}_\ad^i = \rd x^m \bar{\psi}_m{}_\ad^i =  2 E_\ad^i \doubar
 ~.
\end{align}
Here we have defined the double bar projection of a superform as $\O\doubar \equiv \O|_{\theta = \rd \theta = 0}$. On the other hand, a single bar next to a superfield denotes the usual bar projection $X| \equiv X|_{\q=0}$.
The remaining component one-forms are defined as 
\bea
&
A:=\Phi\doubar~,\quad
\phi^{kl} := \Theta{}^{kl} \doubar~, \quad
b := B\doubar ~, \quad
\omega^{cd} := \Omega{}^{cd} \doubar ~, 
\\
&{\phi}_\g^k := 2 \,{{\mathfrak F}}{}_\g^k\doubar~, \quad
\bar{\phi}^\gd_k := 2 \,{{\mathfrak F}}{}^\gd_k\doubar~, \quad
{\mathfrak{f}}{}_c := {\mathfrak{F}}{}_c\doubar~. 
\eea
The covariant matter fields $W_{ab}$, $D$, and $(\Sigma^{\a i},\bar{\Sigma}_{\ad i})$ (we denote these component fields with the same symbols as the super-Weyl superfield and its descendants) arise as some of the components of the multiplet described by the super-Weyl tensor. In particular, it holds that
\bsubeq\label{TWSSb}
\bea
W_{ab}(x)&:=&W_{ab}(z)\loco
~,\qquad
D=
\frac{1}{12}\de^{\a \b}W_{\a\b}\loco
=
\frac{1}{12}\deb^{\ad \bd}\bar{W}_{\ad\bd}\loco
~,
\\
&&\Sigma^{\a i}=\frac{1}{3}\de_\b^i W^{\a\b}\loco
~,~~~
\bar{\Sigma}_{\ad i}=-\frac{1}{3}\deb^\bd_i\bar{W}_{\ad\bd}\loco
~.
\eea
\esubeq
The local superconformal transformations of the gauge fields listed above can be straightforwardly derived by taking the $\q=0$ projection of the superspace transformations \eqref{sugra-group}. The transformations of $W_{ab}$, $D$, and $(\Sigma^{\a i},\bar{\Sigma}_{\ad i})$ can be obtained by applying the transformation rule for covariant superfields, eq.~\eqref{trans-primary}, and the definition of the descendant fields in eq.~\eqref{TWSSb}. 

By taking the double bar projection of the superspace covariant derivative one-form $\de$, eq.~\eqref{eq:covD}, and by appropriately interpreting the projected spinor covariant derivatives $\nabla_{\a}^i\loco$ and $\bar{\nabla}^{\ad}_i\loco$
as the generators of $Q$-supersymmetry,\footnote{Given a covariant superfield $U$, and its lowest component $\cU = U\vert$, one defines $Q_\a^i\cU =\nabla_\a^i| \cU := (\nabla_\a^i U) \loco$ and $\bar{Q}^\ad_i\cU =\bar{\nabla}^\ad_i| \cU := (\bar{\nabla}^\ad_i U) \loco$. The other generators $X_{\underline{a}}$ act on $\cU$ as $X_{\underline{a}}\cU :=(X_{\underline{a}} U)\loco$.}
one obtains the component vector covariant derivative \eqref{conf-derivative-components}, where again we use the same symbols for the superspace and component structures.
With these reduction rules, the algebra of component covariant derivatives acting on a covariant field is also completely determined
by the geometry of conformal superspace. All the component torsions and curvatures are simply the $\theta=0$ projections of the superspace ones. 
The resulting algebra coincides with \eqref{soft-algebra-components} subject to the following conventional constraints
\bsubeq
\bea
R(P)_{ab}{}^c
&=&
0~,
\\
R(Q)_{ab}{}_j{\s}^b
&=&
-\frac{3}{4}\S_{j}\s_a
~,~~~~~~
R(\bar{Q})_{ab}{}^{j}{\tilde\s}^b
=
\frac{3}{4}\bar\S^{j}\tilde\s_a~,
\\
R(M)^c{}_{a}{}_{cb}
&=&
R(\mathbb D)_{ab}
+3\eta_{ab}D
-\eta^{cd}
W^-_{ac}W^+_{bd}
~.
\eea
\esubeq
We later will present the expressions for the superconformal curvatures needed in this paper but refer the reader to \cite{Butter:2011sr,Butter:2012xg} for more detail and their relation to the results presented in \cite{deWit:1979dzm}.

In presenting the multiplet we restrict
to all local superconformal transformations except local 
translations (covariant general coordinate transformations).
These transformations are identified with $\d$ and are defined by the following
operator
\bea
\delta
= 
\xi^\a_iQ_\a^i
+\bar{\xi}_\ad^i\bar{Q}^\ad_i
+\frac{1}{2}\lambda^{ab}M_{ab} 
+ \lambda^{ij}J_{ij} 
+ \lambda_{\mathbb{D}}\mathbb{D} 
+ \ri\lambda_Y Y 
+ \lambda_a K^a 
+ \eta_{\alpha}^i S^\alpha_i 
+ \bar{\eta}^{\dot{\alpha}}_i\bar{S}_{\dot{\alpha}}^i
~.
\eea
The local superconformal transformation of 
the fundamental fields of the standard Weyl multiplet
are then given in the Appendix \ref{local-superconformal-transformation}.
The composite Lorentz and $S$-supersymmetry connections are,
respectively,
\bea
\o_{abc}
&=&
\o(e)_{a}{}_{bc}
-2\eta_{a[b}b_{c]}
-\frac{\ri}{2}\big(
\psi_{a}{}_j\s_{[b}\bar{\psi}_{c]}{}^j
+\psi_{[b}{}_j\s_{c]}\bar{\psi}_a{}^j
+\psi_{[b}{}_j\s_{|a|}\bar{\psi}_{c]}{}^j
\big)
~,
\label{composite-Lorentz}
\eea
and
\bsubeq\label{composite-S-connection}
\bea 
\phi_{m}{}_\b^j
&=&
\frac{\ri}{4} \left(
\sigma^{bc} \sigma_m - \frac{1}{3} \sigma_m \tilde{\sigma}^{bc} 
\right)_{\b \bd}
\overline{\Psi}_{bc}{}^{\bd j}
+ \frac{1}{3} W_{mb}^- \psi^b{}_{\b}^j
- \frac{1}{3} W_{mb}^- (\sigma^{bc} \psi_{c}{}^j)_{\b}   
+ \frac{\ri}{4} (\sigma_{m} \bar\S^{j})_\b
~,
\\
\bphi_{m}{}^\bd_j 
&=& 
\frac{\ri}{4} \left(
\tsigma^{bc} \tsigma_m - \frac{1}{3} \tsigma_m \sigma^{bc} \right)^{\bd\beta}
\Psi_{bc}{}_{\beta j}
-\frac{1}{3} W_{mb}^+ \bar\psi^b{}^{\bd}_{j}
+ \frac{1}{3} W_{mb}^+ (\tsigma^{bc} \bar\psi_{c\, j})^{\bd}
- \frac{\ri}{4} (\tsigma_{m} \S_{j})^\bd
~.~~~~~~~~~~
\eea
\esubeq
The field $\o(e)_{a}{}_{bc}$ in \eqref{composite-Lorentz} 
is the usual torsion-free Lorentz connection 
given in terms of the anholonomy tensor 
${\cal{C}}_{mn}{}^a(e)$ as
\bea
\o(e)_{a}{}_{bc}
=
-\cC_{a}{}_{[bc]}(e)
+\hf\cC_{b}{}_{ca}(e)
~,~~~
{\cal{C}}_{mn}{}^a(e):=2\pa_{[m}e_{n]}{}^a
~,~
{\cal{C}}_{ab}{}^c(e)
:=
e_a{}^me_b{}^n{\cal{C}}_{mn}{}^c(e)
~,~~~~~
\eea
while the fields 
$\big(\Psi_{ab}{}^\g_k,\,{\overline{\Psi}}_{ab}{}^k_{\dot{\g}}\big)$
are the gravitini field strengths
\bea
\Psi_{ab}{}^\g_k 
= 
2{e_a}^m{e_b}^n\cD_{[m}\psi_{n]}{}^\g_k
~,~~~~~~
\overline{\Psi}_{ab}{}_\gd^k = 2{e_a}^m{e_b}^n\cD_{[m}\bar{\psi}_{n]}{}_\gd^k
~.
\eea
Also, the derivatives $\cD_a$ 
are\footnote{In many cases, such as 
eqs.~\eqref{nabla-on-W-and-Sigma},
we do not explicitly write the expressions for
the $\cD_a$ derivatives acting 
on different fields. 
However, it is straightforward to find
the results by use of the 
chiral and dilatation weights in Table 
\ref{chiral-dilatation-weights}
together with the Lorentz and $\rm SU(2)_R$ 
representations of fields 
as well as by using the action of
$M_{ab}$ and $J_{ij}$ generators
corresponding to the notations of 
\cite{Butter:2012xg}.} 
\begin{align}
\cD_a 
=
e_a{}^m\cD_m
=
e_a{}^m\Big( \partial_m 
- \frac{1}{2}\omega_m{}^{cd}M_{cd} 
- \phi_m{}^{ij}J_{ij} 
- \ri A_m Y 
- b_m\mathbb{D}\Big)
~.
\end{align}
Note that $\big(R(Q)_{ab}{}^\g_k,\,R(\bar{Q})_{ab}{}_\gd^k\big)$
are the $Q$-supersymmetry curvatures and satisfy
\bsubeq
\bea
R(Q)_{ab}{}^\g_k
&=& \frac{1}{2} \Psi_{ab}{}^\g_k
- \ri (\bphi_{[a\, k} \,\tsigma_{b]})^{\g}
+\frac{\ri}{4}(\bpsi_{[a\,}{}_k \tsigma_{b]}\sigma^{cd})^{\g} W^+_{cd} 
~,
\\
R(\bar{Q})_{ab}{}_\gd^{k}
&=&
\frac{1}{2} \overline{\Psi}_{ab}{}_\gd^{k}
-\ri(\phi_{[a}{}^{k}\sigma_{b]})_{\gd}
-\frac{\ri}{4}(\psi_{[a}{}^{k}\sigma_{b]}\tsigma^{cd})_\gd W^-_{cd}
~,
\eea
\esubeq
while $R(Y)_{ab}$ and $R(J)_{ab}{}^{kl}$ are
\bsubeq
\bea
R(Y)_{ab} 
&=&
2 e_a{}^me_b{}^n  \partial_{[m} A_{n]}
- \frac{\ri}{2} \psi_{[a}{}_j\phi_{b]}{}^j
+ \frac{\ri}{2} \bpsi_{[a}{}^j\bphi_{b]}{}_j
+ \frac{3}{8} \psi_{[a}{}_j \sigma_{b]} \bar\S{}^j
+ \frac{3}{8} \bpsi_{[a}{}^j \tsigma_{b]} \S{}_j
~,~~~~~~
\\
R(J)_{ab}{}^{kl} 
&= &
2 e_a{}^m e_b{}^n\partial_{[m} \phi_{n]}{}^{kl}
-2\phi_{a}{}^{(k}{}_p
\phi_{b}{}^{ l) p}
+2\psi_{[a}{}^{(k}
\phi_{b]}{}^{ l)}
-2\bpsi_{[a}{}^{(k}
\bphi_{b]}{}^{l)}
\non\\
&&
- \frac{3\ri}{2} \psi_{[a}{}^{(k} \sigma_{b]}\bar\S^{l)}
- \frac{3\ri}{2} \bpsi_{[a}{}^{(k} \tsigma_{b]} \S^{l)}
~.
\eea
\esubeq
We refrain from presenting the expression for the composite special conformal connection $\mathfrak{f}_{m}{}_a$ and other superconformal
curvatures. We will however use $e_a{}^m\mathfrak{f}_{m}{}^a$ which 
is given by 
\bea
{\mathfrak{f}_a}^a 
&=&
- \frac{1}{12}R
+D 
- \frac{1}{24}\varepsilon^{mnpq}
({\bar{\psi}_m}^j\tilde{\sigma}_n\cD_p\psi_{q}{}_{j})
 +\frac{1}{24}\varepsilon^{mnpq}
 (\psi_{mj} \sigma_n \cD_p {\bar{\psi}_q}{}^j)
 \non\\
&&
-\frac{\ri}{8}\psi_{aj}\sigma^a \bar{\Sigma}^j 
+ \frac{\ri}{8}{\bar{\psi}_a}{}^j\tilde{\sigma}^a\Sigma_j
- \frac{1}{12}W^{ab+}({\bar{\psi}_a}{}^j\bar{\psi}_{bj})
+\frac{1}{12}W^{ab-}(\psi_{aj}{\psi_b}{}^j)
~,
\label{composite-f}
\eea
where $R=e_a{}^me_b{}^n R_{mn}{}^{ab}$ is the scalar curvature constructed from the Lorentz curvature
\bea
R_{mn}{}^{cd} 
=
2\pa_{[m}\o_{n]}{}^{cd}
-2\o_{[m}{}^{ce}\o_{n]}{}_e{}^{d}
~.
\eea
Also do remember that the spin connection $\o_m{}^{cd}$ is a composite field of the vielbein, the gravitini, and the dilatation 
connection, eq.~\eqref{composite-Lorentz}.

It is important to note that the transformations 
\eqref{transf-standard-Weyl} 
form an algebra that
closes off shell on a local extension of SU$(2,2|2)$. 
We will not need the explicit form of the algebra here, 
though it is straightforward to derive 
using the results of \cite{deWit:1979dzm} and 
\cite{Butter:2011sr,Butter:2012xg}.
To conclude this subsection, 
for convenience, we include Table \ref{chiral-dilatation-weights} 
which provides a summary of the non-trivial chiral and dilatation
weights of the fields and local gauge parameters 
of the standard Weyl multiplet. 
\begin{table}[hbt!]
\begin{center}
\begin{tabular}{ |c||c| c| c| c| c| c| c| c|c|c|c|} 
 \hline
& $e_m{}^a$
&$\psi_m{}_i$, $\xi_i$
&  $\bar\psi_m{}^i$, $\bar{\xi}^i$
& $\phi_m{}^i$, $\eta^i$ 
& $\bar\phi_m{}_i$, $\bar\eta_i$ 
&$\mathfrak{f}_{m}{}_c$
& $W^+_{ab}$
& $W^-_{ab}$ & 
$\Sigma^{i}$ 
& $\bar\Sigma_{i}$
&$D$
\\ 
\hline
 \hline
$\mathbb{D}$
&$-1$
&$-1/2$
&$-1/2$
&$1/2$
&$1/2$
&$1$
&$1$
&$1$
&$3/2$
&$3/2$
&$2$
\\
$Y$
&$0$
&$-1$
&$1$
&$1$
&$-1$
&$0$
&$-2$
&$2$
&$-1$
&$1$
&$0$
\\
\hline
\end{tabular}
\caption{\footnotesize{Summary of the non-trivial 
dilatation and chiral weights in the standard Weyl 
multiplet.}\label{chiral-dilatation-weights}}
\end{center}
\end{table}

\subsection{The abelian vector multiplet}\label{VectorMultiplet}

\subsubsection{One-form geometry of the abelian vector multiplet and its descendents}\label{VectorMultipletGeometry}
The field strength two-form $F$ of an Abelian vector multiplet 
is given in terms of its one-form potential
$V = \rd z^M V_M= E^A V_A$ by $F = \rd V=\hf E^B\wedge E^A F_{AB}$, or equivalently,
\begin{align}
F_{AB} &= 2 \nabla_{[A} V_{B\}} - T_{AB}{}^C V_C ~.
\end{align}
Due to the existence of the one-form potential the field strength must satisfy the Bianchi identity
\be
\rd F = 0 \implies \nabla_{[A} F_{BC\}} - T_{[AB}{}^{D} F_{|D| C\}} = 0 ~.
\ee
At mass dimension-1 we impose the constraints 
\begin{align}\label{2formConstraint}
F_\a^i{}_\b^j = - 2 \ve^{ij}\ve_{\a\b} \overline{W} \ , \quad F^\ad_i{}^\bd_j = 2 \ve_{ij} \ve^{\ad\bd} W \ , \quad F_\a^i{}^\bd_j = 0 \ ,
\end{align}
where $W$ is a primary superfield with dimension 1 and $\rm U(1)_R$ weight $-2$,
\be \label{Wprimary}
K_A W = 0 \ , \quad \mathbb{D} W = W \ , \quad Y W = -2 W \ .
\ee
Then the Bianchi identities may be solved giving
\begin{subequations}\label{2formCurvs}
\begin{align}
F_a{}_\b^j =& \frac{\ri}{2} (\s_a)_\b{}^\gd \bar{\nabla}_\gd^j \overline{W} \ , \qquad 
F_a{}^\bd_j = - \frac{\ri}{2} (\s_a)_\g{}^\bd \nabla^\g_j W \ , \\  \label{2formCurvsb}
F_{ab} =& - \frac{1}{8} (\s_{ab})_{\a\b} ( \nabla^{\a \b} W + 4 W^{\a\b} \overline{W})
+ \frac{1}{8} (\tilde{\s}_{ab})_{\ad\bd}  (\bar{\nabla}^{\ad \bd} \overline{W} + 4 \overline{W}^{\ad\bd} W) \ .
\end{align}
\end{subequations}
The Bianchi identities also require $W$ to be a reduced chiral superfield,
\begin{align}\label{BianchiYM}
\bar{\nabla}_{\ad}^i W = 0~,\qquad
\nabla^{ij} W = \bar{\nabla}^{ij} \overline{W} ~.
\end{align}
Note that we have introduced the notation
\be \nabla^{ij} := \nabla^{\g (i}  \nabla_\g^{j)} \ , \quad \bar{\nabla}^{ij} := \bar{\nabla}_\gd^{(i}  \bar{\nabla}^{\gd j)}
~,~~~~~~
\nabla^{\a\b} := \nabla^{(\a k}  \nabla_k^{\b)} \ , \quad \bar{\nabla}^{\ad\bd} := \bar{\nabla}_k^{(\ad}  \bar{\nabla}^{\bd) k} \ .
\ee
Acting with spinor covariant derivatives on $W$ gives the following independent descendants:
\bsubeq
\begin{align}
    & \l{}_\a^i :=\nabla_\a^i W ~, \quad  \bar{\l}^{\ad}_{i} := \bar{\nabla}^\ad_i \overline{W} ~, \quad X^{ij} := \nabla^{ij} W = \bar{\nabla}^{ij} \overline{W} ~,  \\
    & F_{ab} := - \frac{1}{8} (\s_{ab})_{\a\b} ( \nabla^{\a \b} W + 4 W^{\a\b} \overline{W})
+ \frac{1}{8} (\tilde{\s}_{ab})_{\ad\bd}  (\bar{\nabla}^{\ad \bd} \overline{W} + 4 \overline{W}^{\ad\bd} W) ~, \label{eq:Fabog}\\
    & F_{\a \b} := \hf (\s^{ab})_{\a\b} F_{ab} = - \frac{1}{8} ( \nabla_{\a \b} W + 4 W_{\a\b} \overline{W}) ~ , \\
    & \bar{F}^{\ad \bd} := - \hf (\tilde{\s}^{ab})^{\ad\bd} F_{ab} = - \frac{1}{8}  (\bar{\nabla}^{\ad \bd} \overline{W} + 4 \overline{W}^{\ad\bd} W) ~.
\end{align}
\esubeq
These superfields satisfy the following tower relations that are particularly useful in analysing the structure of invariants:
\bsubeq
\begin{align}
       &
       \nabla_\a^i \l{}_\b^j =  \hf \e_{\a \b} X^{i j} + 4 \e^{i j} F_{\a \b} + 2 \e^{i j} W_{\a \b} \overline{W} ~ , \\
       &
       \bar{\nabla}^\ad_i \l{}_\b^j = -2 \ri \d^j_i \nabla_{\b}{}^{\ad} W ~ , \qquad
       {\nabla}_\a^i \bar{\l}{}^\bd_j = -2 \ri \d_j^i \nabla_{\a}{}^{\bd} \overline{W} ~ , \\
       &
       \bar{\nabla}^\ad_i \bar{\l}{}^\bd_j = \hf \e^{\ad \bd} X_{i j} + 4 \e_{i j} \bar{F}^{\ad \bd} + 2 \e_{i j} \overline{W}^{\ad \bd} W ~ , \\
       & {\nabla}_\a^i X^{j k} = - 4 \ri \e^{i (j} \nabla_{\a}{}^{\ad} \bar{\l}_{\ad}^{k)} ~ , \quad \bar{\nabla}^\ad_i X^{j k} = 4 \ri \d_{i}^{(j} \nabla_{\a}{}^{\ad}\l^{k) \a} ~ , \\
       & {\nabla}_\g^i F_{\a \b} = \e_{\g (\a} \Sigma_{\b)}^i \overline{W} - \hf W_{\a \b}{}_{\g}^{\, i} \overline{W} + \frac{1}{2} \ri \e_{\g (\a} \nabla_{\b)}{}^{\ad} \bar{\l}^i_\ad ~ , \\
       & \bar{{\nabla}}^\gd_k F_{\a \b} = \frac{\ri}{2} \nabla_{(\a}{}^{\gd} \l_{\b) k} - \hf W_{\a \b} \bar{\l}^{\gd}_k ~ , \\
        & {\nabla}_\g^i \bar{F}^{\ad \bd} = \frac{\ri}{2} \nabla_{\g}{}^{(\ad} \bar{\l}^{\bd) k} - \hf \overline{W}^{\ad \bd} {\l}_{\g}^k ~ , \\
       & \bar{{\nabla}}^\gd_k \bar{F}^{\ad \bd}= - \e^{\gd (\ad} \bar{\Sigma}^{\bd)}_k W -\hf \overline{W}^{\ad \bd}{}^{\gd}_k W  + \frac{1}{2} \ri \e^{\gd (\ad} \nabla_{\a}{}^{\bd)} \l^\a_k ~.
\end{align}
\esubeq
These descendant superfields transform under $S$-supersymmetry as given in the Appendix \ref{S-transformation} of our work.

\subsubsection{The abelian vector multiplet in components}

We define the component fields of the abelian vector multiplet as follows
\begin{align}
\phi := W| \ , \quad \l{}_\a^i := \l_\a^i | = \nabla_\a^i W| \ , \quad X^{ij} := X^{i j}|= \nabla^{ij} W |
\ ,\quad 
F_{ab} := F_{ab}\vert 
\ .
\end{align}
See \cite{Fayet:1975yi,Grimm:1977xp,deWit:1979dzm,deWit:1984rvr,deWit:1984wbb,Cremmer:1984hj} for seminal works on the $\cN=2$ vector multiplet. The reality of $X^{ij}$ follows from the Bianchi identity. The remaining component field being the gauge connection $v_m$ is given by the lowest component of the corresponding superspace connection, $v_m = V_m\vert$. It is worth underlining that the following definition for $F_{a b}$ from \eqref{eq:Fabog} can be directly projected to components
\bea
{F}_{ab} &=& - \frac{1}{8} (\s_{ab})_{\a\b} ( \nabla^{\a \b} W + 4 W^{\a\b} \overline{W})|
+ \frac{1}{8} (\tilde{\s}_{ab})_{\ad\bd}  (\bar{\nabla}^{\ad \bd} \overline{W} + 4 \overline{W}^{\ad\bd} W)| ~.
\eea
The component two-form field strength is constructed from a
projection of the superspace two-form,
\begin{align}
f_{mn} = F_{mn} \lc = 2 \partial_{[m} V_{n]}\lc = 2 \partial_{[m} v_{n]}~.
\end{align}
Making use of the identity
\begin{align}\label{FmnFAB}
F_{mn} = E_m{}^A E_n{}^B F_{AB} (-)^{ab} ,
\end{align}
and projecting to its lowest component, we may solve for $F_{ab}\vert$ to give 
\begin{align}\label{covF}
 F_{ab} := F_{ab}\vert = \; & e_a{}^m e_b{}^n {f}_{mn}
	- \frac{\ri}{2} (\s_{[a})_\a{}^\ad \psi_{b]}{}^\a_k \bar{\l}^k_\ad
	+ \frac{\ri}{2} (\tilde{\s}_{[a})_\ad{}^\a \l^k_\a \psib_{b]}{}^\ad_k \non\\
	& - \hf \psi_{a}{}^\g_k \psi_{b}{}^k_\g \bar{\phi}
	+ \hf \psib_{a}{}^\gd_k \psib_{b}{}^k_\gd \phi ~ .
\end{align}
In the superconformal tensor calculus' language, $F_{ab}$ is
referred to as the supercovariant field strength (as it transforms covariantly under any local superconformal transformations) whereas
$f_{mn}=2\pa_{[m}v_{n]}$ is the conventional field strength.
By construction $F_{ab} $ satisfies the Bianchi identity 
\bea
\de_{[a}F_{bc]} 
=
-\frac{ \ri}{2} R(Q)_{[ab}{}_j \s_{c]}\bar{\l}^{j} 
+\frac{ \ri}{2}R(\bar{Q})_{[ab}{}^j \ts_{c]}\l_{j}
~.
\eea
The local superconformal transformations of the fundamental fields of the vector multiplet fields in a standard Weyl multiplet background are given in Appendix \ref{local-superconformal-transformation}.

\subsection{The linear multiplet}\label{linearMultiplet}
\subsubsection{Two-form geometry of the linear multiplet}\label{linearMultipletGeometry}

The field strength three-form $H$ is given in terms of its two-form gauge
potential $B = \frac{1}{2} E^B\wedge  E^A B_{AB}$ by
\begin{align}
H = \rd B = \frac{1}{3!} E^C \wedge E^B \wedge E^A H_{ABC}~,\qquad
H_{ABC} = 3 \nabla_{[A} B_{BC\}} - 3 T_{[AB}{}^{D} B_{|D|C\}}~. \label{defH}
\end{align}
The field strength remains invariant under gauge transformations
$\d B = \rd V$ with $V$ a one-form gauge parameter.
The existence of the gauge potential requires that the Bianchi identity
\be
\rd H = 0 \implies \nabla_{[A} H_{BCD\}} - \frac{3}{2} T_{[AB}{}^{E} H_{|E|CD\}} = 0 ~,
\ee
be satisfied. As with the gauge one-form, we must impose constraints to reduce the multiplet.
At mass dimension-$\frac{3}{2}$ they consist of 
\be \label{3formConstraint}
H_\a^i{}_\b^j{}_\g^k = H^\ad_i{}^\bd_j{}^\gd_k
= H_\a^i{}_\b^j{}^\gd_k = H^\ad_i{}^\bd_j{}_\g^k = 0 \ .
\ee
The Bianchi identities for $H$ can then be solved.
The solution is
\begin{subequations}\label{3formCurvs}
\begin{align}
H_a{}_\a^i{}_\b^j &= 0 \ , \quad H_a{}^\ad_i{}^\bd_j = 0 \ , \quad 
H_a{}^i_\a{}^\ad_j = \hf (\s_a)_\a{}^\ad \cG^i{}_j \ , \quad \\
H_{ab}{}_\a^i &= \frac{\ri}{6} (\s_{ab})_\a{}^\b \nabla_\b^k \cG^i{}_k \ , \quad 
H_{ab}{}^\ad_i = \frac{\ri}{6} (\tilde{\s}_{ab})^\ad{}_\bd \bar{\nabla}^\bd_k \cG^k{}_i \ , \\
H_{abc} &= \frac{\ri}{96} \ve_{abcd} (\s^d)^\a{}_\bd [\nabla_\a^i , 
\bar{\nabla}^\bd_j] \cG^j{}_i = \ve_{abcd}  H^d \ ,
\end{align}
\end{subequations}
where $\cG^{ij}$ is a real symmetric conformally primary dimension-2 superfield, i.e.,
\begin{align}\label{Gweight}
K_A \cG^{ij} = 0 \ , \quad \mathbb{D} \cG^{ij} = 2 \cG^{ij} \ , \quad Y \cG^{ij} = 0 \ , \quad
(\cG^{ij})^* = \cG_{ij} = \veps_{ik} \veps_{jl} \cG^{kl}~.
\end{align}
The superfield $\cG^{ij}$ also obeys the constraint
\begin{align}\label{Gconstraint}
\bar{\nabla}^{(i}_\ad \cG^{jk)} = \nabla^{(i}_\a \cG^{jk)} = 0~,
\end{align}
which defines the $\cN=2$ linear multiplet. By acting with spinor covariant derivatives on $\cG^{ij}$ gives the following descendants:
\bsubeq
\begin{align}
\c_{\a i} &:= \frac{1}{3} \nabla_\a^j \cG_{i j} ~, \quad\quad 
	\bar{\c}^{\ad i} := \frac{1}{3} \bar{\nabla}^\ad_j \cG^{ij}~, \\
F &:=  \frac{1}{12} \nabla^{ij} \cG_{ij}~, \quad\quad
\bar{F} :=  \frac{1}{12} \bar{\nabla}^{ij} \cG_{ij} \ ,
\\
H_{abc} & := \frac{\ri}{96} \ve_{abcd} (\s^d)^\a{}_\bd [\nabla_\a^i , \bar{\nabla}^\bd_j] \cG^j{}_i = \ve_{abcd} H^d \ ,
\\
 H_{a} &
=\frac{\ri}{96} (\s_a)^\a{}_\bd [\nabla_\a^i , \bar{\nabla}^\bd_j] \cG^j{}_i =  \frac{1}{6}\ve_{abcd}H^{bcd}\ .
\end{align}
\esubeq
These superfields satisfy the following tower relations and are particularly useful in analysing the structure of invariants:
\bsubeq
\begin{align}
    \de^i_\a \cG^{j k}  &= -2 \e^{i (j} \chi^{k )}_\a~, \quad \deb^\ad_i \cG_{j k}  = -2 \e_{i (j} \bar{\chi}_{k )}^\ad~, \\
    \de^i_\a \chi_{\b j} &= \d^i_j \e_{\a \b} F~,  \quad \deb^\ad_i \chi_{\b j} = -4\ri \e_{i j} H_a (\s^a)_\b{}^\ad -  \ri \de_\b{}^\ad \cG_{i j}~, \\
    \deb^\ad_i \bar{\chi}^{\bd j} &= \d_i^j \e^{\ad \bd} \bar{F}~, \quad \de_{i \b} \bar{\chi}^\ad_j  = -4 \ri \e_{i j} H_a (\s^a)_\b{}^\ad -  \ri \de_\b{}^\ad \cG_{i j}~, \\
    \de^i_\a F &= 0~, \quad
    \deb^\ad_i F = 2 \ri \de_\a{}^\ad \chi^\a_i  - 2 \overline{W}^\ad{}_\gd \bar{\chi}^\gd_i - 6 \overline{\S}^{\ad j} G_{i j} ~, \\
    \deb^\ad_i \bar{F} &= 0~, \quad \de^i_\a \bar{F} = 2 \ri \de_\a{}^\ad \bar{\chi}^i_\ad  + 2 W_{\a \g} \chi^{\g i} - 6 \S_{\a}^{j} G^i{}_{j}~, \\
    \de^i_\a H_a &=\frac{1}{2} (\s_{a b})_\a{}^\b \de^b \chi^{ i}_\b  - \frac{\ri}{8} (\s_a)_{\a \bd}
	\left[\overline{W}^{\bd \gd}  \bar{\chi}^i_\gd    +3 \overline{\S}^{\bd}_{l} G^{l i}\right]~, \\
    \deb_{\ad i} H_a &= - \frac{1}{2} (\tilde{\s}_{a b})_{\ad\bd} \de^b \bar{\chi}_i^\bd + \frac{\ri}{8} (\tilde{\s}_a)_{\ad\b} \left[ W^{\b\g} \chi_{\g i} +3 \S^{\b l} G_{l i} \right]~.
\end{align}
\esubeq
These descendant superfields transform under $S$-supersymmetry as given in the Appendix \ref{S-transformation} of our work.

It is possible to construct a superfield which automatically obeys the
above constraints \eqref{Gconstraint} by imposing constraints on the
two-form $B_{AB}$ itself. It holds that
\begin{align}
B_\a^i{}_\b^j = -2\ve^{ij} \ve_{\a\b} \bar{\Psi} \ , \quad
B^\ad_i{}^\bd_j = 2\ve_{ij} \ve^{\ad\bd} \Psi \ , \quad
B_\a^i{}^\bd_j = 0 \ ,
\label{potential-linear-i}
\end{align}
where $\Psi$ is a chiral superfield, $\bar{\nabla}^\ad_i \Psi = 0$,
of dimension 1 and $\rm U(1)_R$ weight -2, but otherwise arbitrary
\be 
K_A \Psi = 0 \ , \quad \mathbb{D} \Psi = \Psi\ , \quad Y \Psi = -2 \Psi \ .
\ee
Constraints of this kind are quite natural since the gauge transformation
$\delta B = \rd \tilde{V} = \tilde{F}$ amounts to
\be\label{gaugeT}
\d \Psi =  \tilde{W} ~,
\ee
with $\tilde{W}$ a generic vector multiplet chiral field strength satisfying the Bianchi identities of eq. \eqref{BianchiYM}. We can then proceed to solve \eqref{defH} for the full two-form $B$:
\begin{subequations} \label{eq_B_two_form}
\begin{align}
B_a{}_\a^i &= \frac{\ri}{2}(\s_a)_\a{}^\ad \bar{\nabla}_\ad^i \bar{\Psi} \ , \ \ 
B_a{}^\ad_i =-\frac{\ri}{2} (\s_a)_\a{}^\ad \nabla^\a_i \Psi \ , \\
B_{ab} &= - \frac{1}{8} (\s_{ab})^{\a\b} (\nabla_{\a\b} \Psi +4W_{\a\b} \bar{\Psi}) 
+ \frac{1}{8} (\tilde{\s}_{ab})_{\ad\bd} (\bar{\nabla}^{\ad\bd} \bar{\Psi} +4 \overline{W}^{\ad\bd} \Psi) \ .
\end{align}
\end{subequations}
Inserting this solution into the definition of $H$ leads to an expression for
the linear multiplet in terms of an unconstrained chiral prepotential 
\begin{align} \label{eq_linear_mult_chiral}
\cG^{ij} &= -\frac{\ri}{2}\big(\nabla^{ij} \Psi -\bar{\nabla}^{ij} \bar{\Psi}\big)
={\rm Im}[\de^{ij}\Psi]~,
~~~~~~
\nabla^{ij} \Psi -\bar{\nabla}^{ij} \bar{\Psi}
=
2\ri \,\cG^{ij} ~.
\end{align}
One can check that $\cG^{ij}$ indeed obeys \eqref{Gweight} and \eqref{Gconstraint}
and is invariant under \eqref{gaugeT}.

\subsubsection{Linear multiplet in components}

The components of the linear multiplet are defined as follows --- see  \cite{FS2,deWit:1980gt,deWit:1980lyi,deWit:1983xhu,N=2tensor,Siegel:1978yi,Siegel80,SSW,SSW-2,deWit:1982na,KLR,LR3} for seminal works.
The linear multiplet is described by a real primary superfield $\cG^{ij}$ of
dimension 2 \eqref{Gweight} satisfying the constraint \eqref{Gconstraint}.
The corresponding 3-form field strength $H$ in superspace is given by
\eqref{3formConstraint} and \eqref{3formCurvs}.
Within the superfield $\cG^{ij}$ are the matter components
of the linear multiplet: a real isotriplet field $G^{ij}$,
a fermion $\chi_{\a i}$, and a complex scalar $F$:
\begin{subequations}\label{component-linear-111}
\begin{alignat}{2}
G^{ij} &:= \cG^{ij}|~, \\
\c_{\a i} &:= \frac{1}{3} \nabla_\a^j \cG_{i j}| ~, \quad\quad 
	\bar{\c}^{\ad i} := \frac{1}{3} \bar{\nabla}^\ad_j \cG^{ij}|~, \\
F &:=  \frac{1}{12} \nabla^{ij} \cG_{ij}|~, \quad\quad
\bar{F} :=  \frac{1}{12} \bar{\nabla}^{ij} \cG_{ij}| \ ,
\\
H_{abc} & := \frac{\ri}{96} \ve_{abcd} (\s^d)^\a{}_\bd [\nabla_\a^i , \bar{\nabla}^\bd_j] \cG^j{}_i |= \ve_{abcd} H^d \ ,
\\
 H_{a} &
=\frac{\ri}{96} (\s_a)^\a{}_\bd [\nabla_\a^i , \bar{\nabla}^\bd_j] \cG^j{}_i |=  \frac{1}{6}\ve_{abcd}H^{bcd}\ .
\end{alignat}
\end{subequations}
The remaining component field, the two-form, is given by $b_{mn} := B_{mn}\lc$.
Owing to the superspace identity
\begin{align}
\nabla_\a^i \nabla_\b^j \cG^{kl} &= - \frac{1}{6} \ve_{\a\b} \ve^{i (k } \ve^{l ) j} \nabla^{pq} \cG_{pq} \ ,
\end{align}
there are no other independent component fields. The local superconformal transformations of the fundamental fields of the linear multiplet in a standard Weyl multiplet background are given in Appendix \ref{local-superconformal-transformation}.

The covariant conservation equation for $ H_a$ is
\bea
\nabla^a  H_{a} 
&=& 
\frac{3}{8}
\S^{i}\chi_{i}
+\frac{3}{8}
\bar{\S}_{i}\bar{\chi}^{i}
~.
\label{covariant-current}
\eea
The constraint locally implies the existence of a gauge two-form potential, $b_{mn}=-b_{nm}$, 
and its exterior derivative $h_{mnp}:=3\pa_{[m}b_{np]}$. The solution of \eqref{covariant-current} is
\bea
 H_a 
=  \frac{1}{6} \ve_a{}^{bcd} \Big(h_{bcd}
-\frac{3 \ri}{4}\psi_b{}_i \s_{cd}\chi^i 
-\frac{3 \ri}{4}
\psib_b{}^i\tilde{\s}_{cd}\bar{\chi}_i
-\frac{3}{4} (\psi_b{}^i\s_c\psib_d{}^j) G_{ij} \Big)
~,
\label{solution-H-hdwm}
\eea
where
$h_{abc}=e_a{}^me_b{}^ne_c{}^ph_{mnp}$.
The local superconformal transformations of $b_{mn}$ are 
\bea
\d b_{mn}
&=&
\frac{\ri}{2} \xi_i\s_{mn} \chi^i
+ \frac{\ri}{2}
\bar{\xi}^i \tilde{\s}_{m n} \bar{\chi}_i  +\frac{1}{2}\Big(
\psib_{[m}{}^i\s_{n]}\xi^j
-\psi_{[m}{}^i\s_{n]}\bar{\xi}^j\Big)
G_{ij}
+2\pa_{[m}l_{n]}
~,
~~~~~~
\label{d-bmn}
\eea
where we have also included 
the vector gauge transformation
$\d_l b_{mn}=2\pa_{[m}l_{n]}$
that leaves 
$h_{mnp}$
and
$ H^a$ invariant. In constructing the superspace three-form $H_{mnp} = 3 \partial_{[m} B_{np]}$
we can make use of the superspace identity
\begin{align}
H_{mnp} = E_m{}^A E_n{}^B E_p{}^C H_{ABC} (-)^{ab+ac+bc} ~.
\end{align}
Projecting this equation to the lowest component, and defining
\begin{align}
h_{mnp} := H_{mnp}\lc = 3 \partial_{[m} b_{np]}
~,~~~
h_{abc} := e_a{}^me_b{}^ne_c{}^ph_{mnp} ~,
\end{align}
it is easy to show that
\bea
\label{h_eqn}
h_{mnp} &=& \eps_{mnpq} H^a\lc e_a{}^q
	+ \frac{3 \ri}{4} (\sigma_{[mn})_\a{}^\b \psi_{p]}{}_j^\a \chi^j_\b
	+\frac{3 \ri}{4} (\ts_{[mn})^\ad{}_\bd \bar\psi_{p]}{}^j_\ad \bar\chi^\bd_j 
 \non\\
 &&
-\frac{3}{4} (\sigma_{[m})_{\a\ad} \psi_{n}{}^{\a i} \bar\psi_{p]}{}^{\ad j} G_{ij} ~,
\eea
or, equivalently,
\bea \label{eq:bigHtoLittleh}
 H^a&:=& H^a|  =   \frac{1}{6} \veps^{abcd} H_{bcd}\lc  \nonumber
        \\
	& = & \frac{1}{6} \ve^{abcd} \Big(h_{bcd}
	-\frac{3 \ri}{4} (\s_{cd})_\a{}^\b \psi_b{}^\a_k \chi_\b^k 
-\frac{3 \ri}{4} (\tilde{\s}_{cd})^\ad{}_\bd \psib_b{}^k_\ad \bar{\chi}^\bd_k
+\frac{3}{4}  (\s_b)_\a{}^\bd \psi_c{}^\a_k \psib_d{}^l_\bd G^k{}_l \Big) \ .~~~~~~~~~
\eea
In the paper, we will denote the Hodge dual of a three-form component field $h_{abc}$ with 
\bea \label{eq:hHodge}
\tilde{h}^a=\frac{1}{6}\ve^{abcd}h_{bcd}
~.
\eea
We also keep using the same notation for the superfield $H^a$ and the covariant component field $H^a|$, but we hope the reader will understand what we refer to depending on the context.

We have emphasized that the construction of the two-form multiplet
is completely geometrical, but it is worth noting that, as
discussed in Subsection \ref{linearMultipletGeometry}, the two-form multiplet can be
encoded in a chiral superfield $\Psi$. 
By making use of the gauge transformations \eqref{gaugeT},
one can choose for the components of $B_{AB}$ that $B_{\a\b}|=\bar{B}^{\ad\bd}|=0$ and $B_{a\b}|=\bar{B}^{a\bd}|=0$ while $B_{ab}| $ remains unconstrained by imposing the component constraints\footnote{The third constraint is not actually necessary to eliminate the other components of the two-form, but it does substantially simplify the component evaluation later.}
\begin{align}\label{eq_psi_constraints}
\Psi \lc = 0~, \quad
\nabla_\a^i \Psi \lc = 0~, \quad
\nabla^{ij} \Psi \lc = -\bar\nabla^{ij} \bar\Psi \lc~.
\end{align}
One may easily construct $b_{mn}$ using
$b_{mn} = e_m{}^a e_n{}^b B_{ab}\lc$.
As usual, the supersymmetry transformation laws of the component
fields may be derived by using the constraints.

\subsection{On-shell hypermultiplet 
and hyper-dilaton Weyl multiplet}
\label{hyper+HDWM}

In this subsection, we review the construction of the 4D, $\cN=2$ hyper-dilaton Weyl multiplet of \cite{Gold:2022bdk}. This plays a central work in our work.

A single on-shell hypermultiplet is comprised of $4+4$ degrees of freedom
described by a Lorentz scalar field $q^{i\underline{i}}$ 
and spinor fields 
$(\rho_\a^{\underline{i}}\,,\bar{\rho}^\ad_{\underline{i}})$
---
see \cite{Fayet:1975yi,FS2,deWit:1980gt,deWit:1980lyi,deWit:1983xhu,deWit:1984rvr}
together with \cite{deWit:1999fp,Freedman:2012zz,Lauria:2020rhc}
and references therein for superconformal approaches
to systems of on-shell hypermultiplets. 
The index $\underline{i}=\underline{1},\underline{2}$ is a SU(2) flavour 
index, and the fields satisfy the following reality conditions
\bea
(q^{i\underline{i}})^*=q_{i\underline{i}}
~,~~~~~~
(\rho_\a^{\underline{i}})^*=\bar{\rho}_{\ad \underline{i}}
~.
\eea
They also satisfy the following dilatation and
chiral weight identities
\bsubeq
\bea
\mathbb{D}q^{i\underline{i}}
=q^{i\underline{i}}
~,~~~
\mathbb{D}\rho_\a^{\underline{i}}
&=&\frac{3}{2}\rho_\a^{\underline{i}}
~,~~~
\mathbb{D}\bar{\rho}_{\ad\underline{i}}
=\frac{3}{2}\bar{\rho}_{\ad\underline{i}}
~,
\\
Yq^{i\underline{i}}
=0
~,~~~
Y\rho_\a^{\underline{i}}
&=&\rho_\a^{\underline{i}}
~,~~~
Y\bar{\rho}_{\ad\underline{i}}
=-\bar{\rho}_{\ad\underline{i}}
~.
\eea
\esubeq
The multiplet, which has the field $q^{i\underline{i}}$ 
as its superconformal primary,
is characterised by the following local superconformal transformations
\cite{deWit:1980gt,deWit:1980lyi,deWit:1983xhu,deWit:1984rvr,deWit:1999fp,Freedman:2012zz,Lauria:2020rhc}
\bsubeq\label{hyper-susy-hdwm}
\bea
\delta q^{i\underline{i}} 
&=& 
\frac{1}{2}\xi^{ i}\rho^{\underline{i}}
-\frac{1}{2}\bar{\xi}^{i} \bar{\rho}^{\underline{i}}
+\lambda^{i}{}_{k}q^{k\underline{i}} 
+ \lambda_{\mathbb{D}}q^{i\underline{i}}
~,
\label{d-qii-hdwm}
\\
\delta \rho_{\a}^{\underline{i}} 
&=&
-4\ri(\sigma^a\bar{\xi}_k)_\a 
\nabla_a q^{k\underline{i}}
 +\hf \lambda_{ab} (\s^{ab} \rho^{\underline{i}})_{\a}    
 + \ri\lambda_Y\rho_{\a}^{\underline{i}}
+ \frac{3}{2}\lambda_{\mathbb{D}}\rho_{\a}^{\underline{i}} 
+ 8\eta_\a^k q_k{}^{\underline{i}}
~,
\label{d-rho-hdwm}
\\
\delta \bar{\rho}^{\ad}_{\underline{i}} 
&=&
~\,\,\,4\ri(\tilde{\sigma}^a{\xi}^k)^{\ad}\nabla_a q_{k{\underline{i}}}
 +\hf \lambda_{ab} (\tilde{\s}^{ab}\bar{\rho}_{\underline{i}} )^{\ad}  
 - \ri\lambda_Y\bar{\rho}^{\ad}_{\underline{i}}
+ \frac{3}{2}\lambda_{\mathbb{D}}\bar{\rho}^{\ad}_{\underline{i}} 
- 8\bar{\eta}^\ad_k q^k{}_{\underline{i}}
~,
\label{d-b-rho-hdwm}
\eea
\esubeq
where 
\bea
\nabla_aq^{i\underline{i}}
=
\cD_a q^{i\underline{i}}
 -\frac{1}{4}\psi_a{}^{i}
 \rho^{\underline{i}}
 + \frac{1}{4}\bar{\psi}_a{}^{i} 
 \bar{\rho}^{\underline{i}}
 ~.
 \label{DD'q-hdwm}
\eea

In conformal superspace, the multiplet is described by a dimension one primary superfield $Q^{i\underline{i}}$, neutral under U(1)$_{\rm R}$, and satisfying the following analyticity constraint
\bea
\de_\a^{(i}Q^{j)\underline{j}}
=\deb_\ad^{(i}Q^{j)\underline{j}}
~.
\label{analyticity-1}
\eea
The transformation rules in \eqref{hyper-susy-hdwm} simply derive from the equation above together with self-consistency of the conformal superspace algebra of covariant derivatives and the following definition of the component fields
\bea
q^{i\underline{i}}
:=Q^{i\underline{i}}|
~,~~~
\rho_\a^{\underline{i}}
:=\de_{\a}^{i} Q_i{}^{\underline{i}}|
~,~~~
\bar{\rho}_\ad^{\underline{i}}
=\deb_{\ad}^{i} Q_i{}^{ \underline{i}}|
~.
\label{analyticity-2}
\eea

In contrast with the standard Weyl 
multiplet described in a previous 
subsection,
the algebra of the local superconformal
transformations \eqref{hyper-susy-hdwm} 
closes 
only when equations of motion for the 
fields are imposed,
see for example \cite{deWit:1999fp,Lauria:2020rhc} for a detailed analysis.
In our notations, the covariant equations of motion
of $q^{i\underline{i}}$
and $(\rho_\a^{\underline{i}},\,\bar{\rho}^{\ad}_{\underline{i}})$
are:
\bsubeq\label{on-shell-hyper-hdwm}
\bea
\left(\nabla_{a}
\rho^{\underline{i}}\,
\s^a\right)_{\dot{\alpha}}
&=&
~\,\,\, \frac{\ri}{2}
(\bar{\rho}^{\underline{i}}\,
\tilde{\s}^{cd})_{\ad}
W^-_{cd}
+6\ri\bar{\Sigma}_{\dot{\alpha} k}q^{k\underline{i}}
 ~,\\
\left(
\nabla_a
\bar{\rho}_{\underline{i}}\,
\tilde{\s}^a\right)^{\alpha}
&=& 
-\frac{\ri}{2}
(\rho_{\underline{i}}\,\s^{cd})^{\alpha}
W^+_{cd}
+6\ri\Sigma^{\a k}q_{k\underline{i}}
~,\\
\Box q^{i\underline{i}} 
&=& 
 -\frac{3}{2}Dq^{i\underline{i}} 
 ~,~~~~~~
\Box:=\de^a\de_a
~.
\label{boxq-0-hdwm}
\eea
\esubeq
The expressions for 
$\nabla_{a}\rho_{\alpha}^{\underline{i}}$, 
$\nabla_{a}\bar{\rho}^{\dot{\alpha}}_{\underline{i}}$, 
and $\square q^{i\underline{i}}$ in terms of the derivatives 
$\cD_a$ are given by
\bsubeq
\bea
\nabla_a \rho_\a^{\underline{i}}
&=& 
\cD_a\rho_\a^{\underline{i}} 
+2\ri{(\sigma^b{\bar{\psi}_a}{}_k)}_{\a}\left( 
\cD_b q^{k\underline{i}} 
-\frac{1}{4}{\psi_b}^{k}
\rho^{\underline{i}} 
+\frac{1}{4}{\bar{\psi}_b}{}^{k}
\bar{\rho}^{\underline{i}}
\right) 
+4{\phi_a}_{\a k}q^{k\underline{i}}
~,
\\\nabla_a \bar{\rho}^\ad_{\underline{i}}
&=& 
\cD_a\bar{\rho}^\ad_{\underline{i}} 
-2\ri{(\tilde{\sigma}^b{{\psi}_a}^{{ k}})}^{\dot{\a}}\left( 
\cD_b q_{k\underline{i}} 
- \frac{1}{4}
{\psi_b}_k\rho_{{\underline{i}}} 
+\frac{1}{4}
{\bar{\psi}_b}{}_k\bar{\rho}_{ {\underline{i}}}
\right) 
- 4{\bar{\phi}_a}{}^{\ad k}q_{k{\underline{i}}}
~,
\\
\square q^{i\underline{i}} 
&=& 
\cD^a\cD_a q^{i\underline{i}}  
- 2 {\mathfrak{f}_a}^a  q^{i \underline{i}}
- \frac{1}{4}\rho^{\underline{i}} \cD_a{\psi^a}^{i}
+ \frac{1}{4}\bar{\rho}^{\underline{i}} \cD_a {\bar{\psi}^a}{}^{ i} 
- \frac{1}{2}{\psi^a}^{ i}\cD_a\rho^{\underline{i}}
+ \frac{1}{2} {\bar{\psi}^a}{}^{ i}\cD_a\bar{\rho}^{\underline{i}}
\nonumber\\
&&
- \frac{\ri}{4}\phi_a{}^i \sigma^a \bar{\rho}^{\underline{i}} 
+ \frac{\ri}{4}\bar{\phi}_a^i \tilde{\sigma}^a \rho^{\underline{i}} 
+\ri
({\psi_a}^{(i}\sigma^b {\bar{\psi}^{k)a}})
\cD_b q_k{}^{\underline{i}} 
+ \frac{3\ri}{4}({\psi_a}^{ i} \sigma^a \bar{\Sigma}_{l}) q^{l \underline{i}} 
+\frac{3\ri}{4} ({\bar{\psi}}_a^i \tilde{\sigma}^a\Sigma_l)q^{l\underline{i}} 
\non\\
&&
-\frac{\ri}{16}  ({\bar{\psi}}_a{}^i 
\tilde{\sigma}^a\s^{cd} 
\rho^{\underline{i}})
W^+_{cd}
- \frac{\ri}{16} ({\psi_a}^{ i}\sigma^a \tilde{\s}^{cd} \bar{\rho}^{\underline{i}}) W^-_{cd}
-({\psi_a}^{i}{\phi^a}{}_{k})
q^{k\underline{i}}
+ ({\bar{\psi}}_a{}^i \bar{\phi}^a{}_k)q^{k\underline{i}} 
 \nonumber \\
&& 
- \frac{\ri}{4}
({\psi_a}^{(i}
\sigma^b
{\bar{\psi}^{k)}{}^a })
({\psi_{b}{}_{k}}\rho^{\underline{i}}) 
+ \frac{\ri}{4}({\psi_a}^{(i}\sigma^c{\bar{\psi}^{k)}}{}^{a})({\bar{\psi}_c}{}_{k}\bar{\rho}^{\underline{i}}) ~.
\eea
\esubeq

It is important to stress that eqs.~\eqref{on-shell-hyper-hdwm}
are typically read as equations of motion for the hypermultiplet fields, see for example, \cite{deWit:1980gt,deWit:1980lyi,deWit:1983xhu,deWit:1984rvr,deWit:1999fp,Freedman:2012zz,Lauria:2020rhc}. They certainly are dynamical equations for 
 $q^{i\underline{i}}$ and $(\rho_\a^{\underline{i}}\,,\bar{\rho}^\ad_{\underline{i}})$
in a flat background (with no central charges as in our case) 
where all conformal supergravity fields are set to zero
\cite{Fayet:1975yi,FS2}. 
For this reason, 
the multiplet is typically referred to as the on-shell 
hypermultiplet. However, such an interpretation is not necessary in a curved background described by the 
standard Weyl multiplet. 
In fact, the eqs. \eqref{on-shell-hyper-hdwm} can be interpreted as algebraic equations for the 
standard Weyl multiplet that determine
the fields $(\S^{\a i},\bar{\S}_{\ad i})$ and $D$ in terms of 
$q^{i\underline{i}}$ and $(\rho_\a^{\underline{i}}\,,\bar{\rho}^\ad_{\underline{i}})$
together with the other independent fields of the standard Weyl multiplet.
If we assume that $q^{i\underline{i}}$ is an invertible matrix, which is equivalent to imposing
\bea
q^2:=q^{i\underline{i}}q_{i\underline{i}}=\ve_{ij}\ve_{\underline{i}\underline{j}}q^{i\underline{i}}q^{j\underline{j}}
=2\det{q^{i\underline{i}}}\ne0
~,
\eea
then the following relations hold
\bsubeq\label{compositeSSbD-hdwm}
\bea
{\Sigma}^{{\alpha} i} 
&=&
2q^{-2}q^{i\underline{i}}\Bigg[
- \frac{\ri}{2}
(\cD_a {\bar{\rho}}_{\underline{i}}
\,\tsigma^a)^\a
 +(\psi_a{}^j \s^b \tilde{\s}^a)^\a \left( 
 \cD_b q_{j\underline{i}} 
 -\frac{1}{4}{{\psi}_b}{}_j
 \rho_{ \underline{i}}
 +\frac{1}{4}
 {\bar\psi_b}{}_{j}\bar\rho_{\underline{i}}
 \right)
 \non\\
&&~~~~~~~~~\,\,
+ \frac{2}{3}
\big(\Psi_{ab}{}_j \s^{ab}\big)^{\a}
q^j{}_{\underline{i}} 
+\frac{1}{4}
(\rho_{\underline{i}}\s^{cd})^{\a}
W^+_{cd} 
+\frac{\ri}{6}
({\bar{\psi}}_a{}_{j}
\ts^a\s^{cd})^{\a} 
q^j{}_{\underline{i}} W^+_{cd}
\,\Bigg] 
~,
\label{sigma-hdwm}
\\
\bar{\Sigma}_{\dot{\alpha} i} 
&=&
2q^{-2}q_{i\underline{i}}\Bigg[
- \frac{\ri}{2}
(\cD_a \rho^{\underline{i}}
\,\sigma^a)_\ad 
 -({\bar{\psi}_a}{}_j \tilde{\s}^b \sigma^a)_\ad \left( 
 \cD_b q^{j\underline{i}} 
 - \frac{1}{4}
 {\psi_b}{}^{j}\rho^{\underline{i}}
 +\frac{1}{4}
 {\bar{\psi}_b}{}^{j}
 \bar{\rho}^{\underline{i}}\right)\nonumber \\
&&~~~~~~~~~~
-\frac{2}{3}
\big(
\overline{\Psi}_{ab}{}^j \ts^{ab}
\big)_{\ad}
q_j{}^{\underline{i}} 
-\frac{1}{4}
(\bar\rho^{\underline{i}}\ts^{cd})_{\ad}
W^-_{cd} 
+\frac{\ri}{6}
(\psi_a{}^{j}
\s^a\ts^{cd})_{\ad} 
q_j{}^{\underline{i}} W^-_{cd}
\,
\Bigg] 
~,
\label{sigma-bar-hdwm}
\\
D 
&=& 
q^{-2}q_{i\underline{i}}\Bigg[
~\cD^a\cD_a q^{i\underline{i}}  
+\frac{1}{6}R\,q^{i\underline{i}} 
-\frac{\ri}{8}({\bar{\psi}}_a{}^i \tilde{\sigma}^a\s^{cd}
\rho^{ \underline{i}})W^+_{cd}
- \frac{\ri}{2}\phi_a{}^i \sigma^a \bar{\rho}^{\underline{i}} 
- \frac{1}{2}\rho^{\underline{i}} \cD_a{\psi^a}^{i}
\nonumber\\
&&
~~~~~~~~\,\,
- {\psi^a}^{ i}\cD_a\rho^{\underline{i}}
+ 2 ({\psi_a}^{i}{\phi^a}^{j})
{q_j}^{\underline{i}}
+ \frac{3\ri}{2}
({\psi_a}^{ i} \sigma^a \bar{\Sigma}_{j}) 
q^{j \underline{i}}
+ \frac{\ri}{2}({\psi_{aj}}\sigma^a \bar{\Sigma}^j)q^{i\underline{i}} 
\nonumber \\
&&~~~~~~~~~
+ \frac{1}{6}\varepsilon^{mnpq}({{\bar{\psi}_{m}}^j} \tilde{\sigma}_n\cD_p\psi_{qj})q^{i\underline{i}}
+ \frac{1}{3}W^{ab+}({{\bar{\psi}_a}^j}{\bar{\psi}_{bj}})q^{i\underline{i}}
+\ri({\psi_a}^{(i}\sigma^b {\bar{\psi}^{j)a}}) \cD_b q_j{}^{\underline{i}} \nonumber \\
&&~~~~~~~~~
- \frac{\ri}{2} ({\psi_a}^{(i}\sigma^b{\bar{\psi}^{j)}{}^a }) ({\psi_{b}{}_{j}}\rho^{\underline{i}})\Bigg] 
+ \text{c.c.}
~.
\label{D_def-hdwm}
\eea
\esubeq
In the expression for $D$, 
eq.~\eqref{D_def-hdwm},
remember that
$(\S^i,\bar\S_i)$
and 
$(\phi^i,\bar\phi_i)$, together with the 
spin connection $\o_m{}^{cd}$,
are composite fields.
Note that so far we have only used one of the four 
equations that are equivalent
to \eqref{boxq-0-hdwm} to solve for $D$ in eq.~\eqref{D_def-hdwm}.
It is simple to show that the remaining independent
three equations are equivalent to the following 
\be
\de^a (q^{i(\underline{i}}\de_aq_i{}^{\underline{j})})=0
~.
\label{2-4-constraints-hdwm}
\ee
As we are going to explain in detail below, 
this equation is solved by turning 
the $\rm SU(2)_R$ connection $\phi_m{}^{kl}$
into a composite field.

As a next step in the construction of the hyper-dilaton Weyl multiplet, we note that, accompanied to an on-shell hypermultiplet
there is always a triplet of composite linear multiplets
\cite{FS2,deWit:1980gt,deWit:1980lyi,deWit:1983xhu}. The following composite fields define a triplet of linear multiplets
\cite{deWit:1984rvr}
\bsubeq\label{composite-linear-hdwm}
\bea
G_{ij}{}^{\underline{i}\underline{j}}
&=&
q_{(i}{}^{\underline{i}}q_{j)}{}^{\underline{j}}
=q_{i}{}^{(\underline{i}}q_{j}{}^{\underline{j})}
~,
~~~~~~~~~~~~~~~~~~~~~~~~~
(G_{ij}{}^{\underline{i}\underline{j}})^*
=
G^{ij}{}_{\underline{i}\underline{j}}
~,\\
\chi_{\alpha i}{}^{\underline{i}\underline{j}}
&=& 
\frac{1}{2} q_{i}{}^{(\underline{i}}\rho_{\alpha}^{\underline{j})}
~,
~~~
\bar\chi^{\ad i}{}_{\underline{i}\underline{j}}
=
-\frac{1}{2} q^{i}{}_{(\underline{i}}\bar{\rho}^{\ad}_{\underline{j})}
~,
~~~~~~~~~
(\chi_{\alpha i}{}^{\underline{i}\underline{j}})^*
=
\bar\chi_\ad^i{}_{\underline{i}\underline{j}}
~,\\
F^{\underline{i}\underline{j}} 
&=& 
\frac{1}{8}\rho^{(\underline{i}}\rho^{\underline{j})}
~,~~~
\bar{F}_{\underline{i}\underline{j}} 
= 
\frac{1}{8}\bar\rho_{(\underline{i}}
\bar\rho_{\underline{j})}
~,~~~~~~~~~~~~~~~~~
(F^{\underline{i}\underline{j}})^*
=
\bar F_{\underline{i}\underline{j}}
~,
\label{composite-F-hdwm}
\\
 H^a{}^{\underline{i}\underline{j}}
&=& -\frac{1}{4}q^{i(\underline{i}} \nabla^a{q_i}^{\underline{j})}
+\frac{\ri}{32}
\rho^{(\underline{i}}\sigma^a\bar{\rho}^{\underline{j})}
~,~~~~~~~~~~~~~
( H^a{}^{\underline{i}\underline{j}})^*
= H^a{}_{\underline{i}\underline{j}}
~.
\label{linear-H-hdwm}
\eea
\esubeq
These fields all transform according to linear multiplet transformation
\eqref{linear-multiplet} and
 each of the previous fields is symmetric in $\underline{i}$ and $\underline{j}$.
Within the previous composite fields, the field 
$ H^a{}^{\underline{i}\underline{j}}$ 
is particularly interesting.
In fact, eq.~\eqref{linear-H-hdwm}
together with \eqref{solution-H-hdwm}
represent the solution to the
constraint \eqref{2-4-constraints-hdwm} 
and can be used to express the 
$\rm SU(2)_R$ connection 
$\phi_m{}^{ij}$ as a composite field.
By introducing the derivative 
\bea
\mathbf{D}_a 
&=& 
{e_a}^m\left( \partial_m - \frac{1}{2}{\omega_m}^{cd}M_{cd} 
- b_m\mathbb{D}\right) 
= \cD_a + {e_a}^m{\phi_m}^{ij}J_{ij} + \ri A_m Y 
~,
\label{boldDa}
\eea
and by using eq.~\eqref{DD'q-hdwm},
eq.~\eqref{linear-H-hdwm} can be rearranged for the $\rm SU(2)_R$ gauge 
connection as follows
\bea
 \phi_a{}^{ij}  
 &=& 
 4q^{-4}q^{(i}{}_{\underline{i}}q^{j)}{}_{\underline{j}}
 \Bigg[\, 
 q^{k\underline{i}} \mathbf{D}_a q_k{}^{\underline{j}}
 - \frac{1}{4}q^{k\underline{i}} 
 ({\psi_a}{}_k \rho^{\underline{j}}) 
 + \frac{1}{4}q^{k\underline{i}}
 ({\bar{\psi}_a}{}_k \bar{\rho}^{\underline{j}})
 - \frac{\ri}{8}
 \rho^{\underline{i}}
 \s_a
 \bar{\rho}^{\underline{j}}
 +4{ H_a{}^{\underline{i}\underline{j}}}\,\Bigg] 
 ~,~~~~~~~~~
 \label{compositeSU2-hdwm}
\eea
with
\bea
H_a{}^{\underline{i}\underline{j}}
=
{ \tilde{h}_a{}^{\underline{i}\underline{j}}}
+ \frac{1}{6} \ve_{abcd} \Big(
-\frac{3 \ri}{4} (\s^{cd})_\a{}^\b \psi^b{}^\a_k \chi_\b^k{}^{\underline{i}\underline{j}}
-\frac{3 \ri}{4} (\tilde{\s}^{cd})^\ad{}_\bd \psib^b{}^k_\ad \bar{\chi}^\bd_k{}^{\underline{i}\underline{j}}
+\frac{3}{4}  (\s^b)_\a{}^\bd \psi^c{}^\a_k \psib^d{}^l_\bd G^k{}_l{}^{\underline{i}\underline{j}} \Big)
~, \non \\
\eea
which plugged back into \eqref{compositeSU2-hdwm} along with \eqref{composite-linear-hdwm} gives
\bea
\phi^{a}{}^{ij}  
 &=& 
 2 {q}^{-2} q^{(i}\,_{\underline{i}} \textbf{D}^{a}{q^{j) \underline{i}}}  
- \frac{1}{2}\psi^{a  \alpha (i} {q}^{-2} q^{j)}\,_{\underline{i}} \rho^{\underline{i}}_{\alpha}
- \frac{1}{2}\overline{\psi}^{a}{}^{(i}_{\dot{\alpha}} {q}^{-2} q^{j) \underline{i}} \overline{\rho}_{\underline{i}}^{\dot{\alpha}} 
+16{q}^{-4} q^{i}\,_{\underline{i}} q^{j}\,_{\underline{j}} \tilde{h}^{a \underline{i} \underline{j}}
\non 
\\ &&
  - \frac{1}{2}{\rm i} (\sigma^{a}) ^{\alpha}\,_{\dot{\alpha}} {q}^{-4} q^{(i}\,_{\underline{i}} q^{j) \underline{j}} \rho^{\underline{i}}_{\alpha} \overline{\rho}_{\underline{j}}^{\dot{\alpha}} 
- \frac{1}{2}\epsilon^{a b c d}   (\sigma_{b})^{\alpha}{}_{\dot{\alpha}} \psi_{c}{}^{(i}_{ \alpha} \overline{\psi}_{d}{}^{j) \dot{\alpha}}  
\non 
\\ &&
  + {q}^{-2} (\sigma^{a b})^{\beta \alpha} \psi_{b}{}_{ \beta}^{i} q^{j}\,_{\underline{i}} \rho^{\underline{i}}_{\alpha}+{q}^{-2} (\tilde{\sigma}^{a b})_{\dot{\beta} \dot{\alpha}} \overline{\psi}_{b}{}^{\dot{\beta} (i} q^{j) \underline{i}} \overline{\rho}_{\underline{i}^{\dot{\alpha}}}
 ~.
\eea
The existence of the linear multiplets \eqref{composite-linear-hdwm} is crucial in the analysis of deformations of vector multiplets that we will perform in our paper. Note that in terms of superfields, the composite linear multiplet is defined by 
\bea
\cG_{ij}{}^{\underline{i}\underline{j}}
&=&
Q_{(i}{}^{\underline{i}}Q_{j)}{}^{\underline{j}}
=Q_{i}{}^{(\underline{i}}Q_{j}{}^{\underline{j})}
~,
~~~~~~
\de_\a^{(i}\cG^{jk)}{}_{\underline{i}\underline{j}}
=\deb_\ad^{(i}\cG^{jk)}{}_{\underline{i}\underline{j}}
=0
~,
\label{composite-linear-triplet}
\eea
where all component fields in \eqref{composite-linear-hdwm} arise from the equations \eqref{component-linear-111} together with \eqref{analyticity-1} and \eqref{analyticity-2}.

This concludes the definition of the hyper-dilaton Weyl multiplet. The result of the analysis is a representation of the off-shell local 4D, $\cN = 2$ superconformal algebra in terms of the following independent fields: 
${e_m}^a$, $b_m$, $A_{m}$, $W_{ab}$, $q^{i\underline{i}}$, $b_{mn}{}^{\underline{i}\underline{j}}$,
$(\psi_m{}_i,\bar\psi_m{}^i)$, and $(\rho^{\underline{i}},\bar\rho_{\underline{i}})$. The multiplet has precisely the same number of off-shell
degrees of freedom as the standard Weyl multiplet,
$24+24$. 
Table \ref{dof2-hdwm} summarises the counting of degrees of freedom, 
underlining the symmetries acting on the fields. 
\begin{table}[hbt!]
\begin{center}
\begin{tabular}{ |c c c c c c c c |c c c c|} 
 \hline
${e_m}^a$ & ${\omega_m}^{ab}$ & $b_m$ & ${\mathfrak{f}_m}{}_a$ & $\phi_{m}{}^{ij}$ & $A_{m}$ & $\psi_m{}_i$ & $\phi_m{}^i$ & $W_{ab}$ & $\rho^{\underline{i}}$ & $q^{i\underline{i}}$ & $b_{mn}{}^{\underline{i}\underline{j}}$\\ 
$16B$ & $0$ & $4B$ & $0$ & $0$ & $4B$ & $32F$ & $0$ & $6B$ & $8F$ & $4B$ & $18B$\\
\hline
$P_a$ & $M_{ab}$ & $\mathbb D$ & $K_a$ & $J^{ij}$ & $Y$ & $Q$ & $S$ & {} & ${}$ & ${}$ & $\lambda_m{}^{\underline{i}\underline{j}}$-sym\\
$-4B$ & $-6B$ & $-1B$ & $-4B$ & $-3B$ & $-1B$ & $-8F$ & $-8F$ & {} & {} & {} & $-9B$\\
\hline
\multicolumn{12}{|c|}{Result: $24+24$ degrees of freedom}\\
\hline
\end{tabular}
\caption{\footnotesize{Degrees of freedom and symmetries of the hyper-dilaton Weyl multiplet. Row one gives all the fields in the multiplet. Row two gives the number of independent components of these fields -- composite connections are counted with zero degrees of freedom. 
Row three gives the gauge symmetries. Note that the parameter $\lambda_m{}^{\underline{i}\underline{j}}$ 
describes the vector symmetry associated with the gauge two-forms
$b_{mn}{}^{\underline{i}\underline{j}}$ 
with field strength
three-forms $h_{mnp}{}^{\underline{i}\underline{j}}$ 
and $ H^a{}^{\underline{i}\underline{j}}$.
Row four gives the number of gauge degrees of freedom to be subtracted when counting the total degrees of freedom. 
Row five gives the resulting number of degrees of freedom. }\label{dof2-hdwm}}
\end{center}
\end{table}
Note that with the ingredients provided so far, it is 
a straightforward
exercise to obtain
the locally superconformal 
transformations of the fundamental fields
of the hyper-dilaton Weyl multiplet
written only in terms of fundamental fields.
These are given by
\eqref{d-vielbein}--\eqref{d-b-gravitino},
\eqref{d-U1}--\eqref{d-W},
\eqref{d-bmn}, and
\eqref{d-qii-hdwm}--\eqref{d-b-rho-hdwm}
after using the appropriate identities 
for all the composite fields 
$\o_m{}^{cd}$, 
${\mathfrak{f}_m}{}_a$,
$\phi_{m}{}^{ij}$, 
$(\phi_{m}{}_{i},\bar\phi_{m}{}^{i})$, 
$(\Sigma^{\a i},\bar\Sigma_{\ad i})$,
and
$D$
respectively given by eqs.~\eqref{composite-Lorentz},
\eqref{compositeSU2-hdwm},
\eqref{composite-S-connection},
and
\eqref{compositeSSbD-hdwm}.

It is important to underline 
that the local gauge transformations of the 
hyper-dilaton Weyl 
multiplet form an algebra that
closes off-shell on a local extension of SU$(2,2|2)$, the 4D $\cN=2$ superconformal group. 
In fact, by construction the resulting algebra
is identical to the one of the standard Weyl multiplet transformations
\eqref{transf-standard-Weyl} 
(see \cite{deWit:1979dzm} and 
\cite{Butter:2011sr,Butter:2012xg}
for detail on the local algebra),
with the only important subtlety being that
the structure functions will have more composite fields.
We also stress that the existence of the triplet of composite linear multiplets in eqs.~\eqref{composite-linear-hdwm} is a key ingredient to engineer FI-type terms in supergravity-matter couplings based on the hyper-dilaton Weyl multiplet.

\subsubsection{Transformation check}

As a side, here we comment more on the consistency of the composite linear multiplet constructed out of the on-shell hypermultiplet. 

It is necessary to use the on-shell condition on the hypermultiplet to prove that the composite multiplet with its lowest component being $G_{ij}{}^{\underline{i}\underline{j}}$ is a linear multiplet. It is straightforward to show that its descendant fields $\chi_{\alpha i}{}^{\underline{i}\underline{j}}$,
$\bar\chi^{\ad i}{}_{\underline{i}\underline{j}}$, $F^{\underline{i}\underline{j}}$, and
$\bar{F}_{\underline{i}\underline{j}}$ transform correctly as linear multiplet fields using only the hypermultiplet supersymmetry transformation rules. No equations of motion are needed. However, for the composite three-form field, one does need to use an equation of motion.

The local supersymmetry transformation of the composite three-form of the linear multiplet in flat space\footnote{For simplicity we restrict to a flat geometry where the derivatives should be $\de_a\to\pa_a$, but it is straightforward to extend this analysis to a Weyl multiplet background.} is given by
\begin{align}
\d  H^a{}^{\underline{i}\underline{j}}
&= -\frac{1}{4} \d q^{i(\underline{i}} \nabla^a{q_i}^{\underline{j})} -\frac{1}{4}q^{i(\underline{i}} \d \nabla^a{q_i}^{\underline{j})}
+\frac{\ri}{32}
\d \rho^{(\underline{i}}\sigma^a\bar{\rho}^{\underline{j})}
+\frac{\ri}{32}
\rho^{(\underline{i}}\sigma^a \d \bar{\rho}^{\underline{j})}
\nonumber \\
&= \frac{1}{8}  \xi^{\a i} q_{i}{}^{(\underline{i}} \nabla^a\rho_{\a}^{\underline{j})} 
+\frac{1}{4}
(\s^{a b})_{\a \g}{\xi}^{\g i} \rho^{\a (\underline{i}} \nabla_b q_{i}{}^{\underline{j})}
+ \textrm{c.c.} ~ .
\end{align}
We use the on-shell hypermultiplet equation of motion to bring the first term into the desired form. In flat space, it is as follows
\bea
\left(\nabla_{a}
\rho^{\underline{i}}\,
\s^a\right)_{\dot{\alpha}}
=
0 \nonumber
 \qquad 
 \implies \qquad 
\nabla^{a} \rho_{\a}^{ \underline{i}}\,  = 2 (\s^{a b})_{\a \b}   \nabla_{b} \rho^{\b \underline{i}}
\, .
\eea
Inserting this back into the transformation rule we have
\bea
\d  H^a{}^{\underline{i}\underline{j}}
&=&  \frac{1}{4}   (\s^{a b})_{\a \b} \xi^{\a i}  (\nabla_{b} \rho^{\b (\underline{i}}  q_{i}{}^{\underline{j})}   
+ \rho^{\b (\underline{i}} \nabla_b q_{i}{}^{\underline{j})})
+ \textrm{c.c.} \nonumber
\\ 
&=& \frac{1}{2} \xi_{i} \s_{a b} \de^b \chi^{i}{}^{\underline{i} \underline{j}}   
+ \textrm{c.c.} ~,
\eea
as required.
Besides working as a consistency check, the previous calculation indicates that we must interpret the on-shell hypermultiplet as matter fields of the hyper-dilaton Weyl multiplet when using the composite linear multiplet to construct supersymmetric invariants and deformations. This will be used in following sections.

\section{The deformed abelian vector multiplet} \label{sec:deformedVec}

In this section, we first revisit how the electric-magnetic duality is implemented in superspace for 4D, $\cN=2$ vector multiplets and how deformed off-shell abelian vector multiplets arise from this duality in the presence of an (electric) Fayet-Iliopoulos term.

\subsection{EM duality in $\cN=2$ superspace}

The purpose of this subsection is to review and motivate the magnetic deformations of vector multiplets, which we will study in more detail in the next subsection. 
We refer the reader to \cite{IZ1,IZ2} for a more extensive discussion of the duality in flat superspace.

We start from a  gauge invariant $\cN=2$ superfield strength $W$ which is chiral
\be
\overline{D}^i_{\dot{\alpha}}W=0 ~,
\ee
and satisfies the additional constraint
\be
D^{ij} W - \bar{D}^{ij} \overline{W} = 0~, \qquad
D^{ij}:=D^{\a i}D_\a^j
~,~~~
\bar{D}_{ij}=\bar{D}_{\ad i}\bar{D}^\ad_j ~,
\ee
where 
\bea 
D_\a^i=
\frac{\pa}{\pa\q^\a_i}+\ri(\s^b)_{\a\bd}\bar{\q}^{\bd i}\pa_b
~,~~~~~~
\bar{D}^\ad_i
=
\frac{\pa}{\pa\bar{\q}_\ad^i}+\ri(\tilde{\s}^b)^{\ad\b}{\q}_{\b i}\pa_b
~,
\eea
are the flat $\cN=2$ superspace spinor covariant derivatives.
The constraints on $W$ can be solved through the Mezincescu 
prepotential \cite{Mezincescu:1979af}
\be
W = \frac{1}{4}\bar{\D} D^{ij} V_{ij} ~,
\ee
where $V_{ij}$ satisfies $V_{ij}=V_{ji}$, $(V_{ij})^*=V^{ij}$, while
\bea
\bar\D:=\frac{1}{48}\bar{D}^{ij}\bar{D}_{ij}=-\frac{1}{48}\bar{D}^{\ad\bd}\bar{D}_{\ad\bd}
~,~~~
\bar{D}_{\ad\bd}=\bar{D}_{\ad k}\bar{D}_\bd^k
~,
\eea
is the $\cN=2$ chiral projecting operator such that
\bea
\int\rd^4 x\rd^4\q\rd^4\bar{\q}\, \cL
=\int\rd^4 x\rd^4\q\,\bar\D \cL
=\int\rd^4 x\rd^4\bar{\q}\, \D \cL
~.
\label{d8=d4D4}
\eea

The dynamics of a free abelian vector multiplet are described by the superspace Lagrangian
\bea
S^0_v=-{\rm Im}\left[\hf\int\rd^4 x\rd^4\q \tau W^2\right]
~,~~~~~~
\tau:=\frac{i}{g^2}+\vartheta
~.
\label{electricS}
\eea
In the case of a self-interacting theory, the previous model can be lifted to
\bea
S_v
=
-{\rm Im}\Big{[}\int\rd^4 x\rd^4\q F(W)\Big{]}
~,
\label{electricS-2}
\eea
where $F(W)$ is an arbitrary function of $W$ which is the special-K\"ahler geometry holomorphic prepotential.
An $\cN=2$ (electric) FI term is defined by
\bea
S_{\rm FI}=\int\rd^4 x\rd^4\q\rd^4\bar{\q} \,\xi^{ij}V_{ij}
~,
\eea
with $\xi^{ij}$ being a triplet of real constants. The theory described by $S_e=S_v+S_{\rm FI}$ is referred to as electrically deformed \cite{Antoniadis:1995vb,IZ1,IZ2}.

The magnetic dual of $S_e$ is described in terms of the Lagrangian 
\bsubeq\label{flat_Sm}
\bea
S_{m}=
-{\rm Im}\Big{[}\int\rd^4 x\rd^4\q\, \hat{F}(\bW)\Big{]}
~,
\label{Sm}
\eea
where, however, the superfield $\bW$ satisfies a modified reduced chiral constraint as follows
\bea
D^{ij}\bW-\bar{D}^{ij}\overline{\bW}=2\ri\zeta^{ij}
~,
\eea
\esubeq 
with $\zeta^{ij}$ being a real triplet of constants \cite{IZ1,IZ2}. Here, we have used a hat to denote the function of the deformed vector multiplet as it can be proven to be related by a duality transformation to $F(W)$ of the electric $S_e$ action.
The duality between $S_e$ and $S_m$ can be implemented through the action
\bea
S_{\rm duality}=
-{\rm Im}\left[\int\rd^4 x\rd^4\q\, \hat{F}(\U)\right]
-\frac{\ri}{8}\int\rd^4 x\rd^4\q\rd^4\bar{\q} \,U_{ij}\Big[D^{ij}\Upsilon-\bar{D}^{ij}\overline{\Upsilon}-2\ri\zeta^{ij}
\Big]
~,~~~~~~
\label{EMdualityS}
\eea
where $U_{ij}$ is an unconstrained real ${(U_{ij})}^*=U^{ij}$ superfield and $\Upsilon$ is an arbitrary 
(and long) chiral superfield. $\Upsilon$ can be represented by using the chiral projecting operator and an arbitrary complex prepotential superfield as
\bea
\Upsilon=\bar\D\Psi
~.
\label{UDb4P}
\eea
When integrating out $U_{ij}$ and renaming $\U=\bW$ in the previous action,
one obtains the $\cN=2$ vector multiplet 
action deformed by a magnetic FI term in eq.~\eqref{flat_Sm}.
After integration by parts and using \eqref{d8=d4D4} one can rewrite \eqref{EMdualityS} as
\bsubeq
\bea
S_{\rm duality}
&=&
-{\rm Im}\left[\int\rd^4 x\rd^4\q\,\Big(  \hat{F}(\Upsilon)
-\Upsilon W_U
\Big)\right]
-\frac{1}{4}\int\rd^4 x\rd^4\q\rd^4\bar{\q} \,\zeta^{ij}U_{ij}
~,
\label{EMdualityS-2}
\eea
\esubeq
with
\bea
W_U=\frac{1}{4}\bar\D D^{ij}U_{ij}
~.
\eea
The variation of the previous action with respect to $\Upsilon$, after using \eqref{UDb4P} and integrating by parts, 
is
\bea
\delta S_{\rm duality}=
\frac{\ri}{2}\int\rd^4 x\rd^4\q\rd^4\bar{\q}\,\delta\Psi\Big(  \frac{\pa \hat{F}(\Upsilon)}{\pa \Upsilon}-W_U\Big)
~,
\eea
implying that on shell it holds
\bea
\frac{\pa \hat{F}(\Upsilon)}{\pa\U}=W_U~,~~~W_U=\frac{1}{4}\bar\D D^{ij}U_{ij} ~,
\eea
which turns \eqref{EMdualityS-2} into
\bea
S_{e}=
-{\rm Im}\left[\int\rd^4 x\rd^4\q\,F(W_U)\right]
+\int\rd^4 x\rd^4\q\rd^4\bar{\q} \,\widetilde{\xi}^{ij}U_{ij}
~,~~~~~~
\widetilde{\xi}^{ij}=-\frac{1}{4}\zeta^{ij} ~.
\eea
This is equivalent to \eqref{electricS-2}
plus a standard FI term
if we define
\bea
F(W_U):=
\hat{F}[\Upsilon(W_U)]
-\Upsilon(W_U) W_U
~,~~~~~~
\frac{\pa F(W_U)}{\pa W_U}
=-\Psi
~.
\eea
This is a usual Legendre transform
of the special-K\"ahler holomorphic prepotential $F$ and its dual $\hat{F}$, and $\Upsilon(W_U)$ is an implicit solution (which we assume to exist) of
$\frac{\pa \hat{F}(\Upsilon)}{\pa\U}=W_U$
satisfying 
\bea
\frac{\pa \U(W_U)}{\pa W_U}=\Big[\frac{\pa W_U}{\pa \U}\Big]^{-1}
=[\tau(\U)]^{-1}
\equiv
-\widetilde{\tau}(W_U)
~,
\eea
where
\bea
\tau(\U)=\frac{\pa^2 \hat{F}(\U)}{(\pa \U)^2}
~,~~~~~~
\widetilde{\tau}(W_U)=\frac{\pa^2 \hat{F}(W_U)}{(\pa W_U)^2}
~.
\eea
These are standard results for the EM duality of a vector multiplet. They show that electric and magnetic FI terms are interchanged with the duality. The same arguments are well known to generalise to several (abelian) vector multiplets.
An important comment is that for flat supersymmetry one can consider a vector multiplet that has both electric and magnetic deformations while still having preserved off-shell supersymmetry \cite{Antoniadis:1995vb,IZ1,IZ2} 
\bsubeq
\bea
&&S=
-{\rm Im}\left[\int\rd^4 x\rd^4\q\,F(\bW)\right]
+\int\rd^4 x\rd^4\q\rd^4\bar{\q} \,\xi^{ij}V_{ij}
~,
\\
&&
D^{ij}\bW-\bar{D}^{ij}\overline{\bW}=2\ri\zeta^{ij}
~,~~~
\bW=\frac{1}{4}\bar{\D} D^{ij} V_{ij}+\frac{\ri}{2}\theta_{ij}\zeta^{ij}
~,
\eea
\esubeq
where $\theta_{ij}:=\theta^\a_i\theta_{\a j}$. The presence of both an electric and a magnetic deformation is the key to obtaining partial supersymmetry breaking in flat superspace with a single physical vector multiplet \cite{Antoniadis:1995vb}. We stress that this is a feature of the globally supersymmetric case.

Note that the previous derivation can straightforwardly be lifted to conformal supergravity defined in conformal superspace. The key ingredient is to realise that the electric and magnetic deformations will turn into linear multiplets. This is potentially straightforward, though there will be various subtleties related to the choices of a conformal supergravity background and compensators which we will discuss in the coming sections. For instance, given a set of vector multiplets in a hyper-dilaton Weyl background, an electric FI-type deformation will be associated to the following full conformal superspace invariant
\bea
\int\rd^4 x\rd^4\q\rd^4\bar{\q} \,E\,\cG_\xi{}_I^{ij}V^I_{ij}
~,~~~~~~
\cG_\xi{}_I^{ij}:=\xi_I^{\underline{i}\underline{j}}
Q^i{}_{\underline{i}}Q^j{}_{\underline{j}}
~,
\eea
where $Q^{i\underline{i}}$ is the on-shell hypermultiplet in conformal superspace, while $\xi_I^{\underline{i}\underline{j}}=\xi_I^{\underline{j}\underline{i}}$ is a triplet of real constant.

Let us now proceed by introducing the modification of the magnetic deformation of an abelian vector multiplet in a conformal supergravity background.

 \subsection{Deformed abelian vector multiplet in conformal superspace}

Consider the following deformation of abelian vector multiplets \cite{Kuzenko:2013gva,Kuzenko:2015rfx}
\bea
\de^{ij}\bW-\deb^{ij}\overline{\bW}=2\ri\cG^{ij} 
~,\label{4.1}
\eea
where $\bW$ is only required to be covariantly chiral. The constraint now holds in conformal supergravity where the $\de_A$ derivatives are the conformal superspace ones that we introduced before.
This multiplet can be thought of as a deformation of a standard vector multiplet described by $W$ by means of the shift
\bea
\bW= W+\Psi ~,
\eea
where $\Psi$ is the prepotential of $\cG^{ij}$ and is a chiral superfield, $\bar{\nabla}^\ad_i \Psi = 0$,
of dimension 1 and $\rm U(1)_R$ weight -2, but otherwise arbitrary --- see eqs. \eqref{potential-linear-i}--\eqref{eq_linear_mult_chiral}. 

It is straightforward to deduce that
\begin{align}
    \de^{ij}\bW-\deb^{ij}\bar{\bW}=\de^{ij}\Psi-\deb^{ij}\bar{\Psi} = 2\ri\cG^{ij}~,
\end{align}
where $\cG^{ij}$ is the linear multiplet superfield. Acting with spinor covariant derivatives on $\bW$ gives the following independent descendants:
\bsubeq
\begin{align}
    & \bl{}_\a^i =\nabla_\a^i \bW \ , \quad  \bar{\bl}^{\ad}_{i} = \bar{\nabla}^\ad_i \bar{\bW} \ , \\
    &  \bX^{ij} = \hf (\nabla^{ij} \bW + \bar{\nabla}^{ij} \bar{\bW}) \ , \quad \cG^{ij} = - \frac{\ri}{2} (\de^{ij}\bW-\deb^{ij}\overline{\bW}) \ ,  \\
    & \bF_{ab} = - \frac{1}{8} (\s_{ab})^{\a\b} (\nabla_{\a\b} \bW +4W_{\a\b} \bar{\bW}) 
+ \frac{1}{8} (\tilde{\s}_{ab})_{\ad\bd} (\bar{\nabla}^{\ad\bd} \bar{\bW} +4 \overline{W}^{\ad\bd} \bW)  \ , \\
    & \bF_{\a \b} = \hf (\s^{ab})_{\a\b} \bF_{ab} = - \frac{1}{8} ( \nabla_{\a \b} \bW + 4 W_{\a\b} \bar{\bW}) , \\
    & \bar{\bF}^{\ad \bd} = - \hf (\tilde{\s}^{ab})^{\ad\bd} \bF_{ab} = - \frac{1}{8}  (\bar{\nabla}^{\ad \bd} \bar{\bW} + 4 \overline{W}^{\ad\bd} \bW)~, \\
    & \c_{\a i} = \frac{1}{3} \nabla_\a^j \cG_{i j} ~, \quad\quad 
	\bar{\c}^{\ad i} = \frac{1}{3} \bar{\nabla}^\ad_j \cG^{ij}~, \\
& F =  \frac{1}{12} \nabla^{ij} \cG_{ij}~, \quad\quad
\bar{F} =  \frac{1}{12} \bar{\nabla}^{ij} \cG_{ij} \ ,
\\
& {H}_{abc} = \frac{\ri}{96} \ve_{abcd} (\s^d)^\a{}_\bd [\nabla_\a^i , \bar{\nabla}^\bd_j] \cG^j{}_i = \ve_{abcd} {H}^d \ ,
\\
 & {H}_{a} 
=\frac{\ri}{96} (\s_a)^\a{}_\bd [\nabla_\a^i , \bar{\nabla}^\bd_j] \cG^j{}_i =  \frac{1}{6}\ve_{abcd}{H}^{bcd}~.
\end{align}
\esubeq
These superfields satisfy the following tower relations that are particularly useful in analyzing the structure of invariants:
\bsubeq
\begin{align}
       &
       \nabla_\a^i \bl{}_\b^j =  \hf \e_{\a \b} \bX^{i j} + \frac{\ri}{2} \e_{\a \b} \cG^{i j}+ 4 \e^{i j} \bF_{\a \b} + 2 \e^{i j} W_{\a \b} \bar{\bW} \ , \\
       &
       \bar{\nabla}^\ad_i \bl{}_\b^j = -2 \ri \d^j_i \nabla_{\b}{}^{\ad} \bW \ , \\
       &
       {\nabla}_\a^i \bar{\bl}{}^\bd_j = -2 \ri \d_j^i \nabla_{\a}{}^{\bd} \bar{\bW} \ , \\
       &
       \bar{\nabla}^\ad_i \bar{\bl}{}^\bd_j = \hf \e^{\ad \bd} \bX^{i j} - \frac{\ri}{2} \e^{\ad \bd} \cG^{i j} + 4 \e_{i j} \bar{\bF}^{\ad \bd} + 2 \e_{i j} \overline{W}^{\ad \bd} \bW \ , \\
       & {\nabla}_\a^i \bX^{j k} = - 2 \ri \e^{i (j} \chi_{\a}^{k)}- 4 \ri \e^{i (j} \nabla_{\a}{}^{\ad} \bar{\bl}_{\ad}^{k)} \, ,  \\
       &  \bar{\nabla}^\ad_i \bX_{j k} =  2 \ri \e_{i (j} \bar{\chi}^{\ad}_{k)} - 4 \ri \e_{i (j} \nabla_{\a}{}^{\ad}\bl_{k)}^{ \a} \ , \\
       & {\nabla}_\g^i \bF_{\a \b} = \e_{\g (\a} \Sigma_{\b)}^i \bar{\bW} - \hf W_{\a \b}{}_{\g}^{\, i} \bar{\bW} + \frac{1}{2} \ri \e_{\g (\a} \chi_{\b)}^i  + \frac{1}{2} \ri \e_{\g (\a} \nabla_{\b)}{}^{\ad} \bar{\bl}^i_\ad \ , \\
       & \bar{{\nabla}}^\gd_k \bF_{\a \b} = \frac{\ri}{2} \nabla_{(\a}{}^{\gd} \bl_{\b) k} - \hf W_{\a \b} \bar{\bl}^{\gd}_k \ , \\
        & {\nabla}_\g^i \bar{\bF}^{\ad \bd} = \frac{\ri}{2} \nabla_{\g}{}^{(\ad} \bar{\bl}^{\bd) k} - \hf \overline{W}^{\ad \bd} {\bl}_{\g}^k \ , \\
       & \bar{{\nabla}}^\gd_k \bar{\bF}^{\ad \bd} =  - \e^{\gd (\ad} \bar{\Sigma}^{\bd)}_k \bW -\hf \overline{W}^{\ad \bd}{}^{\gd}_k \bW - \frac{1}{2} \ri \e^{\gd (\ad} \bar{\chi}^{\bd)}_k + \frac{1}{2} \ri \e^{\gd (\ad} \nabla_{\a}{}^{\bd)} \bl^\a_k~.
\end{align}
\esubeq
The tower of $S$-supersymmetry transformations is identical to the cases of a standard vector multiplet and a linear multiplet, up to appropriately renaming some descendant with bold symbols. As a result, the local superconformal transformation of the fundamental component fields 
of the deformed vector multiplet fields in a standard Weyl multiplet background are
\bsubeq\label{transf-deformedvector}
\bea
\d {\boldsymbol{\phi}}
&=&
\xi_i \bl^i + \l_\mathbb{D} {\boldsymbol{\phi}} - 2 \ri \l_Y {\boldsymbol{\phi}}
~,
\\ 
\d \overline{\boldsymbol{\phi}}
&=&
\bar{\xi}^i \bar{\bl}_i + \l_{\mathbb{D}} \overline{\boldsymbol{\phi}} + 2 \ri \l_Y \overline{\boldsymbol{\phi}}
~,
\\
\d\bl_\alpha^i
&=&
2 (\s^{a b} \xi^i)_\a \bF_{a b} + (\s^{a b}\xi^i)_\a W^{+}_{a b} \overline{\boldsymbol{\phi}}
- \frac{1}{2} \xi_\a{}_j \bX^{ij} - \frac{\ri}{2} \xi_\a{}_j G^{ij} + 2 \ri (\s^a \bar{\xi}^i)_\a \de_a \boldsymbol{\phi}\nonumber \\
&&+ \frac{1}{2}\l^{a b}(\s_{a b} \bl^i)_\a  + \l^i{}_j \bl^j_\a + \frac{3}{2} \l_{\mathbb{D}} \bl^i_\a - \ri \l_Y \bl^i_\a + 4 \eta^i_\a {\boldsymbol{\phi}}
~,
\label{transf-boldlambda}
\\
\d\bar{\bl}^\ad_i
&=&
-2 (\tilde{\s}^{a b} \bar{\xi}_i)^\ad \bF_{a b} - (\tilde{\s}^{a b}\bar{\xi}_i)^\ad W^{-}_{a b} {\boldsymbol{\phi}}
- \frac{1}{2} \bar{\xi}^\ad{}^j \bX_{ij} + \frac{\ri}{2} \bar{\xi}^\ad{}^j G_{ij} + 2 \ri (\tilde{\s}^a \xi_i)^\ad \de_a \overline{\boldsymbol{\phi}} \nonumber \\
&&+ \frac{1}{2}\l^{a b}(\tilde{\s}_{a b} \bar{\bl}_i)^\ad  - \l_i{}^j \bar{\bl}_j^\ad + \frac{3}{2} \l_{\mathbb{D}} \bar{\bl}_i^\ad+ \ri \l_Y \bar{\bl}^\ad_i + 4 \bar{\eta}^\ad_i \overline{\boldsymbol{\phi}}
~,
\label{transf-boldlambda-bar}
\\
\d\boldsymbol{X}^{ij}
&=&
 2 \ri \xi^{\a (i} \chi_{\a}^{j)}+ 4 \ri \xi^{\a (i} \nabla_{\a}{}^{\ad} \bar{\bl}_{\ad}^{j)}
- 2 \ri \bar{\xi}_\ad^{(i} \bar{\chi}^{\ad j)} + 4 \ri  \bar{\xi}_\ad^{(i}\nabla_{\a}{}^{\ad}\bl^{j)  \a} \nonumber \\
&& + 2 \lambda^{(i}{}_{k} \boldsymbol{X}^{j) k}
+ 2 \lambda_{\mathbb{D}} \boldsymbol{X}^{ij}
~,
\label{transf-boldXij-bar}
\\
 \d\boldsymbol{F}^{a b}  &=& 
\Bigg[
-\ri \xi_k \s_{[a}\de_{b]}\bar{\bl}^k 
+ \Big(\xi_k \s_{ab} \S^k - \frac{1}{2} \xi^{\a}_k (\s_{ab})^{\b \g} W_{\a \b\g}{}^{k}\Big) \bar{\boldsymbol{\phi}} 
-\hf (\xi_k\bl^k) W_{ab}^- \non \\
&&
~~~~~ + \frac{\ri}{2} \xi_k \s_{ab} \chi^{k}+ 2 \eta^k \s_{ab} \bl_k 
+ {\rm c.c.} \Bigg] 
+2 \lambda_{\mathbb{D}} \boldsymbol{F}_{ab} 
- 2 \lambda_{[a}{}^{c} \boldsymbol{F}_{b]c}
~,
\label{transf-boldF-bar}
\eea
\esubeq
while all the transformations of the descendants of $\bW$ and $\bar\bW$ associated to $\cG^{ij}$ are exactly the same transformations as the linear multiplet component fields $G^{ij}$, $\chi_{\a i}$, $\bar{\chi}^{\ad i}$, $F$, and $\bar{F}$ given in \eqref{linear-multiplet}.

Note that, in a hyper-dilaton Weyl background, thanks to the existence of the composite triplet of linear multiplets, see, e.g., eq.~\eqref{composite-linear-triplet}, it is natural to consider the deformations of a set of vector multiplets associated to the following deformed constraints
\bea
\de^{ij}\bW^I-\deb^{ij}\overline{\bW}^I=2\ri\cG^I_\zeta{}^{ij} ~,
~~~~~~
\cG^I_\zeta{}^{ij} 
:=
\zeta^I_{\underline{i}\underline{j}}
Q^{i\underline{i}}Q^{j\underline{j}}
~,
\eea
where $Q^{i\underline{i}}$ is the on-shell hypermultiplet in conformal superspace, while $\zeta^I_{\underline{i}\underline{j}}=\zeta^I_{\underline{j}\underline{i}}$ is a triplet of real constants which plays a similar role to the global magnetic FI terms. This will be one of the ingredients that we use in the coming sections.

We proceed next with the definition of several locally superconformal action principles both in superspace and components.

\section{Superconformal actions} \label{sec:actions}

In this section, we review the local superconformal action principles that we use to engineer the supergravity-matter systems studied in the rest of the paper. This includes the abelian vector multiplet action and the linear multiplet action with magnetic and electric deformations, respectively.

\subsection{Chiral action principle}

We introduce here the chiral action involving an integral
over the chiral subspace
\be S = S_c + {\rm c.c.} \ , \quad S_c = \int \rd^8 z \,\cE \, \cL_c \ ,
\quad \rd^{8}z := \rd^4 x \, \rd^4 \q \ ,
\ee
where $\cL_c$ is covariantly chiral, $\bar\nabla^\ad_i \cL_c = 0$, and $\cE$ is a
suitably chosen chiral measure \cite{Muller:ChiralActions,Muller,Kuzenko:2008ry,Butter:2011sr}.
The Lagrangian $\cL_c$ must be a conformally primary Lorentz and
$\rm SU(2)_R$ chiral scalar with conformal dimension two and $\rm U(1)_R$ weight $-4$:
\begin{align}
\mathbb{D} \cL_c = 2 \cL_c \ , \quad Y \cL_c = -4 \cL_c \ , \quad J^{ij} \cL_c = M_{ab} \cL_c = K_a \cL_c = S_\a^i \cL_c = \bar{S}^\ad_i \cL_c = 0 \ .
\end{align}
Any action involving an integral over the full superspace may be converted
to one over the chiral subspace by the rule \cite{Butter:2011sr}
\be \int \rd^{12}z \,E \,\cL = \int \rd^8 z \,\cE \,\bar{\nabla}^4 \cL \ , \quad \rd^{12}z := \rd^4 x \, \rd^4 \q \, \rd^4 \bar{\q} , \quad
\bar{\nabla}^4 := \frac{1}{48} \bar{\nabla}^{ij} \bar{\nabla}_{ij} \ .
\ee

The chiral action in components \cite{Butter:2011sr}, and in our notation that follow the ones of \cite{Butter:2012xg}, takes the form of the following density formula
\begin{align} 
S_c =& \int \rd^4x \,e \Bigg( \frac{1}{48} \nabla^{ij} \nabla_{ij} -  \frac{\ri}{12} \psib_d{}^l_\dd (\tilde{\s}^d)^{\dd \a} \nabla_\a^q \nabla_{lq} 
+ \frac{\ri}{2} \psib_d{}_\dd^l (\s^d)_{\a \ad} \overline{W}^{\ad \dd} \nabla^\a_l + \overline{W}^{\ad\bd} \overline{W}_{\ad\bd} \non\\
&+\frac{1}{4} \psib_c{}_\gd^k \psib_d{}^l_\dd \Big( (\tilde{\s}^{cd})^{\gd \dd} \nabla_{kl} - \frac{1}{2} \ve^{\gd \dd} \ve_{kl} (\s^{cd})_{\b \g} \nabla^{\b \g} 
- 4 \ve^{\gd\dd} \ve_{kl} (\tilde{\s}^{cd})_{\ad\bd} \overline{W}^{\ad\bd} \Big) \non\\
&- \frac{1}{4} \ve^{abcd} (\tilde{\s}_a)^{\bd \a} \psib_b{}_\bd^j \psib_c{}_\gd^k \psib_d{}^\gd_k \nabla_{\a j} 
- \frac{\ri}{4} \ve^{abcd} \psib_a{}_{\ad}^i \psib_b{}^{\ad}_i \psib_c{}_{\bd}^j \psib_d{}^{\bd}_j \Bigg) \cL_c |~. 
\label{CA}
\end{align}
Efficient ways to obtain this result make use of either a normal coordinate expansion in superspace, see \cite{Kuzenko:2008ry}, or alternatively by using the superform approach to constructing supersymmetric invariants, see \cite{Gates:2009xt,Butter:2012ze}.
We stress that the component action \eqref{CA} is the primary building block for the superconformal invariant actions considered throughout this work wherein a consistent choice for $\cL_c$ in conformal superspace satisfying the above properties determines its structure.

\subsection{Deformed abelian vector multiplet action}

Let us now consider the general, superconformal chiral action $\cL_c = \cF(\bW^I)$
of $n$ deformed abelian vector multiplets
$\bW^I$. Note that $\cF(\bW^I)$ must be homogeneous of degree two in $\bW^I$,
\be
    \bW^I \cF_I \equiv \bW^I \frac{\partial}{\partial \bW^I} \cF = 2 \cF ~,
\ee
with
\be
    \de^{ij}\bW^I-\deb^{ij}\overline{\bW}^I=2\ri\cG^{I,ij} ~.
\ee
Recall that the above follows from the fact that the deformed abelian vector multiplet can be defined by the shift
\be
    \bW^I=W^I+\Psi^I ~,
\ee
where $\Psi^I$ are the prepotential for the linear multiplets $\cG^{I,ij}$
\bea
    \cG^{I,ij}=-\frac{\ri}{2}\big(\de^{ij}\Psi^I-\deb^{ij}\overline{\Psi}^I\big) ~.
\eea
Here we do not specify whether the linear multiplets are composite (as for the hyper-dilaton Weyl case that we will study later on) or fundamental.
The model is manifestly invariant under the gauge transformation
\be
    \hat{\d}\bW^I=0 ~,
\ee
which is apparent from 
\bea
    \hat{\d}\Psi^I=\hat{W}^I
    ~,~~~~~~
    \hat{\d} W^I=-\hat{W}^I ~,
\eea
for a vector multiplet field strength $\hat{W}^I$ satisfying 
\bea
\deb^\ad_i \hat{W}^I=0
~,~~~
\de^{ij}\hat{W}^I=\deb^{ij}\overline{\hat{W}}^I
~.
\eea

By the component action principle of eq.~\eqref{CA}, this action in components was generated computationally by the computer algebra software, \textit{Cadabra} \cite{Peeters1, Peeters2, Peeters3}. Note that a specific code repository \cite{Gold:2024nbw} has been developed in parallel to this work that automatically generates 4D, $\cN=2$ actions in components. It has been directly applied here and to all further results in this paper. After further cleaning up by hand, we obtain the following result in components
\begin{align}
    S_c = \int d^4x  \,e	  \Bigg[ &
    \cF_I \Box \bar{\bm\phi}^I   
    - 2 \cF_I \overline{W}_{\dot{\alpha}\dot{\gamma}} {\bar{\bF}}^{I\ad\gd} - \cF \overline{W}_{\dot{\alpha}\dot{\beta}} \overline{W}^{\ad\bd} + 3\cF_I D \bar{\phi}^{I}+\frac{3}{2} \cF_I \bar{\Sigma}^{k}_{\ad} \bar{\bl}^{\dot{\alpha}}_{k} 
    \nonumber\\ 
    & -  \frac{\ri}{2} \mathcal{F}_{I J} \bl^{I \alpha j} \nabla_{\alpha \dot{\alpha}} \bar{\boldsymbol{\lambda}}_{j}^{J \dot{\alpha}}  + \frac{1}{32} \mathcal{F}_{I J} \bM^{I i j} \bM_{i j}^{J}-2 \mathcal{F}_{I J} {\bF}^{I \alpha \beta} {\bF}_{\alpha \beta}^{J} 
    \nonumber\\ 
    & -2 \mathcal{F}_{I J} \bar{\boldsymbol{\phi}}^{I} W^{\alpha \beta} {\bF}_{\alpha \beta}^{J} -\frac{1}{2} \mathcal{F}_{I J} \bar{\boldsymbol{\phi}}^{I} \bar{\boldsymbol{\phi}}^{J} W^{\alpha \beta} W_{\alpha \beta} 
    + \frac{1}{16} \mathcal{F}_{I J K}\left(\bl^{I i} \bl^{J j}\right) \bM_{i j}^{K}
    \nonumber \\
    & + \frac{1}{2} \mathcal{F}_{I J K} \bl^{I \alpha k} \bl_{k}^{J \beta} {\bF}_{\alpha \beta}^{K} +\frac{1}{4} \mathcal{F}_{I J K} \bar{\boldsymbol{\phi}}^{I} \bl^{J \alpha k} \bl_{k}^{K \beta} W_{\alpha \beta}  
    \nonumber \\
    &+ \frac{1}{48}\cF_{I J K L} \bl^{\alpha i, L} \bl^{j, K}_\alpha \bl^{\beta, J}_i \bl^K_{\beta j}
    -\frac{1}{2} \mathcal{F}_{I}\left(\bar{\psi}_{m}^{j} \tilde{\sigma}^{m} \sigma^{b}\right)_{\dot{\alpha}} \nabla_{b} \bar{\boldsymbol{\lambda}}_{j}^{I \dot{\alpha}}
    \nonumber \\
    & -\frac{\ri}{8} \mathcal{F}_{I J}\left(\bar{\psi}_{m}^{i} \tilde{\sigma}^{m} \bl^{I j}\right) \bM_{i j}^{J} - \ri \mathcal{F}_{I J}\left(\bar{\psi}_{m}^{k} \tilde{\sigma}^{m}\right)^{\alpha} \bl_{k}^{I \beta} {\bF}_{\alpha \beta}^{J}  
    \nonumber \\
    & - \frac{\ri}{2} \mathcal{F}_{I J} \bar{\boldsymbol{\phi}}^{I}\left(\bar{\psi}_{m}^{k} \tilde{\sigma}^{m}\right)^{\alpha} \bl_{k}^{J \beta} W_{\alpha \beta} - \frac{\ri}{12} \mathcal{F}_{I J K}\left(\bar{\psi}_{m}^{i} \tilde{\sigma}^{m} \bl^{I j}\right)\left(\bl_{i}^{J} \bl_{j}^{K}\right) 
    \nonumber \\ 
    &+ \frac{\ri}{2} \mathcal{F}_{I} \bar{\psi}_{b \dot{\gamma}}^{j} \overline{W}_{\dot{\beta}}{}^{\dot{\gamma}} (\tilde{\sigma}^{b})^{\dot{\beta} \alpha} \bl_{\alpha j}^{I} - \frac{1}{4} \mathcal{F}_{I}\left(\bar{\psi}_{m}^{i} \tilde{\sigma}^{m n} \bar{\psi}_{n}^{j}\right) \bM_{i j}^{I} \nonumber \\  
    &- \frac{1}{4} \mathcal{F}_{I J}\left(\bar{\psi}_{m}^{i} \tilde{\sigma}^{m n} \bar{\psi}_{n}^{j}\right)\left(\bl_{i}^{I} \bl_{j}^{J}\right) + \mathcal{F}_{I}\left(\bar{\psi}_{m} \bar{\psi}_{n}\right)\left(\sigma^{m n}\right)^{\alpha \beta} {\bF}_{\alpha \beta}^{I} 
    \nonumber\\ 
    &+ \frac{1}{2} \mathcal{F}_{I} \bar{\boldsymbol{\phi}}^{I}\left(\bar{\psi}_{m} \bar{\psi}_{n}\right)\left(\sigma^{m n}\right)^{\alpha \beta} W_{\alpha \beta} + \frac{1}{8} \mathcal{F}_{I J}\left(\bar{\psi}_{m} \bar{\psi}_{n}\right)\left(\bl^{I k} \sigma^{m n} \bl_{k}^{J}\right) 
    \nonumber \\  
    &- \mathcal{F}\left(\bar{\psi}_{m} \bar{\psi}_{n}\right)\left(\tilde{\sigma}^{m n}\right)^{\dot{\alpha} \dot{\beta}} \overline{W}_{\dot{\alpha} \dot{\beta}} + \frac{1}{4} \mathcal{F}_{I} \varepsilon^{m n p q}\left(\bar{\psi}_{m} \bar{\psi}_{n}\right)\left(\bar{\psi}_{p}^{i} \tilde{\sigma}_{q} \bl_{i}^{I}\right) 
    \nonumber \\
    & - \frac{\ri}{4} \mathcal{F} \varepsilon^{m n p q}\left(\bar{\psi}_{m} \bar{\psi}_{n}\right)\left(\bar{\psi}_{p} \bar{\psi}_{q}\right) 
    \nonumber \\
    &+ \frac{\ri}{2} \cF_I F^I 
    + \frac{\ri}{2} \cF_{I J} \left(\bl^{ j,I} \chi_{ j}^J\right)
    + \frac{1}{2} \mathcal{F}_{I}\left(\bar{\psi}_{m}^{j} \tilde{\sigma}^{m} \chi^I_{j}\right) \Bigg] ~, 
    \label{deformed-vectoraction}
\end{align}
where we have introduced the following complex triplet of scalar fields
\begin{align}
    \textbf{M}^I_{ij}= {\bf X}^I_{ij} +\ri G^I_{ij}
    ~,~~~~~~
     \overline{\textbf{M}}^I_{ij}= {\bf X}^I_{ij} -\ri G^I_{ij}
     ~.
     \label{shifted-auxiliary-1}
\end{align}
Note that the effect of the deformation is simply a shift by an imaginary unit times a linear multiplet and that the final line in \eqref{deformed-vectoraction} is comprised of terms that originated from the deformation. Otherwise, this action is equivalent to the chiral component action of the abelian vector multiplets without deformations, $\cL_c=\cF(W^I)$, as found in previous literature, see, e.g., \cite{Butter:2012xg}, up to making various fields appropriately bold through the shift by a linear multiplet.
It is useful to use the following properties in components in the action \eqref{deformed-vectoraction}
\bsubeq
\begin{align} 
    \nabla_{b} \bar{\boldsymbol{\lambda}}_{j}^{I \dot{\alpha}} & 
    = 
    \cD_{b}  \bar{{\bl}}_{j}^{I \dot{\alpha}} 
    - 2 \bar{\phi}_{b}{}^{\ad}_j\bar{\boldsymbol{\phi}}^I   
    + \frac{1}{4} \bar{\psi}_a{}^{\ad k} \overline{\bM}_{jk}^I
    +\bar{\psi}_{a}{}_{\gd j}(2 {\bF}^{I \gd\ad}+W^{\gd \ad} {\boldsymbol{\phi}}^I)
    \non \\
    &~~~    - \ri \psi_{a\g j} \Big(
    \cD^{\ad\g} \bar{\boldsymbol{\phi}}^I  + \frac{1}{2} (\sigma^c)^{\ad\g} (\bar{\psi}_{ck} {{\bar{\bl}}^{Ik}}) \Big)
    ~,
     \\
    \square \bar{\bm\phi}^{I} & =   
    \cD^{a} \cD_{a} \bar{\boldsymbol{\phi}}^I 
    + \frac{1}{2} \cD_b(\bar{\psi}^{b}_k {\bar{\bl}}^{Ik})+\frac{\ri}{2}\left(\phi_{m}^{j} \sigma^{m} \bar{{\bl}}_{j}^{I}\right) +\frac{\ri}{4}\left(\psi_{m j} \sigma^{m}\right)_{\dot{\alpha}} \overline{W}^{\dot{\alpha} \dot{\beta}} \bar{\bl}_{\dot{\beta}}^{I j} 
    \non \\ 
    &~~~ -\frac{3 \ri}{4}\left(\psi_{m j} \sigma^{m} \bar{\Sigma}^{j}\right) \bar{\boldsymbol{\phi}}^{I}-\frac{1}{2} \bar{\psi}_{a}^{j} \nabla^{a} \bar{\boldsymbol{\lambda}}_{j}^{I} -2 \mathfrak{f}_{a}{}^{a} \bar{\bm\phi}^{I} 
    ~.
\end{align}
\esubeq
As a final note, we underline that after covariant vector derivatives are degauged as seen above, due to the component gauge fixing conditions \eqref{eq_psi_constraints}, the bold, ``deformed" fields of this vector multiplet may be thought of as equal to their non-bold, ``non-deformed" counterparts in components with one exception being
\begin{align}
    \bF_{ab}^{I} \vert & = F_{ab}^{I} \vert + B_{ab}^{I} \vert
    ~,
\end{align}
though one should keep in mind that the triplet of scalar auxiliary fields receives an imaginary shift.

Before we proceed, it is useful to make a comment on the effect that the deformation has on the theory's scalar potential. 
A relevant term is given by $\frac{1}{32} \mathcal{F}_{I J} \bM^{I i j} \bM_{i j}^{J}+{\rm c.c.}$ from \eqref{deformed-vectoraction}. The quadratic term in $X_{ij}^I$ is the one that, if the auxiliary fields acquire a vev (typically through an electric gauging by a standard FI term), could ubiquitously lead to a contribution to the vacuum energy. Now, the effect of the ``magnetic'' deformation is similar, but leads to different signs due to the imaginary unit. the difference is that the contribution is already in the Lagrangian without having to integrate out any auxiliary field. Depending on the structure of the model and its holomorphic prepotential, potentially the deformation can lead to Minkowski, anti de Sitter (AdS) or even de Sitter (dS) vacua.

\subsection{Standard $BF$ action/electric FI term}

Let us now consider the supersymmetric $BF$ action \cite{Butter:2012xg,Butter:2010jm}
\be
S_{\text{standard FI}} = - 2 \ri \int \rd^8z\,\cE \,\Psi_I  W^I + \text{c.c.}
 = \int \rd^{12} z \,E \,G^{ij}_I  V_{ij}^I
 ~,
\label{eq:S_standard_FI-superspace}
 \ee
 where
 \be
G^{ij}_I
:=
-\frac{\ri}{2}\left(\de^{ij}\Psi_I-\deb^{ij}\bar{\Psi}_I\right)
~,~~~~~~
W^I:=\frac{1}{4}\bar{\Delta}\de^{ij}V_{ij}^I
 ~.
 \label{eq:S_standard_FI-superspace-2}
\ee
This is defined as a locally superconformal completion of a $BF$ term and emerges as an appropriate product of a linear and a (undeformed) vector multiplet. Here we do not specify whether the linear multiplets, nor vector multiplets are composite. In all cases the previous action proves to be locally superconformal invariant.

The component action for $S_{\text{standard FI}}$ can be obtained by using eq.~(\ref{CA}) in the first definition in \eqref{eq:S_standard_FI-superspace}. The result is
\bsubeq \label{eq:S_standard_FI}
\begin{align}
    S_{\text{standard FI}} = &\int d^4 x e\Bigg[ F_I \phi^I + \chi^\a_i{}_{I} \lambda_\a^{i I} + \frac{1}{8} G^{i j}_I X_{i j}^I 
    - \e^{m n p q} b_{m n}{}_{I} f^I_{p q} \non \\
    - & \frac{\ri}{2}\bar{\psi}_d{}_{\dd}^l(\tilde{\sigma}^d)^{\dd \a} \left[2\chi_{\a l}{}_{I}\phi^I 
    + G_{l q}{}_{I}\lambda^{q I}_\a \right] 
    + \psib_c{}^k_{\gd} \psib_d{}^l_{\dd}(\tilde{\sigma}^{c d})^{\gd \dd}G_{k l}{}_{I}\phi^I\Bigg] +\text{c.c.} ~,
\end{align}
or equivalently
\begin{align}
    S_{\text{standard FI}} = &\int d^4 x e\Bigg[ F_I \phi^I + \chi^\a_i{}_{I} \lambda_\a^{i I} + \frac{1}{8} G^{i j}_I X_{i j}^I 
    +\frac{2}{3} \e^{m n p q} h_{m np}{}_{I} v^I_{q} \non \\
    - & \frac{\ri}{2}\bar{\psi}^l_{d \dd}(\tilde{\sigma}^d)^{\dd \a} \left[2\chi_{\a l}{}_{I}\phi^I 
    + G_{l q}{}_{I}\lambda^{qI}_\a \right] 
    + \psib^k_{c \gd} \psib^l_{d \dd}(\tilde{\sigma}^{c d})^{\gd \dd}G_{k l}{}_{I}\phi^I\Bigg] +\text{c.c.} ~.
\end{align}
\esubeq
Notably, and consistently, the previous action is invariant under the defining shift symmetry of the linear multiplet prepotentials by vector multiplets,
\bea
{\tilde\d}\Psi_I ={\tilde{W}}_I
~.
\eea
This is manifestly an invariance, assuming that ${\tilde\d}W^I=0$ and
\begin{align}
    \deb^\ad_i {\tilde{W}}_I=0~,~~~~~~
    \de^{ij}{\tilde{W}}_I=\deb^{ij}\overline{{\tilde{W}}}_I
    ~.
\end{align}
Here we have used the symbol $\tilde{W}_I$ to distinguish the closed,  super two-form field strength vector multiplet gauge parameter from the physical vector multiplet $W^I$. The invariance can be trivially seen when looking at the second equation in \eqref{eq:S_standard_FI-superspace} and by noticing that $G_I^{ij}$ is identically zero if $\Psi_I$ is replaced with the vector multiplet, $\tilde{W}_I$, in \eqref{eq:S_standard_FI-superspace-2}.
Moreover, the action \eqref{eq:S_standard_FI-superspace} is also invariant under the following gauge transformations of the vector multiplets prepotentials
\bea
\d_\Lambda V^I_{ij}
=
\de^\a_k\L^I{}_\a^{ijk}
+\deb_{\ad k}\bar{\L}^I{}^\ad{}^{ijk}
~,~~~~~~
\L^I{}_\a^{ijk}=\L^I{}_\a^{(ijk)}
~,~~~
\bar{\L}^I{}^\ad_{ijk}=(\L^I{}^\a{}^{ijk})^*
~,
\eea
for a set of complex gauge parameter superfields $\L^I{}_\a^{ijk}$ being arbitrary up to the algebraic and reality conditions stated above. This transformation leaves the field strengths $W^I$ in \eqref{eq:S_standard_FI-superspace-2} invariant and, after superspace integration by parts and using $\de_\a^{(i}G_I^{jk)}=0$, $\deb_\ad^{(i}G_I^{jk)}=0$, one can directly show the invariance of the second form of \eqref{eq:S_standard_FI-superspace}.
The invariances under $\tilde{\d}$ and $\d_\L$ manifest themselves in the component action \eqref{eq:S_standard_FI} by the fact that the only term transforming would be  $\e^{m n p q} b_{mn}{}_{I} f^I_{p q} $, equivalent to $\e^{m n p q} h_{mnp}{}_{I} v^I_{q}$, which transform as total derivatives under the $\tilde{\d}$ and ${\d}_\L$ variations.

We did stress that in the global case, it is possible to simultaneously turn on an electric and a magnetic FI term preserving (and deforming) supersymmetry off shell \cite{Antoniadis:1995vb,IZ1,IZ2}. This is a fundamental ingredient in engineering global partial supersymmetry breaking by the use of vector multiplets. It is natural to ask whether the same is possible in the local off-shell superconformal setting that we have described above. In contrast to the global case, due to the gauge symmetries of the linear multiplets involved in the two types of deformations, in the local case, it does not seem possible to have the two FI-type deformations turned on at the same time. Let us comment more on this.

Suppose that we consider magnetically deformed vector multiplets
\bea
\bW^I=W^I+\Psi^I
~,~~~
\de^{ij} \bW^I-\deb^{ij}\overline{\bW}^I
=2\ri\cG^I{}^{ij}
~,
\eea
which possess the gauge transformation
\bea
\hat{\d}\bW^I=0 ~,~~~~~~
    \hat{\d}\Psi^I=\hat{W}^I
    ~,~~~
    \hat{\d}W^I=-\hat{W}^I ~,
\eea
for a vector multiplet field strength $\hat{W}^I$ satisfying 
\bea
\deb^\ad_i \hat{W}^I=0
~,~~~
\de^{ij} \hat{W}^I=\deb^{ij}\overline{\hat{W}}^I
~.
\eea
We do use $\hat{\d}$ to distinguish from $\tilde{\d}$ and also to distinguish $\hat{W}^I$ from $\tilde{W}^I$ and $W^I$. Assuming $\Psi^I$ and $\Psi_I$ are unrelated, the possible candidates for an electric FI-type deformation would be
\be
- 2 \ri \int \rd^8 z \,\cE \,\Psi_I  W^I + \text{c.c.}
 = \int \rd^{12} z \,E \,G^{ij}_I  V_{ij}^I
~,~~~~~~
 W^I:=\frac{1}{4}\bar{\Delta}\de^{ij}V_{ij}^I
 ~,
 \label{no-curved-e-FI-1}
\ee
which, is invariant under the $\tilde{\d}\Psi_I=\tilde{W}_I$ transformation, but, with $\hat{\d}\Psi_I=0$, it is not invariant under $\hat{\d}$ transformations,
and
\be
- 2 \ri \int \rd^8 z \,\cE \,\Psi_I  \bW^I + \text{c.c.}
~,
\label{no-curved-e-FI-2}
\ee
which, is invariant under the $\hat{\d}$ transformation but not $\tilde{\d}$.
Note that the previous no-go argument holds also in cases where the electric and magnetic deformations are defined in terms of the same $\Psi$ building block. This is for instance the case of the hyper-dilaton Weyl composite linear multiplet $\cG_{ij}{}^{\underline{i}\underline{j}}$ that leads to $\cG_\xi{}_I^{ij}=\xi_I^{\underline{i}\underline{j}}\cG^{ij}{}_{\underline{i}\underline{j}}$ and $\cG_\zeta^I{}^{ij}=\zeta^I_{\underline{i}\underline{j}}\cG^{ij\underline{i}\underline{j}}$. In this case, assuming the existence of a potential $\Psi^{\underline{i}\underline{j}}$ for $\cG^{ij\underline{i}\underline{j}}$, the $\tilde{\d}$ and $\hat{\d}$ transformations would coincide with a single one generated by $\underline{\delta}\Psi^{\underline{i}\underline{j}}=\underline{W}^{\underline{i}\underline{j}}$ for a triplet of vector multiplets with field strengths $\underline{W}^{\underline{i}\underline{j}}$. The reader can check that by choosing $\Psi_\xi{}_I:=\xi_I^{\underline{i}\underline{j}}\Psi_{\underline{i}\underline{j}}$
and $\Psi_\zeta^I:=\zeta^I_{\underline{i}\underline{j}}\Psi^{\underline{i}\underline{j}}$, together with $\bW^I=W^I+\Psi_\zeta^I$, both \eqref{no-curved-e-FI-1} and \eqref{no-curved-e-FI-2} are in general not invariant under $\underline{\d}$ transformations. Given the discussion above, in this paper, we will consider the existence of off-shell ``electric'' and ``magnetic'' deformations as mutually exclusive. Despite this difference compared to the off-shell global case, we will see that, also due to the presence of the compensating vector multiplet in supergravity, there is still enough freedom to obtain an off-shell model exhibiting local partial supersymmetry breaking.

\section{Deformed $\mathcal{N}=2$ supergravity in a hyper-dilaton Weyl multiplet background} \label{section:gauged}

The action for a deformed abelian vector multiplet in a hyper-dilaton Weyl multiplet background can be derived by substituting the expressions \eqref{compositeSSbD-hdwm} and \eqref{compositeSU2-hdwm} into \eqref{deformed-vectoraction}. Because the special conformal ${{\mathfrak f}}_m{}_{c}$ and $S$-supersymmetry $({\phi}_m{}_{\a}^{ i},\bar{{\phi}}_m{}^{\ad}_{i})$ connections depend on $D$ and $(\Sigma^{\a i},\bar{\S}_{\ad i})$ and because the $\rm SU(2)_R$ connection $\phi_{m}{}^{ij}$ is composite in the hyper-dilaton Weyl background, we degauge the derivative $\nabla_a$ to $\textbf{D}_a$ as defined in eq.~\eqref{boldDa}. Also, note that the linear multiplet fields are composite of hypermultiplet fields, eq.~\eqref{composite-linear-hdwm} and $G_\zeta{}^{ij}{}^I=\zeta^I_{\underline{i}\underline{j}}G^{ij\underline{i}\underline{j}}$, $\chi_{\a i}^I=\zeta^I_{\underline{i}\underline{j}}\chi_{\a i}{}^{\underline{i}\underline{j}}$, $\bar{\chi}^{\ad i}{}^I=\zeta^I_{\underline{i}\underline{j}}\bar{\chi}^{\ad i}{}^{\underline{i}\underline{j}}$, and $F{}^I=\zeta^I_{\underline{i}\underline{j}}F{}^{\underline{i}\underline{j}}$. With this in mind, the bosonic part of the action follows:\footnote{In the discussion in this section, we should use $\boldsymbol{\phi}^I$ rather than ${\textbf{W}}^{I}$ since we are projecting to components several supersymmetric invariants. However, since we will obtain covariant superfields equations of motions, we continue to use ${\textbf{W}}^{I}$ with the hope that it will be clear from the context whether we denote the superfield or its lowest component field.}
\bea
\label{deformed_vector_hdw_action_boson}
  \cL_{c, \textrm{bosons}}
  &=&~
  \frac{1}{2}\mathcal{F}_{I} \overline{\textbf{W}}^{I} R
  +\frac{1}{2}{\rm i} \mathcal{F}_{I} F_\zeta^{I} 
  +\frac{1}{4}\mathcal{F} W_{a b} W^{a b}  
  - \frac{1}{8}{\rm i} \mathcal{F} \epsilon_{a b c d} W^{a b} W^{c d}  
  - \frac{1}{2}\mathcal{F}_{I} W_{a b} \textbf{F}^{a b I} 
  \non\\
  &&
  +\frac{1}{4}{\rm i} \mathcal{F}_{I} \epsilon_{a b c d} W^{a b} \textbf{F}^{c d I}  
  - \frac{1}{2}\mathcal{F}_{I J} \textbf{F}_{a b}\,^{I} \textbf{F}^{a b J} 
  - \frac{1}{4}{\rm i} \mathcal{F}_{I J} \epsilon_{a b c d} \textbf{F}^{a b I} \textbf{F}^{c d J}  
  - \frac{1}{32}\mathcal{F}_{I J} G_\zeta{}_{i j}^{I} G_\zeta^{i j J} 
  \non\\
  &&
  +\frac{1}{32}\mathcal{F}_{I J} \textbf{X}_{i j}\,^{I} \textbf{X}^{i j J} 
  +\mathcal{F}_{I} \textbf{D}_{a}{\textbf{D}^{a}{\overline{\textbf{W}}^{I}}} 
  -4\mathcal{F}_{I} \overline{\textbf{W}}^{I} A_{a} A^{a}  
  - \frac{1}{4}\mathcal{F}_{I} W_{a b} W^{a b} \textbf{W}^{I} 
  \non\\
  &&
  +\frac{1}{8}{\rm i} \mathcal{F}_{I} \epsilon_{a b c d} W^{a b} W^{c d} \textbf{W}^{I} 
  - \frac{1}{2}\mathcal{F}_{I J} W_{a b} \overline{\textbf{W}}^{I} \textbf{F}^{a b J} 
  - \frac{1}{4}{\rm i} \mathcal{F}_{I J} \epsilon_{a b c d} W^{a b} \overline{\textbf{W}}^{I} \textbf{F}^{c d J} 
  \non\\
  &&
  +\frac{1}{16}{\rm i} \mathcal{F}_{J I} \textbf{X}_{i j}\,^{J} G_\zeta^{i j I} 
  - \frac{1}{8}\mathcal{F}_{I J} W_{a b} W^{a b} \overline{\textbf{W}}^{I} \overline{\textbf{W}}^{J} 
  - \frac{1}{16}{\rm i} \mathcal{F}_{I J} \epsilon_{a b c d} W^{a b} W^{c d} \overline{\textbf{W}}^{I} \overline{\textbf{W}}^{J} 
\non\\
  &&
  -4{\rm i} \mathcal{F}_{I} A_{a} \textbf{D}^{a}{\overline{\textbf{W}}^{I}} 
-2{\rm i} \mathcal{F}_{I} \overline{\textbf{W}}^{I} \textbf{D}_{a}{A^{a}} 
-64\mathcal{F}_{I} \overline{\textbf{W}}^{I} {q}^{-4} H_{a \underline{i} \underline{j}} H^{a \underline{i} \underline{j}} 
\non\\
  &&
+\mathcal{F}_{I} \overline{\textbf{W}}^{I} {q}^{-2}
\left(2q_{i \underline{i}} \textbf{D}_{a}{\textbf{D}^{a}{q^{i \underline{i}}}} 
+\textbf{D}_{a}{q^{i \underline{i}}} \textbf{D}^{a}{q_{i \underline{i}}} 
-2 {q}^{-2} q_{i \underline{i}} q_{j \underline{j}} \textbf{D}_{a}{q^{i \underline{j}}} \textbf{D}^{a}{q^{j \underline{i}}}
\right)
~.
\eea

Note that implicit fermions exist here and can be seen by converting $H^{a \underline{i} \underline{j}}$, $F^{a b}$, and $\textbf{F}^{a b}$ to $\tilde{h}^{a \underline{i} \underline{j}}$, $f^{a b}$, and $\textbf{f}^{a b}$, respectively, by eqs. \eqref{eq:bigHtoLittleh}, \eqref{eq:hHodge}, and  \eqref{covF}. This notably includes the coupling of $\tilde{h}^{a \underline{i} \underline{j}}$ to two gravitini. Otherwise, the fermionic counterpart of the action is given in Section II of the supplementary file. The component action for $S_{\text{standard FI}}$ in the hyper-dilaton Weyl background will remain the same as in the standard Weyl multiplet background \eqref{eq:S_standard_FI}, as it does not depend on the composite fields $D$, $(\Sigma^{\a i},\bar{\S}_{\ad i})$, and $\phi_{m}{}^{ij}$.   Keep also in mind that some of the fields in the previous bosonic Lagrangian are composite and include fermions. For example, $F^I_\zeta$ in a hyper-dilaton Weyl background is purely quadratic in fermions while the $\textbf{D}_{a}$ derivative is defined in terms of $\o_m{}^{cd}$ which contains the torsion quadratic in gravitini.

\subsection{Equations of motion}

The goal of this section is to obtain superconformal primary equations of motion that describe gauged $\cN=2$ deformed supergravity based on a hyper-dilaton Weyl multiplet and defined by the action
\be
    \mathcal{S} = \mathcal{S}_{c} + \textrm{c.c.} + \mathcal{S}_{\textrm{standard FI}} ~.
    \label{action-sec-5}
\ee
In all the expressions in this section, we will formally allow for arbitrary ``electric'' ($\cG_\xi{}^{ij}_I$) and ``magnetic'' (${\cG}_\zeta{}_{ij}^I$)  deformations but, as discussed before, the reader should keep in mind that we consider them to be mutually exclusive. Given a fixed value of the index $I$, we allow for either of the two to be turned on, but not both at the same time.

We obtain the equations of motion by the variation of the action \eqref{action-sec-5} in components with respect to the auxiliary fields, i.e., the highest dimension independent fields, of each multiplet. The resulting equations of motion then describe the primary fields, i.e., the bottom components, of the multiplets of the equations of motion that arise from the variation of the full superfields. It is then straightforward to reinterpret them as the primary superfields of the equations of motion. See \cite{Gold:2023dfe,Gold:2023ykx} for a recent analysis in a five-dimensional setting.

In components, the EOM for the vector multiplet is obtained by varying the action with respect to the auxiliary field $\textbf{X}^{ij I}$. The $\textbf{X}^{ij I}$-dependent terms in the action are
\bea
    \cL_X &=& \frac{1}{16} \left(\cF_{I J K} \boldsymbol{\lambda}^{\a I}_{i} \boldsymbol{\lambda}^{J}_{\a j} - \overline{\cF}_{I J K} \bar{\boldsymbol{\lambda}}^{\ad I}_{i} \bar{\boldsymbol{\lambda}}^{J}_{\ad j}\right)  \textbf{X}^{i j K} + \frac{1}{32} N_{I J} \textbf{X}^I_{i j} \textbf{X}^{i j J} 
    \non \\
    &&+ \frac{1}{16} \ri \left(\cF_{I J} - \bar{\cF}_{I J}\right) \textbf{X}^I_{i j} G_\zeta^{i j J} + \frac{1}{4} G_\xi{}_I^{i j} \textbf{X}^I_{i j} ~,
\eea
where $G_\zeta^{i j I}$ and ${G}_\xi{}^{i j}_{I}$ are the magnetic and electric deformations, respectively. Thus the equations of motion follow (one for each selection of $I$):
\bea\label{EOMXij}
    0 =  \left(\cF_{I J K} \boldsymbol{\lambda}^{\a J}_{i} \boldsymbol{\lambda}^{K}_{\a j} 
    + \overline{\cF}_{I J K} \bar{\boldsymbol{\lambda}}^{J}_{\ad i} \bar{\boldsymbol{\lambda}}^{\ad K}_{j}\right)   +  N_{I J} \textbf{X}^J_{i j}+ \ri \left(\cF_{I J} - \bar{\cF}_{I J}\right)  G_\zeta{}_{i j}^{J} + 4 G_\xi{}_{i j I} ~, 
\eea
where we have defined the special K\"ahler metric
\bea
N_{I J} = \cF_{I J} + \bar{\cF}_{I J}
\label{special-K-Metric}
~.
\eea

Next, we find the Euler-Lagrange equations of motion for the auxiliary fields $W_{\a \b}$ and $\overline{W}^{\ad \bd}$. It is worth pointing out that in superspace the equations of motion derived by varying prepotentials are manifestly covariant. Hence, one expects the same to be true once the superspace results are reduced to component fields. However, in the component approach of finding the EOMs, the component action computed from eq.~\eqref{CA} includes hundreds of terms when fermions are considered, and it is not manifestly covariant due to the presence of naked gravitini. Although the action lacks manifest covariance, it has recently been explicitly demonstrated in components that for any supergravity theory, there exist covariant equations of motion that are equivalent to the regular field equations \cite{Ferrara:2017yhz,Vanhecke:2017chr}. These covariant equations are obtained by covariantising the regular field equations, resulting in a multiplet of field equations \cite{Ferrara:2017yhz,Vanhecke:2017chr}.

To find the covariant equations of motion for $W_{\a \b}$, $\Wb^{\ad \bd}$, we can directly use the above action as degauging is already completed. Collecting all terms up to all orders in fermions with $W_{\a \b}$ and $\Wb^{\ad \bd}$, we have
\bea
\mathcal{L}_{W_{\alpha \lambda}}&=&-2\mathcal{F}_{I J}  W_{\alpha \lambda} \overline{\textbf{W}}^{I} \textbf{F}^{\alpha \lambda}\,^{J} - \frac{1}{2}\mathcal{F}_{I J}  W_{\alpha \lambda} W^{\alpha \lambda} \overline{\textbf{W}}^{I} \overline{\textbf{W}}^{J} 
+ \frac{1}{4} \mathcal{F}_{I J K} \epsilon_{i j} W_{\alpha \lambda} \overline{\textbf{W}}^{I} \boldsymbol{\lambda}^{i \alpha}{}^{J} \boldsymbol{\lambda}^{j \lambda}{}^{K} \non 
\\ &&
+\mathcal{F}_{I J}   W_{\alpha \lambda} \overline{\textbf{W}}^{I} q_{i \underline{i}}  \boldsymbol{\lambda}^{i \a}{}^{J} \rho^{\underline{i} \l} {q}^{-2}
+\frac{1}{4}\mathcal{F}_{I}  \epsilon_{\underline{i} \underline{j}} W_{\alpha \lambda} \overline{\textbf{W}}^{I} \rho^{\underline{i} \a} \rho^{\underline{j}\l} {q}^{-2} \non 
\\ &&
+\bar{\mathcal{F}}_{I}  {W}_{{\alpha} {\lambda}} {\boldsymbol{\lambda}}^{\,{\a} i}{}^{I} q_{i \underline{i}} {\rho}^{{\l} \underline{i}} {q}^{-2}-2\bar{\mathcal{F}}_{I} {W}_{{\alpha} {\lambda}} {\textbf{F}}^{{\alpha} {\lambda} I}-\bar{\mathcal{F}}  {W}_{{\alpha} {\lambda}} {W}^{{\alpha} {\lambda}} + \textrm{c.c.} ~. \non 
\eea
The Euler-Lagrange equations of motion for the auxiliary fields $W_{\a \b}$, $\Wb^{\ad \bd}$ follow
\bea\label{EOMWab}
     N_{I J}  W^{\alpha \lambda} \overline{\textbf{W}}^{I} \overline{\textbf{W}}^{J} &=&-2N_{I J}  \overline{\textbf{W}}^{I} \textbf{F}^{\alpha \lambda}\,^{J}  
+ \frac{1}{4} \mathcal{F}_{I J K} \epsilon_{i j} \overline{\textbf{W}}^{I} \boldsymbol{\lambda}^{i J (\alpha}\boldsymbol{\lambda}^{ \lambda) j}{}^{K} \non 
\\ &&
+ (\bar{\mathcal{F}}_{IJ} -\mathcal{F}_{I J})    \overline{\textbf{W}}^{I} q_{i \underline{i}} \boldsymbol{\lambda}^{i J (\a}  \rho^{ \l)\underline{i} } {q}^{-2}
+\frac{1}{4}\mathcal{F}_{I}  \epsilon_{\underline{i} \underline{j}} \overline{\textbf{W}}^{I} \rho^{\underline{i} (\a} \rho^{\l)\underline{j}} {q}^{-2} ~,
\eea
together with its complex conjugate.
Note that the EOM for $W_{\a \b}$ and $\Wb^{\ad \bd}$ are manifestly covariant and do not depend on the gravitini as expected. 

Next, to prove that the equation of motion for $A_{m}$ is covariant, we need to perform integration by parts, which makes it essential to degauge the covariant derivative $\textbf{D}_{a}$ with respect to $M_{a b}$ in the above action and insert the composite expression for the spin connection $\omega_{m}{}^{ab}$ in terms of $\omega(e)_{m}{}^{ab}$ together with bilinear terms in gravitini, eq.~\eqref{composite-Lorentz}. Once these steps are carried out, the terms involving $A_a = e_a{}^m A_m$ in the Lagrangian take the following form:
\begin{align}
    \mathcal{L}_{A_a}=& {}-{\rm i} \mathcal{F}_{I J}  \epsilon^{\alpha \beta} \epsilon_{i j} \psi^{a}{}^{i}_{\alpha} \overline{\textbf{W}}^{I} \boldsymbol{\lambda}^{j}_{\beta}{}^{J} A_{a}-\mathcal{F}_{I J} \epsilon^{\alpha \beta} \epsilon_{\dot{\alpha} \dot{\beta}} \epsilon_{i j} (\sigma_{a})_{\alpha}{}^{\dot{\alpha}} \overline{\textbf{W}}^{I} \boldsymbol{\lambda}^{i}_{\beta}{}^{J} q^{j \underline{i}} \overline{\rho}_{\underline{i}}\,^{\dot{\beta}} A^{a} {q}^{-2} 
    \non \\ &
    +\frac{1}{2}\mathcal{F}_{I J} \epsilon^{\alpha \beta} \epsilon_{\dot{\alpha} \dot{\beta}} (\sigma_{a})_{\alpha}{}^{\dot{\alpha}} \boldsymbol{\lambda}^{i}_{\beta}{}^{I} \overline{\boldsymbol{\lambda}}_{\,i}^{\,\dot{\beta}}{}^{J} A^{a}
    +{\rm i} \mathcal{F}_{I}  \epsilon_{\dot{\alpha} \dot{\beta}} \epsilon^{i j} \bar{\psi}^{a}{}_{i}^{\dot{\alpha}} \overline{\boldsymbol{\lambda}}_{\,j}^{\,\dot{\beta}}{}^{I} A_{a}
    \non \\ &
    +\mathcal{F}_{I} \epsilon^{\alpha \beta} \epsilon_{\dot{\alpha} \dot{\beta}} \epsilon_{\underline{i} \underline{j}} (\sigma_{a})_{\alpha}{}^{\dot{\alpha}} \overline{\boldsymbol{\lambda}}_{\,i}^{\,\dot{\beta}}{}^{I} q^{i \underline{i}} \rho^{\underline{j}}\,_{\beta} A^{a} {q}^{-2}
    +\frac{1}{2}\mathcal{F}_{I} \epsilon^{\alpha \beta} \epsilon_{\dot{\alpha} \dot{\beta}} (\sigma_{a})_{\alpha}{}^{\dot{\alpha}} \overline{\textbf{W}}^{I} \rho^{\underline{i}}\,_{\beta} \overline{\rho}_{\underline{i}}\,^{\dot{\beta}} A^{a} {q}^{-2}
    \non \\ &
    -4{\rm i} \mathcal{F}_{I}  A^{a} e_{a}{}^{m}\partial_{m}{\overline{\textbf{W}}^{I}} -4\mathcal{F}_{I}  \overline{\textbf{W}}^{I} A^{a} A_{a}-2{\rm i} \mathcal{F}_{I} \overline{\textbf{W}}^{I} e_{a}{}^{m}\partial_{m}{A^{a}} + \textrm{c.c.}~.
\end{align}
The Euler-Lagrange equation of the auxiliary fields $A^{a}$ give:
\bea\label{EOMAa}
   8 N A_{a} &=&- \ri \mathcal{F}_{I J} \psi_{a}{}^{i}_{\alpha} \overline{\textbf{W}}^{I} \boldsymbol{\lambda}^{\a}_{i}{}^{J} - \mathcal{F}_{I J} (\sigma_{a})_{\alpha}{}^{\dot{\alpha}} \overline{\textbf{W}}^{I} \boldsymbol{\lambda}^{i \a J} q_i{}^{\underline{i}} \overline{\rho}_{\underline{i} \ad} {q}^{-2} + \frac{1}{2}\mathcal{F}_{I J}  (\sigma_{a})_{\alpha}{}^{\dot{\alpha}} \boldsymbol{\lambda}^{i \a}{}^{I} \overline{\boldsymbol{\lambda}}_{\,i \ad}^{J} 
    \non \\
    && + \ri \mathcal{F}_{I}  \bar{\psi}_{a}{}_{i}^{\dot{\alpha}} \overline{\boldsymbol{\lambda}}^{\,i}_{\,\dot{\a}}{}^{I} + \mathcal{F}_{I} (\sigma_{a})_{\alpha}{}^{\dot{\alpha}} \overline{\boldsymbol{\lambda}}_{\,i \ad}^{I} q^{i \underline{i}} \rho_{\underline{i}}^{\a} {q}^{-2}
    \non \\
    &&+\frac{1}{2}\mathcal{F}_{I}(\sigma_{a})_{\alpha}{}^{\dot{\alpha}} \overline{\textbf{W}}^{I} \rho^{\underline{i} \a} \overline{\rho}_{\underline{i} \ad} {q}^{-2}  - 2 \ri \mathcal{F}_{I J} \textbf{W}^{J}  \cD'_a {\overline{\textbf{W}}^{I}}+ 2 \ri \mathcal{F}_{I J}  \overline{\textbf{W}}^{I} \cD'_a \textbf{W}^{J} + \textrm{c.c.}  ~,~~~~~~
\eea
where we have defined 
\bea
N = N_{IJ} \textbf{W}^{I} \overline{\textbf{W}}^{J}
~,
\label{special-K-N}
\eea
and the covariant derivative $\cD'_a$ contains only the Lorentz connection (without gravitini torsion) and the inverse vielbein, i.e.,
\begin{align}
\cD'_a 
=
 e_{a}{}^{m}\Big(\partial_m
- \frac{1}{2}\omega_m{}^{cd}(e)M_{cd}\Big) ~.
\end{align}
Finally, we uplift the derivative to the superconformal covariant derivative. This will absorb leftover gravitini terms to get the covariant equation of motion for $A^{a}$
\bea
    &&  N_{I J} (\sigma_{a})_{\alpha}{}^{\dot{\alpha}} \left( \overline{\textbf{W}}^{I} \boldsymbol{\lambda}_i^{\a J}  \overline{\rho}_{\underline{i} \ad}+\textbf{W}^{I}\overline{\boldsymbol{\lambda}}_{\,i \ad}^{J} \rho_{\underline{i}}^{\a}\right) q^{i \underline{i}}{q}^{-2} + \frac{1}{2}N_{I J}  (\sigma_{a})_{\alpha}{}^{\dot{\alpha}} \boldsymbol{\lambda}^{i \a}{}^{I} \overline{\boldsymbol{\lambda}}_{\,i \ad}^{J} 
    \non \\
    &&+\frac{1}{2} N(\sigma_{a})_{\alpha}{}^{\dot{\alpha}}  \rho^{\underline{i} \a} \overline{\rho}_{\underline{i} \ad} {q}^{-2}  - 2 \ri N_{I J}  \textbf{W}^{I} \nabla^{a}{\overline{\textbf{W}}^{J}}+ 2 \ri N_{I J}  \overline{\textbf{W}}^{I}\nabla_{a}\textbf{W}^{J} = 0~.
\eea

With these covariant equations of motion computed, we are in a position to integrate out the auxiliary fields prerequisite to going on shell. This is explored in the following section in the context of a SU(1,1)/U(1) model leading to partial supersymmetry breaking. We leave for future work a general analysis of the on-shell action for the model described by \eqref{action-sec-5} in a hyper-dilaton Weyl setup.

\section{Off-shell model with on-shell partial-susy breaking} \label{sec:SU11U1}

In this section, we move to present a new off-shell model for partial supersymmetry breaking engineered by using off-shell deformed vector multiples in a hyper-dilaton Weyl background. Before presenting the details of the construction, it is worth stressing some of the key features of local supersymmetry breaking that guide our analysis.

For simplicity, we seek for a model of local partial supersymmetry breaking on a Minkowski vacuum. Hence, once auxiliary fields are integrated out, we want an on-shell theory possessing a Minkowski solution and no (effective) cosmological constant. Partial supersymmetry breaking emerges once the on-shell transformations of the fermions, which we collectively denote here as $f$, all possess a shit symmetric term schematically of the form 
\bea
\d_\xi f
=
\bM\xi  
+\cdots
~.
\eea
Here $\xi=(\xi^\a_i,\bar{\xi}_\ad^i)$ refers to the supersymmetry transformation parameters and $\bM$ is a (field dependent) matrix. If $\bM$ is degenerate, $\det \bM=0$, but non-trivial, then part of local supersymmetry is spontaneously broken. As a consequence of this fact, the fermionic mass terms are all parametrised in terms of $\bM$, hence with part of the spectrum remaining massless. For instance, in the case of local $\cN=2\to\cN=1$ susy breaking, one gravitino acquires a mass-like term while one remains massless.

As reviewed in the introduction, finding models based on vector multiples, potentially coupled to appropriate hypermultiplets, that possess local partial supersymmetry breaking is a non-trivial task --- see, for example, 
\cite{Cecotti:1984rk,Cecotti:1984fn,Cecotti:1984wn,Cecotti:1985sf,Ferrara:1995gu,Ferrara:1995xi,Fre:1996js,Louis:2012ux,Louis1,Louis2,Andrianopoli:2015wqa,Antoniadis:2018blk,Abe:2019svc,Abe:2019vzi,Lauria:2020rhc}. Three of the features of our new construction given in this section for supergravity with partial supersymmetry breaking are: (i) our model is manifestly off shell, which, to the best of our knowledge, is a first explicit example; (ii) the spectrum of the on-shell theory, which includes a triplet of gauged two-forms, differs from previous examples described in the literature; (iii) though based on electric and magnetic deformations of vector multiples, our model is not based on a standard gauging procedure, and in fact the fermions, e.g.~the gravitini, are not charged under any of the U(1) symmetries of the vector multiples (a generic feature of working with an hyper-dilaton multiplet).
Let us now move to the description of our construction to see these properties unfolding.

\subsection{${\rm SU(1,1)/U(1)}$ model} 

We consider two vector multiples, and the holomorphic prepotential 
\begin{align}\label{SU(1,1)/U(1)model}
    \cF = c \phi \boldsymbol{\phi} ~,
\end{align}
where $\phi$ is a compensator, $\boldsymbol{\phi}$ is a deformed physical vector multiplet, and $c$ is a real, nonzero constant which we decide to leave as a free normalisation parameter. 
This model is directly inspired by the well-known SU(1,1)/U(1) special K\"ahler sigma model, which, in a different set up, is known to lead to partial supersymmetry breaking. In particular, as described in the introduction, \eqref{SU(1,1)/U(1)model} arises from the SU(1,1)/U(1)  special-K\"ahler sigma model after performing a duality transformation in the geometry used in \cite{Cecotti:1985sf,Ferrara:1995gu} ending up into a symplectic frame where a holomorphic prepotential exists and is given by \eqref{SU(1,1)/U(1)model}.

As shown in \cite{Cecotti:1985sf,Ferrara:1995gu}, within the context of $\cN=2$ supergravity in the standard Weyl multiplet background, the minimal matter content required for partial supersymmetry breaking includes a physical vector multiplet and a hypermultiplet. In this scenario, the vector multiplet parametrizes the SU(1,1)/U(1) special Kähler manifold, while the physical hypermultiplet parametrizes the SO(4,1)/SO(4) quaternionic
manifold. This model was further generalized in \cite{Fre:1996js} by coupling the standard Weyl multiplet to $n+1$ vector multiplets and $m$ hypermultiplets in a set-up that inherently has supersymmetry closing (partially) on shell.

Considering the minimal field content mentioned above, it is natural to argue that in a hyper-dilaton Weyl multiplet background, only a single physical vector multiplet plus a compensator, which would parametrize the special Kähler manifold SU(1,1)/U(1), might be sufficient to achieve partial supersymmetry breaking. In fact,  the hypermultiplet is already a part of the hyper-dilaton Weyl multiplet itself, though in an off-shell setting, and in a new type of matter content. We will see in this section that this intuition is correct.

Note that instead of using numbered indices for the vector multiplets, we employ a bold symbol to denote the physical ``1" multiplet, while the unbold symbol represents the compensating ``0" multiplet,  which, after taking derivatives of the holomorphic prepotential and using eqs.~\eqref{special-K-Metric} and \eqref{special-K-N}, implies
\begin{align}
    &\cF_0 = c \boldsymbol{\phi}, ~~ \cF_1 =  c \phi, ~~\cF_{0 0} = 0, ~~ \cF_{1 0} = \cF_{0 1} = c , ~~ \cF_{1 1} = 0~, \non \\
    & N_{0 0} = 0, ~~ N_{1 0} = N_{0 1} = 2c , ~~ N_{1 1} = 0, ~~ N = 2 c (\phi \bar{\boldsymbol{\phi}} + \boldsymbol{\phi}  \bar{\phi})~.
\end{align}
We also choose the electric and magnetic deformations to be $G_\xi{}_{I}^{ij}=(\xi_{\underline{i}\underline{j}} q^{i\underline{i}} q^{j\underline{j}}, 0)$ and $G_\zeta{}_{ij}^I = (0, \zeta_{\underline{i}\underline{j}} q_i{}^{\underline{i}} q_j{}^{\underline{j}})$, respectively. This means that the compensator is ``electrically deformed'', which generically induces a negative contribution to the vacuum energy, while the physical vector multiplet is ``magnetically'' deformed, which generically induces a positive contribution to the vacuum energy. By tuning appropriately $\xi_{\underline{i}\underline{j}}$ and $\zeta_{\underline{i}\underline{j}}$ we will find zero vacuum energy and partial supersymmetry breaking.

For simplicity, we consider the case where $c\ne0$ is real. The off-shell component action for the SU(1,1)/U(1) model can be obtained by first substituting eq.~\eqref{SU(1,1)/U(1)model} and its derivatives into the general, deformed off-shell action of eqs.~\eqref{deformed_vector_hdw_action_boson} and the results given in the supplementary file (for bosons and fermions, respectively) along with their complex conjugates, and then adding the standard FI action of eq.~\eqref{eq:S_standard_FI}. The bosonic part of the action takes the following form:
\begin{align}\label{SU(1,1)/U(1)_action_hdw_boson}
    \cL_{\textrm{bosons}} ={}&~~~\frac{N}{2} R-2c F_{a b} \textbf{F}^{a b}+\frac{1}{8}c X_{i j} \textbf{X}^{i j} -4 N A_{a} A^{a} 
    \non \\ &
    -c (\overline{\boldsymbol{\phi}} F^{a b} + \overline{\phi} \textbf{F}^{a b}) W_{a b} - \frac{1}{2}{\rm i} c \epsilon_{a b c d }  (\overline{\boldsymbol{\phi}} F^{a b} + \overline{\phi} \textbf{F}^{a b}) W^{c d}
    \non \\ &
    -c ({\boldsymbol{\phi}} F^{a b} + {\phi} \textbf{F}^{a b}) W_{a b}+ \frac{1}{2}{\rm i} c \epsilon_{a b c d }  ({\boldsymbol{\phi}} F^{a b} + {\phi} \textbf{F}^{a b}) W^{c d}
    \non \\ &
    +c \boldsymbol{\phi} \textbf{D}_{a}{\textbf{D}^{a}{\overline{\phi}}} +c \overline{\boldsymbol{\phi}} \textbf{D}_{a}{\textbf{D}^{a}{\phi}}+c \phi \textbf{D}_{a}{\textbf{D}^{a}{\overline{\boldsymbol{\phi}}}}+c \overline{\phi} \textbf{D}_{a}{\textbf{D}^{a}{\boldsymbol{\phi}}}%
    \non \\ &
    - \frac{1}{2}c W_{a b} W^{a b} (\overline{\phi} \overline{\boldsymbol{\phi}} + \phi \boldsymbol{\phi}) - \frac{1}{4}{\rm i} c \epsilon_{a b c d} W^{a b} W^{c d} (\overline{\phi} \overline{\boldsymbol{\phi}} - \phi \boldsymbol{\phi})
    \non \\ &
    -4{\rm i} c (\boldsymbol{\phi}- \overline{\boldsymbol{\phi}}) A_{a} \textbf{D}^{a}{\overline{\phi}}-4{\rm i} c (\phi - \overline{\phi}) A_{a} \textbf{D}^{a}{\overline{\boldsymbol{\phi}}}
    \non \\ &
    -64 N {q}^{-4} H_{a \underline{i} \underline{j}} H^{a \underline{i} \underline{j}}
    +N {q}^{-2} \textbf{D}_{a}{q_{i \underline{i}}} \textbf{D}^{a}{q^{i \underline{i}}}+2 N {q}^{-2} q_{i \underline{i}} \textbf{D}_{a}{\textbf{D}^{a}{q^{i \underline{i}}}}
    \non \\ &
    -2 N {q}^{-4} q_{i \underline{i}} q_{j \underline{j}} \textbf{D}_{a}{q^{i \underline{j}}} \textbf{D}^{a}{q^{j \underline{i}}} 
    \non \\ & 
+ \frac{1}{4} \xi_{\underline{i} \underline{j}} q_i{}^{\underline{i}} q_j{}^{\underline{j}} X^{i j} +\frac{4}{3}  \xi_{\underline{i} \underline{j}} \e^{m n p q} h_{m np}{}^{{\underline{i} \underline{j}}} v_{q} ~.
\end{align}

Note that, once again, implicit fermions exist from the conversion of $H^{a \underline{i} \underline{j}}$, $F^{a b}$, and $\textbf{F}^{a b}$ to $\tilde{h}^{a \underline{i} \underline{j}}$, $f^{a b}$, and $\textbf{f}^{a b}$, respectively, by eqs.~\eqref{eq:bigHtoLittleh}, \eqref{eq:hHodge}, and  \eqref{covF}. Otherwise, the fermionic counterpart to this action can be found in Section III of the supplementary file. The Euler-Lagrange equations of motion for the vector multiplet auxiliary fields $X^{ij}$ and $\textbf{X}^{ij}$ can be obtained by substituting \eqref{SU(1,1)/U(1)model} into the general equation of motion \eqref{EOMXij} and are given by the following two equations
\bsubeq\label{EOMXijSU(1,1)/U(1)model}
\begin{align}
     X_{ij} &= 0~, \\
    \bX_{ij} &= - \frac{2}{c} \xi_{\underline{i} \underline{j}} q_{i}{}^{\underline{i}} q_{j}{}^{\underline{j}}
     ~.
\end{align}
The second equation leads to the following expression for the shifted auxiliary field of eq.~\eqref{shifted-auxiliary-1}
\begin{align}
    \bM_{ij} := \left(- \frac{2}{c} \xi_{\underline{i} \underline{j}} 
     +\ri \zeta_{\underline{i} \underline{j}} \right)q_i{}^{\underline{i}} q_j{}^{\underline{j}}~.
\end{align}
\esubeq

The Euler-Lagrange equations of motion for the remaining auxiliary fields $W_{\alpha \beta}$, $\bar{W}^{\dot{\alpha} \dot{\beta}}$, and $A_a$ can once again be obtained by substituting \eqref{SU(1,1)/U(1)model} into the general equations of motion for \eqref{EOMWab}, its complex conjugate, and \eqref{EOMAa}, respectively. They are all given by the following:
\bsubeq \label{EOM_Wab_AaSU(1,1)/U(1)model}
\bea
     4 c \bar{\phi}\bar{\boldsymbol{\phi}} {W}^{\a\b}  
     &= & 
     -\frac{3}{2} c ( \phi\bar{\boldsymbol{\phi}} +  \bar{\phi} \boldsymbol{\phi})  q^{-2}  {\rho}_{\underline{j}}^{(\a} {\rho}^{\b) \underline{j}} 
    - 4 c (\bar{\phi}{\bF}^{\a\b} +\bar{\boldsymbol{\phi}} {F}^{\a\b})~, \label{eq:WabEOM}~\\
      4 c \phi \boldsymbol{\phi} \overline{W}^{\ad\bd} 
      &= &
      -\frac{3}{2} c ( \phi\bar{\boldsymbol{\phi}} +  \bar{\phi} \boldsymbol{\phi})  q^{-2}  \bar{\rho}^{\underline{j} (\ad} \bar{\rho}^{\bd)}_{ \underline{j}} 
    - 4 c (\phi {\bar{\bF}}^{\ad\bd} + \boldsymbol{\phi} {\bar{F}}^{\ad\bd})     \label{eq:bWabEOM} ~,
     \\
       8 N A_{a}
       & =& 
       ~4 \ri c  \left(
       \bar{\phi}\cD'_{a}{\boldsymbol{\phi}} +\bar{\boldsymbol{\phi}}\cD'_{a}{\phi}
-{\phi}  \cD'_{a}{\bar{\boldsymbol{\phi}}}
- {\boldsymbol{\phi}}  \cD'_{a}{\bar{\phi}}\right) 
       \non\\
       &&
    - 2 \ri c \psi_{a}{}^{i}_{\alpha} (\bar{\phi} \boldsymbol{\lambda}^{\a}_{i}{} +\bar{\boldsymbol{\phi}} {\lambda}^{\a}_{i}{}) 
      + 2 \ri c  \bar{\psi}_{a}{}_{i}^{\dot{\alpha}} ({\phi} \overline{\boldsymbol{\lambda}}^{\,i}_{\,\dot{\a}} + {\boldsymbol{\phi}} \overline{{\lambda}}^{\,i}_{\,\dot{\a}} )
      \non\\
      &&
+ c  (\sigma_{a})_{\alpha}{}^{\dot{\alpha}} ({\lambda}^{i \a} \overline{\boldsymbol{\lambda}}_{\,i \ad} + \boldsymbol{\lambda}^{i \a}{} \overline{{\lambda}}_{\,i \ad} )
   +c (\sigma_{a})_{\alpha}{}^{\dot{\alpha}}  ( \phi\bar{\boldsymbol{\phi}} +  \bar{\phi} \boldsymbol{\phi}) \rho^{\underline{i} \a} \overline{\rho}_{\underline{i} \ad} {q}^{-2} 
    \non  \\
    &&
    + 2 c (\sigma_{a})_{\alpha}{}^{\dot{\alpha}} (\bar{\phi} \boldsymbol{\lambda}^{i \a}{} +\bar{\boldsymbol{\phi}} {\lambda}^{i \a}{}) q_{i \underline{i}} \overline{\rho}_{\ad}^{\underline{i}} {q}^{-2} 
       \non \\
       &&
  + 2 c (\sigma_{a})_{\alpha}{}^{\dot{\alpha}} ({\phi} \overline{\boldsymbol{\lambda}}_{i \dot{\a}} + {\boldsymbol{\phi}} \overline{{\lambda}}_{i \dot{\a}} ) q^{i \underline{i}} \rho_{\underline{i}}^{\a} {q}^{-2}
    ~.
    \label{eq:AaEOM}
\eea
\esubeq
Note that from the action in eq. \eqref{SU(1,1)/U(1)_action_hdw_boson} we see that the scalar potential 
\begin{align}
    S_{\rm potential}
    = \int d^4x  \,e\Bigg[& 
     \frac{1}{16} c X^{i j} (\textbf{M}_{i j}+\overline{\textbf{M}}_{i j} )
+\frac{1}{4} \xi_{\underline{i} \underline{j}} q_{i}{}^{\underline{i}} q_j{}^{\underline{j}} X^{i j}
\Bigg]=0 ~,
\end{align}
is zero on the vacuum where $X^{ij}=0$. Hence, we have zero cosmological constant. By analysing in the following subsection the supersymmetry variation of the fermions, we will see that
the condition $\det{\bM}=0$, together with the assumption that the rank of the matrix $\bM$ is one, will ensure local partial supersymmetry breaking in Minkowski space-time, for any scalar fields configurations.

\subsection{Gauge fixing and fermion shifts}
\label{section-6.2}

In this section, we give the explicit expressions for the gauge fixing that lead to Poincar\'e supergravity. In particular, we will gauge fix all superconformal structure group transformations except local $Q$-supersymmetry and Lorentz. For the dilatations,
we aim to have a standard kinetic term for gravity. Hence, we collect the terms with the scalar curvature and obtain the following gauge condition for dilatation
\bsubeq\label{gauge-conditions}
\bea\label{D-gauge-condition}
    {{\text{$\mathbb{D}$-{gauge:}}}} \qquad & ( \phi\bar{\boldsymbol{\phi}} +  \bar{\phi} \boldsymbol{\phi}) = - \frac{1}{2c} ~.
\eea
The $S$-gauge can be obtained by simply taking the $Q$-supersymmetry transformation of the $\mathbb{D}$-gauge 
\bea\label{S-gauge-condition}
     {\text{$S$-gauge:}} \qquad & \l^{ i}_{\a}\bar{\boldsymbol{\phi}} +  \bar{\phi} \boldsymbol{\l}^{i}_{\a} = 0 ~.
\eea
With this choice, the $\mathbb{D}$-gauge is invariant under $Q$-supersymmetry.
Next, the consistent gauge choice for the $\rm U(1)_R$ symmetry \cite{Lauria:2020rhc} is
\bea\label{U(1)-gauge-condition}
    \textrm{U(1)}\textrm{-gauge:} \qquad \phi = \bar{\phi}= y ~,
\eea
which clearly imposes the compensating vector multiplet field to be real. A characterising feature of the hyper-dilaton Weyl multiplet is that it contains an $\rm SU(2)_R$ compensator being the $q_{i\underline{i}}$ fields. We then impose
\bea\label{SU(2)-gauge-condition}
    \textrm{SU(2)-gauge}: \qquad q_{i\underline{i}} = \e_{i \underline{i}} e^{-U} ~,
\eea
which gauge fixes $\rm SU(2)_R$. Lastly, we take the standard choice of gauge fixing condition
\begin{equation} \label{K-gauge-condition}
    K\textrm{-gauge}: \qquad b_m = 0 ~,
\end{equation}
\esubeq
to fix special conformal symmetry.

The transformation rules of the resulting Poincar\'e supergravity multiplet 
\cite{Muller_hyper:1986ts}
are those that preserve the previous gauge conditions of eqs.~\eqref{gauge-conditions}.
To preserve the gauge condition
\eqref{D-gauge-condition} we need to impose 
$\l_{\mathbb D}\equiv0$. 
Because $Q$-supersymmetry does not preserve the gauge, it is necessary 
to accompany these transformations with appropriate $S$-supersymmetry, $\rm U(1)_R$, special conformal, and $\rm SU(2)_R$
compensating transformations.
To preserve \eqref{S-gauge-condition}, by examining the transformations of eqs.~\eqref{transf-vector}
and \eqref{transf-deformedvector}, it is straightforward to show that any $Q$-supersymmetry transformation
has to be accompanied by a compensating  $S$-supersymmetry transformation with the following parameter
\bea
 \eta^i_\a =   &&  \frac{c}{2} \Big(2 (\s^{a b} \xi^i)_\a (F_{a b}\bar{\boldsymbol{\phi}} +\bar{\phi}  \bF_{a b}) + 2 (\s^{a b}\xi^i)_\a W^{ +}_{a b} (\bar{\phi}\bar{\boldsymbol{\phi}} ) - \frac{1}{2} \xi_\a{}_j (X^{ij}\bar{\boldsymbol{\phi}} + \bar{\phi}  \bM^{ij})\non \\
   &&
 + 2 \ri (\s^a \bar{\xi}^i)_\a (\de_a \phi\bar{\boldsymbol{\phi}} + \bar{\phi}  \de_a {\boldsymbol{\phi}} )  +  \l^{i}_{\a} (\bar{\xi}^j \bar{\bl}_j ) +  (\bar{\xi}^j \bar{\l}_j) \boldsymbol{\l}^{i}_{\a}\Big) ~.
\eea
To preserve the gauge condition \eqref{U(1)-gauge-condition}, by examining the transformations of eqs.~\eqref{transf-vector}
and \eqref{transf-deformedvector}, it is straightforward to show that any $Q$-supersymmetry transformation
has to be accompanied by a compensating  $\rm U(1)_R$-symmetry transformation with parameter
\bea
    \l_Y  = -\frac{\ri}{4 y}(\xi_i \l^{i} - \bar{\xi}^i \bar{\l}_i) ~.
\eea
A similar analysis shows that to preserve the gauge condition \eqref{K-gauge-condition} one needs to enforce non-trivial compensating 
special conformal $K$-transformations with a parameter $\l^a(\xi)$. However, because all the other supergravity fields
are conformal primaries (though not necessarily superconformal primaries) that do not transform under special conformal boosts, 
in practice, we will never have to worry about inserting the compensating $\l^a(\xi)$ parameter (whose expression is quite involved)
in any Poincar\'e supergravity transformations.
The last gauge fixing condition that is not preserved is 
\eqref{SU(2)-gauge-condition}.
It is straightforward to check that we can consistently have $\d q^{(i\underline{i})}=0$ by implementing a 
compensating $\rm SU(2)_R$ transformation with the following parameter
\bea
\lambda^{ij}(\xi)
= 
-\,\frac{\re^U}{2}\Big[
\xi^{(i}\rho^{j)} 
- \bar{\xi}^{(i}\bar{\rho}^{j)}
\Big]
~,
\eea
where 
$\rho^{i}=\d^i_{\underline{i}}\rho^{\underline{i}}$
and 
$\bar\rho_{i}=\d_i^{\underline{i}}\bar\rho_{\underline{i}}$.

The local super-Poincar\'e transformations of the fermionic fields after gauge fixing are given by:
\bsubeq
\bea
\d\lambda_\alpha^i
&=&
2 (\s^{a b} \xi^i)_\a F_{a b} + (\s^{a b}\xi^i)_\a W^{+}_{a b} \bar{\phi} 
- \frac{1}{2} \xi_\a{}_j X^{ij} + \frac{1}{2}\l^{a b}(\s_{a b} \l^i)_\a \nonumber \\
&&   + 2 \ri (\s^a \bar{\xi}^i)_\a (\cD'_a {\phi} + 2 \ri A_{a} {\phi} - \hf \psi_{a}{}^{\b}_{i} \l_{\b}^{ i}) + \frac{\re^U}{2}( \xi^{(i}\rho^{j)} - \bar{\xi}^{(i}\bar{\rho}^{j)}
) \l_{\a j} -\frac{1}{4 y}(\xi_k \l^{ k} - \bar{\xi}^k \bar{\l}_k) \l^i_\a \non \\
&& + 2c \left[2 (\s^{a b} \xi^i)_\a (F_{a b}\bar{\boldsymbol{\phi}} +\bar{\phi}  \bF_{a b}) + 2 (\s^{a b}\xi^i)_\a W^{ +}_{a b} (\bar{\phi}\bar{\boldsymbol{\phi}} ) - \frac{1}{2} \xi_\a{}_j (X^{ ij}\bar{\boldsymbol{\phi}} + \bar{\phi}  \bM^{i j})\right] {\phi} \non \\
   &&
+ 2c \left[ 2 \ri (\s^a \bar{\xi}^i)_\a \left(\big(\cD'_a {\phi} + 2 \ri A_{a} {\phi} - \hf \psi_{a}{}^{\b}_{i} \l_{\b}^{ i}\big)\bar{\boldsymbol{\phi}} 
+ \bar{\phi}  \big(\cD'_a \boldsymbol{\phi} + 2 \ri A_{a} \boldsymbol{\phi} - \hf \psi_{a}{}^{\b}_{i} \bl_{\b}^{ i}\big) \right) \right] {\phi}  \non 
\\
&&
+  2c \left[ \l^{ i}_{\a} (\bar{\xi}^j \bar{\bl}_j ) +  (\bar{\xi}^j \bar{\l}_j) \boldsymbol{\l}^{ i}_{\a}\right] {\phi} 
 ~, \\
\delta \psi_m{}^\alpha_i  
&=& 
\Big(2\partial_m \xi^\alpha_i  
+ {\omega}_m{}^{ab}(\xi_i\sigma_{ab})^\alpha 
+2\phi_m{}_{i}{}^{j}\xi^{\alpha}_{j} 
+ 2\ri A_m\xi^\alpha_i 
+ b_m\xi^\alpha_i
\Big)
-\frac{\ri}{2}(\bar{\xi}_{i}\tilde{\sigma}_m\s^{cd})^{\alpha}W^+_{cd}
\non\\
&&
-\frac{1}{2}\lambda^{ab}({\psi}_m{}_i\sigma_{ab})^\alpha 
-\,\frac{\re^U}{2}(
\xi_{(i}\rho_{j)} 
- \bar{\xi}_{(i}\bar{\rho}_{j)}
) \,\psi_m{}^{\alpha j}  - \frac{1}{4 y}(\xi_i \l^{ i} - \bar{\xi}^i \bar{\l}_i)\,\psi_m{}^\alpha_i
\non \\
&&
-\ri c ({\sigma}_m)^{\a}{}_{\ad}  \Big[ -2 (\tilde{\s}^{a b} \bar{\xi}_i)^\ad  ( \phi\bF_{a b} + F_{a b}  \boldsymbol{\phi}  + W^{-}_{a b} \phi \boldsymbol{\phi} ) 
    - \frac{1}{2} \bar{\xi}^\ad{}^j (X_{ij}  \boldsymbol{\phi}  + \phi \overline{\bM}_{ij}) \non \\ &&
    + 2 \ri (\tilde{\s}^a \xi_i)^\ad \big((\phi (\cD'_a\bar{\boldsymbol{\phi}} - 2 \ri A_{a}\bar{\boldsymbol{\phi}} - \hf \bpsi_{a}{}_{\bd}^{i} \bar{\bl}^{ \bd}_{i}) + (\cD'_a \bar{\phi}- 2 \ri A_{a} \bar{\phi}- \hf \bpsi_{a}{}_{\bd}^{i} \bar{\l}^{ \bd}_{i}) \boldsymbol{\phi}\big) \non 
    \\
    &&
    + (\xi_i \l^{ i}  ) \bar{\bl}^{\ad}_{i}   
    + \bar{\l}^{\ad}_{i}  (\xi_i \bl^{i})\Big] 
~, \\
    \delta \rho_{\a}^{\underline{i}} 
    &=& -4\ri(\sigma^a\bar{\xi}_k)_\a \nabla_a q^{k\underline{i}} +\hf \lambda_{ab} (\s^{ab} \rho^{\underline{i}})_{\a}    
     + \frac{1}{4 y}(\xi_i \l^{ i} - \bar{\xi}^i \bar{\l}_i) \rho_{\a}^{\underline{i}} \non 
     \\&& 
    + 4 c  \Big[2 (\s^{a b} \xi^i)_\a (F_{a b}\bar{\boldsymbol{\phi}} +\bar{\phi}  \bF_{a b}) + 2 (\s^{a b}\xi^i)_\a W^{ +}_{a b} (\bar{\phi}\bar{\boldsymbol{\phi}} ) - \frac{1}{2} \xi_\a{}_j (X^{ij}\bar{\boldsymbol{\phi}} + \bar{\phi}  \bM^{i j})\non \\
   &&
 + 2 \ri (\s^a \bar{\xi}^i)_\a \big((\cD'_a {\phi} + 2 \ri A_{a} {\phi} - \hf \psi_{a}{}^{\b}_{i} \l_{\b}^{ i})\bar{\boldsymbol{\phi}} + \bar{\phi}  (\cD'_a \boldsymbol{\phi} + 2 \ri A_{a} \boldsymbol{\phi} - \hf \psi_{a}{}^{\b}_{i} \bl_{\b}^{ i}) \big)  \non 
 \\
 && +  \l^{ i}_{\a} (\bar{\xi}^j \bar{\bl}_j ) +  (\bar{\xi}^j \bar{\l}_j) \boldsymbol{\l}^{ i}_{\a}\Big] q_i{}^{\underline{i}} ~,
\eea
\esubeq
together with their complex conjugates.
Note that we have only given the transformation of $\l_\a^i$ since the $S$-supersymmetry gauge condition \eqref{S-gauge-condition} implies that the other gaugino is not independent. 
In principle, we should substitute all the auxiliary fields equations of motion and gauge conditions, but we are mainly interested in the shift terms (where we ignore higher fermionic terms and tensorial structures, as, for example, $F_{ab}$, $W_{ab}$, etc.) as these are the ones to investigate the supersymmetry breaking pattern in the model. The resulting equations are
\bsubeq
\bea
\d\l_\alpha^{ i}
&=&
-  c y^2  \bM^{i j} \xi_\a{}_j   + \cdots
~,
\\
\delta \psi_m{}^\alpha_i  
&=& 
\frac{c y\ri}{2} ({\sigma}_m)^{\a}{}_{\ad} 
     \overline{\bM}_{ij}\bar{\xi}^\ad{}^j 
    + \cdots
~,
\\
\delta \rho_{\a}^{\underline{i}} 
    &=&
    -2  c y  q^{i\underline{i}} \bM_{i j}\xi_\a^j
    + \cdots~.
\eea
\esubeq
We see immediately that all these terms are proportional to a single complex matrix, $\bM_{ ij}$. 
To achieve partial supersymmetry breaking and zero vacuum energy, we impose a vanishing determinant of this matrix, $\bM_{i j}$. This implies the following expression is zero
\begin{align}
    \bM_{ij} \bM^{ ij} = \frac{1}{c^2} \xi_{\underline{i} \underline{j}} \xi^{\underline{i} \underline{j}} - \frac{1}{4} \zeta_{\underline{i} \underline{j}} \zeta^{\underline{i} \underline{j}} - \frac{1}{c} \ri \xi_{\underline{i} \underline{j}}\zeta^{\underline{i} \underline{j}}   = 0~,
\end{align}
thereby giving the two independent conditions on the magnetic and electric deformation parameters
\bsubeq
\begin{align}
    &\frac{4}{c^2} \xi_{\underline{i} \underline{j}} \xi^{\underline{i} \underline{j}} - \zeta_{\underline{i} \underline{j}} \zeta^{\underline{i} \underline{j}} = 0 ~, \\
    &  \xi_{\underline{i} \underline{j}}\zeta^{\underline{i} \underline{j}} = 0 ~.
\end{align}
\esubeq
It follows that for generic solutions of the previous equations, $\det{\bM}=0$ while the rank of the matrix $\bM$ is one.
This is irrespective of the values of the scalars in the model.
In this case, we have one supersymmetry preserved and one broken on a Minkowski vacuum, as only one of the two supersymmetry transformations has a local shift term.

\subsection{On-shell theory and fermionic mass matrix} \label{subsec:onshell}

The on-shell component action for the ${\rm SU(1,1)/U(1)}$ model can be derived by substituting the equations of motion for all auxiliary fields as given in eqs.~\eqref{EOMXijSU(1,1)/U(1)model} and \eqref{EOM_Wab_AaSU(1,1)/U(1)model} into the off-shell action of eq.~\eqref{SU(1,1)/U(1)_action_hdw_boson}. This is followed by the imposition of gauge-fixing conditions as seen in eqs.~\eqref{gauge-conditions} of the previous subsection. We give here some details of this process.

After imposing the U(1)$_\textrm{R}$ gauge condition, the $\mathbb{D}$-gauge simplifies to the following:
\begin{equation}  \label{eq:D+U1}
    y (\bar{\boldsymbol{\phi}} + \boldsymbol{\phi}) = -\frac{1}{2c} ~.
\end{equation}
This implies that the real part of $\boldsymbol{\phi}$ is nonzero being a unique characteristic of this sigma model in the symplectic frame chosen. We should also then interpret this condition as requiring $y$ to be a function of the real part of the physical vector multiplet field while its imaginary part remains independent. That is,
\begin{equation} \label{eq:y}
    y := y(\boldsymbol{\phi},\bar{\boldsymbol{\phi}}) = -\frac{1}{2c\left(\boldsymbol{\phi}+\bar{\boldsymbol{\phi}}\right)} = -\frac{1}{4c \, \textrm{Re}\boldsymbol{\phi}}  ~.
\end{equation}
One should therefore interpret $y$ as in eq.~\eqref{eq:y} in all equations that follow. Next, one algebraically solve for $W_{\a \b}$ and $\bar{W}_{\ad \bd}$ in terms of the other independent fields by using \eqref{eq:WabEOM} and \eqref{eq:bWabEOM}.
After imposing the U(1)$_\textrm{R}$ gauge condition, we will have the inverse vector multiplet field and its conjugate, $\boldsymbol{\phi}^{-1}$ and $\bar{\boldsymbol{\phi}}^{-1}$, in our action as a direct result of integrating out these auxiliary fields. Remember that eq.~\eqref{eq:D+U1} requires the real part of $\boldsymbol{\phi}$ to be nonzero, so their presence is not problematic. This condition both on the compensator and physical fields is a feature of the SU(1,1)/U(1) target space sigma model in the symplectic frame that we have chosen which admits the holomorphic prepotential \eqref{SU(1,1)/U(1)model}.
It is also straightforward to obtain the algebraic expression for $A_a$ in terms of the independent component fields by solving \eqref{eq:AaEOM} after imposing all the gauge fixings.
Finally, also note that the $S$-gauge \eqref{S-gauge-condition} can be applied in the following way
\begin{equation}
    \boldsymbol{\l^i_\a}= -y^{-1} \bar{\boldsymbol{\phi}} \l^i_{\a} ~,
\end{equation}
thereby removing $\boldsymbol{\l}^i_\a$ from the final result. At the end of all these substitutions, the bosonic part of the on-shell action is
\begin{align} \label{eq:onshellbosons}
    \cL_{\textrm{bosons}}=&{} - \frac{1}{2}R+\frac{1}{2}c {y}^{-1} \bar{\boldsymbol{\phi}} f_{a b} f^{a b}+\frac{1}{4}{\rm i} c {y}^{-1} \epsilon_{a b c d} \bar{\boldsymbol{\phi}} f^{a b} f^{c d}  +\frac{1}{2}c y {\bar{\boldsymbol{\phi}}}^{-1} \textbf{f}_{a b} \textbf{f}^{a b}
    \non \\ & 
   +\frac{1}{4}{\rm i} c y \epsilon_{a b c d} {\bar{\boldsymbol{\phi}}}^{-1} \textbf{f}^{a b} \textbf{f}^{c d}+\frac{1}{2}c {y}^{-1} \boldsymbol{\phi} f_{a b} f^{a b} 
    - \frac{1}{4}{\rm i} c {y}^{-1} \epsilon_{a b c d} \boldsymbol{\phi} f^{a b} f^{c d}
    \non \\ & 
    +\frac{1}{2}c y {\boldsymbol{\phi}}^{-1} \textbf{f}_{a b} \textbf{f}^{a b}  - \frac{1}{4}{\rm i} c y \epsilon_{a b c d} {\boldsymbol{\phi}}^{-1} \textbf{f}^{a b} \textbf{f}^{c d}
   +c y \mathcal{D}'_{a}\mathcal{D}'^{a}
    \left({\boldsymbol{\phi}}+{\bar{\boldsymbol{\phi}}}\right)
    \non \\ & 
    +{c}^{2} {y}^{2} \mathcal{D}'_{a}{\bar{\boldsymbol{\phi}}} \mathcal{D}'^{a}{\bar{\boldsymbol{\phi}}}-2{c}^{2} {y}^{2} \mathcal{D}'_{a}{\boldsymbol{\phi}} \mathcal{D}'^{a}{\bar{\boldsymbol{\phi}}}+{c}^{2} {y}^{2} \mathcal{D}'_{a}{\boldsymbol{\phi}} \mathcal{D}'^{a}{\boldsymbol{\phi}}
    \non \\ & 
+16{e}^{4U} {\tilde{h}}^{a i j} {\tilde{h}}_{a i j} 
-2\mathcal{D}'_{a}{U} \mathcal{D}'^{a}{U}+2\mathcal{D}'_{a}{\mathcal{D}'^{a}{U}} +\frac{4}{3} \e^{m n p q} \xi_{i j}  h_{m np}{}^{i j} v_{q}~.
\end{align}
Note that the Lagrangian has no scalar potential, as in the analogue model described in \cite{Cecotti:1985sf,Ferrara:1995gu}.
The rest of the Lagrangian, including all fermionic terms, is given in Section IV of the supplementary file. Here we only present the fermionic mass terms, meaning quadratic terms in the fermions that do not have any derivative coupling or coupling to fields other than scalars. These terms take the form
\begin{align}\label{eq:onshell2ferms}
    \cL_{\textrm{fermions~mass~terms}}=& 
    -\frac{1}{16}c y \overline{\bM}_{ij} e^{2U}    \rho^{i \alpha} \rho^{j}\,_{\alpha} - \frac{1}{16} c y{\bM}^{ij} e^{2U} \overline{\rho}_{i \dot{\alpha}} \overline{\rho}_{j}\,^{\dot{\alpha}}  \non 
    \non \\ & 
      + \frac{1}{4}c {e}^{U} \overline{\bM}_{i j} \lambda^{j \alpha} \rho^{i}\,_{\alpha}
     +\frac{1}{4} c {e}^{U} \bM^{i j} \bar{\lambda}_{j \dot{\alpha}} \overline{\rho}_{i}\,^{\dot{\alpha}} 
    \non \\ &
    - \frac{1}{4} {\rm i}c  \bM_{i j} (\sigma_{a})^{\alpha}{}_{\dot{\alpha}} \psi^{a}{}^{j}_{\alpha} \bar{\lambda}^{i \dot{\alpha}}-\frac{1}{4} {\rm i} c  \overline{\bM}_{i j} (\sigma_{a})^{\alpha}{}_{\dot{\alpha}} \bar{\psi}^{a}{}^{j \dot{\alpha}} \lambda^{i}\,_{\alpha} 
    \non \\ &
   +\frac{1}{4}{\rm i} c y {e}^{U} \bM_{i j} (\sigma_{a})^{\alpha}{}_{\dot{\alpha}} \psi^{a}{}^{j}_{\alpha} \overline{\rho}^{i \dot{\alpha}} + \frac{1}{4} {\rm i} c y {e}^{U} \overline{\bM}_{i j} (\sigma_{a})^{\alpha}{}_{\dot{\alpha}} \bar{\psi}^{a}{}^{j \dot{\alpha}} \rho^{i}\,_{\alpha}
    \non \\ &
     - \frac{1}{2} c y  \bM_{i j} (\sigma_{ab})^{\alpha\beta} \psi^{a}{}^{i}_{\alpha} \psi^{b}{}^{j}_{\beta} - \frac{1}{2} c y \overline{\bM}^{i j} (\tilde{\sigma}_{ab})_{\dot{\alpha}\dot{\beta}} \bar{\psi}^{a}{}_{i}^{\dot{\alpha}} \bar{\psi}^{b}{}_{j}^{\dot{\beta}}
 ~.
\end{align}
They are all proportional to either $\bM_{ij}$ or its complex conjugate $\overline{\bM}^{ij}$. This is expected from the analysis in subsection \ref{section-6.2}, where we did show that in the on-shell theory, for any scalar field configurations, half of the local supersymmetry is spontaneously broken while half is preserved.  Imposing the zero determinant (but rank one) condition that we discussed in the previous subsections implies that half of the fermions remain massless, in agreement with local partial supersymmetry breaking in a Minkowski vacuum. 

We repeat that the complete action up to all orders in fermions is given in Section IV of the supplementary file accompanying our paper.
This notably includes the coupling between two gravitini and the three-form fields $h_{abc}{}^{\underline{i}\underline{j}}$ that appear through their Hodge duals, $\tilde{h}_{a}{}^{\underline{i}\underline{j}}$. 
 Remember that, roughly, in the hyper-dilaton Weyl multiplet $\tilde{h}_{a}{}^{\underline{i}\underline{j}}$ takes the place of the ${\rm SU(2)_R}$ connection $\phi_a{}^{ij}$ which within the supergravity-matter systems engineered in terms of the standard Weyl multiplet (where matter fields have ubiquitous couplings with $\phi_a{}^{ij}$) becomes a linear combination of the vector multiplet gauge connections together with terms arising from the hypermultiplet moment map of the quaternion-K\"ahler geometry. In the hyper-dilaton Weyl case, there is no gauging and only the coupling with $\tilde{h}_{a}{}^{\underline{i}\underline{j}}$ appears in place of FI-type terms. It would be interesting in the future to explore in more detail such property of this off-shell engineering of 4D $\cN=2$ supergravity-matter systems.


\section{Conclusion and outlook} \label{sec:conclusion}

In this paper, we have elaborated on the deformation of off-shell vector multiplets in supergravity, both in components and superspace. In a superconformal framework, the deformations are associated with (composite) linear multiplets. Analogue to the globally supersymmetric case where an interplay of electric and magnetic deformations can lead to (partial) breaking of $\cN=2$ global supersymmetry for systems of vector multiplets \cite{Fayet:1974jb,Fayet:1975yi,Antoniadis:1995vb,IZ1,IZ2}, the aim of our work was to explore the off-shell engineering of local partial supersymmetry breaking. To construct new off-shell models, we made use of superconformal tensor calculus techniques where the multiplet of conformal supergravity was chosen to be the hyper-dilaton Weyl multiplet introduced in 2022 in \cite{Gold:2022bdk}, while general off-shell vector multiplets were deformed with what proves to be local analogous to global electric and magnetic FI terms. The hyper-dilaton Weyl multiplet was chosen since it naturally contains a triplet of composite linear multiples, and one can easily engineer non-parallel deformations, a prerequisite to obtaining partial supersymmetry breaking, in a fashion very similar to the global case of \cite{Antoniadis:1995vb,IZ1,IZ2}. As a proof of concept, in this work, we did show that by considering the SU(1,1)/U(1) special-K\"ahler sigma model, originally employed in \cite{Cecotti:1985sf,Ferrara:1995gu}, however working in a symplectic frame which admits a holomorphic prepotential given by \eqref{SU(1,1)/U(1)model}, and with both electric and magnetic deformations appropriately turned on, we obtain local partial supersymmetry breaking.

It is inspiring that an off-shell model with partial supersymmetry breaking can be engineered by the hyper-dilaton Weyl multiplet and off-shell deformations, as there is potential in extending this simple example to more complicate supergravity-matter couplings. Our set-up is related to the off-shell work of M\"uller from 1986  \cite{Muller_hyper:1986ts} that has not been appreciated so far. We expect our results can be extended in various directions.

First of all, it would be interesting to extend the analysis of Section \ref{sec:SU11U1} to general special-K\"ahler target spaces leading to scalar potentials and also with more physical vector multiplets. In principle, one could revisit all the analyses performed in the past, see, e.g., \cite{Ferrara:1995gu,Ferrara:1995xi,Fre:1996js,Louis1,Louis:2012ux,Louis2,Andrianopoli:2015wqa,Antoniadis:2018blk}, by employing the off-shell setting offered by vector multiplets in a hyper-dilaton Weyl background. This might provide a new description of sectors of compactified string theories with fluxes and their various patterns of supersymmetry breaking, see, e.g., \cite{Louis:2012ux} and references therein.

Several features of our analysis resemble the global partial supersymmetry breaking APT model. It is natural to expect that one can obtain this theory as the global limit of our construction, as it was done with a different setting in \cite{Ferrara:1995xi} and more recently revisited in \cite{Antoniadis:2018blk} for the case of a single physical vector multiplet. We aim to look at the global limit of our construction in the future.

It would also be intriguing to understand if and how the dilaton and triplet of gauge two-forms sector of our models, and M\"uller's Poincar\'e supergravity, is mapped to the quaternion-K\"ahler hypermultiplet target spaces that characterise the work in \cite{Ferrara:1995gu,Ferrara:1995xi,Fre:1996js,Antoniadis:2018blk}. For example, at least at the level of the on-shell Lagrangians, the ${\rm SO(4,1)/SO(4)}$ hypermultiplet sector could arise by taking our model and dualising the triplet of physical gauge two-forms into scalars which could then organise with the dilaton to parametrise the conventional target space with two isometries. For general models based on the hyper-dilaton Weyl multiplet, it also remains unclear how various fermions would become charged in accordance with the standard gauging in $\cN=2$ supergravity, see, e.g., \cite{Lauria:2020rhc,DAuria:1990qxt,Andrianopoli:1996vr,Andrianopoli:1996cm,DallAgata:2003sjo,Trigiante:2016mnt,VanProeyen:2004xt}. It is certainly an interesting option to further analyse, in $\cN=2$ supergravity, this and the surprising and intriguing mechanism that leads to scalar potentials without gauging the ${\rm SU(2)_R}$ symmetry in the hyper-dilaton and M\"uller frameworks. Moreover, perhaps similar, so far missed, structures might be found also in $\cN>2$ extended supergravities.

As repeatedly mentioned, the fact that we have an off-shell construction allows us to straightforwardly extend our model without changing the local transformations of the multiplets. For example, it would be possible to add higher-derivative actions to these models based on the hyper-dilaton Weyl and the hyper-dilaton Poincar\'e multiplets. Higher-derivative supergravity naturally arise in the low-energy description of string theory but, despite its importance, is still poorly understood. Other types of dilaton Weyl multiplets have been key to the construction of several off-shell higher-derivative supergravities in $4\leq{\rm D}\leq 6$ dimensions, see, e.g., \cite{BSS1,CVanP,Bergshoeff:2012ax,OP131,OP132,OzkanThesis,Butter:2014xxa,NOPT-M17,Butter:2018wss,Mishra:2020jlc}. One can look at this problem starting from a hyper-dilaton Weyl multiplet coupled to systems of vector multiplets with electric and magnetic FI-type terms. Among higher-derivative couplings, it would also be interesting to study the interplay between the FI-type terms used in our paper and the new $\cN=2$ FI term introduced in \cite{Antoniadis:2019hbu}.

To conclude, it would be interesting to engineer other off-shell constructions for local partial supersymmetry breaking. In fact, though we do value the simplicity of the deformations in our hyper-dilaton Weyl set-up (which works well for constructions with vector multiplets but no arbitrary sector for physical charged hypermultiplets), an approach based on the standard Weyl multiplet and different types of off-shell matter multiplets would be welcome. 
In our paper, we have avoided the option of admitting the (composite) linear multiplets to be charged under central charge transformations, see for example \cite{deWit:1984rvr}. It would be interesting to understand in detail if and how off-shell magnetic deformations could be implemented when charged under the action of a gauged central charge in a standard Weyl multiplet background.
It is also worth mentioning that the most general $\cN=2$ supergravity-matter couplings are expected to be engineered off shell by coupling the standard Weyl multiplet to matter multiplets defined by harmonic or projective superfields \cite{Galperin:1984av,Galperin:1987ek,Galperin:1987em,Galperin:2001seg,KLR,LR3,LR:SYM,Kuzenko:SPH,Kuzenko:2008ep,Kuzenko:2008qz,Kuzenko:2009zu,Butter:2011sr,Butter:2012xg,Butter:2014gha,Butter:2014xua,Butter:2015nza}, which can include an infinite number of auxiliary fields. Though in this setup the electric gauging has been studied, to the best of our knowledge, the off-shell magnetic one has not. If one wanted to construct some composite multiplet similar to the linear ones but quadratic in a hypermultiplet with an infinite number of auxiliary fields, then an off-shell ``magnetic'' deformation would be associated with an extended vector multiplet field strength that should also have an infinite amount of component fields. This might lead to some new extended vector multiplet, and possibly new implementations of a central charge in the supergravity algebra. 
If understood, all these off-shell extensions might lead to new mechanisms of partial supersymmetry breaking, possibly even engineered with only (extended) vector multiplets and no physical hypers, in line with some of the on-shell results obtained in \cite{Altendorfer:1999mn,Altendorfer:1999qb}. We hope to come back to these various questions in the future.

\vspace{0.3cm}
\noindent
{\bf Acknowledgements:}\\
We thank W.~Kitchin for collaboration at the early stage of this project and J.~Hutomo for collaboration on related projects.
G.\,T.-M. is also grateful to I.~Antoniadis, J.-P.~Derendinger, L.~Girardello, F.~Farakos, H.~Jiang, S.~Kuzenko, and A.~Van Proeyen for several discussions and collaborations on topics related to this work.
We are also grateful to F.~Farakos and S.~Kuzenko for the careful reading and the useful feedback on our manuscript.
This work has been supported by the Australian Research Council (ARC) Future Fellowship FT180100353, ARC Discovery
Project DP240101409, and the Capacity Building Package of the University of Queensland.
G.\,G. and S.\,K. have been supported by postgraduate scholarships at the University of Queensland.
We acknowledge the kind hospitality and financial support at the MATRIX Program “New Deformations of Quantum Field and Gravity
Theories,” that took place in Creswick (Australia) between 22 January and 2 February 2024, and the meeting ``Integrability in low-supersymmetry theories,'' held in Trani between 22 July and 2 Aug 2024 and funded by the COST Action CA22113 by INFN and by Salento University, where part of this work was performed.

\appendix


\section{Notations and conventions} \label{NC}

Our notations and conventions follow mostly those in \cite{Buchbinder-Kuzenko}. We briefly summarize them here.

We use two-component notation where
dotted and undotted spinor indices are raised and lowered by $\ve$ tensors
\be
\psi_\a = \ve_{\a\b} \psi^\b \ , \quad \bar{\chi}^\ad = \ve^{\ad\bd} \bar{\chi}_\bd \ ,
\ee
obeying
\be \ve_{\a\b} = - \ve_{\b\a} \ , \quad \ve_{\a\b} \ve^{\b\g} = \d_\a^\g \ , \quad \ve_{\ad\bd} \ve^{\bd\gd} = \d_\ad^\gd \ , \quad \ve^{12} = 1 \ . \non\\
\ee
Similarly ${\rm SU}(2)$ indices are raised and lowered by $\ve_{ij}$ and $\ve^{ij}$ having the same properties as $\ve_{\a\b}$.
Spinor indices are contracted as
\begin{align}
\psi \chi &:= \psi^\a \chi_\a \ , \quad \bar{\psi} \bar{\chi} = \bar{\psi}_\ad \bar{\chi}^\ad~.
\end{align}
For spinors which are also isospinors, we define
\begin{align}
\psi \chi &= \psi^\a_i \chi_\a^i \ , \quad \bar{\psi} \bar{\chi} = \bar{\psi}_\ad^i \bar{\chi}^\ad_i \ .
\end{align}

The metric is $\eta_{ab} = \textrm{diag}(-1, 1,1,1)$. The sigma matrices are defined as
\be
(\s^a)_{\a\ad} = (1, \vec{\s}) \ , \quad (\tilde{\s}^a)^{\ad\a} = \ve^{\ad\bd} \ve^{\a\b} (\s^a)_{\b\bd} = (1 , - \vec{\s}) ~,
\ee
and have the properties
\begin{align}
(\s_a)_{\a\bd} (\tilde{\s}_b)^{\bd\b} = - \eta_{ab} \d^\b_\a - 2 (\s_{ab})_\a{}^\b \ , \\
(\tilde{\s}_a)^{\ad\b} (\s_b)_{\b\bd} = - \eta_{ab} \d^\ad_\bd - 2 (\tilde{\s}_{ab})^\ad{}_\bd \ ,
\end{align}
together with the following useful identities
\begin{subequations}
\begin{align*}
(\s^a)_{\a \ad} (\s_a)_{\b \bd} &= - 2 \ve_{\a\b} \ve_{\ad \bd}~, \\
(\s_{ab})_{\a\b} (\s^{ab})_{\g \d} &= -2 \ve_{\g (\a} \ve_{\b) \d}~, \\
(\tilde{\s}_{ab})_{\ad\bd} (\tilde{\s}^{ab})_{\gd \dd} &= -2 \ve_{\gd (\ad} \ve_{\bd) \dd}~, \\
(\s_{ab})_{\a\b} (\tilde{\s}^{ab})_{\gd \dd} &= 0~, \\
\tr(\s_{ab} \s_{cd}) &= (\s_{ab})_\a{}^\b (\s_{cd})_\b{}^\a = - \eta_{a[c} \eta_{d] b} - \frac{\ri}{2} \ve_{abcd}~, \\
\tr(\tilde{\s}_{ab} \tilde{\s}_{cd}) &= (\tilde{\s}_{ab})_\ad{}^\bd (\tilde{\s}_{cd})_\bd{}^\ad = - \eta_{a[c} \eta_{d] b} + \frac{\ri}{2} \ve_{abcd}~, \\
(\s^{a})_{\a}{}^{\ad} (\s_{ab})_{\b\g} &= \e_{\a (\b} (\s_b)_{\g)}{}^{\ad}~,\\
(\s^{a})_{\a}{}^{\ad} (\tilde{\s}_{ab})^{\bd\gd} &= \e^{\ad (\bd} (\s_b)_{\a}{}^{\gd)}~,\\
(\s_{[a})_{\a \bd} (\s_{bc]})_{\g\d} &= \frac{\ri}{3} \ve_{abcd} \ve_{\a (\g} (\s^d)_{\d)\bd}~, \\
\ve^{abcd} \ve_{a' b' c' d'} &= - 4! \d^a_{[a'} \d^b_{b'} \d^c_{c'} \d^d_{d']}~, \\
(\s_{ab} \s_c)_\a{}^\ad &= (\s_{ab})_\a{}^\b (\s_c)_\b{}^\ad = - \eta_{c[a} (\s_{b]})_\a{}^\ad - \frac{\ri}{2} \ve_{abcd} (\s^d)_\a{}^\ad~, \\
(\tilde{\s}_{ab} \s_c)^\ad{}_\a &= (\tilde{\s}_{ab})^\ad{}_\bd (\s_c)_\a{}^\bd = - \eta_{c[a} (\s_{b]})_\a{}^\ad + \frac{\ri}{2} \ve_{abcd} (\s^d)_\a{}^\ad~, \\
\ve_{abcd} (\s^{cd})_{\a \b} &= -2 i (\s_{ab})_{\a\b} \ , \qquad \ve_{abcd} (\tilde{\s}^{cd})_{\ad \bd} = 2 i (\tilde{\s}_{ab})_{\ad\bd} \ .
\end{align*}
\end{subequations}
The antisymmetric tensor is
\be \ve^{0123} = - \ve_{0123} = 1 \ ,
\ee
and (anti-)symmetrization includes a normalization factor, for example
\be V_{[ab]} = \frac{1}{2!} (V_{ab} - V_{ba}) \ , \quad \psi_{(\a\b)} = \frac{1}{2!} (\psi_{\a\b} + \psi_{\b\a}) \ .
\ee
For superform indices, we introduce graded antisymmetrization, e.g.,
\begin{align}
V_{[AB\}} = \frac{1}{2!} (V_{AB} - (-)^{ab} V_{BA})~.
\end{align}
When an index is not included, we separate it with vertical bars, e.g.,
\begin{align}
T_{[AB}{}^D F_{|D|C\}} &= \frac{1}{3!} \Big(T_{AB}{}^D F_{DC}
	- (-)^{ab} T_{BA}{}^D F_{DC}
	+ (-)^{ca+cb} T_{CA}{}^D F_{DB}
	\eol & \quad
	- (-)^{cb} T_{AC}{}^D F_{DB}
	+ (-)^{ab+ac} T_{BC}{}^D F_{DA}
	- (-)^{ab+ac+cb} T_{CB}{}^D F_{DA}\Big)~.
\end{align}

A vector $V_a$ can be rewritten with spinor indices as
\be
V_{\a\bd} = (\s^a)_{\a\bd} V_a \ , \quad V_a = - \hf (\tilde{\s}_a)^{\bd\a} V_{\a\bd} \ .
\ee
A real antisymmetric tensor, $F_{ab} = - F_{ba}$ is converted to spinor indices as
\be
F_{\a\b} = \hf (\s^{ab})_{\a\b} F_{ab} \ , \quad \bar{F}_{\ad\bd} = - \hf (\tilde{\s}^{ab})_{\ad\bd} F_{ab} \ ,\quad
F_{ab} = (\s_{ab})^{\a\b} F_{\a\b} - (\tilde{\s}_{ab})_{\ad\bd} \bar{F}^{\ad\bd} \ .
\ee

\section{Conformal superspace} \label{Appendix-B}

In this appendix, we collect results about conformal superspace relevant to our discussion in the paper. The Lorentz generators obey
\begin{align}
[M_{ab}, M_{cd} ] &= 2 \eta_{c [a} M_{b] d} - 2 \eta_{d[a} M_{b] c}~,\quad
[M_{ab}, \nabla_c ] = 2 \eta_{c [a} \nabla_{b]}~, \non \\
[M_{ab}, \nabla_\a^i] &= (\s_{ab})_\a{}^\b \nabla_\b^i ~,\quad
[M_{ab}, \bar\nabla^\ad_i] = (\tilde{\s}_{ab})^\ad{}_\bd \bar\nabla^\bd_i~.
\end{align}
The $\rm SU(2)_R$, $\rm U(1)_R$, and dilatation generators obey
\begin{align}
[J_{ij}, J_{kl}] &= -\ve_{k(i} J_{j)l} - \ve_{l(i} J_{j) k}~, \quad
[J_{ij}, \nabla_\a^k] = - \d^k_{(i} \nabla_{\a j)} ~,\quad
[J_{ij}, \bar\nabla^\ad_k] = - \ve_{k (i} \bar\nabla^{\ad}_{j)}~, \non \\
[Y, \nabla_\a^i] &= \nabla_\a^i ~,\quad [Y, \bar\nabla^\ad_i] = - \bar\nabla^\ad_i~,  \non \\
[\mathbb{D}, \nabla_a] &= \nabla_a ~, \quad
[\mathbb{D}, \nabla_\a^i] = \hf \nabla_\a^i ~, \quad
[\mathbb{D}, \bar\nabla^\ad_i] = \hf \bar\nabla^\ad_i ~.
\end{align}
The special superconformal generators $K^A$ transform in the obvious
way under Lorentz and $\rm SU(2)_R$ rotations,
\begin{align}
[M_{ab}, K_c] &= 2 \eta_{c [a} K_{b]} ~, \quad
[M_{ab} , S^\g_i] = - (\s_{ab})_\b{}^\g S^\b_i ~, \quad
[M_{ab} , \bar S_\gd^i] = - (\ts_{ab})^\bd{}_\gd \bar S_\bd^i~, \non \\
[J_{ij}, S^\g_k] &= - \ve_{k (i} S^\g_{j)} ~, \quad
[J_{ij}, \bar{S}^k_\gd] = - \d^k_{(i} \bar{S}_{\gd j)}~,
\end{align}
while their transformation under $\rm U(1)_R$ and dilatations is opposite
that of $\nabla_A$:
\begin{align}
[Y, S^\a_i] &= - S^\a_i ~, \quad
[Y, \bar{S}^i_\ad] = \bar{S}^i_\ad~, \non \\
[\mathbb{D}, K_a] &= - K_a ~, \quad
[\mathbb{D}, S^\a_i] = - \hf S^\a_i ~, \quad
[\mathbb{D}, \bar{S}_\ad^i] = - \hf \bar{S}_\ad^i ~.
\end{align}
Among themselves, the generators $K^A$ obey the algebra
\begin{align}
\{ S^\a_i , \bar{S}^j_\ad \} &= 2 \ri \d^j_i (\s^a)^\a{}_\ad K_a~.
\end{align}

Finally, the algebra of $K^A$ with $\nabla_B$ is given by
\begin{align}
[K^a, \nabla_b] &= 2 \delta^a_b \mathbb{D} + 2 M^{a}{}_b ~,\non \\
\{ S^\a_i , \nabla_\b^j \} &= 2 \d^j_i \d^\a_\b \mathbb{D} - 4 \d^j_i M^\a{}_\b 
- \d^j_i \d^\a_\b Y + 4 \d^\a_\b J_i{}^j ~,\non \\
\{ \bar{S}^i_\ad , \bar{\nabla}^\bd_j \} &= 2 \d^i_j \d^\bd_\ad \mathbb{D} 
+ 4 \d^i_j \overline{M}_\ad{}^\bd + \d^i_j \d_\ad^\bd Y - 4 \d_\ad^\bd J^i{}_j ~,\non \\
[K^a, \nabla_\b^j] &= -\ri (\s^a)_\b{}^\bd \bar{S}_\bd^j \ , \quad [K^a, \bar{\nabla}^\bd_j] = 
-\ri ({\s}^a)^\bd{}_\b S^\b_j ~, \non \\
[S^\a_i , \nabla_b] &= \ri (\s_b)^\a{}_\bd \bar{\nabla}^\bd_i \ , \quad [\bar{S}^i_\ad , \nabla_b] = 
\ri ({\s}_b)_\ad{}^\b \nabla_\b^i \ ,
\end{align}
where all other (anti-)commutations vanish.

The covariant derivatives obey (anti-)commutation relations of the form
\begin{align} [\nabla_A, \nabla_B\} 
         &= -T_{AB}{}^C \nabla_C - \hf R_{AB}{}^{cd} M_{cd} - R_{AB}{}^{kl} J_{kl}
	\eol & \quad
	- \ri R_{AB}(Y) Y - R_{AB} (\mathbb{D}) \mathbb{D} - R_{AB}{}^C K_C \ , \label{CDA1}
\end{align}
where $T_{AB}{}^C$ is the torsion, and $R_{AB}{}^{cd}$, $R_{AB}{}^{kl}$, $R_{AB}(Y)$, $R_{AB} (\mathbb{D})$ 
and $R_{AB}{}^C$ are the curvatures. Some of the components of the torsion and curvature must be constrained. Following 
\cite{Butter:2011sr}, the spinor derivative torsions and curvatures are chosen to obey
\begin{align}\label{fullConstraints}
\{ \nabla_\a^i , \nabla_\b^j \} = - 2 \ve^{ij} \ve_{\a\b} \bar{\cW}~,\quad
\{ \bar{\nabla}^\ad_i , \bar{\nabla}^\bd_j \} = 2 \ve_{ij} \ve^{\ad\bd} \cW~,\quad
\{ \nabla_\a^i , \bar{\nabla}^\bd_j \} = -2 \ri \d^i_j \nabla_\a{}^\bd ~,
\end{align}
where $\cW$ is some operator valued in the superconformal algebra.
In \cite{Butter:2011sr}, it was shown how to constrain $\cW$ entirely in terms
of a superfield $W_{\a\b}$ so that the component structure reproduces
$\cN=2$ conformal supergravity. In our notation, the constraints lead to
\begin{subequations}\label{CSGAlgebra}
\begin{align}
\{ \nabla_\a^i , \nabla_\b^j \} &= 2 \ve^{ij} \ve_{\a\b} \overline{W}_{\gd\dd} \overline{M}^{\gd\dd} + \hf \ve^{ij} \ve_{\a\b} \bar{\nabla}_{\gd k} \overline{W}^{\gd\dd} \bar{S}^k_\dd - \hf \ve^{ij} \ve_{\a\b} \nabla_{\g\dd} \overline{W}^\dd{}_\gd K^{\g \gd}~, \\
\{ \bar{\nabla}^\ad_i , \bar{\nabla}^\bd_j \} &= - 2 \ve_{ij} \ve^{\ad\bd} W^{\g\d} M_{\g\d} + \frac{1}{2} \ve_{ij} \ve^{\ad\bd} \nabla^{\g k} W_{\g\d} S^\d_k - \frac{1}{2} \ve_{ij} \ve^{\ad\bd} \nabla^{\g\gd} W_{\g}{}^\d K_{\d \gd}~, \\
\{ \nabla_\a^i , \bar{\nabla}^\bd_j \} &= - 2 \ri \d_j^i \nabla_\a{}^\bd~, \\
[\nabla_{\a\ad} , \nabla_\b^i ] &= - \ri \ve_{\a\b} \overline{W}_{\ad\bd} \bar{\nabla}^{\bd i} - \frac{\ri}{2} \ve_{\a\b} \bar{\nabla}^{\bd i} \overline{W}_{\ad\bd} \mathbb{D} - \frac{\ri}{4} \ve_{\a\b} \bar{\nabla}^{\bd i} \overline{W}_{\ad\bd} Y + \ri \ve_{\a\b} \bar{\nabla}^\bd_j \overline{W}_{\ad\bd} J^{ij}
	\eol & \quad
	- \ri \ve_{\a\b} \bar{\nabla}_\bd^i \overline{W}_{\gd\ad} \overline{M}^{\bd \gd} - \frac{\ri}{4} \ve_{\a\b} \bar{\nabla}_\ad^i \bar{\nabla}^\bd_k \overline{W}_{\bd\gd} \bar{S}^{\gd k} + \frac{1}{2} \ve_{\a\b} \nabla^{\g \bd} \overline{W}_{\ad\bd} S^i_\g
	\eol & \quad
	+ \frac{\ri}{4} \ve_{\a\b} \bar{\nabla}_\ad^i \nabla^\g{}_\gd \overline{W}^{\gd \bd} K_{\g \bd}~, \\
[ \nabla_{\a\ad} , \bar{\nabla}^\bd_i ] &=  \ri \d^\bd_\ad W_{\a\b} \nabla^{\b}_i + \frac{\ri}{2} \d^\bd_\ad \nabla^{\b}_i W_{\a\b} \mathbb{D} - \frac{\ri}{4} \d^\bd_\ad \nabla^{\b}_i W_{\a\b} Y + \ri \d^\bd_\ad \nabla^{\b j} W_{\a\b} J_{ij}
	\eol & \quad
	+ \ri \d^\bd_\ad \nabla^{\b}_i W^\g{}_\a M_{\b\g} + \frac{\ri}{4} \d^\bd_\ad \nabla_{\a i} \nabla^{\b j} W_\b{}^\g S_{\g j} - \hf \d^\bd_\ad \nabla^\b{}_\gd W_{\a\b} \bar{S}^{\gd}_i
	\eol & \quad
	+ \frac{\ri}{4} \d^\bd_\ad \nabla_{\a i} \nabla^\g{}_\gd W_{\b\g} K^{\b\gd} ~.
\end{align}
\end{subequations}
The complex superfield $W_{\a\b} = W_{\b\a}$ and its complex conjugate
${\overline{W}}_{\ad \bd} := \overline{W_{\a\b}}$ are superconformally primary,
$K_A W_{\a\b} = 0$, and obey the additional constraints
\begin{align}
\bar{\nabla}^\ad_i W_{\b\g} = 0~,\qquad
\nabla_{\a\b} W^{\a\b} &= \bar{\nabla}^{\ad\bd} \overline{W}_{\ad\bd} ~,
\end{align}
where we introduce the notation
\begin{align}
\nabla_{\a\b} := \nabla_{(\a}^k \nabla_{\b) k} \ , \quad
\bar{\nabla}^{\ad\bd} := \bar\nabla^{(\ad}_k \bar\nabla^{\bd) k} \ .
\end{align}
Despite the appearance of the $S$-supersymmetry and special conformal $K_a$ generators,
the algebra of covariant derivatives \eqref{CSGAlgebra} is significantly
simpler to work with than the corresponding algebras of $\rm SU(2)$ \cite{Grimm, Kuzenko:2008ep}
or $\rm U(2)$ superspace \cite{Howe1, Howe:1981gz , Kuzenko:2009zu}.

\section{$S$-supersymmetry transformations}\label{S-transformation}
The $S$-supersymmetry transformations of the descendant superfields of the super-Weyl tensor are given as follows:
\bsubeq 
\bea 
    && S_\d^k W_{\a\b\g}{}^{i}  = 8 \e^{k i} \e_{\d(\a} W_{\b \g)} ~, \quad
    \bar{S}^\dd_k W_{\a\b\g}{}^{i} = 0 ~, 
    \\ &&
     \bar{S}^\dd_k \overline{W}^{\ad \bd}{}^{\gd}_i = 8 \e_{k i} \e^{\dd (\ad} \overline{W}^{\bd \gd)}  ~,  \quad
     S_\d^k \overline{W}^{\ad \bd}{}^{\gd}_i  = 0
     ~,\\
    && S_\a^k \S_{\b}^{i}  = - \frac{4}{3} \e^{k i} W_{\a \b} ~, \quad
    \bar{S}^\ad_k \S_{\b}^{i} = 0 ~, 
    \\ &&
    \bar{S}^\ad_k \bar{\S}^{\bd}_i =  \frac{4}{3} \e_{k i} \overline{W}^{\ad \bd} ~,  
     \quad
    S_\a^k \bar{\S}^{\bd}_i  = 0 ~, \\
    && S_\l^k  W_{\a \b \g \d}  = 24  \e_{\l(\a} W_{\b \g \d)}{}^k ~, \quad
    \bar{S}^\ld_k  W_{\a \b \g \d} = 0 ~, 
     \\ 
     &&
    \bar{S}^\ld_k \overline{W}^{\ad \bd \gd \dd} = 24 \e^{\ld (\ad} \overline{W}^{\bd \gd \dd)}{}_{k} ~, \quad
     S_\l^k \overline{W}^{\ad \bd \gd \dd}  = 0 ~, \\
    && S_\l^k  \S_{\a \b}{}^{ij}  = \frac{28}{3} \e^{k (i} \e_{\l (\a} \S^{j)}_{\b)} + \frac{4}{3} \e^{k (i} W_{\a \b \l}{}^{j)} ~, \quad
    \bar{S}^\ld_k  \S_{\a \b}{}^{ij} = 0 ~, 
    \\ &&
     \bar{S}^\ld_k \bar{\S}^{\ad \bd}{}_{ij} = \frac{28}{3} \e_{k (i} \e^{\ld (\ad} \bar{\S}_{j)}^{\bd)} -\frac{4}{3} \e_{k (i} \overline{W}^{\ad \bd \ld}{}_{j)} ~, \quad
     S_\l^k \bar{\S}^{\ad \bd}{}_{ij}  = 0 ~,\\
    && S_\l^k  \S_{\a \b}  = \frac{32}{3} \e_{\l (\a} \S^{k}_{\b)} - \frac{4}{3} W_{\a \b \l}{}^{k} ~, \quad
    \bar{S}^\ld_k  \S_{\a \b} = 0 ~, 
     \\ &&
    \bar{S}^\ld_k \bar{\S}^{\ad \bd} = \frac{32}{3} \e^{\ld(\ad} \bar{\S}_{k}^{\bd)} + \frac{4}{3} \overline{W}^{\ad \bd \ld}{}_{k} ~, \quad
     S_\l^k \bar{\S}^{\ad \bd}  = 0 ~, \\
    && S_\l^k  D  = 0 ~, \quad
    \bar{S}^\ld_k  D = 0 ~, \\
    && S_\l^l   \S_{\a \b \g}{}^k  = \frac{2}{3} \e^{l k} W_{\a \b \g \l} + 3 \e^{l k} \e_{\l (\a} \S_{\b \g)} - 12 \e_{\l (\a} \S_{\b \g)}{}^{l k}~, \quad
    \bar{S}^\ld_l   \S_{\a \b \g}{}^k = 0 ~, 
    \\ &&
    \bar{S}^\ld_l \bar{\S}^{\ad \bd \gd}{}_k =  - \frac{2}{3} \e_{l k} \overline{W}^{\ad \bd \gd \ld} + 3 \e_{l k} \e^{\ld (\ad} \bar{\S}^{\bd \gd)} -12 \e^{\ld (\ad} \bar{\S}^{\bd \gd)}{}_{l k}  ~, \quad
     S_\l^l \bar{\S}^{\ad \bd \gd}{}_k  = 0 ~.
\eea 
\esubeq

The $S$-supersymmetry transformations of the descendants of the abelian vector multiplet are given as follows:
\begin{align}
    & S^\a_i \lambda^{j}_{\b} = 4 \d^\a_\b \d^j_i W~, \quad  \bar{S}^i_\ad \lambda^{j}_{\a} = 0~,    \quad S^\a_i \bar{\lambda}^{\ad}_{i} = 0~, \quad  \bar{S}^i_\ad \bar{\lambda}^{\bd}_{j} = 4 \d^\bd_\ad \d^i_j \overline{W}~, \\
    & S^\g_i F_{\a \b} = \hf \d^{\g}_{(\a} \l_{\b) i}~, \quad  \bar{S}^i_\ad F_{\a \b} = 0~,  \quad S^\g_i \bar{F}^{\ad \bd} = 0~, \quad  \bar{S}^i_\gd \bar{F}^{\ad \bd} = \hf \d^{(\ad}_{\gd} \l^{\bd) i}~, \\
    & S^\a_i X^{kl} = 0, \quad  \bar{S}^i_\ad X^{kl} = 0~.
\end{align}

The $S$-supersymmetry transformations of the descendants of the tensor multiplet are given as follows:
\begin{align}
    S^\a_i \chi_{\b j} &= 4 \d^\a_\b G_{i j}~, \quad \bar{S}^i_\ad \bar{\chi}^{\bd j} = 4 \d_\ad^\bd G^{i j}~, \quad S^\a_i \bar{\chi}^{\bd j} = 0~, \quad \bar{S}^i_\ad \chi_{\b j} =  0~, \\
    S^\a_i F &= -4 \chi^\a_i~, \quad  \bar{S}^i_\ad \bar{F} = -4 \bar{\chi}^i_\ad~, \quad  S^\a_i \bar{F} = 0~, \quad \bar{S}_\ad^i F = 0~, \\
    S^\a_i H_a &= \frac{3\ri}{4} (\s_a)^\a{}_\bd\bar{\chi}^\bd_i~, \quad \bar{S}^i_\ad H_a = -\frac{3 \ri}{4} (
  \bar{\s}_a)_\ad{}^\b \chi^i_\b~.
\end{align}

\section{Local superconformal transformations}\label{local-superconformal-transformation}
The local superconformal transformations of 
the fundamental fields of the standard Weyl multiplet
are given by
\bsubeq\label{transf-standard-Weyl}
\bea
\delta e_m{}^a 
&=& 
\ri\,\xi_i\sigma^a\bar{\psi}_m{}^i
+\ri\,\bar{\xi}^i \tilde\sigma^a\psi_m{}_{i}
- \lambda_{\mathbb{D}}{e}_m{}^a 
+ \lambda^{a}{}_{b}e_m{}^b
~,
\label{d-vielbein}
\\
\delta \psi_m{}^\alpha_i  
&=& 
\Big(2\partial_m \xi^\alpha_i  
+ {\omega}_m{}^{ab}(\xi_i\sigma_{ab})^\alpha 
+2\phi_m{}_{i}{}^{j}\xi^{\alpha}_{j} 
+ 2\ri A_m\xi^\alpha_i 
+ b_m\xi^\alpha_i
\Big)
-\frac{\ri}{2}(\bar{\xi}_{i}\tilde{\sigma}_m\s^{cd})^{\alpha}W^+_{cd}
\non\\
&&
-\frac{1}{2}\lambda^{ab}({\psi}_m{}_i\sigma_{ab})^\alpha 
- \lambda_{i}{}^{j} \,\psi_m{}^{\alpha}_{j}
- \ri\lambda_{Y}\,\psi_m{}^\alpha_i
- \frac{1}{2}\lambda_{\mathbb{D}}\,{\psi}_m{}^\alpha_i
+2\ri(\bar{\eta}_{i}\tilde{\sigma}_m)^{\alpha} 
~,
\label{d-gravitino}
\\
\delta 
\bar{\psi}_m{}_\ad^{i}  
&=&
\Big(
2\partial_m \bar{\xi}_\ad^{i}  
+{\omega}_m{}^{ab}(\bar{\xi}^{i} \tilde{\sigma}_{ab})_{\ad} 
- 2\phi_m{}^{i}{}_{j} \bar{\xi}_{\dot{\alpha}}^j 
- 2\ri A_m\bar{\xi}_\ad^{i} 
+ b_m \bar{\xi}_\ad^{i}
\Big)
+\frac{\ri}{2} (\xi^{i}{\sigma}_m\tilde{\s}^{cd})_{\dot{\alpha}} W^-_{cd}
\non
\\
&&
-\frac{1}{2}\lambda^{ab}({\bar{\psi}}_m{}^{i}\tilde{\sigma}_{ab})_{\ad}  
+ \lambda^{i}{}_{j} \,\bar{\psi}_m{}_{\dot{\alpha}}^j 
+ \ri\lambda_{Y} \,\bar{\psi}_m{}_\ad^{i}
- \frac{1}{2}\lambda^{\mathbb{D}}\,{\bar{\psi}}_m{}_\ad^{i}
+ 2\ri (\eta^{i}{\sigma}_m)_{\dot{\alpha}}
~,
\label{d-b-gravitino}
\\
 \delta  \phi_m{}^{ij} &=&
\Big(\partial_m\lambda^{ij} 
  -2\phi_m{}^{(i}{}_k \lambda^{j)k}
  \Big)
+\frac{3\ri}{2} \,\xi^{(i}{\sigma}_m\bar{\Sigma}^{j)}
 +\frac{3\ri}{2}\,\bar{\xi}^{(i}\tilde{\sigma}_m\Sigma^{j)}
 - \phi_m{}^{(i} \xi^{j)} 
+ \bar{\phi}_m{}^{(i}\bar{\xi}^{j)}
\non\\
&& 
 + 2 \psi_m{}^{(i}\eta^{j)} 
 - 2 \bar{\psi}_m{}^{(i} \bar{\eta}^{j)} 
~,
\label{d-SU2}
\\
\delta A_m 
&=& 
\partial_m\lambda_Y  
- \frac{3}{8}\,\xi_i{\sigma}_m\bar{\Sigma}^{i}
- \frac{3}{8}\,\bar{\xi}^i\tilde{\sigma}_m\Sigma_{i}
+ \frac{\ri}{2}\,\xi_i\phi_m{}^i
- \frac{\ri}{2}\,\bar{\xi}^i\bar{\phi}_m{}_i
- \frac{\ri}{2}\,\psi_m{}_i \eta^i
+\frac{\ri}{2}\,\bar{\psi}_m{}^i\bar{\eta}_i 
~,~~~~~~
\label{d-U1}
\\
\delta b_m 
&=& \partial_m \lambda_{\mathbb{D}} 
-\frac{3\ri}{4}\,\xi_i \sigma_m\bar{\Sigma}^{i} 
+ \frac{3\ri}{4}\,\bar{\xi}^i\tilde{\sigma}_m\Sigma_{i}
+\xi_i   \phi_m{}^i
+\bar{\xi}^i\bar{\phi}_m{}_i
- \psi_m{}_i\eta^i  
-\bar{\psi}_m{}^i\bar{\eta}_i 
- 2\lambda_m
~,
\label{d-dilatation}
\\
 \delta W_{ab} &=& 
-4\xi_{k}R(Q)_{ab}{}^k
-4\bar\xi^{k}R(\bar{Q})_{ab}{}_k
-2\lambda_{[a}{}^{c}W_{b]c}
  +\lambda_{\mathbb{D}} W_{ab}
  -2\ri \lambda_{Y} W^+_{ab}
  +2\ri \lambda_{Y} W^-_{ab}
  ~,~~~~~~~
  \label{d-W}
  \\
\delta  D
&=& 
-\ri \xi^{k} \s^a \de_{a}\bar{\Sigma}_k 
- \ri \bar{\xi}_k \tilde{\s}^a \de_a \Sigma^{k} 
+ 2 \l_{\mathbb{D}} D
~,
\label{d-D}
\\
\delta \Sigma^{\a i}
&=&      
\xi^{\a i }  D 
+ \frac{4\ri}{3} (\xi^{ i} \s^{ab})^\a R(Y)_{ab} 
+\frac{2}{3} (\xi_{j} \s^{ab})^\a R(J)_{ab}{}^{ j i} 
- \frac{\ri}{3} (\bar{\xi}^{i} \tilde{\sigma}^a\s^{cd})^\a
\nabla_{a}W^+_{cd}
\non\\ 
&& 
-\frac{1}{2} \lambda^{a b} (\Sigma^i \sigma_{ab})^{ \a} 
+\lambda^{i}{}_{ j}  \Sigma^{\a j} 
+\frac{3}{2}\lambda_{\mathbb{D}}
\Sigma^{\a i}
- \ri \lambda_{Y}\Sigma^{\a i}
+\frac{2}{3} (\eta^{i}\s^{cd})^{\alpha}W^+_{cd}
~,
\label{d-Sigma}
\\
\delta  \bar{\Sigma}_{\ad i}
&=& 
-\bar{\xi}_{\ad i} D 
+ \frac{4\ri}{3} (\bar{\xi}_{ i} \tilde{\s}^{ab})_{\ad} R(Y)_{ab} 
+ \frac{2}{3} (\bar{\xi}^{j} \tilde{\s}^{ab})_\ad R(J)_{ a b j i} 
- \frac{\ri}{3}(\xi_{ i} \s^a\tilde\s^{cd})_\ad \nabla_{a}W^-_{cd} 
\nonumber \\ 
&&  
- \frac{1}{2} \lambda^{a b}   (\bar{\Sigma}_i \tilde{\s}_{ab})_{\ad} 
- \lambda_{i}{}^{j}  \bar{\Sigma}_{\ad j}
+\frac{3}{2}\lambda_{\mathbb{D}} \bar{\Sigma}_{\ad i}
+ \ri\lambda_{Y} \bar{\Sigma}_{\ad i} 
+\frac{2}{3} (\bar{\eta}_{i}\tilde{\s}^{cd})_\ad W^-_{cd}
~,
\label{d-b-Sigma}
\eea
\esubeq
where
\bsubeq\label{nabla-on-W-and-Sigma}
\bea
\nabla_{a}W_{bc} &=& 
\cD_{a}W_{bc}
+2 \psi_{a}{}_{k}  R(Q)_{bc}{}^k
+2 \bar{\psi}_{a}{}^{k} R(\bar{Q})_{bc}{}_k 
~,\\
\de_a \Sigma^{\a i}
&=& 
\cD_a\Sigma^{\a i} 
-\frac{1}{2} \psi_{a}{}^{\a i}  D 
- \frac{2\ri}{3} (\psi_{a}{}^{i}\s^{cd})^\a  R(Y)_{cd}
- \frac{1}{3}(\psi_{a}{}_{j}\s^{cd})^\a R(J)_{cd}{}^{j i}  
\nonumber\\
&&
+ \frac{\ri}{6} (\bar{\psi}_{a}{}^{i}  \tilde{\s}^b\s^{cd})^\a \de_{b} W^+_{cd}
+ (\phi_{a}{}^{i}\s^{cd})^\a\, W^+_{cd}
~,
\\
\de_a \bar{\Sigma}_{\ad i}
&=& 
\cD_a\bar{\Sigma}_{\ad i} 
+\frac{1}{2} \bar\psi_{a}{}_{\ad i}  D 
- \frac{2\ri}{3}  (\bar\psi_{a}{}_{i}\tilde{\s}^{cd})_\ad  R(Y)_{cd}
- \frac{1}{3}(\bar\psi_{a}{}^{j}\tilde{\s}^{cd})_\ad R(J)_{cd}{}_{j i}  
\nonumber\\
&&
+ \frac{\ri}{6} (\psi_{a}{}_{i}\s^b\tilde{\s}^{cd})_\ad \de_{b} W^-_{cd} 
+  (\bar\phi_{a}{}_{i}\tilde{\s}^{cd})_\ad\, W^-_{cd}
~.
\eea
\esubeq

The local superconformal transformations of the fundamental fields of the vector multiplet fields in a standard Weyl multiplet background are
\bsubeq\label{transf-vector}
\bea
\d \phi
&=&
\xi_i \l^i + \l_\mathbb{D} \phi - 2 \ri \l_Y \phi
~,
\\ 
\d \bar{\phi}
&=&
\bar{\xi}^i \bar{\l}_i + \l_{\mathbb{D}} \bar{\phi} + 2 \ri \l_Y \bar{\phi}
~,
\\
\d\lambda_\alpha^i
&=&
2 (\s^{a b} \xi^i)_\a F_{a b} + (\s^{a b}\xi^i)_\a W^{+}_{a b} \bar{\phi} 
- \frac{1}{2} \xi_\a{}_j X^{ij} + 2 \ri (\s^a \bar{\xi}^i)_\a \de_a \phi \nonumber \\
&&+ \frac{1}{2}\l^{a b}(\s_{a b} \l^i)_\a  + \l^i{}_j \l^j_\a + \frac{3}{2} \l_{\mathbb{D}} \l^i_\a - \ri \l_Y \l^i_\a + 4 \eta^i_\a \phi
~,
\label{transf-lambda}
\\
\d\bar{\lambda}^\ad_i
&=&
-2 (\tilde{\s}^{a b} \bar{\xi}_i)^\ad F_{a b} - (\tilde{\s}^{a b}\bar{\xi}_i)^\ad W^{-}_{a b} \phi 
- \frac{1}{2} \bar{\xi}^\ad{}^j X_{ij} + 2 \ri (\tilde{\s}^a \xi_i)^\ad \de_a \bar{\phi} \nonumber \\
&&+ \frac{1}{2}\l^{a b}(\tilde{\s}_{a b} \bar{\l}_i)^\ad  - \l_i{}^j \bar{\l}_j^\ad + \frac{3}{2} \l_{\mathbb{D}} \bar{\l}_i^\ad+ \ri \l_Y \bar{\l}^\ad_i + 4 \bar{\eta}^\ad_i \bar{\phi}
~, 
\label{transf-lambda-bar}
\\
\d X^{ij}
&=& -4 \ri \xi^{(i} \s^a \de_a \bar{\lambda}^{j)} - 4 \ri \bar{\xi}^{(i} \tilde{\s}^a \de_a \lambda^{j)} 
+ 2 \lambda^{(i}{}_k X^{j)k} 
+ 2 \lambda_{\mathbb{D}} X^{ij}
~,
\\
\d F_{ab}
&=&
\Bigg[
-\ri \xi_k \s_{[a}\de_{b]}\bar{\lambda}^k 
+ 2\big(\xi_k R(Q)_{ab}{}^k\big) \bar{\phi} 
-\hf (\xi_k\lambda^k) W_{ab}^- 
+ 2 \eta^k \s_{ab} \lambda_k 
+ {\rm c.c.} \Bigg] 
\non\\
&& 
~~
+2 \lambda_{\mathbb{D}} F_{ab} 
- 2 \lambda_{[a}{}^{c} F_{b]c}
~,
\\
\d v_m
&=& (\xi_k \psi_{m}{}^k) \bar{\phi} 
-(\bar{\xi}^k \bar{\psi}_{m k}) \phi 
+\partial_m \l_V
~,
\label{dvm}
\eea
\esubeq
where
\bsubeq
\bea
\de_a \phi
&=&
\cD_a \phi - \frac{1}{2}\psi_a{}_i \l^i
~,
\\
\de_a \bar{\phi}
&=&
\cD_a \bar{\phi} 
- \frac{1}{2}\bar{\psi}_a{}^i \bar{\l}_i
~,
\\
\de_a \l^i_\a
&=&
\cD_{a} \l^i_\a 
-(\s^{cd}\psi_a{}^{i})_\a\Big( F^+_{cd}
+ \frac{1}{2}  W^+_{cd} \bar{\phi} \Big)
+\frac{1}{4} \psi_{a}{}_\a{}_j X^{ij} 
\non\\
&&
-\ri (\s^b \bar{\psi}_{a}{}^i)_\a \de_b \phi - 2 \phi_a{}_\a^i \phi
~,
\\
\de_a \bar\l_i^\ad
&=& 
\cD_{a} \bar{\l}^\ad_i 
+ (\ts^{cd}\bar{\psi}_a{}_i)^{\ad}\Big( 
F^-_{cd}
+\frac{1}{2} W^-_{cd} \phi\Big)
+\frac{1}{4} \bar{\psi}_a{}^{\ad j} X_{ij} 
\non\\ 
&& 
- \ri (\tilde{\s}^b \psi_a{}_i)^\ad \de_b \bar{\phi} - 2 \bar{\phi}_a{}^\ad_i \bar{\phi}
~,
\eea
\esubeq
and we have also included in \eqref{dvm}
the gauge field transformation
parametrised by the local real parameter $\l_V$.

The local superconformal transformations of the fundamental fields of the tensor multiplet in a standard Weyl multiplet background are given by \cite{deWit:1980lyi}
\bsubeq
\label{linear-multiplet}
\bea
\d G_{ij}
&=&
2 \xi_{(i} \chi_{j)}
+ 2\bar\xi_{ (i}  \bar{\chi}_{j )}
- 2\lambda_{(i}{}^{k}  G_{j) k} 
+2 \lambda_{\mathbb{D}} G_{ i j}
~,
\\
\d \chi_{\a i}
&=&
-\xi_{\a i} F 
-4\ri H_a (\s^a \bar{\xi}_{i})_\a 
+ \ri   
(\s^a \bar{\xi}^{j})_\a 
\de_{a} G_{i j}
+ 4 \eta_{\a}^{j}   G_{j i}
\nonumber\\ 
&& 
+\frac{1}{2} \lambda^{a b}
(\s_{a b}\chi_{i})_{\a} 
-\lambda_i{}^j \chi_{j \a}
+\frac{5}{2}\lambda_{\mathbb{D}}\chi_{\a i} +\ri \lambda_{Y} \chi_{\a i} 
~,
\\
\d \bar{\chi}^{\ad i}
&=&
-\bar\xi^{\ad i} \bar{F}
+4\ri H_a (\ts^a {\xi}^{i})^\ad 
+\ri(\ts^a \xi_{j})^\ad 
\de_{a} G^{i j}
+ 4 \bar\eta^{\ad}_{j}G^{j i}
\nonumber\\ 
&& 
+\frac{1}{2} \lambda^{a b}
(\ts_{a b}\bar\chi^{i})^{\ad} 
+\lambda^i{}_j \bar{\chi}^{j\ad}
+\frac{5}{2}\lambda_{\mathbb{D}}
\bar\chi^{\ad i} 
-\ri \lambda_{Y} \bar\chi^{\ad i} 
~,
\\
\d F
&=&
- 2 \ri \bar{\xi}^{i} \tilde{\s}^a \de_a \chi_i
+(\bar{\xi}^{i} \ts^{cd}\bar{\chi}_i)
W^-_{cd}
- 6 (\bar{\xi}^{i}\bar{\Sigma}^{ j})
G_{i j} 
+ 4\eta^{i} \chi_{i}
\nonumber\\
&&
+3 \lambda_{\mathbb{D}} F 
+ 2 \ri \lambda_{Y} F 
~,\\
\d \bar{F}
&=&
- 2 \ri{\xi}_{i}\s^a \de_a \bar\chi^i
-({\xi}_{i} \s^{cd}\chi^{i})W^+_{cd}
+6(\xi^{i}\Sigma^{j})G_{i j} 
+4\bar\eta_{i} \bar\chi^{i}
\nonumber\\
&&
+3\lambda_{\mathbb{D}} \bar{F} 
- 2 \ri \lambda_{Y} \bar{F} 
~,
\\
\d  H_{a}
&=&
\frac{1}{2} \xi_{i} \s_{a b} \de^b \chi^{ i}
-\frac{\ri}{16} 
(\xi_{i} \s_a\ts^{cd} \bar{\chi}^{i})
W^-_{cd}
- \frac{3\ri}{8} 
(\xi_{i} \s_a\bar{\Sigma}_j) 
G^{ij} 
\non\\
&&
-\frac{1}{2}\bar{\xi}^{i}\tilde{\s}_{a b} \de^b \bar{\chi}_i
- \frac{\ri}{16} 
(\bar\xi^{i}\tilde{\s}_a\s^{cd}\chi_{i})
W^+_{cd}
-\frac{3\ri}{8}
(\bar\xi^{i}\tilde{\s}_a\Sigma^{j}) 
G_{ij}
\non\\
&&
+ \lambda_a{}^b H_b
+3\lambda_{\mathbb{D}}  H_a 
-\frac{3\ri}{4} 
\eta^{i}\s_a \bar{\chi}_i
+\frac{3 \ri}{4}
\bar{\eta}_{i} \tilde{\s}_a \chi^i
~,
\eea
\esubeq
where
\bsubeq
\bea
\nabla_{a} G_{i j}
&=&
\cD_{a} G_{i j} - \psi_{a (i} \chi_{j)} -\bar{\psi}_{a (i} \bar{ \chi}_{j)} 
~,
\\
\nabla_{a} \chi_{\a i}
&=&
\cD_{a} \chi_{\a i} 
+\frac{1}{2} \psi_{a \a i}  F 
+2\ri (\s^b \bar{\psi}_{a i})_\a   
 H_b 
-\frac{\ri}{2}
(\s^b \bar{\psi}_{a}{}^{j})_\a
\de_{b}G_{ij} 
-2 \phi_a{}_\a^j  G_{i j}
~,
\\
\nabla_{a} \bar\chi^{\ad i}
&=&
\cD_{a}\bar\chi^{\ad i} 
+\frac{1}{2}\bar\psi_{a}{}^{ \ad i}  
\bar F 
-2\ri (\ts^b {\psi}_{a}{}^{i})_\a   
 H_b 
-\frac{\ri}{2}
(\ts^b\psi_{a}{}_{j})^\ad
\de_{b}G^{ij} 
-2 \bar\phi_a{}^\ad_j  G^{i j}
~.
\eea
\esubeq


\begin{footnotesize}

\end{footnotesize}

\end{document}



\renewcommand{\thefootnote}{\arabic{footnote}}
\setcounter{footnote}{0}

\begin{flushright}
September, 2024
\end{flushright}
\vspace{5mm}

\begin{center}
{\Large \bf 
Supplementary file for
``On 4D, $\cN=2$ deformed vector multiplets and partial supersymmetry breaking in off-shell supergravity''
}
\end{center}

\begin{center}

{\bf
Gregory Gold${}^{a}$,
Saurish Khandelwal${}^{a}$,\\
and Gabriele Tartaglino-Mazzucchelli${}^{a}$
} \\
\vspace{5mm}

\footnotesize{
{\it 
School of Mathematics and Physics, University of Queensland,
\\
 St Lucia, Brisbane, Queensland 4072, Australia}
}
\vspace{2mm}
~\\
\texttt{g.gold@uq.edu.au; 
s.khandelwal@uq.edu.au;
\\ g.tartaglino-mazzucchelli@uq.edu.au}\\
\vspace{2mm}

\end{center}


\bigskip\hrule
\tableofcontents{}
\vspace{1cm}
\bigskip\hrule

\allowdisplaybreaks

\section{Overview of the structure of this file: Fermionic component actions}
\label{Cadabra-action}

Here we provide the fermionic parts to component actions corresponding to those seen in the main body as generated computationally by \textit{Cadabra} \cite{Peeters1,Peeters2,Peeters3} --- see \cite{Gold:2024nbw} for information on the types of codes we use and detail of the repository where they are stored. First, we give the fermionic terms of the general gauged off-shell $\cN=2$ supergravity component action for deformed abelian vector multiplets in a hyper-dilaton Weyl multiplet background as discussed in Section 5 of the main paper. Second, we give the fermionic terms of the off-shell action with the specific choice of the SU(1,1)/U(1) model as discussed in Section 6.1. Lastly, we give the remaining fermionic terms of the same SU(1,1)/U(1) model but on shell as discussed in Section 6.3. Note that given their computationally generated origin, further reduction and simplification of these expressions can likely be achieved.

\section{General off-shell vector multiplet action}

Here, we give the entire fermionic Lagrangian for the general off-shell vector multiplet action of eq.~(5.1) in the paper divided as follows
\bea
\cL^{{\rm off-shell}}
=\cL^{{\rm off-shell}}_{c,0}
+\cL^{{\rm off-shell}}_{c,2}
+\cL^{{\rm off-shell}}_{c,4}
+{\rm c.c.}
~,
\eea
where the subscript of the Lagrangian refers the order in explicit fermions. Note that implicit fermions exist here and can be seen by converting $H^{a \underline{i} \underline{j}}$, $F^{a b}$, and $\textbf{F}^{a b}$ to $\tilde{h}^{a \underline{i} \underline{j}}$, $f^{a b}$, and $\textbf{f}^{a b}$, respectively, by eqs.~(2.67), (2.68), and (2.44) in the main paper as well by the fact that the $\textbf{D}_{a}$ derivative in eq.~(2.82) has the gravitini torsion in the spin connection.

\subsection{$\cL^{{\rm off-shell}}_{c,0}$}

\begin{adjustwidth}{1em}{4cm}${}\frac{1}{2}\mathcal{F}_{I} \overline{\textbf{W}}^{I} R+\frac{1}{2}{\rm i} \mathcal{F}_{I} F^{I}+\frac{1}{4}\mathcal{F} W_{a b} W^{a b} - \frac{1}{8}{\rm i} \mathcal{F} \epsilon_{a b c d} W^{a b} W^{c d} - \frac{1}{2}\mathcal{F}_{I} W_{a b} \textbf{F}^{a b I}+\frac{1}{4}{\rm i} \mathcal{F}_{I} \epsilon_{a b c d} W^{a b} \textbf{F}^{c d I} - \frac{1}{2}\mathcal{F}_{I J} \textbf{F}_{a b}\,^{I} \textbf{F}^{a b J} - \frac{1}{4}{\rm i} \mathcal{F}_{I J} \epsilon_{a b c d} \textbf{F}^{a b I} \textbf{F}^{c d J} - \frac{1}{32}\mathcal{F}_{I J} G_{i j}\,^{I} G^{i j J}+\frac{1}{32}\mathcal{F}_{I J} \textbf{X}_{i j}\,^{I} \textbf{X}^{i j J}+\mathcal{F}_{I} \textbf{D}_{a}{\textbf{D}^{a}{\overline{\textbf{W}}^{I}}}-4\mathcal{F}_{I} \overline{\textbf{W}}^{I} A_{a} A^{a} - \frac{1}{4}\mathcal{F}_{I} W_{a b} W^{a b} \textbf{W}^{I}+\frac{1}{8}{\rm i} \mathcal{F}_{I} \epsilon_{a b c d} W^{a b} W^{c d} \textbf{W}^{I} - \frac{1}{2}\mathcal{F}_{I J} W_{a b} \overline{\textbf{W}}^{I} \textbf{F}^{a b J} - \frac{1}{4}{\rm i} \mathcal{F}_{I J} \epsilon_{a b c d} W^{a b} \overline{\textbf{W}}^{I} \textbf{F}^{c d J}+\frac{1}{16}{\rm i} \mathcal{F}_{J I} \textbf{X}_{i j}\,^{J} G^{i j I} - \frac{1}{8}\mathcal{F}_{I J} W_{a b} W^{a b} \overline{\textbf{W}}^{I} \overline{\textbf{W}}^{J} - \frac{1}{16}{\rm i} \mathcal{F}_{I J} \epsilon_{a b c d} W^{a b} W^{c d} \overline{\textbf{W}}^{I} \overline{\textbf{W}}^{J}%
-4{\rm i} \mathcal{F}_{I} A_{a} \textbf{D}^{a}{\overline{\textbf{W}}^{I}}-2{\rm i} \mathcal{F}_{I} \overline{\textbf{W}}^{I} \textbf{D}_{a}{A^{a}}-64\mathcal{F}_{I} \overline{\textbf{W}}^{I} {q}^{-4} H_{a \underline{i} \underline{j}} H^{a \underline{i} \underline{j}}+\mathcal{F}_{I} \epsilon_{\underline{i} \underline{j}} \epsilon_{i j} \overline{\textbf{W}}^{I} {q}^{-2} \textbf{D}_{a}{q^{i \underline{i}}} \textbf{D}^{a}{q^{j \underline{j}}}+2\mathcal{F}_{I} \overline{\textbf{W}}^{I} {q}^{-2} q_{i \underline{i}} \textbf{D}_{a}{\textbf{D}^{a}{q^{i \underline{i}}}}-2\mathcal{F}_{I} \overline{\textbf{W}}^{I} {q}^{-4} q_{i \underline{i}} q_{j \underline{j}} \textbf{D}_{a}{q^{i \underline{j}}} \textbf{D}^{a}{q^{j \underline{i}}}$\end{adjustwidth}

\subsection{$\cL^{{\rm off-shell}}_{c,2}$}
\begin{adjustwidth}{1em}{4cm}${}\frac{1}{2}\mathcal{F}_{I} (\sigma_{a})^{\alpha}{}_{\dot{\alpha}} \overline{\psi}^{a}{}_{i}^{\dot{\alpha}} \chi^{i}_{\alpha}{}^{I}+\frac{1}{2}\mathcal{F}_{I J} (\sigma_{a})^{\alpha}{}_{\dot{\alpha}} \boldsymbol{\lambda}^{i}_{\alpha}{}^{I} \overline{\boldsymbol{\lambda}}_{\,i}^{\,\dot{\alpha}}{}^{J} A^{a} - \frac{1}{2}{\rm i} \mathcal{F}_{I J} \boldsymbol{\lambda}_{i}\,^{\alpha I} \chi^{i}_{\alpha}{}^{J} - \frac{1}{4}\mathcal{F}_{I J K} (\sigma_{ab})^{\alpha\beta} \boldsymbol{\lambda}_{i \alpha}\,^{I} \boldsymbol{\lambda}^{i}_{\beta}{}^{J} \textbf{F}^{a b K}+\frac{1}{16}\mathcal{F}_{I J K} \boldsymbol{\lambda}_{i}\,^{\alpha I} \boldsymbol{\lambda}_{j \alpha}\,^{J} \textbf{X}^{i j K}+\frac{1}{2}{\rm i} \mathcal{F}_{I J} (\sigma_{a})^{\alpha}{}_{\dot{\alpha}} \boldsymbol{\lambda}^{i}_{\alpha}{}^{I} \textbf{D}^{a}{\overline{\boldsymbol{\lambda}}_{\,i}^{\,\dot{\alpha}}{}^{J}}+\frac{1}{2}\mathcal{F}_{I} (\sigma_{a})^{\alpha}{}_{\dot{\alpha}} \overline{\textbf{W}}^{I} {q}^{-2} \rho^{\underline{i}}\,_{\alpha} \overline{\rho}_{\underline{i}}\,^{\dot{\alpha}} A^{a}+\frac{1}{2}\mathcal{F}_{I} (\tilde{\sigma}_{ab})_{\dot{\beta}\dot{\alpha}} W^{a b} \overline{\boldsymbol{\lambda}}_{\,i}^{\,\dot{\beta}}{}^{I} {q}^{-2} q^{i \underline{i}} \overline{\rho}_{\underline{i}}\,^{\dot{\alpha}} - \frac{1}{4}\mathcal{F}_{J I} (\sigma_{a})^{\alpha}{}_{\dot{\alpha}} \overline{\psi}^{a}{}_{j}^{\dot{\alpha}} \boldsymbol{\lambda}_{i \alpha}\,^{J} G^{i j I} - \frac{1}{8}\mathcal{F}_{I J K} (\sigma_{ab})^{\alpha\beta} W^{a b} \overline{\textbf{W}}^{I} \boldsymbol{\lambda}_{i \alpha}\,^{J} \boldsymbol{\lambda}^{i}_{\beta}{}^{K}+\frac{1}{2}{\rm i} \mathcal{F}_{I} (\tilde{\sigma}_{ab})_{\dot{\alpha}\dot{\beta}} \overline{\psi}^{a}{}_{i}^{\dot{\alpha}} \overline{\psi}^{b}{}_{j}^{\dot{\beta}} G^{i j I}+\frac{1}{16}{\rm i} \mathcal{F}_{J K I} \boldsymbol{\lambda}_{i}\,^{\alpha J} \boldsymbol{\lambda}_{j \alpha}\,^{K} G^{i j I}+\frac{1}{2}\mathcal{F} \overline{\psi}_{a}\,^{i}\,_{\dot{\alpha}} \overline{\psi}_{b i}\,^{\dot{\alpha}} W^{a b} - \frac{1}{4}{\rm i} \mathcal{F} \epsilon_{c d a b} \overline{\psi}^{c i}\,_{\dot{\alpha}} \overline{\psi}^{d}{}_{i}^{\dot{\alpha}} W^{a b}+\frac{1}{16}\mathcal{F}_{I} \overline{\textbf{W}}^{I} {q}^{-4} \rho^{\underline{i} \alpha} \rho^{\underline{j}}\,_{\alpha} \overline{\rho}_{\underline{i} \dot{\alpha}} \overline{\rho}_{\underline{j}}\,^{\dot{\alpha}}+\frac{1}{2}\mathcal{F}_{I} \overline{\boldsymbol{\lambda}}^{i}\,_{\dot{\alpha}}\,^{I} \textbf{D}_{a}{\overline{\psi}^{a}{}_{i}^{\dot{\alpha}}} - \frac{1}{2}\mathcal{F}_{I} \overline{\psi}_{a}\,^{i}\,_{\dot{\alpha}} \textbf{D}^{a}{\overline{\boldsymbol{\lambda}}_{\,i}^{\,\dot{\alpha}}{}^{I}}+\mathcal{F}_{I} (\sigma_{b})^{\alpha}{}_{\dot{\alpha}} \psi_{a}\,^{i}\,_{\alpha} \overline{\psi}^{b}{}_{i}^{\dot{\alpha}} \overline{\textbf{W}}^{I} A^{a}-\mathcal{F}_{I} (\sigma_{b})^{\alpha}{}_{\dot{\alpha}} \psi^{b}{}^{i}_{\alpha} \overline{\psi}_{a i}\,^{\dot{\alpha}} \overline{\textbf{W}}^{I} A^{a}%
-\mathcal{F}_{I} (\sigma_{a})^{\alpha}{}_{\dot{\alpha}} \overline{\boldsymbol{\lambda}}_{\,i}^{\,\dot{\alpha}}{}^{I} {q}^{-2} q^{i}\,_{\underline{i}} \rho^{\underline{i}}\,_{\alpha} A^{a} - \frac{1}{8}\mathcal{F}_{I} (\sigma_{ab})^{\alpha\beta} W^{a b} \overline{\textbf{W}}^{I} {q}^{-2} \rho_{\underline{i} \alpha} \rho^{\underline{i}}\,_{\beta}+\frac{1}{2}\mathcal{F}_{I} \overline{\psi}_{a}\,^{i}\,_{\dot{\alpha}} \overline{\psi}_{b i}\,^{\dot{\alpha}} \textbf{F}^{a b I}+\frac{1}{4}{\rm i} \mathcal{F}_{I} \epsilon_{c d a b} \overline{\psi}^{c i}\,_{\dot{\alpha}} \overline{\psi}^{d}{}_{i}^{\dot{\alpha}} \textbf{F}^{a b I}-\mathcal{F}_{I} (\tilde{\sigma}_{ba})_{\dot{\beta}\dot{\alpha}} \overline{\psi}^{b i \dot{\beta}} \textbf{D}^{a}{\overline{\boldsymbol{\lambda}}_{\,i}^{\,\dot{\alpha}}{}^{I}}-\mathcal{F}_{I} (\tilde{\sigma}_{cd})_{\dot{\rho}\dot{\alpha}} (\tilde{\sigma}_{ab})^{\dot{\rho}}{}_{\dot{\beta}} \overline{\psi}^{c i \dot{\alpha}} \overline{\psi}^{d}{}_{i}^{\dot{\beta}} \textbf{F}^{a b I}+\frac{1}{8}\mathcal{F}_{I} (\tilde{\sigma}_{ab})_{\dot{\alpha}\dot{\beta}} W^{a b} \overline{\textbf{W}}^{I} {q}^{-2} \overline{\rho}^{\underline{i} \dot{\alpha}} \overline{\rho}_{\underline{i}}\,^{\dot{\beta}} - \frac{1}{48}\mathcal{F}_{I J K L} \boldsymbol{\lambda}_{i}\,^{\alpha I} \boldsymbol{\lambda}^{i \beta J} \boldsymbol{\lambda}_{j \alpha}\,^{K} \boldsymbol{\lambda}^{j}_{\beta}{}^{L} - \frac{1}{4}{\rm i} \mathcal{F}_{I} (\sigma_{a})^{\alpha}{}_{\dot{\alpha}} \overline{\textbf{W}}^{I} {q}^{-2} \overline{\rho}_{\underline{i}}\,^{\dot{\alpha}} \textbf{D}^{a}{\rho^{\underline{i}}\,_{\alpha}}+\frac{1}{4}{\rm i} \mathcal{F}_{I} (\sigma_{b})^{\alpha}{}_{\dot{\alpha}} \overline{\psi}_{a i}\,^{\dot{\alpha}} W^{a b} \boldsymbol{\lambda}^{i}_{\alpha}{}^{I} - \frac{1}{8}\mathcal{F}_{I} \epsilon^{c}\,_{d a b} (\sigma_{c})^{\alpha}{}_{\dot{\alpha}} \overline{\psi}^{d}{}_{i}^{\dot{\alpha}} W^{a b} \boldsymbol{\lambda}^{i}_{\alpha}{}^{I} - \frac{1}{4}{\rm i} \mathcal{F}_{I} (\sigma_{a})^{\alpha}{}_{\dot{\alpha}} \overline{\textbf{W}}^{I} {q}^{-2} \rho^{\underline{i}}\,_{\alpha} \textbf{D}^{a}{\overline{\rho}_{\underline{i}}\,^{\dot{\alpha}}}+\frac{1}{2}\mathcal{F}_{I J} \epsilon^{c}\,_{d a b} (\sigma_{c})^{\alpha}{}_{\dot{\alpha}} \overline{\psi}^{d}{}_{i}^{\dot{\alpha}} \boldsymbol{\lambda}^{i}_{\alpha}{}^{I} \textbf{F}^{a b J} - \frac{1}{2}\mathcal{F}_{I} (\tilde{\sigma}_{cd})_{\dot{\rho}\dot{\alpha}} (\tilde{\sigma}_{ab})^{\dot{\rho}}{}_{\dot{\beta}} \overline{\psi}^{c i \dot{\alpha}} \overline{\psi}^{d}{}_{i}^{\dot{\beta}} W^{a b} \textbf{W}^{I}-8\mathcal{F}_{I} \overline{\psi}_{a i \dot{\alpha}} \overline{\textbf{W}}^{I} {q}^{-4} q^{i}\,_{\underline{i}} \overline{\rho}_{\underline{j}}\,^{\dot{\alpha}} H^{a \underline{i} \underline{j}}+\frac{1}{2}\mathcal{F}_{I J} \psi_{a i}\,^{\alpha} \boldsymbol{\lambda}^{i}_{\alpha}{}^{I} \textbf{D}^{a}{\overline{\textbf{W}}^{J}}+\mathcal{F}_{I J} (\sigma_{ba})^{\beta\alpha} \psi^{b}\,_{i \beta} \boldsymbol{\lambda}^{i}_{\alpha}{}^{I} \textbf{D}^{a}{\overline{\textbf{W}}^{J}}+\mathcal{F}_{I J} (\sigma_{a})^{\alpha}{}_{\dot{\alpha}} \overline{\textbf{W}}^{I} \boldsymbol{\lambda}_{i \alpha}\,^{J} {q}^{-2} q^{i \underline{i}} \overline{\rho}_{\underline{i}}\,^{\dot{\alpha}} A^{a}+{\rm i} \mathcal{F}_{I} \overline{\psi}_{a}\,^{i}\,_{\dot{\alpha}} \overline{\boldsymbol{\lambda}}_{\,i}^{\,\dot{\alpha}}{}^{I} A^{a}+\frac{1}{2}{\rm i} \mathcal{F}_{I} (\sigma_{b})^{\alpha}{}_{\dot{\alpha}} \psi_{a}\,^{i}\,_{\alpha} \overline{\psi}^{b}{}_{i}^{\dot{\alpha}} \textbf{D}^{a}{\overline{\textbf{W}}^{I}}%
 - \frac{1}{2}{\rm i} \mathcal{F}_{I} (\sigma_{b})^{\alpha}{}_{\dot{\alpha}} \psi^{b}{}^{i}_{\alpha} \overline{\psi}_{a i}\,^{\dot{\alpha}} \textbf{D}^{a}{\overline{\textbf{W}}^{I}} - \frac{1}{2}\mathcal{F}_{I} \epsilon^{b}\,_{c d a} (\sigma_{b})^{\alpha}{}_{\dot{\alpha}} \psi^{c}{}^{i}_{\alpha} \overline{\psi}^{d}{}_{i}^{\dot{\alpha}} \textbf{D}^{a}{\overline{\textbf{W}}^{I}}-{\rm i} \mathcal{F}_{I} (\sigma_{a})^{\alpha}{}_{\dot{\alpha}} \overline{\boldsymbol{\lambda}}_{\,i}^{\,\dot{\alpha}}{}^{I} {q}^{-2} q^{i}\,_{\underline{i}} \textbf{D}^{a}{\rho^{\underline{i}}\,_{\alpha}} - \frac{1}{4}{\rm i} \mathcal{F}_{I J} (\sigma_{b})^{\alpha}{}_{\dot{\alpha}} \overline{\psi}_{a i}\,^{\dot{\alpha}} W^{a b} \textbf{W}^{I} \boldsymbol{\lambda}^{i}_{\alpha}{}^{J}+\frac{1}{8}\mathcal{F}_{I J} \epsilon^{c}\,_{d a b} (\sigma_{c})^{\alpha}{}_{\dot{\alpha}} \overline{\psi}^{d}{}_{i}^{\dot{\alpha}} W^{a b} \textbf{W}^{I} \boldsymbol{\lambda}^{i}_{\alpha}{}^{J}+\frac{1}{2}\mathcal{F}_{I} \overline{\textbf{W}}^{I} {q}^{-2} q^{i \underline{i}} \overline{\rho}_{\underline{i} \dot{\alpha}} \textbf{D}_{a}{\overline{\psi}^{a}{}_{i}^{\dot{\alpha}}}+\frac{1}{2}\mathcal{F}_{I} \overline{\psi}_{a i \dot{\alpha}} \overline{\textbf{W}}^{I} {q}^{-2} \overline{\rho}_{\underline{i}}\,^{\dot{\alpha}} \textbf{D}^{a}{q^{i \underline{i}}}+\mathcal{F}_{I} (\tilde{\sigma}_{ba})_{\dot{\beta}\dot{\alpha}} \overline{\psi}^{b}{}_{i}^{\dot{\beta}} \overline{\textbf{W}}^{I} {q}^{-2} \overline{\rho}_{\underline{i}}\,^{\dot{\alpha}} \textbf{D}^{a}{q^{i \underline{i}}}+\frac{1}{2}\mathcal{F}_{I} \overline{\psi}_{a i \dot{\alpha}} \overline{\textbf{W}}^{I} {q}^{-2} q^{i \underline{i}} \textbf{D}^{a}{\overline{\rho}_{\underline{i}}\,^{\dot{\alpha}}}+\mathcal{F}_{I} (\tilde{\sigma}_{ba})_{\dot{\beta}\dot{\alpha}} \overline{\psi}^{b}{}_{i}^{\dot{\beta}} \overline{\textbf{W}}^{I} {q}^{-2} q^{i \underline{i}} \textbf{D}^{a}{\overline{\rho}_{\underline{i}}\,^{\dot{\alpha}}}-16\mathcal{F}_{I} \overline{\psi}_{a i \dot{\alpha}} \overline{\boldsymbol{\lambda}}_{\,j}^{\,\dot{\alpha}}{}^{I} {q}^{-4} q^{i}\,_{\underline{i}} q^{j}\,_{\underline{j}} H^{a \underline{i} \underline{j}} - \frac{1}{2}\mathcal{F}_{I J} (\sigma_{ab})^{\beta\alpha} W^{a b} \overline{\textbf{W}}^{I} \boldsymbol{\lambda}_{i \beta}\,^{J} {q}^{-2} q^{i}\,_{\underline{i}} \rho^{\underline{i}}\,_{\alpha}-4{\rm i} \mathcal{F}_{I} (\sigma_{a})^{\alpha}{}_{\dot{\alpha}} \overline{\textbf{W}}^{I} {q}^{-4} \rho_{\underline{i} \alpha} \overline{\rho}_{\underline{j}}\,^{\dot{\alpha}} H^{a \underline{i} \underline{j}}-{\rm i} \mathcal{F}_{I J} \psi_{a i}\,^{\alpha} \overline{\textbf{W}}^{I} \boldsymbol{\lambda}^{i}_{\alpha}{}^{J} A^{a}-{\rm i} \mathcal{F}_{I J} (\sigma_{a})^{\alpha}{}_{\dot{\alpha}} \overline{\textbf{W}}^{I} \boldsymbol{\lambda}_{i \alpha}\,^{J} {q}^{-2} q^{i \underline{i}} \textbf{D}^{a}{\overline{\rho}_{\underline{i}}\,^{\dot{\alpha}}}+\mathcal{F}_{I} \overline{\psi}_{a j \dot{\alpha}} \overline{\boldsymbol{\lambda}}_{\,i}^{\,\dot{\alpha}}{}^{I} {q}^{-2} q^{i}\,_{\underline{i}} \textbf{D}^{a}{q^{j \underline{i}}}+4\mathcal{F}_{I} (\tilde{\sigma}_{ba})_{\dot{\beta}\dot{\alpha}} \overline{\psi}^{b}{}_{j}^{\dot{\beta}} \overline{\boldsymbol{\lambda}}_{\,i}^{\,\dot{\alpha}}{}^{I} {q}^{-2} q^{i}\,_{\underline{i}} \textbf{D}^{a}{q^{j \underline{i}}}-\mathcal{F}_{I} \overline{\psi}_{a i \dot{\alpha}} \overline{\boldsymbol{\lambda}}_{\,j}^{\,\dot{\alpha}}{}^{I} {q}^{-2} q^{i}\,_{\underline{i}} \textbf{D}^{a}{q^{j \underline{i}}}+8\mathcal{F}_{I} \psi_{a i}\,^{\alpha} \overline{\textbf{W}}^{I} {q}^{-4} q^{i}\,_{\underline{i}} \rho_{\underline{j} \alpha} H^{a \underline{i} \underline{j}} - \frac{1}{4}\mathcal{F}_{I J} \boldsymbol{\lambda}_{i}\,^{\alpha I} \overline{\boldsymbol{\lambda}}_{j \dot{\alpha}}\,^{J} {q}^{-4} q^{i}\,_{\underline{i}} q^{j \underline{j}} \rho^{\underline{i}}\,_{\alpha} \overline{\rho}_{\underline{j}}\,^{\dot{\alpha}}%
 - \frac{1}{4}\mathcal{F}_{I J} \boldsymbol{\lambda}_{i}\,^{\alpha I} \overline{\boldsymbol{\lambda}}_{j \dot{\alpha}}\,^{J} {q}^{-4} q^{i \underline{j}} q^{j}\,_{\underline{i}} \rho^{\underline{i}}\,_{\alpha} \overline{\rho}_{\underline{j}}\,^{\dot{\alpha}} - \frac{1}{2}{\rm i} \mathcal{F}_{I J} (\sigma_{a})^{\alpha}{}_{\dot{\alpha}} \boldsymbol{\lambda}_{i \alpha}\,^{I} \overline{\boldsymbol{\lambda}}_{\,j}^{\,\dot{\alpha}}{}^{J} {q}^{-2} q^{i}\,_{\underline{i}} \textbf{D}^{a}{q^{j \underline{i}}} - \frac{1}{2}{\rm i} \mathcal{F}_{I J} (\sigma_{a})^{\alpha}{}_{\dot{\alpha}} \boldsymbol{\lambda}_{j \alpha}\,^{I} \overline{\boldsymbol{\lambda}}_{\,i}^{\,\dot{\alpha}}{}^{J} {q}^{-2} q^{i}\,_{\underline{i}} \textbf{D}^{a}{q^{j \underline{i}}} - \frac{1}{2}\mathcal{F}_{I} \overline{\textbf{W}}^{I} {q}^{-2} q_{i \underline{i}} \rho^{\underline{i} \alpha} \textbf{D}_{a}{\psi^{a}{}^{i}_{\alpha}}+\frac{1}{2}\mathcal{F}_{I} \psi_{a i}\,^{\alpha} \overline{\textbf{W}}^{I} {q}^{-2} q^{i}\,_{\underline{i}} \textbf{D}^{a}{\rho^{\underline{i}}\,_{\alpha}}+\mathcal{F}_{I} (\sigma_{ba})^{\beta\alpha} \psi^{b}\,_{i \beta} \overline{\textbf{W}}^{I} {q}^{-2} q^{i}\,_{\underline{i}} \textbf{D}^{a}{\rho^{\underline{i}}\,_{\alpha}} - \frac{1}{2}\mathcal{F}_{I} \psi_{a i}\,^{\alpha} \overline{\textbf{W}}^{I} {q}^{-2} \rho_{\underline{i} \alpha} \textbf{D}^{a}{q^{i \underline{i}}}-\mathcal{F}_{I} (\sigma_{ba})^{\beta\alpha} \psi^{b}\,_{i \beta} \overline{\textbf{W}}^{I} {q}^{-2} \rho_{\underline{i} \alpha} \textbf{D}^{a}{q^{i \underline{i}}}+\frac{1}{4}\mathcal{F}_{I} \epsilon^{b}\,_{c a d} (\sigma_{b})^{\alpha}{}_{\dot{\alpha}} \psi^{c}{}^{i}_{\alpha} \overline{\textbf{W}}^{I} \textbf{D}^{a}{\overline{\psi}^{d}{}_{i}^{\dot{\alpha}}} - \frac{1}{4}\mathcal{F}_{I} \epsilon^{b}\,_{c a d} (\sigma_{b})^{\alpha}{}_{\dot{\alpha}} \overline{\psi}^{c}{}_{i}^{\dot{\alpha}} \overline{\textbf{W}}^{I} \textbf{D}^{a}{\psi^{d}{}^{i}_{\alpha}}+16\mathcal{F}_{I J} \psi_{a i}\,^{\alpha} \overline{\textbf{W}}^{I} \boldsymbol{\lambda}_{j \alpha}\,^{J} {q}^{-4} q^{i}\,_{\underline{i}} q^{j}\,_{\underline{j}} H^{a \underline{i} \underline{j}}-8{\rm i} \mathcal{F}_{I J} (\sigma_{a})^{\alpha}{}_{\dot{\alpha}} \boldsymbol{\lambda}_{i \alpha}\,^{I} \overline{\boldsymbol{\lambda}}_{\,j}^{\,\dot{\alpha}}{}^{J} {q}^{-4} q^{i}\,_{\underline{i}} q^{j}\,_{\underline{j}} H^{a \underline{i} \underline{j}}-\mathcal{F}_{I J} \psi_{a j}\,^{\alpha} \overline{\textbf{W}}^{I} \boldsymbol{\lambda}_{i \alpha}\,^{J} {q}^{-2} q^{i}\,_{\underline{i}} \textbf{D}^{a}{q^{j \underline{i}}}-4\mathcal{F}_{I J} (\sigma_{ba})^{\beta\alpha} \psi^{b}\,_{j \beta} \overline{\textbf{W}}^{I} \boldsymbol{\lambda}_{i \alpha}\,^{J} {q}^{-2} q^{i}\,_{\underline{i}} \textbf{D}^{a}{q^{j \underline{i}}}+\mathcal{F}_{I J} \psi_{a i}\,^{\alpha} \overline{\textbf{W}}^{I} \boldsymbol{\lambda}_{j \alpha}\,^{J} {q}^{-2} q^{i}\,_{\underline{i}} \textbf{D}^{a}{q^{j \underline{i}}}-{\rm i} \mathcal{F}_{I} (\sigma_{b})^{\alpha}{}_{\dot{\alpha}} \psi_{a i \alpha} \overline{\psi}^{b}{}_{j}^{\dot{\alpha}} \overline{\textbf{W}}^{I} {q}^{-2} q^{i}\,_{\underline{i}} \textbf{D}^{a}{q^{j \underline{i}}}+{\rm i} \mathcal{F}_{I} (\sigma_{b})^{\alpha}{}_{\dot{\alpha}} \psi^{b}\,_{i \alpha} \overline{\psi}_{a j}\,^{\dot{\alpha}} \overline{\textbf{W}}^{I} {q}^{-2} q^{i}\,_{\underline{i}} \textbf{D}^{a}{q^{j \underline{i}}}-\mathcal{F}_{I} \epsilon^{b}\,_{c d a} (\sigma_{b})^{\alpha}{}_{\dot{\alpha}} \psi^{c}\,_{i \alpha} \overline{\psi}^{d}{}_{j}^{\dot{\alpha}} \overline{\textbf{W}}^{I} {q}^{-2} q^{i}\,_{\underline{i}} \textbf{D}^{a}{q^{j \underline{i}}}+{\rm i} \mathcal{F}_{I} (\sigma_{b})^{\alpha}{}_{\dot{\alpha}} \psi_{a j \alpha} \overline{\psi}^{b}{}_{i}^{\dot{\alpha}} \overline{\textbf{W}}^{I} {q}^{-2} q^{i}\,_{\underline{i}} \textbf{D}^{a}{q^{j \underline{i}}}-{\rm i} \mathcal{F}_{I} (\sigma_{b})^{\alpha}{}_{\dot{\alpha}} \psi^{b}\,_{j \alpha} \overline{\psi}_{a i}\,^{\dot{\alpha}} \overline{\textbf{W}}^{I} {q}^{-2} q^{i}\,_{\underline{i}} \textbf{D}^{a}{q^{j \underline{i}}}%
-\mathcal{F}_{I} \epsilon^{b}\,_{c d a} (\sigma_{b})^{\alpha}{}_{\dot{\alpha}} \psi^{c}\,_{j \alpha} \overline{\psi}^{d}{}_{i}^{\dot{\alpha}} \overline{\textbf{W}}^{I} {q}^{-2} q^{i}\,_{\underline{i}} \textbf{D}^{a}{q^{j \underline{i}}} - \frac{1}{4}\mathcal{F}_{I} \epsilon^{b}\,_{c d e} (\sigma_{b})^{\alpha}{}_{\dot{\alpha}} \psi^{c}{}^{i}_{\alpha} \overline{\textbf{W}}^{I} \overline{\bold{\Psi}}^{d e}{}_i^\ad - \frac{1}{2}\mathcal{F}_{I} (\tilde{\sigma}_{cd})_{\dot{\alpha}\dot{\beta}} \overline{\boldsymbol{\lambda}}^{i \dot{\alpha} I} \overline{\bold{\Psi}}^{c d}{}_i^\bd + \frac{1}{2} \mathcal{F}_{I} (\tilde{\sigma}_{cd})_{\dot{\alpha}\dot{\beta}} \overline{\textbf{W}}^{I} {q}^{-2} q^{i \underline{i}} \overline{\rho}_{\underline{i}}\,^{\dot{\alpha}} \overline{\bold{\Psi}}^{c d}{}_i^\bd - \mathcal{F}_{I J} (\sigma_{cd})^{\alpha\beta} \overline{\textbf{W}}^{I} \boldsymbol{\lambda}_{i \alpha}\,^{J} \bold{\Psi}^{c d}{}^i_\b+ \frac{1}{2} \mathcal{F}_{I} \epsilon^{b}\,_{c d e} (\sigma_{b})^{\alpha}{}_{\dot{\alpha}} \overline{\psi}^{c}{}_{i}^{\dot{\alpha}} \overline{\textbf{W}}^{I} \bold{\Psi}^{d e}{}^i_\a - \frac{1}{2}\mathcal{F}_{I} (\sigma_{cd})^{\alpha\beta} \overline{\textbf{W}}^{I} {q}^{-2} q_{i \underline{i}} \rho^{\underline{i}}\,_{\alpha} \bold{\Psi}^{c d}{}^i_\b$\end{adjustwidth}

where,
 \bea
\bold{\Psi}_{ab}{}^\g_k 
= 
2{e_a}^m{e_b}^n\textbf{D}_{[m}\psi_{n]}{}^\g_k
~,~~~~~~
\overline{\bold{\Psi}}_{ab}{}_\gd^k = 2{e_a}^m{e_b}^n\textbf{D}_{[m}\bar{\psi}_{n]}{}_\gd^k
~.
\eea

\subsection{$\cL^{{\rm off-shell}}_{c,4}$}

\begin{adjustwidth}{1em}{4cm}${}\frac{1}{4}\mathcal{F}_{I J} (\tilde{\sigma}_{ab})_{\dot{\alpha}\dot{\beta}} \overline{\psi}^{a}{}_{i}^{\dot{\alpha}} \overline{\psi}^{b}{}_{j}^{\dot{\beta}} \boldsymbol{\lambda}^{i \alpha I} \boldsymbol{\lambda}^{j}_{\alpha}{}^{J}+\frac{1}{8}\mathcal{F}_{I} \psi_{a}\,^{i \alpha} \overline{\psi}^{a}\,_{i \dot{\alpha}} \overline{\textbf{W}}^{I} {q}^{-2} \rho^{\underline{i}}\,_{\alpha} \overline{\rho}_{\underline{i}}\,^{\dot{\alpha}}+\frac{1}{4}\mathcal{F}_{I J} \psi_{a i}\,^{\alpha} \overline{\psi}^{a j}\,_{\dot{\alpha}} \boldsymbol{\lambda}^{i}_{\alpha}{}^{I} \overline{\boldsymbol{\lambda}}_{\,j}^{\,\dot{\alpha}}{}^{J}+\frac{1}{4}\mathcal{F}_{I} (\sigma_{ab})^{\beta\alpha} \psi^{a}{}^{i}_{\beta} \overline{\psi}^{b}\,_{i \dot{\alpha}} \overline{\textbf{W}}^{I} {q}^{-2} \rho^{\underline{i}}\,_{\alpha} \overline{\rho}_{\underline{i}}\,^{\dot{\alpha}} - \frac{1}{4}\mathcal{F}_{I} (\tilde{\sigma}_{ab})_{\dot{\beta}\dot{\alpha}} \psi^{a i \alpha} \overline{\psi}^{b}{}_{i}^{\dot{\beta}} \overline{\textbf{W}}^{I} {q}^{-2} \rho^{\underline{i}}\,_{\alpha} \overline{\rho}_{\underline{i}}\,^{\dot{\alpha}} - \frac{1}{4}\mathcal{F}_{I} \psi_{a}\,^{j \alpha} \overline{\psi}^{a}\,_{j \dot{\alpha}} \overline{\boldsymbol{\lambda}}_{\,i}^{\,\dot{\alpha}}{}^{I} {q}^{-2} q^{i}\,_{\underline{i}} \rho^{\underline{i}}\,_{\alpha}+\frac{1}{4}\mathcal{F}_{I} \psi_{a}\,^{j \alpha} \overline{\psi}^{a}\,_{i \dot{\alpha}} \overline{\boldsymbol{\lambda}}_{\,j}^{\,\dot{\alpha}}{}^{I} {q}^{-2} q^{i}\,_{\underline{i}} \rho^{\underline{i}}\,_{\alpha} - \frac{1}{4}\mathcal{F}_{I} \overline{\psi}_{a}\,^{j}\,_{\dot{\beta}} \overline{\psi}^{a}\,_{j \dot{\alpha}} \overline{\boldsymbol{\lambda}}_{\,i}^{\,\dot{\beta}}{}^{I} {q}^{-2} q^{i \underline{i}} \overline{\rho}_{\underline{i}}\,^{\dot{\alpha}}+\frac{1}{4}\mathcal{F}_{I} \overline{\psi}_{a}\,^{j}\,_{\dot{\alpha}} \overline{\psi}^{a}\,_{i \dot{\beta}} \overline{\boldsymbol{\lambda}}_{\,j}^{\,\dot{\beta}}{}^{I} {q}^{-2} q^{i \underline{i}} \overline{\rho}_{\underline{i}}\,^{\dot{\alpha}}+\frac{1}{2}\mathcal{F}_{I J} (\sigma_{ab})^{\beta\alpha} \psi^{a}\,_{i \beta} \overline{\psi}^{b j}\,_{\dot{\alpha}} \boldsymbol{\lambda}^{i}_{\alpha}{}^{I} \overline{\boldsymbol{\lambda}}_{\,j}^{\,\dot{\alpha}}{}^{J}+\frac{1}{8}\mathcal{F}_{I J} (\sigma_{ab})^{\alpha\beta} \overline{\psi}^{a j}\,_{\dot{\alpha}} \overline{\psi}^{b}{}_{j}^{\dot{\alpha}} \boldsymbol{\lambda}_{i \alpha}\,^{I} \boldsymbol{\lambda}^{i}_{\beta}{}^{J} - \frac{1}{4}{\rm i} \mathcal{F} \epsilon_{a b c d} \overline{\psi}^{a i}\,_{\dot{\alpha}} \overline{\psi}^{b}{}_{i}^{\dot{\alpha}} \overline{\psi}^{c j}\,_{\dot{\beta}} \overline{\psi}^{d}{}_{j}^{\dot{\beta}}+\frac{1}{12}{\rm i} \mathcal{F}_{I J K} (\sigma_{a})^{\alpha}{}_{\dot{\alpha}} \overline{\psi}^{a}{}_{j}^{\dot{\alpha}} \boldsymbol{\lambda}_{i \alpha}\,^{I} \boldsymbol{\lambda}^{i \beta J} \boldsymbol{\lambda}^{j}_{\beta}{}^{K}+\mathcal{F}_{I} (\tilde{\sigma}_{ab})_{\dot{\beta}\dot{\alpha}} \psi^{a j \alpha} \overline{\psi}^{b}{}_{j}^{\dot{\beta}} \overline{\boldsymbol{\lambda}}_{\,i}^{\,\dot{\alpha}}{}^{I} {q}^{-2} q^{i}\,_{\underline{i}} \rho^{\underline{i}}\,_{\alpha}-\mathcal{F}_{I} (\tilde{\sigma}_{ab})_{\dot{\rho}\dot{\beta}} \overline{\psi}^{a j \dot{\rho}} \overline{\psi}^{b}\,_{j \dot{\alpha}} \overline{\boldsymbol{\lambda}}_{\,i}^{\,\dot{\beta}}{}^{I} {q}^{-2} q^{i \underline{i}} \overline{\rho}_{\underline{i}}\,^{\dot{\alpha}} - \frac{1}{16}\mathcal{F}_{I} \psi_{a i}\,^{\alpha} \psi^{a i \beta} \overline{\textbf{W}}^{I} {q}^{-2} \rho_{\underline{i} \alpha} \rho^{\underline{i}}\,_{\beta} - \frac{1}{16}\mathcal{F}_{I} \overline{\psi}_{a}\,^{i}\,_{\dot{\alpha}} \overline{\psi}^{a}\,_{i \dot{\beta}} \overline{\textbf{W}}^{I} {q}^{-2} \overline{\rho}^{\underline{i} \dot{\alpha}} \overline{\rho}_{\underline{i}}\,^{\dot{\beta}}+\frac{1}{8}\mathcal{F}_{I} \overline{\psi}_{a i \dot{\beta}} \overline{\psi}^{a}\,_{j \dot{\alpha}} \overline{\textbf{W}}^{I} {q}^{-4} q^{i \underline{i}} q^{j \underline{j}} \overline{\rho}_{\underline{i}}\,^{\dot{\alpha}} \overline{\rho}_{\underline{j}}\,^{\dot{\beta}}+\frac{1}{4}\mathcal{F}_{I} \epsilon^{a}\,_{b c d} (\sigma_{a})^{\alpha}{}_{\dot{\beta}} \psi^{b}{}^{j}_{\alpha} \overline{\psi}^{c i}\,_{\dot{\alpha}} \overline{\psi}^{d}{}_{j}^{\dot{\beta}} \overline{\boldsymbol{\lambda}}_{\,i}^{\,\dot{\alpha}}{}^{I}%
 - \frac{1}{4}\mathcal{F}_{I} \epsilon^{a}\,_{b c d} (\sigma_{a})^{\alpha}{}_{\dot{\beta}} \overline{\psi}^{b j}\,_{\dot{\alpha}} \overline{\psi}^{c}{}_{j}^{\dot{\alpha}} \overline{\psi}^{d}{}_{i}^{\dot{\beta}} \boldsymbol{\lambda}^{i}_{\alpha}{}^{I} - \frac{1}{4}\mathcal{F}_{I J} \psi_{a i}\,^{\alpha} \overline{\psi}^{a}\,_{j \dot{\alpha}} \overline{\textbf{W}}^{I} \boldsymbol{\lambda}^{j}_{\alpha}{}^{J} {q}^{-2} q^{i \underline{i}} \overline{\rho}_{\underline{i}}\,^{\dot{\alpha}}+\frac{1}{4}\mathcal{F}_{I J} \psi_{a}\,^{j \alpha} \overline{\psi}^{a}\,_{j \dot{\alpha}} \overline{\textbf{W}}^{I} \boldsymbol{\lambda}_{i \alpha}\,^{J} {q}^{-2} q^{i \underline{i}} \overline{\rho}_{\underline{i}}\,^{\dot{\alpha}}+\frac{1}{4}{\rm i} \mathcal{F}_{I} (\sigma_{b})^{\alpha}{}_{\dot{\beta}} \psi_{a}\,^{j}\,_{\alpha} \overline{\psi}^{a i}\,_{\dot{\alpha}} \overline{\psi}^{b}{}_{j}^{\dot{\beta}} \overline{\boldsymbol{\lambda}}_{\,i}^{\,\dot{\alpha}}{}^{I} - \frac{1}{4}{\rm i} \mathcal{F}_{I} (\sigma_{a})^{\alpha}{}_{\dot{\beta}} \psi^{a}{}^{j}_{\alpha} \overline{\psi}_{b}\,^{i}\,_{\dot{\alpha}} \overline{\psi}^{b}{}_{j}^{\dot{\beta}} \overline{\boldsymbol{\lambda}}_{\,i}^{\,\dot{\alpha}}{}^{I} - \frac{1}{8}{\rm i} \mathcal{F}_{I J} (\sigma_{a})^{\alpha}{}_{\dot{\beta}} \overline{\psi}^{a}\,_{j \dot{\alpha}} \boldsymbol{\lambda}^{j}_{\alpha}{}^{I} \overline{\boldsymbol{\lambda}}_{\,i}^{\,\dot{\beta}}{}^{J} {q}^{-2} q^{i \underline{i}} \overline{\rho}_{\underline{i}}\,^{\dot{\alpha}} - \frac{1}{4}\mathcal{F}_{I} (\sigma_{ab})^{\rho\alpha} \psi^{a}\,_{i \rho} \psi^{b i \beta} \overline{\textbf{W}}^{I} {q}^{-2} \rho_{\underline{i} \alpha} \rho^{\underline{i}}\,_{\beta} - \frac{1}{4}\mathcal{F}_{I} (\tilde{\sigma}_{ab})_{\dot{\rho}\dot{\alpha}} \overline{\psi}^{a i \dot{\rho}} \overline{\psi}^{b}\,_{i \dot{\beta}} \overline{\textbf{W}}^{I} {q}^{-2} \overline{\rho}^{\underline{i} \dot{\alpha}} \overline{\rho}_{\underline{i}}\,^{\dot{\beta}}+\mathcal{F}_{I J} (\sigma_{ab})^{\beta\alpha} \psi^{a}{}^{j}_{\beta} \overline{\psi}^{b}\,_{j \dot{\alpha}} \overline{\textbf{W}}^{I} \boldsymbol{\lambda}_{i \alpha}\,^{J} {q}^{-2} q^{i \underline{i}} \overline{\rho}_{\underline{i}}\,^{\dot{\alpha}}+\frac{1}{8}{\rm i} \mathcal{F}_{I} (\sigma_{a})^{\alpha}{}_{\dot{\alpha}} \overline{\psi}^{a}\,_{i \dot{\beta}} \overline{\textbf{W}}^{I} {q}^{-4} q^{i \underline{j}} \rho^{\underline{i}}\,_{\alpha} \overline{\rho}_{\underline{j}}\,^{\dot{\alpha}} \overline{\rho}_{\underline{i}}\,^{\dot{\beta}} - \frac{1}{4}\mathcal{F}_{I} \psi_{a i}\,^{\alpha} \overline{\psi}^{a}\,_{j \dot{\alpha}} \overline{\textbf{W}}^{I} {q}^{-4} q^{i \underline{j}} q^{j}\,_{\underline{i}} \rho^{\underline{i}}\,_{\alpha} \overline{\rho}_{\underline{j}}\,^{\dot{\alpha}} - \frac{1}{4}\mathcal{F}_{I} \epsilon^{a}\,_{b c d} (\sigma_{a})^{\alpha}{}_{\dot{\beta}} \psi^{b}{}^{j}_{\alpha} \overline{\psi}^{c}\,_{j \dot{\alpha}} \overline{\psi}^{d}{}_{i}^{\dot{\beta}} \overline{\textbf{W}}^{I} {q}^{-2} q^{i \underline{i}} \overline{\rho}_{\underline{i}}\,^{\dot{\alpha}} - \frac{1}{4}\mathcal{F}_{I J} \psi_{a j}\,^{\beta} \psi^{a j \alpha} \overline{\textbf{W}}^{I} \boldsymbol{\lambda}_{i \beta}\,^{J} {q}^{-2} q^{i}\,_{\underline{i}} \rho^{\underline{i}}\,_{\alpha}+\frac{1}{4}\mathcal{F}_{I J} \psi_{a j}\,^{\alpha} \psi^{a}\,_{i}\,^{\beta} \overline{\textbf{W}}^{I} \boldsymbol{\lambda}^{j}_{\beta}{}^{J} {q}^{-2} q^{i}\,_{\underline{i}} \rho^{\underline{i}}\,_{\alpha}+\frac{1}{4}{\rm i} \mathcal{F}_{I} (\sigma_{a})^{\alpha}{}_{\dot{\alpha}} \overline{\psi}^{a}\,_{i \dot{\beta}} \overline{\boldsymbol{\lambda}}_{\,j}^{\,\dot{\beta}}{}^{I} {q}^{-4} q^{i}\,_{\underline{i}} q^{j \underline{j}} \rho^{\underline{i}}\,_{\alpha} \overline{\rho}_{\underline{j}}\,^{\dot{\alpha}}+\frac{1}{4}{\rm i} \mathcal{F}_{I} (\sigma_{a})^{\alpha}{}_{\dot{\alpha}} \overline{\psi}^{a}\,_{i \dot{\beta}} \overline{\boldsymbol{\lambda}}_{\,j}^{\,\dot{\beta}}{}^{I} {q}^{-4} q^{i \underline{j}} q^{j}\,_{\underline{i}} \rho^{\underline{i}}\,_{\alpha} \overline{\rho}_{\underline{j}}\,^{\dot{\alpha}} - \frac{1}{4}{\rm i} \mathcal{F}_{I} (\sigma_{b})^{\alpha}{}_{\dot{\beta}} \psi_{a}\,^{j}\,_{\alpha} \overline{\psi}^{a}\,_{j \dot{\alpha}} \overline{\psi}^{b}{}_{i}^{\dot{\beta}} \overline{\textbf{W}}^{I} {q}^{-2} q^{i \underline{i}} \overline{\rho}_{\underline{i}}\,^{\dot{\alpha}}+\frac{1}{4}{\rm i} \mathcal{F}_{I} (\sigma_{a})^{\alpha}{}_{\dot{\beta}} \psi^{a}{}^{j}_{\alpha} \overline{\psi}_{b j \dot{\alpha}} \overline{\psi}^{b}{}_{i}^{\dot{\beta}} \overline{\textbf{W}}^{I} {q}^{-2} q^{i \underline{i}} \overline{\rho}_{\underline{i}}\,^{\dot{\alpha}}+\frac{1}{8}{\rm i} \mathcal{F}_{I J} (\sigma_{a})^{\alpha}{}_{\dot{\beta}} \overline{\psi}^{a j}\,_{\dot{\alpha}} \boldsymbol{\lambda}_{i \alpha}\,^{I} \overline{\boldsymbol{\lambda}}_{\,j}^{\,\dot{\beta}}{}^{J} {q}^{-2} q^{i \underline{i}} \overline{\rho}_{\underline{i}}\,^{\dot{\alpha}} - \frac{1}{8}{\rm i} \mathcal{F}_{I J} (\sigma_{a})^{\beta}{}_{\dot{\alpha}} \psi^{a}\,_{j}\,^{\alpha} \boldsymbol{\lambda}^{j}_{\beta}{}^{I} \overline{\boldsymbol{\lambda}}_{\,i}^{\,\dot{\alpha}}{}^{J} {q}^{-2} q^{i}\,_{\underline{i}} \rho^{\underline{i}}\,_{\alpha}%
+\frac{1}{8}{\rm i} \mathcal{F}_{I J} (\sigma_{a})^{\beta}{}_{\dot{\alpha}} \psi^{a j \alpha} \boldsymbol{\lambda}_{i \beta}\,^{I} \overline{\boldsymbol{\lambda}}_{\,j}^{\,\dot{\alpha}}{}^{J} {q}^{-2} q^{i}\,_{\underline{i}} \rho^{\underline{i}}\,_{\alpha}-\mathcal{F}_{I J} (\sigma_{ab})^{\rho\beta} \psi^{a}\,_{j \rho} \psi^{b j \alpha} \overline{\textbf{W}}^{I} \boldsymbol{\lambda}_{i \beta}\,^{J} {q}^{-2} q^{i}\,_{\underline{i}} \rho^{\underline{i}}\,_{\alpha} - \frac{1}{8}{\rm i} \mathcal{F}_{I} (\sigma_{a})^{\alpha}{}_{\dot{\alpha}} \overline{\psi}^{a}\,_{i \dot{\beta}} \overline{\textbf{W}}^{I} {q}^{-4} q^{i}\,_{\underline{i}} \rho^{\underline{i}}\,_{\alpha} \overline{\rho}^{\underline{j} \dot{\alpha}} \overline{\rho}_{\underline{j}}\,^{\dot{\beta}}+\frac{1}{8}{\rm i} \mathcal{F}_{I} (\sigma_{a})^{\alpha}{}_{\dot{\alpha}} \psi^{a}\,_{i}\,^{\beta} \overline{\textbf{W}}^{I} {q}^{-4} q^{i}\,_{\underline{i}} \rho^{\underline{i}}\,_{\alpha} \rho^{\underline{j}}\,_{\beta} \overline{\rho}_{\underline{j}}\,^{\dot{\alpha}} - \frac{1}{8}{\rm i} \mathcal{F}_{I} (\sigma_{a})^{\alpha}{}_{\dot{\alpha}} \psi^{a}\,_{i}\,^{\beta} \overline{\textbf{W}}^{I} {q}^{-4} q^{i \underline{j}} \rho_{\underline{i} \alpha} \rho^{\underline{i}}\,_{\beta} \overline{\rho}_{\underline{j}}\,^{\dot{\alpha}}+\frac{1}{8}\mathcal{F}_{I} \psi_{a i}\,^{\beta} \psi^{a}\,_{j}\,^{\alpha} \overline{\textbf{W}}^{I} {q}^{-4} q^{i}\,_{\underline{i}} q^{j}\,_{\underline{j}} \rho^{\underline{i}}\,_{\alpha} \rho^{\underline{j}}\,_{\beta} - \frac{1}{4}\mathcal{F}_{I} \epsilon^{a}\,_{b c d} (\sigma_{a})^{\alpha}{}_{\dot{\beta}} \psi^{b}\,_{i \alpha} \overline{\psi}^{c j \dot{\beta}} \overline{\psi}^{d}\,_{j \dot{\alpha}} \overline{\textbf{W}}^{I} {q}^{-2} q^{i \underline{i}} \overline{\rho}_{\underline{i}}\,^{\dot{\alpha}}+\frac{1}{4}\mathcal{F}_{I} \epsilon^{a}\,_{b c d} (\sigma_{a})^{\beta}{}_{\dot{\alpha}} \psi^{b}\,_{j \beta} \psi^{c j \alpha} \overline{\psi}^{d}{}_{i}^{\dot{\alpha}} \overline{\textbf{W}}^{I} {q}^{-2} q^{i}\,_{\underline{i}} \rho^{\underline{i}}\,_{\alpha}+\frac{1}{4}\mathcal{F}_{I} \epsilon^{a}\,_{b c d} (\sigma_{a})^{\beta}{}_{\dot{\alpha}} \psi^{b}\,_{i \beta} \psi^{c j \alpha} \overline{\psi}^{d}{}_{j}^{\dot{\alpha}} \overline{\textbf{W}}^{I} {q}^{-2} q^{i}\,_{\underline{i}} \rho^{\underline{i}}\,_{\alpha} - \frac{1}{4}{\rm i} \mathcal{F}_{I} (\sigma_{b})^{\alpha}{}_{\dot{\beta}} \psi_{a i \alpha} \overline{\psi}^{a j}\,_{\dot{\alpha}} \overline{\psi}^{b}{}_{j}^{\dot{\beta}} \overline{\textbf{W}}^{I} {q}^{-2} q^{i \underline{i}} \overline{\rho}_{\underline{i}}\,^{\dot{\alpha}}+\frac{1}{4}{\rm i} \mathcal{F}_{I} (\sigma_{a})^{\alpha}{}_{\dot{\beta}} \psi^{a}\,_{i \alpha} \overline{\psi}_{b}\,^{j \dot{\beta}} \overline{\psi}^{b}\,_{j \dot{\alpha}} \overline{\textbf{W}}^{I} {q}^{-2} q^{i \underline{i}} \overline{\rho}_{\underline{i}}\,^{\dot{\alpha}}+\frac{1}{4}{\rm i} \mathcal{F}_{I} (\sigma_{b})^{\beta}{}_{\dot{\alpha}} \psi_{a j \beta} \psi^{a j \alpha} \overline{\psi}^{b}{}_{i}^{\dot{\alpha}} \overline{\textbf{W}}^{I} {q}^{-2} q^{i}\,_{\underline{i}} \rho^{\underline{i}}\,_{\alpha} - \frac{1}{4}{\rm i} \mathcal{F}_{I} (\sigma_{b})^{\beta}{}_{\dot{\alpha}} \psi_{a i \beta} \psi^{a j \alpha} \overline{\psi}^{b}{}_{j}^{\dot{\alpha}} \overline{\textbf{W}}^{I} {q}^{-2} q^{i}\,_{\underline{i}} \rho^{\underline{i}}\,_{\alpha} - \frac{1}{4}{\rm i} \mathcal{F}_{I} (\sigma_{b})^{\beta}{}_{\dot{\alpha}} \psi_{a j}\,^{\alpha} \psi^{b}{}^{j}_{\beta} \overline{\psi}^{a}{}_{i}^{\dot{\alpha}} \overline{\textbf{W}}^{I} {q}^{-2} q^{i}\,_{\underline{i}} \rho^{\underline{i}}\,_{\alpha} - \frac{1}{4}{\rm i} \mathcal{F}_{I} (\sigma_{b})^{\beta}{}_{\dot{\alpha}} \psi_{a}\,^{j \alpha} \psi^{b}\,_{i \beta} \overline{\psi}^{a}{}_{j}^{\dot{\alpha}} \overline{\textbf{W}}^{I} {q}^{-2} q^{i}\,_{\underline{i}} \rho^{\underline{i}}\,_{\alpha} - \frac{1}{4}{\rm i} \mathcal{F}_{I J} (\sigma_{a})^{\alpha}{}_{\dot{\alpha}} \psi^{a}\,_{i}\,^{\beta} \overline{\textbf{W}}^{I} \boldsymbol{\lambda}_{j \beta}\,^{J} {q}^{-4} q^{i}\,_{\underline{i}} q^{j \underline{j}} \rho^{\underline{i}}\,_{\alpha} \overline{\rho}_{\underline{j}}\,^{\dot{\alpha}} - \frac{1}{4}{\rm i} \mathcal{F}_{I J} (\sigma_{a})^{\alpha}{}_{\dot{\alpha}} \psi^{a}\,_{i}\,^{\beta} \overline{\textbf{W}}^{I} \boldsymbol{\lambda}_{j \beta}\,^{J} {q}^{-4} q^{i \underline{j}} q^{j}\,_{\underline{i}} \rho^{\underline{i}}\,_{\alpha} \overline{\rho}_{\underline{j}}\,^{\dot{\alpha}}$\end{adjustwidth}

\section{Off-Shell SU(1,1)/U(1) sigma model}

Here, we give the entire fermionic Lagrangian for the SU(1,1)/U(1) off-shell model corresponding to eq. (6.3) of the main paper divided as follows
\bea
\cL^{{\rm off-shell,\,SU(1,1)/U(1)}}
=\cL^{{\rm off,\,SU(1,1)/U(1)}}_{0}
+\cL^{{\rm off,\,SU(1,1)/U(1)}}_{2}
+\cL^{{\rm off,\,SU(1,1)/U(1)}}_{4}
~,~~~
\eea
where once again the subscript of the Lagrangian refers the order in explicit fermions. Note again that implicit fermions exist here and can be seen by converting $H^{a \underline{i} \underline{j}}$, $F^{a b}$, and $\textbf{F}^{a b}$ to $\tilde{h}^{a \underline{i} \underline{j}}$, $f^{a b}$, and $\textbf{f}^{a b}$, respectively, by eqs. (2.67), (2.68), and (2.44) in the main paper and that the $\textbf{D}_{a}$ derivative has gravitini hidden in the spin connection.

\subsection{$\cL^{{\rm off,\,SU(1,1)/U(1)}}_{0}$}

\begin{adjustwidth}{1em}{4cm}${}c \phi \overline{\boldsymbol{\phi}} R+c \overline{\phi} \boldsymbol{\phi} R-2c F_{a b} \textbf{F}^{a b}+\frac{1}{8}c X_{i j} \textbf{X}^{i j}-c W_{a b} \overline{\boldsymbol{\phi}} F^{a b} - \frac{1}{2}{\rm i} c \epsilon_{c d a b} W^{c d} \overline{\boldsymbol{\phi}} F^{a b}-c W_{a b} \overline{\phi} \textbf{F}^{a b} - \frac{1}{2}{\rm i} c \epsilon_{a b c d} W^{a b} \overline{\phi} \textbf{F}^{c d}-c W_{a b} \boldsymbol{\phi} F^{a b}+\frac{1}{2}{\rm i} c \epsilon_{c d a b} W^{c d} \boldsymbol{\phi} F^{a b}-c W_{a b} \phi \textbf{F}^{a b}+\frac{1}{2}{\rm i} c \epsilon_{a b c d} W^{a b} \phi \textbf{F}^{c d}+c \boldsymbol{\phi} \textbf{D}_{a}{\textbf{D}^{a}{\overline{\phi}}}+c \overline{\boldsymbol{\phi}} \textbf{D}_{a}{\textbf{D}^{a}{\phi}}+c \phi \textbf{D}_{a}{\textbf{D}^{a}{\overline{\boldsymbol{\phi}}}}-8c \phi \overline{\boldsymbol{\phi}} A_{a} A^{a}+c \overline{\phi} \textbf{D}_{a}{\textbf{D}^{a}{\boldsymbol{\phi}}}-8c \overline{\phi} \boldsymbol{\phi} A_{a} A^{a} - \frac{1}{2}c W_{a b} W^{a b} \overline{\phi} \overline{\boldsymbol{\phi}}%
 - \frac{1}{4}{\rm i} c \epsilon_{a b c d} W^{a b} W^{c d} \overline{\phi} \overline{\boldsymbol{\phi}} - \frac{1}{2}c W_{a b} W^{a b} \phi \boldsymbol{\phi}+\frac{1}{4}{\rm i} c \epsilon_{a b c d} W^{a b} W^{c d} \phi \boldsymbol{\phi}-4{\rm i} c \boldsymbol{\phi} A_{a} \textbf{D}^{a}{\overline{\phi}}+4{\rm i} c \overline{\boldsymbol{\phi}} A_{a} \textbf{D}^{a}{\phi}-4{\rm i} c \phi A_{a} \textbf{D}^{a}{\overline{\boldsymbol{\phi}}}+4{\rm i} c \overline{\phi} A_{a} \textbf{D}^{a}{\boldsymbol{\phi}}-128c \phi \overline{\boldsymbol{\phi}} {q}^{-4} H_{a \underline{i} \underline{j}} H^{a \underline{i} \underline{j}}-128c \overline{\phi} \boldsymbol{\phi} {q}^{-4} H_{a \underline{i} \underline{j}} H^{a \underline{i} \underline{j}}+2c \epsilon_{\underline{i} \underline{j}} \epsilon_{i j} \phi \overline{\boldsymbol{\phi}} {q}^{-2} \textbf{D}_{a}{q^{i \underline{i}}} \textbf{D}^{a}{q^{j \underline{j}}}+4c \phi \overline{\boldsymbol{\phi}} {q}^{-2} q_{i \underline{i}} \textbf{D}_{a}{\textbf{D}^{a}{q^{i \underline{i}}}}+2c \epsilon_{\underline{i} \underline{j}} \epsilon_{i j} \overline{\phi} \boldsymbol{\phi} {q}^{-2} \textbf{D}_{a}{q^{i \underline{i}}} \textbf{D}^{a}{q^{j \underline{j}}}+4c \overline{\phi} \boldsymbol{\phi} {q}^{-2} q_{i \underline{i}} \textbf{D}_{a}{\textbf{D}^{a}{q^{i \underline{i}}}}-4c \phi \overline{\boldsymbol{\phi}} {q}^{-4} q_{i \underline{i}} q_{j \underline{j}} \textbf{D}_{a}{q^{i \underline{j}}} \textbf{D}^{a}{q^{j \underline{i}}}-4c \overline{\phi} \boldsymbol{\phi} {q}^{-4} q_{i \underline{i}} q_{j \underline{j}} \textbf{D}_{a}{q^{i \underline{j}}} \textbf{D}^{a}{q^{j \underline{i}}} +  \frac{1}{4} \xi_{\underline{i} \underline{j}} q^{i \underline{i}}q^{j \underline{j}} X_{i j}     +\frac{4}{3} \xi_{\underline{i} \underline{j}} \e^{m n p q} h_{m np}{}^{\underline{i}\underline{j}} v_{q}$\end{adjustwidth}

\subsection{$\cL^{{\rm off,\,SU(1,1)/U(1)}}_{2}$}
\begin{adjustwidth}{1em}{4.5cm}${}\left(\frac{1}{16}{\rm i} c \zeta_{\underline{i} \underline{j}} + \frac{1}{8} \xi_{\underline{i} \underline{j}}\right)  \phi \rho^{\underline{i} \alpha} \rho^{\underline{j}}\,_{\alpha} - \left(\frac{1}{4}{\rm i} c \zeta_{\underline{i} \underline{j}} + \frac{1}{2} \xi_{\underline{i} \underline{j}}\right) \lambda_{i}\,^{\alpha} q^{i \underline{j}} \rho^{\underline{i}}\,_{\alpha} + \left(\frac{1}{4}{\rm i} c \zeta_{\underline{i} \underline{j}} - \frac{1}{2} \xi_{\underline{i} \underline{j}}\right)  \overline{\lambda}_{i \dot{\alpha}} q^{i \underline{j}} \overline{\rho}^{\underline{i} \dot{\alpha}}+\left(\frac{1}{4} c \zeta_{\underline{i} \underline{j}} - \frac{1}{2} {\rm i} \xi_{\underline{i} \underline{j}}\right) (\sigma_{a})^{\alpha}{}_{\dot{\alpha}} \overline{\psi}^{a}{}_{i}^{\dot{\alpha}} \phi q^{i \underline{j}} \rho^{\underline{i}}\,_{\alpha} + \left(-\frac{1}{16}{\rm i} c \zeta_{\underline{i} \underline{j}} + \frac{1}{8} \xi_{\underline{i} \underline{j}}\right) \overline{\phi} \overline{\rho}^{\underline{i}}_{\dot{\alpha}} \overline{\rho}^{\underline{j} \dot{\alpha}}+\left(\frac{1}{4} c \zeta_{\underline{i} \underline{j}} + \frac{1}{2} {\rm i} \xi_{\underline{i} \underline{j}}\right) (\sigma_{a})^{\alpha}{}_{\dot{\alpha}} \psi^{a}\,_{i \alpha} \overline{\lambda}_{j}\,^{\dot{\alpha}} q^{i \underline{i}} q^{j \underline{j}} - \left(\frac{1}{4} c \zeta_{\underline{i} \underline{j}} - \frac{1}{2} {\rm i} \xi_{\underline{i} \underline{j}}\right) (\sigma_{a})^{\alpha}{}_{\dot{\alpha}} \overline{\psi}^{a}{}_{j}^{\dot{\alpha}} \lambda_{i \alpha} q^{i \underline{i}} q^{j \underline{j}}-\left(\frac{1}{4} c \zeta_{\underline{i} \underline{j}} + \frac{1}{2} {\rm i} \xi_{\underline{i} \underline{j}}\right) (\sigma_{a})^{\alpha}{}_{\dot{\alpha}} \psi^{a}\,_{i \alpha} \overline{\phi} q^{i \underline{j}} \overline{\rho}^{\underline{i}\dot{\alpha}}+\left(\frac{1}{2}{\rm i} c \zeta_{\underline{i} \underline{j}} + \xi_{\underline{i} \underline{j}} \right)(\tilde{\sigma}_{ab})_{\dot{\alpha}\dot{\beta}} \overline{\psi}^{a}{}_{i}^{\dot{\alpha}} \overline{\psi}^{b}{}_{j}^{\dot{\beta}} \phi q^{i \underline{i}} q^{j \underline{j}} +\left(- \frac{1}{2}{\rm i} c \zeta_{\underline{i} \underline{j}} + \xi_{\underline{i} \underline{j}}\right) (\sigma_{ab})^{\alpha\beta} \psi^{a}\,_{i \alpha} \psi^{b}\,_{j \beta} \overline{\phi} q^{i \underline{i}} q^{j \underline{j}}+c (\sigma_{a})^{\alpha}{}_{\dot{\alpha}} \lambda^{i}\,_{\alpha} \overline{\boldsymbol{\lambda}}_{i}\,^{\dot{\alpha}} A^{a}-c (\sigma_{a})^{\alpha}{}_{\dot{\alpha}} \overline{\lambda}_{i}\,^{\dot{\alpha}} \boldsymbol{\lambda}^{i}\,_{\alpha} A^{a}+\frac{1}{2}{\rm i} c (\sigma_{a})^{\alpha}{}_{\dot{\alpha}} \overline{\boldsymbol{\lambda}}_{i}\,^{\dot{\alpha}} \textbf{D}^{a}{\lambda^{i}\,_{\alpha}}+\frac{1}{2}{\rm i} c (\sigma_{a})^{\alpha}{}_{\dot{\alpha}} \overline{\lambda}_{i}\,^{\dot{\alpha}} \textbf{D}^{a}{\boldsymbol{\lambda}^{i}\,_{\alpha}}+\frac{1}{2}{\rm i} c (\sigma_{a})^{\alpha}{}_{\dot{\alpha}} \boldsymbol{\lambda}^{i}\,_{\alpha} \textbf{D}^{a}{\overline{\lambda}_{i}\,^{\dot{\alpha}}}+\frac{1}{2}{\rm i} c (\sigma_{a})^{\alpha}{}_{\dot{\alpha}} \lambda^{i}\,_{\alpha} \textbf{D}^{a}{\overline{\boldsymbol{\lambda}}_{i}\,^{\dot{\alpha}}}+c (\sigma_{a})^{\alpha}{}_{\dot{\alpha}} \phi \overline{\boldsymbol{\phi}} {q}^{-2} \rho^{\underline{i}}\,_{\alpha} \overline{\rho}_{\underline{i}}\,^{\dot{\alpha}} A^{a}+c (\sigma_{a})^{\alpha}{}_{\dot{\alpha}} \overline{\phi} \boldsymbol{\phi} {q}^{-2} \rho^{\underline{i}}\,_{\alpha} \overline{\rho}_{\underline{i}}\,^{\dot{\alpha}} A^{a}+c (\tilde{\sigma}_{ab})_{\dot{\beta}\dot{\alpha}} W^{a b} \boldsymbol{\phi} \overline{\lambda}_{i}\,^{\dot{\beta}} {q}^{-2} q^{i \underline{i}} \overline{\rho}_{\underline{i}}\,^{\dot{\alpha}}%
+c (\tilde{\sigma}_{ab})_{\dot{\beta}\dot{\alpha}} W^{a b} \phi \overline{\boldsymbol{\lambda}}_{i}\,^{\dot{\beta}} {q}^{-2} q^{i \underline{i}} \overline{\rho}_{\underline{i}}\,^{\dot{\alpha}}+\frac{1}{2}c \epsilon^{c}\,_{d a b} (\sigma_{c})^{\alpha}{}_{\dot{\alpha}} \psi^{d}{}^{i}_{\alpha} \overline{\boldsymbol{\lambda}}_{i}\,^{\dot{\alpha}} F^{a b}+\frac{1}{2}c \epsilon^{c}\,_{d a b} (\sigma_{c})^{\alpha}{}_{\dot{\alpha}} \psi^{d}{}^{i}_{\alpha} \overline{\lambda}_{i}\,^{\dot{\alpha}} \textbf{F}^{a b}+\frac{1}{2}c \epsilon^{c}\,_{d a b} (\sigma_{c})^{\alpha}{}_{\dot{\alpha}} \overline{\psi}^{d}{}_{i}^{\dot{\alpha}} \boldsymbol{\lambda}^{i}\,_{\alpha} F^{a b}+\frac{1}{2}c \epsilon^{c}\,_{d a b} (\sigma_{c})^{\alpha}{}_{\dot{\alpha}} \overline{\psi}^{d}{}_{i}^{\dot{\alpha}} \lambda^{i}\,_{\alpha} \textbf{F}^{a b}+\frac{1}{2}c \overline{\boldsymbol{\phi}} \lambda_{i}\,^{\alpha} \textbf{D}_{a}{\psi^{a}{}^{i}_{\alpha}} - \frac{1}{2}c \psi_{a i}\,^{\alpha} \overline{\phi} \textbf{D}^{a}{\boldsymbol{\lambda}^{i}\,_{\alpha}}-c (\sigma_{ba})^{\beta\alpha} \psi^{b}\,_{i \beta} \overline{\phi} \textbf{D}^{a}{\boldsymbol{\lambda}^{i}\,_{\alpha}} - \frac{1}{2}c \psi_{a i}\,^{\alpha} \overline{\boldsymbol{\phi}} \textbf{D}^{a}{\lambda^{i}\,_{\alpha}}+\frac{1}{2}c \psi_{a i}\,^{\alpha} \lambda^{i}\,_{\alpha} \textbf{D}^{a}{\overline{\boldsymbol{\phi}}}+c (\sigma_{ba})^{\beta\alpha} \psi^{b}\,_{i \beta} \lambda^{i}\,_{\alpha} \textbf{D}^{a}{\overline{\boldsymbol{\phi}}}+\frac{1}{2}c \overline{\phi} \boldsymbol{\lambda}_{i}\,^{\alpha} \textbf{D}_{a}{\psi^{a}{}^{i}_{\alpha}} - \frac{1}{2}c \psi_{a i}\,^{\alpha} \psi_{b}\,^{i}\,_{\alpha} \overline{\boldsymbol{\phi}} F^{a b}+\frac{1}{4}{\rm i} c \epsilon_{c d a b} \psi^{c}\,_{i}\,^{\alpha} \psi^{d}{}^{i}_{\alpha} \overline{\boldsymbol{\phi}} F^{a b} - \frac{1}{2}c \psi_{a i}\,^{\alpha} \psi_{b}\,^{i}\,_{\alpha} \overline{\phi} \textbf{F}^{a b}+\frac{1}{4}{\rm i} c \epsilon_{c d a b} \psi^{c}\,_{i}\,^{\alpha} \psi^{d}{}^{i}_{\alpha} \overline{\phi} \textbf{F}^{a b}+\frac{1}{2}c \boldsymbol{\phi} \overline{\lambda}^{i}\,_{\dot{\alpha}} \textbf{D}_{a}{\overline{\psi}^{a}{}_{i}^{\dot{\alpha}}} - \frac{1}{2}c \overline{\psi}_{a}\,^{i}\,_{\dot{\alpha}} \boldsymbol{\phi} \textbf{D}^{a}{\overline{\lambda}_{i}\,^{\dot{\alpha}}}+\frac{1}{2}c \overline{\psi}_{a}\,^{i}\,_{\dot{\alpha}} \overline{\boldsymbol{\lambda}}_{i}\,^{\dot{\alpha}} \textbf{D}^{a}{\phi}+c (\tilde{\sigma}_{ba})_{\dot{\beta}\dot{\alpha}} \overline{\psi}^{b i \dot{\beta}} \overline{\boldsymbol{\lambda}}_{i}\,^{\dot{\alpha}} \textbf{D}^{a}{\phi}%
+\frac{1}{2}c \overline{\psi}_{a}\,^{i}\,_{\dot{\alpha}} \overline{\lambda}_{i}\,^{\dot{\alpha}} \textbf{D}^{a}{\boldsymbol{\phi}}+c (\tilde{\sigma}_{ba})_{\dot{\beta}\dot{\alpha}} \overline{\psi}^{b i \dot{\beta}} \overline{\lambda}_{i}\,^{\dot{\alpha}} \textbf{D}^{a}{\boldsymbol{\phi}} - \frac{1}{2}c \overline{\psi}_{a}\,^{i}\,_{\dot{\alpha}} \phi \textbf{D}^{a}{\overline{\boldsymbol{\lambda}}_{i}\,^{\dot{\alpha}}}+\frac{1}{2}c \phi \overline{\boldsymbol{\lambda}}^{i}\,_{\dot{\alpha}} \textbf{D}_{a}{\overline{\psi}^{a}{}_{i}^{\dot{\alpha}}}-2c (\sigma_{a})^{\alpha}{}_{\dot{\alpha}} \boldsymbol{\phi} \overline{\lambda}_{i}\,^{\dot{\alpha}} {q}^{-2} q^{i}\,_{\underline{i}} \rho^{\underline{i}}\,_{\alpha} A^{a}-2c (\sigma_{a})^{\alpha}{}_{\dot{\alpha}} \phi \overline{\boldsymbol{\lambda}}_{i}\,^{\dot{\alpha}} {q}^{-2} q^{i}\,_{\underline{i}} \rho^{\underline{i}}\,_{\alpha} A^{a}+\frac{1}{8}c \phi \overline{\boldsymbol{\phi}} {q}^{-4} \rho^{\underline{i} \alpha} \rho^{\underline{j}}\,_{\alpha} \overline{\rho}_{\underline{i} \dot{\alpha}} \overline{\rho}_{\underline{j}}\,^{\dot{\alpha}}+\frac{1}{8}c \overline{\phi} \boldsymbol{\phi} {q}^{-4} \rho^{\underline{i} \alpha} \rho^{\underline{j}}\,_{\alpha} \overline{\rho}_{\underline{i} \dot{\alpha}} \overline{\rho}_{\underline{j}}\,^{\dot{\alpha}}+c (\sigma_{cd})^{\rho\alpha} (\sigma_{ab})_{\rho}{}^{\beta} \psi^{c}\,_{i \alpha} \psi^{d}{}^{i}_{\beta} \overline{\boldsymbol{\phi}} F^{a b}+c (\sigma_{cd})^{\rho\alpha} (\sigma_{ab})_{\rho}{}^{\beta} \psi^{c}\,_{i \alpha} \psi^{d}{}^{i}_{\beta} \overline{\phi} \textbf{F}^{a b}-c (\sigma_{ba})^{\beta\alpha} \psi^{b}\,_{i \beta} \overline{\boldsymbol{\phi}} \textbf{D}^{a}{\lambda^{i}\,_{\alpha}} - \frac{1}{4}c (\sigma_{ab})^{\alpha\beta} W^{a b} \phi \overline{\boldsymbol{\phi}} {q}^{-2} \rho_{\underline{i} \alpha} \rho^{\underline{i}}\,_{\beta} - \frac{1}{4}c (\sigma_{ab})^{\alpha\beta} W^{a b} \overline{\phi} \boldsymbol{\phi} {q}^{-2} \rho_{\underline{i} \alpha} \rho^{\underline{i}}\,_{\beta}+\frac{1}{2}c \psi_{a i}\,^{\alpha} \boldsymbol{\lambda}^{i}\,_{\alpha} \textbf{D}^{a}{\overline{\phi}}+c (\sigma_{ba})^{\beta\alpha} \psi^{b}\,_{i \beta} \boldsymbol{\lambda}^{i}\,_{\alpha} \textbf{D}^{a}{\overline{\phi}}+\frac{1}{2}c \overline{\psi}_{a}\,^{i}\,_{\dot{\alpha}} \overline{\psi}_{b i}\,^{\dot{\alpha}} \boldsymbol{\phi} F^{a b}+\frac{1}{4}{\rm i} c \epsilon_{c d a b} \overline{\psi}^{c i}\,_{\dot{\alpha}} \overline{\psi}^{d}{}_{i}^{\dot{\alpha}} \boldsymbol{\phi} F^{a b}+\frac{1}{2}c \overline{\psi}_{a}\,^{i}\,_{\dot{\alpha}} \overline{\psi}_{b i}\,^{\dot{\alpha}} \phi \textbf{F}^{a b}+\frac{1}{4}{\rm i} c \epsilon_{c d a b} \overline{\psi}^{c i}\,_{\dot{\alpha}} \overline{\psi}^{d}{}_{i}^{\dot{\alpha}} \phi \textbf{F}^{a b}-c (\tilde{\sigma}_{ba})_{\dot{\beta}\dot{\alpha}} \overline{\psi}^{b i \dot{\beta}} \boldsymbol{\phi} \textbf{D}^{a}{\overline{\lambda}_{i}\,^{\dot{\alpha}}}%
-c (\tilde{\sigma}_{ba})_{\dot{\beta}\dot{\alpha}} \overline{\psi}^{b i \dot{\beta}} \phi \textbf{D}^{a}{\overline{\boldsymbol{\lambda}}_{i}\,^{\dot{\alpha}}}+2c (\sigma_{a})^{\alpha}{}_{\dot{\alpha}} \overline{\boldsymbol{\phi}} \lambda_{i \alpha} {q}^{-2} q^{i \underline{i}} \overline{\rho}_{\underline{i}}\,^{\dot{\alpha}} A^{a}+2c (\sigma_{a})^{\alpha}{}_{\dot{\alpha}} \overline{\phi} \boldsymbol{\lambda}_{i \alpha} {q}^{-2} q^{i \underline{i}} \overline{\rho}_{\underline{i}}\,^{\dot{\alpha}} A^{a}-c (\tilde{\sigma}_{cd})_{\dot{\rho}\dot{\alpha}} (\tilde{\sigma}_{ab})^{\dot{\rho}}{}_{\dot{\beta}} \overline{\psi}^{c i \dot{\alpha}} \overline{\psi}^{d}{}_{i}^{\dot{\beta}} \boldsymbol{\phi} F^{a b}-c (\tilde{\sigma}_{cd})_{\dot{\rho}\dot{\alpha}} (\tilde{\sigma}_{ab})^{\dot{\rho}}{}_{\dot{\beta}} \overline{\psi}^{c i \dot{\alpha}} \overline{\psi}^{d}{}_{i}^{\dot{\beta}} \phi \textbf{F}^{a b}+\frac{1}{4}c (\tilde{\sigma}_{ab})_{\dot{\alpha}\dot{\beta}} W^{a b} \phi \overline{\boldsymbol{\phi}} {q}^{-2} \overline{\rho}^{\underline{i} \dot{\alpha}} \overline{\rho}_{\underline{i}}\,^{\dot{\beta}}+\frac{1}{4}c (\tilde{\sigma}_{ab})_{\dot{\alpha}\dot{\beta}} W^{a b} \overline{\phi} \boldsymbol{\phi} {q}^{-2} \overline{\rho}^{\underline{i} \dot{\alpha}} \overline{\rho}_{\underline{i}}\,^{\dot{\beta}} - \frac{1}{2}{\rm i} c (\sigma_{a})^{\alpha}{}_{\dot{\alpha}} \phi \overline{\boldsymbol{\phi}} {q}^{-2} \overline{\rho}_{\underline{i}}\,^{\dot{\alpha}} \textbf{D}^{a}{\rho^{\underline{i}}\,_{\alpha}} - \frac{1}{2}{\rm i} c (\sigma_{a})^{\alpha}{}_{\dot{\alpha}} \overline{\phi} \boldsymbol{\phi} {q}^{-2} \overline{\rho}_{\underline{i}}\,^{\dot{\alpha}} \textbf{D}^{a}{\rho^{\underline{i}}\,_{\alpha}} - \frac{1}{2}{\rm i} c (\sigma_{a})^{\alpha}{}_{\dot{\alpha}} \phi \overline{\boldsymbol{\phi}} {q}^{-2} \rho^{\underline{i}}\,_{\alpha} \textbf{D}^{a}{\overline{\rho}_{\underline{i}}\,^{\dot{\alpha}}} - \frac{1}{2}{\rm i} c (\sigma_{a})^{\alpha}{}_{\dot{\alpha}} \overline{\phi} \boldsymbol{\phi} {q}^{-2} \rho^{\underline{i}}\,_{\alpha} \textbf{D}^{a}{\overline{\rho}_{\underline{i}}\,^{\dot{\alpha}}} - \frac{1}{12}c \psi_{a i}\,^{\alpha} \psi_{b}\,^{i}\,_{\alpha} W^{a b} \phi \overline{\boldsymbol{\phi}} - \frac{1}{12}c \psi_{a i}\,^{\alpha} \psi_{b}\,^{i}\,_{\alpha} W^{a b} \overline{\phi} \boldsymbol{\phi}+c (\sigma_{cd})^{\rho\alpha} (\sigma_{ab})_{\rho}{}^{\beta} \psi^{c}\,_{i \alpha} \psi^{d}{}^{i}_{\beta} W^{a b} \overline{\phi} \overline{\boldsymbol{\phi}} - \frac{1}{2}c \psi_{a i}\,^{\alpha} \psi_{b}\,^{i}\,_{\alpha} W^{a b} \overline{\phi} \overline{\boldsymbol{\phi}} - \frac{1}{4}{\rm i} c \epsilon_{c d a b} \psi^{c}\,_{i}\,^{\alpha} \psi^{d}{}^{i}_{\alpha} W^{a b} \overline{\phi} \overline{\boldsymbol{\phi}}-c (\sigma_{ab})^{\beta\alpha} W^{a b} \overline{\boldsymbol{\phi}} \lambda_{i \beta} {q}^{-2} q^{i}\,_{\underline{i}} \rho^{\underline{i}}\,_{\alpha}-c (\sigma_{ab})^{\beta\alpha} W^{a b} \overline{\phi} \boldsymbol{\lambda}_{i \beta} {q}^{-2} q^{i}\,_{\underline{i}} \rho^{\underline{i}}\,_{\alpha}-c (\tilde{\sigma}_{cd})_{\dot{\rho}\dot{\alpha}} (\tilde{\sigma}_{ab})^{\dot{\rho}}{}_{\dot{\beta}} \overline{\psi}^{c i \dot{\alpha}} \overline{\psi}^{d}{}_{i}^{\dot{\beta}} W^{a b} \phi \boldsymbol{\phi}+\frac{1}{2}c \overline{\psi}_{a}\,^{i}\,_{\dot{\alpha}} \overline{\psi}_{b i}\,^{\dot{\alpha}} W^{a b} \phi \boldsymbol{\phi}%
 - \frac{1}{4}{\rm i} c \epsilon_{c d a b} \overline{\psi}^{c i}\,_{\dot{\alpha}} \overline{\psi}^{d}{}_{i}^{\dot{\alpha}} W^{a b} \phi \boldsymbol{\phi}-16c \overline{\psi}_{a i \dot{\alpha}} \phi \overline{\boldsymbol{\phi}} {q}^{-4} q^{i}\,_{\underline{i}} \overline{\rho}_{\underline{j}}\,^{\dot{\alpha}} H^{a \underline{i} \underline{j}}-16c \overline{\psi}_{a i \dot{\alpha}} \overline{\phi} \boldsymbol{\phi} {q}^{-4} q^{i}\,_{\underline{i}} \overline{\rho}_{\underline{j}}\,^{\dot{\alpha}} H^{a \underline{i} \underline{j}}-2{\rm i} c \psi_{a i}\,^{\alpha} \overline{\boldsymbol{\phi}} \lambda^{i}\,_{\alpha} A^{a}-2{\rm i} c \psi_{a i}\,^{\alpha} \overline{\phi} \boldsymbol{\lambda}^{i}\,_{\alpha} A^{a}+\frac{1}{2}{\rm i} c (\sigma_{b})^{\alpha}{}_{\dot{\alpha}} \psi_{a}\,^{i}\,_{\alpha} \overline{\psi}^{b}{}_{i}^{\dot{\alpha}} \overline{\boldsymbol{\phi}} \textbf{D}^{a}{\phi} - \frac{1}{2}{\rm i} c (\sigma_{b})^{\alpha}{}_{\dot{\alpha}} \psi^{b}{}^{i}_{\alpha} \overline{\psi}_{a i}\,^{\dot{\alpha}} \overline{\boldsymbol{\phi}} \textbf{D}^{a}{\phi}+\frac{1}{2}c \epsilon^{b}\,_{c d a} (\sigma_{b})^{\alpha}{}_{\dot{\alpha}} \psi^{c}{}^{i}_{\alpha} \overline{\psi}^{d}{}_{i}^{\dot{\alpha}} \overline{\boldsymbol{\phi}} \textbf{D}^{a}{\phi}+\frac{1}{2}{\rm i} c (\sigma_{b})^{\alpha}{}_{\dot{\alpha}} \psi_{a}\,^{i}\,_{\alpha} \overline{\psi}^{b}{}_{i}^{\dot{\alpha}} \overline{\phi} \textbf{D}^{a}{\boldsymbol{\phi}} - \frac{1}{2}{\rm i} c (\sigma_{b})^{\alpha}{}_{\dot{\alpha}} \psi^{b}{}^{i}_{\alpha} \overline{\psi}_{a i}\,^{\dot{\alpha}} \overline{\phi} \textbf{D}^{a}{\boldsymbol{\phi}}+\frac{1}{2}c \epsilon^{b}\,_{c d a} (\sigma_{b})^{\alpha}{}_{\dot{\alpha}} \psi^{c}{}^{i}_{\alpha} \overline{\psi}^{d}{}_{i}^{\dot{\alpha}} \overline{\phi} \textbf{D}^{a}{\boldsymbol{\phi}}+2{\rm i} c \overline{\psi}_{a}\,^{i}\,_{\dot{\alpha}} \boldsymbol{\phi} \overline{\lambda}_{i}\,^{\dot{\alpha}} A^{a}+2{\rm i} c \overline{\psi}_{a}\,^{i}\,_{\dot{\alpha}} \phi \overline{\boldsymbol{\lambda}}_{i}\,^{\dot{\alpha}} A^{a}+\frac{1}{2}{\rm i} c (\sigma_{b})^{\alpha}{}_{\dot{\alpha}} \psi_{a}\,^{i}\,_{\alpha} \overline{\psi}^{b}{}_{i}^{\dot{\alpha}} \boldsymbol{\phi} \textbf{D}^{a}{\overline{\phi}} - \frac{1}{2}{\rm i} c (\sigma_{b})^{\alpha}{}_{\dot{\alpha}} \psi^{b}{}^{i}_{\alpha} \overline{\psi}_{a i}\,^{\dot{\alpha}} \boldsymbol{\phi} \textbf{D}^{a}{\overline{\phi}} - \frac{1}{2}c \epsilon^{b}\,_{c d a} (\sigma_{b})^{\alpha}{}_{\dot{\alpha}} \psi^{c}{}^{i}_{\alpha} \overline{\psi}^{d}{}_{i}^{\dot{\alpha}} \boldsymbol{\phi} \textbf{D}^{a}{\overline{\phi}}+\frac{1}{2}{\rm i} c (\sigma_{b})^{\alpha}{}_{\dot{\alpha}} \psi_{a}\,^{i}\,_{\alpha} \overline{\psi}^{b}{}_{i}^{\dot{\alpha}} \phi \textbf{D}^{a}{\overline{\boldsymbol{\phi}}} - \frac{1}{2}{\rm i} c (\sigma_{b})^{\alpha}{}_{\dot{\alpha}} \psi^{b}{}^{i}_{\alpha} \overline{\psi}_{a i}\,^{\dot{\alpha}} \phi \textbf{D}^{a}{\overline{\boldsymbol{\phi}}} - \frac{1}{2}c \epsilon^{b}\,_{c d a} (\sigma_{b})^{\alpha}{}_{\dot{\alpha}} \psi^{c}{}^{i}_{\alpha} \overline{\psi}^{d}{}_{i}^{\dot{\alpha}} \phi \textbf{D}^{a}{\overline{\boldsymbol{\phi}}}-2{\rm i} c (\sigma_{a})^{\alpha}{}_{\dot{\alpha}} \boldsymbol{\phi} \overline{\lambda}_{i}\,^{\dot{\alpha}} {q}^{-2} q^{i}\,_{\underline{i}} \textbf{D}^{a}{\rho^{\underline{i}}\,_{\alpha}}%
-2{\rm i} c (\sigma_{a})^{\alpha}{}_{\dot{\alpha}} \phi \overline{\boldsymbol{\lambda}}_{i}\,^{\dot{\alpha}} {q}^{-2} q^{i}\,_{\underline{i}} \textbf{D}^{a}{\rho^{\underline{i}}\,_{\alpha}}-2{\rm i} c (\sigma_{a})^{\alpha}{}_{\dot{\alpha}} \overline{\boldsymbol{\phi}} \lambda_{i \alpha} {q}^{-2} q^{i \underline{i}} \textbf{D}^{a}{\overline{\rho}_{\underline{i}}\,^{\dot{\alpha}}}-2{\rm i} c (\sigma_{a})^{\alpha}{}_{\dot{\alpha}} \overline{\phi} \boldsymbol{\lambda}_{i \alpha} {q}^{-2} q^{i \underline{i}} \textbf{D}^{a}{\overline{\rho}_{\underline{i}}\,^{\dot{\alpha}}}+c \overline{\psi}_{a i \dot{\alpha}} \phi \overline{\boldsymbol{\phi}} {q}^{-2} \overline{\rho}_{\underline{i}}\,^{\dot{\alpha}} \textbf{D}^{a}{q^{i \underline{i}}}+2c (\tilde{\sigma}_{ba})_{\dot{\beta}\dot{\alpha}} \overline{\psi}^{b}{}_{i}^{\dot{\beta}} \phi \overline{\boldsymbol{\phi}} {q}^{-2} \overline{\rho}_{\underline{i}}\,^{\dot{\alpha}} \textbf{D}^{a}{q^{i \underline{i}}}+c \overline{\psi}_{a i \dot{\alpha}} \phi \overline{\boldsymbol{\phi}} {q}^{-2} q^{i \underline{i}} \textbf{D}^{a}{\overline{\rho}_{\underline{i}}\,^{\dot{\alpha}}}+c \overline{\psi}_{a i \dot{\alpha}} \overline{\phi} \boldsymbol{\phi} {q}^{-2} \overline{\rho}_{\underline{i}}\,^{\dot{\alpha}} \textbf{D}^{a}{q^{i \underline{i}}}+2c (\tilde{\sigma}_{ba})_{\dot{\beta}\dot{\alpha}} \overline{\psi}^{b}{}_{i}^{\dot{\beta}} \overline{\phi} \boldsymbol{\phi} {q}^{-2} \overline{\rho}_{\underline{i}}\,^{\dot{\alpha}} \textbf{D}^{a}{q^{i \underline{i}}}+c \overline{\psi}_{a i \dot{\alpha}} \overline{\phi} \boldsymbol{\phi} {q}^{-2} q^{i \underline{i}} \textbf{D}^{a}{\overline{\rho}_{\underline{i}}\,^{\dot{\alpha}}}+c \phi \overline{\boldsymbol{\phi}} {q}^{-2} q^{i \underline{i}} \overline{\rho}_{\underline{i} \dot{\alpha}} \textbf{D}_{a}{\overline{\psi}^{a}{}_{i}^{\dot{\alpha}}}+c \overline{\phi} \boldsymbol{\phi} {q}^{-2} q^{i \underline{i}} \overline{\rho}_{\underline{i} \dot{\alpha}} \textbf{D}_{a}{\overline{\psi}^{a}{}_{i}^{\dot{\alpha}}} - \frac{1}{2}c \lambda_{i}\,^{\alpha} \overline{\boldsymbol{\lambda}}_{j \dot{\alpha}} {q}^{-4} q^{i}\,_{\underline{i}} q^{j \underline{j}} \rho^{\underline{i}}\,_{\alpha} \overline{\rho}_{\underline{j}}\,^{\dot{\alpha}} - \frac{1}{2}c \lambda_{i}\,^{\alpha} \overline{\boldsymbol{\lambda}}_{j \dot{\alpha}} {q}^{-4} q^{i \underline{j}} q^{j}\,_{\underline{i}} \rho^{\underline{i}}\,_{\alpha} \overline{\rho}_{\underline{j}}\,^{\dot{\alpha}}+\frac{1}{2}c \overline{\lambda}_{j \dot{\alpha}} \boldsymbol{\lambda}_{i}\,^{\alpha} {q}^{-4} q^{i}\,_{\underline{i}} q^{j \underline{j}} \rho^{\underline{i}}\,_{\alpha} \overline{\rho}_{\underline{j}}\,^{\dot{\alpha}}+\frac{1}{2}c \overline{\lambda}_{j \dot{\alpha}} \boldsymbol{\lambda}_{i}\,^{\alpha} {q}^{-4} q^{i \underline{j}} q^{j}\,_{\underline{i}} \rho^{\underline{i}}\,_{\alpha} \overline{\rho}_{\underline{j}}\,^{\dot{\alpha}}+2c (\tilde{\sigma}_{ba})_{\dot{\beta}\dot{\alpha}} \overline{\psi}^{b}{}_{i}^{\dot{\beta}} \phi \overline{\boldsymbol{\phi}} {q}^{-2} q^{i \underline{i}} \textbf{D}^{a}{\overline{\rho}_{\underline{i}}\,^{\dot{\alpha}}}+2c (\tilde{\sigma}_{ba})_{\dot{\beta}\dot{\alpha}} \overline{\psi}^{b}{}_{i}^{\dot{\beta}} \overline{\phi} \boldsymbol{\phi} {q}^{-2} q^{i \underline{i}} \textbf{D}^{a}{\overline{\rho}_{\underline{i}}\,^{\dot{\alpha}}}-32c \overline{\psi}_{a i \dot{\alpha}} \boldsymbol{\phi} \overline{\lambda}_{j}\,^{\dot{\alpha}} {q}^{-4} q^{i}\,_{\underline{i}} q^{j}\,_{\underline{j}} H^{a \underline{i} \underline{j}}-32c \overline{\psi}_{a i \dot{\alpha}} \phi \overline{\boldsymbol{\lambda}}_{j}\,^{\dot{\alpha}} {q}^{-4} q^{i}\,_{\underline{i}} q^{j}\,_{\underline{j}} H^{a \underline{i} \underline{j}} - \frac{1}{4}{\rm i} c (\sigma_{a})^{\alpha}{}_{\dot{\beta}} \overline{\psi}^{a}\,_{j \dot{\alpha}} \lambda^{j}\,_{\alpha} \overline{\boldsymbol{\lambda}}_{i}\,^{\dot{\beta}} {q}^{-2} q^{i \underline{i}} \overline{\rho}_{\underline{i}}\,^{\dot{\alpha}}%
+\frac{1}{4}{\rm i} c (\sigma_{a})^{\alpha}{}_{\dot{\beta}} \overline{\psi}^{a}\,_{j \dot{\alpha}} \overline{\lambda}_{i}\,^{\dot{\beta}} \boldsymbol{\lambda}^{j}\,_{\alpha} {q}^{-2} q^{i \underline{i}} \overline{\rho}_{\underline{i}}\,^{\dot{\alpha}}-{\rm i} c (\sigma_{a})^{\alpha}{}_{\dot{\alpha}} \lambda_{i \alpha} \overline{\boldsymbol{\lambda}}_{j}\,^{\dot{\alpha}} {q}^{-2} q^{i}\,_{\underline{i}} \textbf{D}^{a}{q^{j \underline{i}}}-{\rm i} c (\sigma_{a})^{\alpha}{}_{\dot{\alpha}} \lambda_{j \alpha} \overline{\boldsymbol{\lambda}}_{i}\,^{\dot{\alpha}} {q}^{-2} q^{i}\,_{\underline{i}} \textbf{D}^{a}{q^{j \underline{i}}}+{\rm i} c (\sigma_{a})^{\alpha}{}_{\dot{\alpha}} \overline{\lambda}_{j}\,^{\dot{\alpha}} \boldsymbol{\lambda}_{i \alpha} {q}^{-2} q^{i}\,_{\underline{i}} \textbf{D}^{a}{q^{j \underline{i}}}+{\rm i} c (\sigma_{a})^{\alpha}{}_{\dot{\alpha}} \overline{\lambda}_{i}\,^{\dot{\alpha}} \boldsymbol{\lambda}_{j \alpha} {q}^{-2} q^{i}\,_{\underline{i}} \textbf{D}^{a}{q^{j \underline{i}}}-8{\rm i} c (\sigma_{a})^{\alpha}{}_{\dot{\alpha}} \phi \overline{\boldsymbol{\phi}} {q}^{-4} \rho_{\underline{i} \alpha} \overline{\rho}_{\underline{j}}\,^{\dot{\alpha}} H^{a \underline{i} \underline{j}}-8{\rm i} c (\sigma_{a})^{\alpha}{}_{\dot{\alpha}} \overline{\phi} \boldsymbol{\phi} {q}^{-4} \rho_{\underline{i} \alpha} \overline{\rho}_{\underline{j}}\,^{\dot{\alpha}} H^{a \underline{i} \underline{j}}+2c \overline{\psi}_{a j \dot{\alpha}} \boldsymbol{\phi} \overline{\lambda}_{i}\,^{\dot{\alpha}} {q}^{-2} q^{i}\,_{\underline{i}} \textbf{D}^{a}{q^{j \underline{i}}}+8c (\tilde{\sigma}_{ba})_{\dot{\beta}\dot{\alpha}} \overline{\psi}^{b}{}_{j}^{\dot{\beta}} \boldsymbol{\phi} \overline{\lambda}_{i}\,^{\dot{\alpha}} {q}^{-2} q^{i}\,_{\underline{i}} \textbf{D}^{a}{q^{j \underline{i}}}-2c \overline{\psi}_{a i \dot{\alpha}} \boldsymbol{\phi} \overline{\lambda}_{j}\,^{\dot{\alpha}} {q}^{-2} q^{i}\,_{\underline{i}} \textbf{D}^{a}{q^{j \underline{i}}}+2c \overline{\psi}_{a j \dot{\alpha}} \phi \overline{\boldsymbol{\lambda}}_{i}\,^{\dot{\alpha}} {q}^{-2} q^{i}\,_{\underline{i}} \textbf{D}^{a}{q^{j \underline{i}}}+8c (\tilde{\sigma}_{ba})_{\dot{\beta}\dot{\alpha}} \overline{\psi}^{b}{}_{j}^{\dot{\beta}} \phi \overline{\boldsymbol{\lambda}}_{i}\,^{\dot{\alpha}} {q}^{-2} q^{i}\,_{\underline{i}} \textbf{D}^{a}{q^{j \underline{i}}}-2c \overline{\psi}_{a i \dot{\alpha}} \phi \overline{\boldsymbol{\lambda}}_{j}\,^{\dot{\alpha}} {q}^{-2} q^{i}\,_{\underline{i}} \textbf{D}^{a}{q^{j \underline{i}}}+16c \psi_{a i}\,^{\alpha} \phi \overline{\boldsymbol{\phi}} {q}^{-4} q^{i}\,_{\underline{i}} \rho_{\underline{j} \alpha} H^{a \underline{i} \underline{j}}+16c \psi_{a i}\,^{\alpha} \overline{\phi} \boldsymbol{\phi} {q}^{-4} q^{i}\,_{\underline{i}} \rho_{\underline{j} \alpha} H^{a \underline{i} \underline{j}}+c \psi_{a i}\,^{\alpha} \phi \overline{\boldsymbol{\phi}} {q}^{-2} q^{i}\,_{\underline{i}} \textbf{D}^{a}{\rho^{\underline{i}}\,_{\alpha}}+2c (\sigma_{ba})^{\beta\alpha} \psi^{b}\,_{i \beta} \phi \overline{\boldsymbol{\phi}} {q}^{-2} q^{i}\,_{\underline{i}} \textbf{D}^{a}{\rho^{\underline{i}}\,_{\alpha}}+c \psi_{a i}\,^{\alpha} \overline{\phi} \boldsymbol{\phi} {q}^{-2} q^{i}\,_{\underline{i}} \textbf{D}^{a}{\rho^{\underline{i}}\,_{\alpha}}+2c (\sigma_{ba})^{\beta\alpha} \psi^{b}\,_{i \beta} \overline{\phi} \boldsymbol{\phi} {q}^{-2} q^{i}\,_{\underline{i}} \textbf{D}^{a}{\rho^{\underline{i}}\,_{\alpha}}-c \phi \overline{\boldsymbol{\phi}} {q}^{-2} q_{i \underline{i}} \rho^{\underline{i} \alpha} \textbf{D}_{a}{\psi^{a}{}^{i}_{\alpha}}%
-c \overline{\phi} \boldsymbol{\phi} {q}^{-2} q_{i \underline{i}} \rho^{\underline{i} \alpha} \textbf{D}_{a}{\psi^{a}{}^{i}_{\alpha}}-c \psi_{a i}\,^{\alpha} \phi \overline{\boldsymbol{\phi}} {q}^{-2} \rho_{\underline{i} \alpha} \textbf{D}^{a}{q^{i \underline{i}}}-2c (\sigma_{ba})^{\beta\alpha} \psi^{b}\,_{i \beta} \phi \overline{\boldsymbol{\phi}} {q}^{-2} \rho_{\underline{i} \alpha} \textbf{D}^{a}{q^{i \underline{i}}}-c \psi_{a i}\,^{\alpha} \overline{\phi} \boldsymbol{\phi} {q}^{-2} \rho_{\underline{i} \alpha} \textbf{D}^{a}{q^{i \underline{i}}}-2c (\sigma_{ba})^{\beta\alpha} \psi^{b}\,_{i \beta} \overline{\phi} \boldsymbol{\phi} {q}^{-2} \rho_{\underline{i} \alpha} \textbf{D}^{a}{q^{i \underline{i}}}+32c \psi_{a i}\,^{\alpha} \overline{\boldsymbol{\phi}} \lambda_{j \alpha} {q}^{-4} q^{i}\,_{\underline{i}} q^{j}\,_{\underline{j}} H^{a \underline{i} \underline{j}}+32c \psi_{a i}\,^{\alpha} \overline{\phi} \boldsymbol{\lambda}_{j \alpha} {q}^{-4} q^{i}\,_{\underline{i}} q^{j}\,_{\underline{j}} H^{a \underline{i} \underline{j}} - \frac{1}{4}{\rm i} c (\sigma_{a})^{\beta}{}_{\dot{\alpha}} \psi^{a}\,_{j}\,^{\alpha} \lambda^{j}\,_{\beta} \overline{\boldsymbol{\lambda}}_{i}\,^{\dot{\alpha}} {q}^{-2} q^{i}\,_{\underline{i}} \rho^{\underline{i}}\,_{\alpha}+\frac{1}{4}{\rm i} c (\sigma_{a})^{\beta}{}_{\dot{\alpha}} \psi^{a j \alpha} \lambda_{i \beta} \overline{\boldsymbol{\lambda}}_{j}\,^{\dot{\alpha}} {q}^{-2} q^{i}\,_{\underline{i}} \rho^{\underline{i}}\,_{\alpha}+\frac{1}{4}{\rm i} c (\sigma_{a})^{\alpha}{}_{\dot{\beta}} \overline{\psi}^{a j}\,_{\dot{\alpha}} \lambda_{i \alpha} \overline{\boldsymbol{\lambda}}_{j}\,^{\dot{\beta}} {q}^{-2} q^{i \underline{i}} \overline{\rho}_{\underline{i}}\,^{\dot{\alpha}} - \frac{1}{4}{\rm i} c (\sigma_{a})^{\alpha}{}_{\dot{\beta}} \overline{\psi}^{a j}\,_{\dot{\alpha}} \overline{\lambda}_{j}\,^{\dot{\beta}} \boldsymbol{\lambda}_{i \alpha} {q}^{-2} q^{i \underline{i}} \overline{\rho}_{\underline{i}}\,^{\dot{\alpha}}-16{\rm i} c (\sigma_{a})^{\alpha}{}_{\dot{\alpha}} \lambda_{i \alpha} \overline{\boldsymbol{\lambda}}_{j}\,^{\dot{\alpha}} {q}^{-4} q^{i}\,_{\underline{i}} q^{j}\,_{\underline{j}} H^{a \underline{i} \underline{j}}+\frac{1}{4}{\rm i} c (\sigma_{a})^{\beta}{}_{\dot{\alpha}} \psi^{a}\,_{j}\,^{\alpha} \overline{\lambda}_{i}\,^{\dot{\alpha}} \boldsymbol{\lambda}^{j}\,_{\beta} {q}^{-2} q^{i}\,_{\underline{i}} \rho^{\underline{i}}\,_{\alpha} - \frac{1}{4}{\rm i} c (\sigma_{a})^{\beta}{}_{\dot{\alpha}} \psi^{a j \alpha} \overline{\lambda}_{j}\,^{\dot{\alpha}} \boldsymbol{\lambda}_{i \beta} {q}^{-2} q^{i}\,_{\underline{i}} \rho^{\underline{i}}\,_{\alpha}+16{\rm i} c (\sigma_{a})^{\alpha}{}_{\dot{\alpha}} \overline{\lambda}_{j}\,^{\dot{\alpha}} \boldsymbol{\lambda}_{i \alpha} {q}^{-4} q^{i}\,_{\underline{i}} q^{j}\,_{\underline{j}} H^{a \underline{i} \underline{j}}-2c \psi_{a j}\,^{\alpha} \overline{\boldsymbol{\phi}} \lambda_{i \alpha} {q}^{-2} q^{i}\,_{\underline{i}} \textbf{D}^{a}{q^{j \underline{i}}}-8c (\sigma_{ba})^{\beta\alpha} \psi^{b}\,_{j \beta} \overline{\boldsymbol{\phi}} \lambda_{i \alpha} {q}^{-2} q^{i}\,_{\underline{i}} \textbf{D}^{a}{q^{j \underline{i}}}+2c \psi_{a i}\,^{\alpha} \overline{\boldsymbol{\phi}} \lambda_{j \alpha} {q}^{-2} q^{i}\,_{\underline{i}} \textbf{D}^{a}{q^{j \underline{i}}}-2c \psi_{a j}\,^{\alpha} \overline{\phi} \boldsymbol{\lambda}_{i \alpha} {q}^{-2} q^{i}\,_{\underline{i}} \textbf{D}^{a}{q^{j \underline{i}}}-8c (\sigma_{ba})^{\beta\alpha} \psi^{b}\,_{j \beta} \overline{\phi} \boldsymbol{\lambda}_{i \alpha} {q}^{-2} q^{i}\,_{\underline{i}} \textbf{D}^{a}{q^{j \underline{i}}}%
+2c \psi_{a i}\,^{\alpha} \overline{\phi} \boldsymbol{\lambda}_{j \alpha} {q}^{-2} q^{i}\,_{\underline{i}} \textbf{D}^{a}{q^{j \underline{i}}}+\frac{1}{2}c \epsilon^{b}\,_{c a d} (\sigma_{b})^{\alpha}{}_{\dot{\alpha}} \psi^{c}{}^{i}_{\alpha} \phi \overline{\boldsymbol{\phi}} \textbf{D}^{a}{\overline{\psi}^{d}{}_{i}^{\dot{\alpha}}}+\frac{1}{2}c \epsilon^{b}\,_{c a d} (\sigma_{b})^{\alpha}{}_{\dot{\alpha}} \psi^{c}{}^{i}_{\alpha} \overline{\phi} \boldsymbol{\phi} \textbf{D}^{a}{\overline{\psi}^{d}{}_{i}^{\dot{\alpha}}} - \frac{1}{2}c \epsilon^{b}\,_{c a d} (\sigma_{b})^{\alpha}{}_{\dot{\alpha}} \overline{\psi}^{c}{}_{i}^{\dot{\alpha}} \phi \overline{\boldsymbol{\phi}} \textbf{D}^{a}{\psi^{d}{}^{i}_{\alpha}} - \frac{1}{2}c \epsilon^{b}\,_{c a d} (\sigma_{b})^{\alpha}{}_{\dot{\alpha}} \overline{\psi}^{c}{}_{i}^{\dot{\alpha}} \overline{\phi} \boldsymbol{\phi} \textbf{D}^{a}{\psi^{d}{}^{i}_{\alpha}}-2{\rm i} c (\sigma_{b})^{\alpha}{}_{\dot{\alpha}} \psi_{a i \alpha} \overline{\psi}^{b}{}_{j}^{\dot{\alpha}} \phi \overline{\boldsymbol{\phi}} {q}^{-2} q^{i}\,_{\underline{i}} \textbf{D}^{a}{q^{j \underline{i}}}+2{\rm i} c (\sigma_{b})^{\alpha}{}_{\dot{\alpha}} \psi^{b}\,_{i \alpha} \overline{\psi}_{a j}\,^{\dot{\alpha}} \phi \overline{\boldsymbol{\phi}} {q}^{-2} q^{i}\,_{\underline{i}} \textbf{D}^{a}{q^{j \underline{i}}}-2c \epsilon^{b}\,_{c d a} (\sigma_{b})^{\alpha}{}_{\dot{\alpha}} \psi^{c}\,_{i \alpha} \overline{\psi}^{d}{}_{j}^{\dot{\alpha}} \phi \overline{\boldsymbol{\phi}} {q}^{-2} q^{i}\,_{\underline{i}} \textbf{D}^{a}{q^{j \underline{i}}}-2{\rm i} c (\sigma_{b})^{\alpha}{}_{\dot{\alpha}} \psi_{a i \alpha} \overline{\psi}^{b}{}_{j}^{\dot{\alpha}} \overline{\phi} \boldsymbol{\phi} {q}^{-2} q^{i}\,_{\underline{i}} \textbf{D}^{a}{q^{j \underline{i}}}+2{\rm i} c (\sigma_{b})^{\alpha}{}_{\dot{\alpha}} \psi^{b}\,_{i \alpha} \overline{\psi}_{a j}\,^{\dot{\alpha}} \overline{\phi} \boldsymbol{\phi} {q}^{-2} q^{i}\,_{\underline{i}} \textbf{D}^{a}{q^{j \underline{i}}}-2c \epsilon^{b}\,_{c d a} (\sigma_{b})^{\alpha}{}_{\dot{\alpha}} \psi^{c}\,_{i \alpha} \overline{\psi}^{d}{}_{j}^{\dot{\alpha}} \overline{\phi} \boldsymbol{\phi} {q}^{-2} q^{i}\,_{\underline{i}} \textbf{D}^{a}{q^{j \underline{i}}}+2{\rm i} c (\sigma_{b})^{\alpha}{}_{\dot{\alpha}} \psi_{a j \alpha} \overline{\psi}^{b}{}_{i}^{\dot{\alpha}} \phi \overline{\boldsymbol{\phi}} {q}^{-2} q^{i}\,_{\underline{i}} \textbf{D}^{a}{q^{j \underline{i}}}-2{\rm i} c (\sigma_{b})^{\alpha}{}_{\dot{\alpha}} \psi^{b}\,_{j \alpha} \overline{\psi}_{a i}\,^{\dot{\alpha}} \phi \overline{\boldsymbol{\phi}} {q}^{-2} q^{i}\,_{\underline{i}} \textbf{D}^{a}{q^{j \underline{i}}}-2c \epsilon^{b}\,_{c d a} (\sigma_{b})^{\alpha}{}_{\dot{\alpha}} \psi^{c}\,_{j \alpha} \overline{\psi}^{d}{}_{i}^{\dot{\alpha}} \phi \overline{\boldsymbol{\phi}} {q}^{-2} q^{i}\,_{\underline{i}} \textbf{D}^{a}{q^{j \underline{i}}}+2{\rm i} c (\sigma_{b})^{\alpha}{}_{\dot{\alpha}} \psi_{a j \alpha} \overline{\psi}^{b}{}_{i}^{\dot{\alpha}} \overline{\phi} \boldsymbol{\phi} {q}^{-2} q^{i}\,_{\underline{i}} \textbf{D}^{a}{q^{j \underline{i}}}-2{\rm i} c (\sigma_{b})^{\alpha}{}_{\dot{\alpha}} \psi^{b}\,_{j \alpha} \overline{\psi}_{a i}\,^{\dot{\alpha}} \overline{\phi} \boldsymbol{\phi} {q}^{-2} q^{i}\,_{\underline{i}} \textbf{D}^{a}{q^{j \underline{i}}}-2c \epsilon^{b}\,_{c d a} (\sigma_{b})^{\alpha}{}_{\dot{\alpha}} \psi^{c}\,_{j \alpha} \overline{\psi}^{d}{}_{i}^{\dot{\alpha}} \overline{\phi} \boldsymbol{\phi} {q}^{-2} q^{i}\,_{\underline{i}} \textbf{D}^{a}{q^{j \underline{i}}} - \frac{1}{48}c \epsilon^{c d}\,_{e {e_{1}}} \epsilon_{c d a b} \psi^{e}\,_{i}\,^{\alpha} \psi^{{e_{1}}}{}^{i}_{\alpha} W^{a b} \phi \overline{\boldsymbol{\phi}} - \frac{1}{48}c \epsilon^{c d}\,_{e {e_{1}}} \epsilon_{c d a b} \psi^{e}\,_{i}\,^{\alpha} \psi^{{e_{1}}}{}^{i}_{\alpha} W^{a b} \overline{\phi} \boldsymbol{\phi} - \frac{3}{4}c \epsilon^{b}\,_{c d e} (\sigma_{b})^{\alpha}{}_{\dot{\alpha}} \psi^{c}{}^{i}_{\alpha} \overline{\phi} \boldsymbol{\phi} \overline{\bold{\Psi}}^{d e}{}_i^\ad -\frac{3}{2} c (\tilde{\sigma}_{cd})_{\dot{\alpha}\dot{\beta}} \boldsymbol{\phi} \overline{\lambda}^{i \dot{\alpha}}\overline{\bold{\Psi}}^{c d}{}_i^\bd+c (\tilde{\sigma}_{cd})_{\dot{\alpha}\dot{\beta}} \overline{\phi} \boldsymbol{\phi} {q}^{-2} q^{j \underline{i}} \overline{\rho}_{\underline{i}}\,^{\dot{\alpha}} \overline{\bold{\Psi}}^{c d}{}_j^\bd %
 - \frac{3}{4}c \epsilon^{b}\,_{c d e} (\sigma_{b})^{\alpha}{}_{\dot{\alpha}} \psi^{c}{}^{i}_{\alpha} \phi \overline{\boldsymbol{\phi}} \overline{\bold{\Psi}}^{d e}{}_i^\ad -\frac{3}{2}c (\tilde{\sigma}_{cd})_{\dot{\alpha}\dot{\beta}} \phi \overline{\boldsymbol{\lambda}}^{i \dot{\alpha}} \overline{\bold{\Psi}}^{c d}{}_i^\bd + c (\tilde{\sigma}_{cd})_{\dot{\alpha}\dot{\beta}} \phi \overline{\boldsymbol{\phi}} {q}^{-2} q^{j \underline{i}} \overline{\rho}_{\underline{i}}\,^{\dot{\alpha}} \overline{\bold{\Psi}}^{c d}{}_j^\bd - \frac{3}{2} c (\sigma_{cd})^{\alpha\beta} \overline{\phi} \boldsymbol{\lambda}_{i \alpha} \bold{\Psi}^{c d}{}^i_\b +\frac{3}{4}c \epsilon^{b}\,_{c d e} (\sigma_{b})^{\alpha}{}_{\dot{\alpha}} \overline{\psi}^{c}{}_{i}^{\dot{\alpha}} \overline{\phi} \boldsymbol{\phi}  \bold{\Psi}^{d e}{}^i_\a -c (\sigma_{cd})^{\alpha\beta} \overline{\phi} \boldsymbol{\phi} {q}^{-2} q_{j \underline{i}} \rho^{\underline{i}}\,_{\alpha} \bold{\Psi}^{c d}{}^j_\b -\frac{3}{2} c (\sigma_{cd})^{\alpha\beta} \overline{\boldsymbol{\phi}} \lambda_{i \alpha} \bold{\Psi}^{c d}{}^i_\b +\frac{3}{4}c \epsilon^{b}\,_{c d e} (\sigma_{b})^{\alpha}{}_{\dot{\alpha}} \overline{\psi}^{c}{}_{i}^{\dot{\alpha}} \phi \overline{\boldsymbol{\phi}}\bold{\Psi}^{d e}{}^i_\a-c (\sigma_{cd})^{\alpha\beta} \phi \overline{\boldsymbol{\phi}} {q}^{-2} q_{j \underline{i}} \rho^{\underline{i}}\,_{\alpha} \bold{\Psi}^{c d}{}^{j}_{\beta}$\end{adjustwidth}

\subsection{$\cL^{{\rm off,\,SU(1,1)/U(1)}}_{4}$}

\begin{adjustwidth}{1em}{4.5cm}${}\frac{1}{2}c (\sigma_{ab})^{\alpha\beta} \psi^{a}{}^{j}_{\alpha} \psi^{b}{}^{i}_{\beta} \overline{\lambda}_{j \dot{\alpha}} \overline{\boldsymbol{\lambda}}_{i}\,^{\dot{\alpha}}+\frac{1}{2}c (\tilde{\sigma}_{ab})_{\dot{\alpha}\dot{\beta}} \overline{\psi}^{a}{}_{j}^{\dot{\alpha}} \overline{\psi}^{b}{}_{i}^{\dot{\beta}} \lambda^{j \alpha} \boldsymbol{\lambda}^{i}\,_{\alpha}+\frac{1}{2}c \psi_{a j}\,^{\alpha} \overline{\psi}^{a i}\,_{\dot{\alpha}} \lambda^{j}\,_{\alpha} \overline{\boldsymbol{\lambda}}_{i}\,^{\dot{\alpha}} - \frac{1}{2}c \psi_{a i}\,^{\alpha} \overline{\psi}^{a j}\,_{\dot{\alpha}} \overline{\lambda}_{j}\,^{\dot{\alpha}} \boldsymbol{\lambda}^{i}\,_{\alpha}+\frac{1}{2}c (\sigma_{ab})^{\beta\alpha} \psi^{a}\,_{j \beta} \overline{\psi}^{b i}\,_{\dot{\alpha}} \lambda^{j}\,_{\alpha} \overline{\boldsymbol{\lambda}}_{i}\,^{\dot{\alpha}}+\frac{1}{4}c (\tilde{\sigma}_{ab})_{\dot{\beta}\dot{\alpha}} \psi^{a}\,_{j}\,^{\alpha} \psi^{b}{}^{j}_{\alpha} \overline{\lambda}^{i \dot{\beta}} \overline{\boldsymbol{\lambda}}_{i}\,^{\dot{\alpha}} - \frac{1}{2}c (\tilde{\sigma}_{ab})_{\dot{\beta}\dot{\alpha}} \psi^{a}\,_{j}\,^{\alpha} \overline{\psi}^{b i \dot{\beta}} \lambda^{j}\,_{\alpha} \overline{\boldsymbol{\lambda}}_{i}\,^{\dot{\alpha}} - \frac{1}{2}c (\sigma_{ab})^{\beta\alpha} \psi^{a}\,_{i \beta} \overline{\psi}^{b j}\,_{\dot{\alpha}} \overline{\lambda}_{j}\,^{\dot{\alpha}} \boldsymbol{\lambda}^{i}\,_{\alpha}+\frac{1}{4}c (\sigma_{ab})^{\beta\alpha} \overline{\psi}^{a j}\,_{\dot{\alpha}} \overline{\psi}^{b}{}_{j}^{\dot{\alpha}} \lambda_{i \beta} \boldsymbol{\lambda}^{i}\,_{\alpha}+\frac{1}{2}c (\tilde{\sigma}_{ab})_{\dot{\beta}\dot{\alpha}} \psi^{a}\,_{i}\,^{\alpha} \overline{\psi}^{b j \dot{\beta}} \overline{\lambda}_{j}\,^{\dot{\alpha}} \boldsymbol{\lambda}^{i}\,_{\alpha}+\frac{1}{4}c \psi_{a}\,^{i \alpha} \overline{\psi}^{a}\,_{i \dot{\alpha}} \phi \overline{\boldsymbol{\phi}} {q}^{-2} \rho^{\underline{i}}\,_{\alpha} \overline{\rho}_{\underline{i}}\,^{\dot{\alpha}}+\frac{1}{4}c \psi_{a}\,^{i \alpha} \overline{\psi}^{a}\,_{i \dot{\alpha}} \overline{\phi} \boldsymbol{\phi} {q}^{-2} \rho^{\underline{i}}\,_{\alpha} \overline{\rho}_{\underline{i}}\,^{\dot{\alpha}}+\frac{1}{2}c (\sigma_{ab})^{\beta\alpha} \psi^{a}{}^{i}_{\beta} \overline{\psi}^{b}\,_{i \dot{\alpha}} \phi \overline{\boldsymbol{\phi}} {q}^{-2} \rho^{\underline{i}}\,_{\alpha} \overline{\rho}_{\underline{i}}\,^{\dot{\alpha}}+\frac{1}{2}c (\sigma_{ab})^{\beta\alpha} \psi^{a}{}^{i}_{\beta} \overline{\psi}^{b}\,_{i \dot{\alpha}} \overline{\phi} \boldsymbol{\phi} {q}^{-2} \rho^{\underline{i}}\,_{\alpha} \overline{\rho}_{\underline{i}}\,^{\dot{\alpha}} - \frac{1}{2}c (\tilde{\sigma}_{ab})_{\dot{\beta}\dot{\alpha}} \psi^{a i \alpha} \overline{\psi}^{b}{}_{i}^{\dot{\beta}} \phi \overline{\boldsymbol{\phi}} {q}^{-2} \rho^{\underline{i}}\,_{\alpha} \overline{\rho}_{\underline{i}}\,^{\dot{\alpha}} - \frac{1}{2}c (\tilde{\sigma}_{ab})_{\dot{\beta}\dot{\alpha}} \psi^{a i \alpha} \overline{\psi}^{b}{}_{i}^{\dot{\beta}} \overline{\phi} \boldsymbol{\phi} {q}^{-2} \rho^{\underline{i}}\,_{\alpha} \overline{\rho}_{\underline{i}}\,^{\dot{\alpha}} - \frac{1}{2}c \psi_{a}\,^{j \alpha} \overline{\psi}^{a}\,_{j \dot{\alpha}} \phi \overline{\boldsymbol{\lambda}}_{i}\,^{\dot{\alpha}} {q}^{-2} q^{i}\,_{\underline{i}} \rho^{\underline{i}}\,_{\alpha}+\frac{1}{2}c \psi_{a}\,^{j \alpha} \overline{\psi}^{a}\,_{i \dot{\alpha}} \phi \overline{\boldsymbol{\lambda}}_{j}\,^{\dot{\alpha}} {q}^{-2} q^{i}\,_{\underline{i}} \rho^{\underline{i}}\,_{\alpha} - \frac{1}{2}c \psi_{a i}\,^{\alpha} \overline{\psi}^{a}\,_{j \dot{\alpha}} \overline{\boldsymbol{\phi}} \lambda^{j}\,_{\alpha} {q}^{-2} q^{i \underline{i}} \overline{\rho}_{\underline{i}}\,^{\dot{\alpha}}%
 - \frac{1}{2}c \psi_{a i}\,^{\alpha} \overline{\psi}^{a}\,_{j \dot{\alpha}} \overline{\phi} \boldsymbol{\lambda}^{j}\,_{\alpha} {q}^{-2} q^{i \underline{i}} \overline{\rho}_{\underline{i}}\,^{\dot{\alpha}}+\frac{1}{2}c \psi_{a}\,^{j \alpha} \overline{\psi}^{a}\,_{j \dot{\alpha}} \overline{\boldsymbol{\phi}} \lambda_{i \alpha} {q}^{-2} q^{i \underline{i}} \overline{\rho}_{\underline{i}}\,^{\dot{\alpha}}+\frac{1}{2}c \psi_{a}\,^{j \alpha} \overline{\psi}^{a}\,_{j \dot{\alpha}} \overline{\phi} \boldsymbol{\lambda}_{i \alpha} {q}^{-2} q^{i \underline{i}} \overline{\rho}_{\underline{i}}\,^{\dot{\alpha}} - \frac{1}{2}c \psi_{a}\,^{j \alpha} \overline{\psi}^{a}\,_{j \dot{\alpha}} \boldsymbol{\phi} \overline{\lambda}_{i}\,^{\dot{\alpha}} {q}^{-2} q^{i}\,_{\underline{i}} \rho^{\underline{i}}\,_{\alpha}+\frac{1}{2}c \psi_{a}\,^{j \alpha} \overline{\psi}^{a}\,_{i \dot{\alpha}} \boldsymbol{\phi} \overline{\lambda}_{j}\,^{\dot{\alpha}} {q}^{-2} q^{i}\,_{\underline{i}} \rho^{\underline{i}}\,_{\alpha} - \frac{1}{2}c \overline{\psi}_{a}\,^{j}\,_{\dot{\beta}} \overline{\psi}^{a}\,_{j \dot{\alpha}} \boldsymbol{\phi} \overline{\lambda}_{i}\,^{\dot{\beta}} {q}^{-2} q^{i \underline{i}} \overline{\rho}_{\underline{i}}\,^{\dot{\alpha}} - \frac{1}{2}c \overline{\psi}_{a}\,^{j}\,_{\dot{\beta}} \overline{\psi}^{a}\,_{j \dot{\alpha}} \phi \overline{\boldsymbol{\lambda}}_{i}\,^{\dot{\beta}} {q}^{-2} q^{i \underline{i}} \overline{\rho}_{\underline{i}}\,^{\dot{\alpha}}+\frac{1}{2}c \overline{\psi}_{a}\,^{j}\,_{\dot{\alpha}} \overline{\psi}^{a}\,_{i \dot{\beta}} \boldsymbol{\phi} \overline{\lambda}_{j}\,^{\dot{\beta}} {q}^{-2} q^{i \underline{i}} \overline{\rho}_{\underline{i}}\,^{\dot{\alpha}}+\frac{1}{2}c \overline{\psi}_{a}\,^{j}\,_{\dot{\alpha}} \overline{\psi}^{a}\,_{i \dot{\beta}} \phi \overline{\boldsymbol{\lambda}}_{j}\,^{\dot{\beta}} {q}^{-2} q^{i \underline{i}} \overline{\rho}_{\underline{i}}\,^{\dot{\alpha}}+2c (\sigma_{ab})^{\beta\alpha} \psi^{a}{}^{j}_{\beta} \overline{\psi}^{b}\,_{j \dot{\alpha}} \overline{\boldsymbol{\phi}} \lambda_{i \alpha} {q}^{-2} q^{i \underline{i}} \overline{\rho}_{\underline{i}}\,^{\dot{\alpha}}+2c (\sigma_{ab})^{\beta\alpha} \psi^{a}{}^{j}_{\beta} \overline{\psi}^{b}\,_{j \dot{\alpha}} \overline{\phi} \boldsymbol{\lambda}_{i \alpha} {q}^{-2} q^{i \underline{i}} \overline{\rho}_{\underline{i}}\,^{\dot{\alpha}}+2c (\tilde{\sigma}_{ab})_{\dot{\beta}\dot{\alpha}} \psi^{a j \alpha} \overline{\psi}^{b}{}_{j}^{\dot{\beta}} \phi \overline{\boldsymbol{\lambda}}_{i}\,^{\dot{\alpha}} {q}^{-2} q^{i}\,_{\underline{i}} \rho^{\underline{i}}\,_{\alpha}+2c (\tilde{\sigma}_{ab})_{\dot{\beta}\dot{\alpha}} \psi^{a j \alpha} \overline{\psi}^{b}{}_{j}^{\dot{\beta}} \boldsymbol{\phi} \overline{\lambda}_{i}\,^{\dot{\alpha}} {q}^{-2} q^{i}\,_{\underline{i}} \rho^{\underline{i}}\,_{\alpha}-2c (\tilde{\sigma}_{ab})_{\dot{\rho}\dot{\beta}} \overline{\psi}^{a j \dot{\rho}} \overline{\psi}^{b}\,_{j \dot{\alpha}} \boldsymbol{\phi} \overline{\lambda}_{i}\,^{\dot{\beta}} {q}^{-2} q^{i \underline{i}} \overline{\rho}_{\underline{i}}\,^{\dot{\alpha}}-2c (\tilde{\sigma}_{ab})_{\dot{\rho}\dot{\beta}} \overline{\psi}^{a j \dot{\rho}} \overline{\psi}^{b}\,_{j \dot{\alpha}} \phi \overline{\boldsymbol{\lambda}}_{i}\,^{\dot{\beta}} {q}^{-2} q^{i \underline{i}} \overline{\rho}_{\underline{i}}\,^{\dot{\alpha}} - \frac{1}{8}c \psi_{a i}\,^{\alpha} \psi^{a i \beta} \phi \overline{\boldsymbol{\phi}} {q}^{-2} \rho_{\underline{i} \alpha} \rho^{\underline{i}}\,_{\beta} - \frac{1}{8}c \psi_{a i}\,^{\alpha} \psi^{a i \beta} \overline{\phi} \boldsymbol{\phi} {q}^{-2} \rho_{\underline{i} \alpha} \rho^{\underline{i}}\,_{\beta} - \frac{1}{8}c \overline{\psi}_{a}\,^{i}\,_{\dot{\alpha}} \overline{\psi}^{a}\,_{i \dot{\beta}} \phi \overline{\boldsymbol{\phi}} {q}^{-2} \overline{\rho}^{\underline{i} \dot{\alpha}} \overline{\rho}_{\underline{i}}\,^{\dot{\beta}} - \frac{1}{8}c \overline{\psi}_{a}\,^{i}\,_{\dot{\alpha}} \overline{\psi}^{a}\,_{i \dot{\beta}} \overline{\phi} \boldsymbol{\phi} {q}^{-2} \overline{\rho}^{\underline{i} \dot{\alpha}} \overline{\rho}_{\underline{i}}\,^{\dot{\beta}}+\frac{1}{4}c \overline{\psi}_{a i \dot{\beta}} \overline{\psi}^{a}\,_{j \dot{\alpha}} \phi \overline{\boldsymbol{\phi}} {q}^{-4} q^{i \underline{i}} q^{j \underline{j}} \overline{\rho}_{\underline{i}}\,^{\dot{\alpha}} \overline{\rho}_{\underline{j}}\,^{\dot{\beta}}%
+\frac{1}{4}c \overline{\psi}_{a i \dot{\beta}} \overline{\psi}^{a}\,_{j \dot{\alpha}} \overline{\phi} \boldsymbol{\phi} {q}^{-4} q^{i \underline{i}} q^{j \underline{j}} \overline{\rho}_{\underline{i}}\,^{\dot{\alpha}} \overline{\rho}_{\underline{j}}\,^{\dot{\beta}}+\frac{1}{4}c \epsilon^{a}\,_{b c d} (\sigma_{a})^{\beta}{}_{\dot{\alpha}} \psi^{b}\,_{j}\,^{\alpha} \psi^{c}{}^{j}_{\alpha} \psi^{d}{}^{i}_{\beta} \overline{\boldsymbol{\phi}} \overline{\lambda}_{i}\,^{\dot{\alpha}}+\frac{1}{4}c \epsilon^{a}\,_{b c d} (\sigma_{a})^{\beta}{}_{\dot{\alpha}} \psi^{b}\,_{j}\,^{\alpha} \psi^{c}{}^{j}_{\alpha} \psi^{d}{}^{i}_{\beta} \overline{\phi} \overline{\boldsymbol{\lambda}}_{i}\,^{\dot{\alpha}} - \frac{1}{4}c \epsilon^{a}\,_{b c d} (\sigma_{a})^{\beta}{}_{\dot{\alpha}} \psi^{b}\,_{i}\,^{\alpha} \psi^{c}{}^{j}_{\beta} \overline{\psi}^{d}{}_{j}^{\dot{\alpha}} \overline{\boldsymbol{\phi}} \lambda^{i}\,_{\alpha}+\frac{1}{4}c \epsilon^{a}\,_{b c d} (\sigma_{a})^{\alpha}{}_{\dot{\beta}} \psi^{b}{}^{j}_{\alpha} \overline{\psi}^{c i}\,_{\dot{\alpha}} \overline{\psi}^{d}{}_{j}^{\dot{\beta}} \boldsymbol{\phi} \overline{\lambda}_{i}\,^{\dot{\alpha}}+\frac{1}{4}c \epsilon^{a}\,_{b c d} (\sigma_{a})^{\alpha}{}_{\dot{\beta}} \psi^{b}{}^{j}_{\alpha} \overline{\psi}^{c i}\,_{\dot{\alpha}} \overline{\psi}^{d}{}_{j}^{\dot{\beta}} \phi \overline{\boldsymbol{\lambda}}_{i}\,^{\dot{\alpha}} - \frac{1}{4}c \epsilon^{a}\,_{b c d} (\sigma_{a})^{\alpha}{}_{\dot{\beta}} \overline{\psi}^{b j}\,_{\dot{\alpha}} \overline{\psi}^{c}{}_{j}^{\dot{\alpha}} \overline{\psi}^{d}{}_{i}^{\dot{\beta}} \boldsymbol{\phi} \lambda^{i}\,_{\alpha} - \frac{1}{4}c \epsilon^{a}\,_{b c d} (\sigma_{a})^{\beta}{}_{\dot{\alpha}} \psi^{b}\,_{i}\,^{\alpha} \psi^{c}{}^{j}_{\beta} \overline{\psi}^{d}{}_{j}^{\dot{\alpha}} \overline{\phi} \boldsymbol{\lambda}^{i}\,_{\alpha} - \frac{1}{4}c \epsilon^{a}\,_{b c d} (\sigma_{a})^{\alpha}{}_{\dot{\beta}} \overline{\psi}^{b j}\,_{\dot{\alpha}} \overline{\psi}^{c}{}_{j}^{\dot{\alpha}} \overline{\psi}^{d}{}_{i}^{\dot{\beta}} \phi \boldsymbol{\lambda}^{i}\,_{\alpha}+\frac{1}{4}{\rm i} c \epsilon_{a b c d} \psi^{a}\,_{i}\,^{\alpha} \psi^{b}{}^{i}_{\alpha} \psi^{c}\,_{j}\,^{\beta} \psi^{d}{}^{j}_{\beta} \overline{\phi} \overline{\boldsymbol{\phi}} - \frac{1}{4}{\rm i} c \epsilon_{a b c d} \overline{\psi}^{a i}\,_{\dot{\alpha}} \overline{\psi}^{b}{}_{i}^{\dot{\alpha}} \overline{\psi}^{c j}\,_{\dot{\beta}} \overline{\psi}^{d}{}_{j}^{\dot{\beta}} \phi \boldsymbol{\phi} - \frac{1}{4}{\rm i} c (\sigma_{b})^{\beta}{}_{\dot{\alpha}} \psi_{a i}\,^{\alpha} \psi^{a}{}^{j}_{\beta} \overline{\psi}^{b}{}_{j}^{\dot{\alpha}} \overline{\boldsymbol{\phi}} \lambda^{i}\,_{\alpha}+\frac{1}{4}{\rm i} c (\sigma_{b})^{\beta}{}_{\dot{\alpha}} \psi_{a i}\,^{\alpha} \psi^{b}{}^{j}_{\beta} \overline{\psi}^{a}{}_{j}^{\dot{\alpha}} \overline{\boldsymbol{\phi}} \lambda^{i}\,_{\alpha}+\frac{1}{4}{\rm i} c (\sigma_{b})^{\alpha}{}_{\dot{\beta}} \psi_{a}\,^{j}\,_{\alpha} \overline{\psi}^{a i}\,_{\dot{\alpha}} \overline{\psi}^{b}{}_{j}^{\dot{\beta}} \boldsymbol{\phi} \overline{\lambda}_{i}\,^{\dot{\alpha}}+\frac{1}{4}{\rm i} c (\sigma_{b})^{\alpha}{}_{\dot{\beta}} \psi_{a}\,^{j}\,_{\alpha} \overline{\psi}^{a i}\,_{\dot{\alpha}} \overline{\psi}^{b}{}_{j}^{\dot{\beta}} \phi \overline{\boldsymbol{\lambda}}_{i}\,^{\dot{\alpha}} - \frac{1}{4}{\rm i} c (\sigma_{a})^{\alpha}{}_{\dot{\beta}} \psi^{a}{}^{j}_{\alpha} \overline{\psi}_{b}\,^{i}\,_{\dot{\alpha}} \overline{\psi}^{b}{}_{j}^{\dot{\beta}} \boldsymbol{\phi} \overline{\lambda}_{i}\,^{\dot{\alpha}} - \frac{1}{4}{\rm i} c (\sigma_{a})^{\alpha}{}_{\dot{\beta}} \psi^{a}{}^{j}_{\alpha} \overline{\psi}_{b}\,^{i}\,_{\dot{\alpha}} \overline{\psi}^{b}{}_{j}^{\dot{\beta}} \phi \overline{\boldsymbol{\lambda}}_{i}\,^{\dot{\alpha}} - \frac{1}{4}{\rm i} c (\sigma_{b})^{\beta}{}_{\dot{\alpha}} \psi_{a i}\,^{\alpha} \psi^{a}{}^{j}_{\beta} \overline{\psi}^{b}{}_{j}^{\dot{\alpha}} \overline{\phi} \boldsymbol{\lambda}^{i}\,_{\alpha}+\frac{1}{4}{\rm i} c (\sigma_{b})^{\beta}{}_{\dot{\alpha}} \psi_{a i}\,^{\alpha} \psi^{b}{}^{j}_{\beta} \overline{\psi}^{a}{}_{j}^{\dot{\alpha}} \overline{\phi} \boldsymbol{\lambda}^{i}\,_{\alpha} - \frac{1}{2}c (\sigma_{ab})^{\rho\alpha} \psi^{a}\,_{i \rho} \psi^{b i \beta} \phi \overline{\boldsymbol{\phi}} {q}^{-2} \rho_{\underline{i} \alpha} \rho^{\underline{i}}\,_{\beta}%
 - \frac{1}{2}c (\sigma_{ab})^{\rho\alpha} \psi^{a}\,_{i \rho} \psi^{b i \beta} \overline{\phi} \boldsymbol{\phi} {q}^{-2} \rho_{\underline{i} \alpha} \rho^{\underline{i}}\,_{\beta} - \frac{1}{2}c (\tilde{\sigma}_{ab})_{\dot{\rho}\dot{\alpha}} \overline{\psi}^{a i \dot{\rho}} \overline{\psi}^{b}\,_{i \dot{\beta}} \phi \overline{\boldsymbol{\phi}} {q}^{-2} \overline{\rho}^{\underline{i} \dot{\alpha}} \overline{\rho}_{\underline{i}}\,^{\dot{\beta}} - \frac{1}{2}c (\tilde{\sigma}_{ab})_{\dot{\rho}\dot{\alpha}} \overline{\psi}^{a i \dot{\rho}} \overline{\psi}^{b}\,_{i \dot{\beta}} \overline{\phi} \boldsymbol{\phi} {q}^{-2} \overline{\rho}^{\underline{i} \dot{\alpha}} \overline{\rho}_{\underline{i}}\,^{\dot{\beta}} - \frac{1}{2}c \psi_{a j}\,^{\beta} \psi^{a j \alpha} \overline{\boldsymbol{\phi}} \lambda_{i \beta} {q}^{-2} q^{i}\,_{\underline{i}} \rho^{\underline{i}}\,_{\alpha} - \frac{1}{2}c \psi_{a j}\,^{\beta} \psi^{a j \alpha} \overline{\phi} \boldsymbol{\lambda}_{i \beta} {q}^{-2} q^{i}\,_{\underline{i}} \rho^{\underline{i}}\,_{\alpha}+\frac{1}{2}c \psi_{a j}\,^{\alpha} \psi^{a}\,_{i}\,^{\beta} \overline{\boldsymbol{\phi}} \lambda^{j}\,_{\beta} {q}^{-2} q^{i}\,_{\underline{i}} \rho^{\underline{i}}\,_{\alpha}+\frac{1}{2}c \psi_{a j}\,^{\alpha} \psi^{a}\,_{i}\,^{\beta} \overline{\phi} \boldsymbol{\lambda}^{j}\,_{\beta} {q}^{-2} q^{i}\,_{\underline{i}} \rho^{\underline{i}}\,_{\alpha}+\frac{1}{4}{\rm i} c (\sigma_{a})^{\alpha}{}_{\dot{\alpha}} \overline{\psi}^{a}\,_{i \dot{\beta}} \phi \overline{\boldsymbol{\phi}} {q}^{-4} q^{i \underline{j}} \rho^{\underline{i}}\,_{\alpha} \overline{\rho}_{\underline{j}}\,^{\dot{\alpha}} \overline{\rho}_{\underline{i}}\,^{\dot{\beta}}+\frac{1}{4}{\rm i} c (\sigma_{a})^{\alpha}{}_{\dot{\alpha}} \overline{\psi}^{a}\,_{i \dot{\beta}} \overline{\phi} \boldsymbol{\phi} {q}^{-4} q^{i \underline{j}} \rho^{\underline{i}}\,_{\alpha} \overline{\rho}_{\underline{j}}\,^{\dot{\alpha}} \overline{\rho}_{\underline{i}}\,^{\dot{\beta}}-2c (\sigma_{ab})^{\rho\beta} \psi^{a}\,_{j \rho} \psi^{b j \alpha} \overline{\boldsymbol{\phi}} \lambda_{i \beta} {q}^{-2} q^{i}\,_{\underline{i}} \rho^{\underline{i}}\,_{\alpha}-2c (\sigma_{ab})^{\rho\beta} \psi^{a}\,_{j \rho} \psi^{b j \alpha} \overline{\phi} \boldsymbol{\lambda}_{i \beta} {q}^{-2} q^{i}\,_{\underline{i}} \rho^{\underline{i}}\,_{\alpha} - \frac{1}{2}c \psi_{a i}\,^{\alpha} \overline{\psi}^{a}\,_{j \dot{\alpha}} \phi \overline{\boldsymbol{\phi}} {q}^{-4} q^{i \underline{j}} q^{j}\,_{\underline{i}} \rho^{\underline{i}}\,_{\alpha} \overline{\rho}_{\underline{j}}\,^{\dot{\alpha}} - \frac{1}{2}c \psi_{a i}\,^{\alpha} \overline{\psi}^{a}\,_{j \dot{\alpha}} \overline{\phi} \boldsymbol{\phi} {q}^{-4} q^{i \underline{j}} q^{j}\,_{\underline{i}} \rho^{\underline{i}}\,_{\alpha} \overline{\rho}_{\underline{j}}\,^{\dot{\alpha}} - \frac{1}{2}c \epsilon^{a}\,_{b c d} (\sigma_{a})^{\alpha}{}_{\dot{\beta}} \psi^{b}{}^{j}_{\alpha} \overline{\psi}^{c}\,_{j \dot{\alpha}} \overline{\psi}^{d}{}_{i}^{\dot{\beta}} \phi \overline{\boldsymbol{\phi}} {q}^{-2} q^{i \underline{i}} \overline{\rho}_{\underline{i}}\,^{\dot{\alpha}} - \frac{1}{2}c \epsilon^{a}\,_{b c d} (\sigma_{a})^{\alpha}{}_{\dot{\beta}} \psi^{b}{}^{j}_{\alpha} \overline{\psi}^{c}\,_{j \dot{\alpha}} \overline{\psi}^{d}{}_{i}^{\dot{\beta}} \overline{\phi} \boldsymbol{\phi} {q}^{-2} q^{i \underline{i}} \overline{\rho}_{\underline{i}}\,^{\dot{\alpha}}+\frac{1}{2}{\rm i} c (\sigma_{a})^{\alpha}{}_{\dot{\alpha}} \overline{\psi}^{a}\,_{i \dot{\beta}} \boldsymbol{\phi} \overline{\lambda}_{j}\,^{\dot{\beta}} {q}^{-4} q^{i}\,_{\underline{i}} q^{j \underline{j}} \rho^{\underline{i}}\,_{\alpha} \overline{\rho}_{\underline{j}}\,^{\dot{\alpha}}+\frac{1}{2}{\rm i} c (\sigma_{a})^{\alpha}{}_{\dot{\alpha}} \overline{\psi}^{a}\,_{i \dot{\beta}} \boldsymbol{\phi} \overline{\lambda}_{j}\,^{\dot{\beta}} {q}^{-4} q^{i \underline{j}} q^{j}\,_{\underline{i}} \rho^{\underline{i}}\,_{\alpha} \overline{\rho}_{\underline{j}}\,^{\dot{\alpha}}+\frac{1}{2}{\rm i} c (\sigma_{a})^{\alpha}{}_{\dot{\alpha}} \overline{\psi}^{a}\,_{i \dot{\beta}} \phi \overline{\boldsymbol{\lambda}}_{j}\,^{\dot{\beta}} {q}^{-4} q^{i}\,_{\underline{i}} q^{j \underline{j}} \rho^{\underline{i}}\,_{\alpha} \overline{\rho}_{\underline{j}}\,^{\dot{\alpha}}+\frac{1}{2}{\rm i} c (\sigma_{a})^{\alpha}{}_{\dot{\alpha}} \overline{\psi}^{a}\,_{i \dot{\beta}} \phi \overline{\boldsymbol{\lambda}}_{j}\,^{\dot{\beta}} {q}^{-4} q^{i \underline{j}} q^{j}\,_{\underline{i}} \rho^{\underline{i}}\,_{\alpha} \overline{\rho}_{\underline{j}}\,^{\dot{\alpha}} - \frac{1}{2}{\rm i} c (\sigma_{b})^{\alpha}{}_{\dot{\beta}} \psi_{a}\,^{j}\,_{\alpha} \overline{\psi}^{a}\,_{j \dot{\alpha}} \overline{\psi}^{b}{}_{i}^{\dot{\beta}} \phi \overline{\boldsymbol{\phi}} {q}^{-2} q^{i \underline{i}} \overline{\rho}_{\underline{i}}\,^{\dot{\alpha}}%
 - \frac{1}{2}{\rm i} c (\sigma_{b})^{\alpha}{}_{\dot{\beta}} \psi_{a}\,^{j}\,_{\alpha} \overline{\psi}^{a}\,_{j \dot{\alpha}} \overline{\psi}^{b}{}_{i}^{\dot{\beta}} \overline{\phi} \boldsymbol{\phi} {q}^{-2} q^{i \underline{i}} \overline{\rho}_{\underline{i}}\,^{\dot{\alpha}}+\frac{1}{2}{\rm i} c (\sigma_{a})^{\alpha}{}_{\dot{\beta}} \psi^{a}{}^{j}_{\alpha} \overline{\psi}_{b j \dot{\alpha}} \overline{\psi}^{b}{}_{i}^{\dot{\beta}} \phi \overline{\boldsymbol{\phi}} {q}^{-2} q^{i \underline{i}} \overline{\rho}_{\underline{i}}\,^{\dot{\alpha}}+\frac{1}{2}{\rm i} c (\sigma_{a})^{\alpha}{}_{\dot{\beta}} \psi^{a}{}^{j}_{\alpha} \overline{\psi}_{b j \dot{\alpha}} \overline{\psi}^{b}{}_{i}^{\dot{\beta}} \overline{\phi} \boldsymbol{\phi} {q}^{-2} q^{i \underline{i}} \overline{\rho}_{\underline{i}}\,^{\dot{\alpha}} - \frac{1}{4}{\rm i} c (\sigma_{a})^{\alpha}{}_{\dot{\alpha}} \overline{\psi}^{a}\,_{i \dot{\beta}} \phi \overline{\boldsymbol{\phi}} {q}^{-4} q^{i}\,_{\underline{i}} \rho^{\underline{i}}\,_{\alpha} \overline{\rho}^{\underline{j} \dot{\alpha}} \overline{\rho}_{\underline{j}}\,^{\dot{\beta}} - \frac{1}{4}{\rm i} c (\sigma_{a})^{\alpha}{}_{\dot{\alpha}} \overline{\psi}^{a}\,_{i \dot{\beta}} \overline{\phi} \boldsymbol{\phi} {q}^{-4} q^{i}\,_{\underline{i}} \rho^{\underline{i}}\,_{\alpha} \overline{\rho}^{\underline{j} \dot{\alpha}} \overline{\rho}_{\underline{j}}\,^{\dot{\beta}}+\frac{1}{4}{\rm i} c (\sigma_{a})^{\alpha}{}_{\dot{\alpha}} \psi^{a}\,_{i}\,^{\beta} \phi \overline{\boldsymbol{\phi}} {q}^{-4} q^{i}\,_{\underline{i}} \rho^{\underline{i}}\,_{\alpha} \rho^{\underline{j}}\,_{\beta} \overline{\rho}_{\underline{j}}\,^{\dot{\alpha}} - \frac{1}{4}{\rm i} c (\sigma_{a})^{\alpha}{}_{\dot{\alpha}} \psi^{a}\,_{i}\,^{\beta} \phi \overline{\boldsymbol{\phi}} {q}^{-4} q^{i \underline{j}} \rho_{\underline{i} \alpha} \rho^{\underline{i}}\,_{\beta} \overline{\rho}_{\underline{j}}\,^{\dot{\alpha}}+\frac{1}{4}{\rm i} c (\sigma_{a})^{\alpha}{}_{\dot{\alpha}} \psi^{a}\,_{i}\,^{\beta} \overline{\phi} \boldsymbol{\phi} {q}^{-4} q^{i}\,_{\underline{i}} \rho^{\underline{i}}\,_{\alpha} \rho^{\underline{j}}\,_{\beta} \overline{\rho}_{\underline{j}}\,^{\dot{\alpha}} - \frac{1}{4}{\rm i} c (\sigma_{a})^{\alpha}{}_{\dot{\alpha}} \psi^{a}\,_{i}\,^{\beta} \overline{\phi} \boldsymbol{\phi} {q}^{-4} q^{i \underline{j}} \rho_{\underline{i} \alpha} \rho^{\underline{i}}\,_{\beta} \overline{\rho}_{\underline{j}}\,^{\dot{\alpha}}+\frac{1}{4}c \psi_{a i}\,^{\beta} \psi^{a}\,_{j}\,^{\alpha} \phi \overline{\boldsymbol{\phi}} {q}^{-4} q^{i}\,_{\underline{i}} q^{j}\,_{\underline{j}} \rho^{\underline{i}}\,_{\alpha} \rho^{\underline{j}}\,_{\beta}+\frac{1}{4}c \psi_{a i}\,^{\beta} \psi^{a}\,_{j}\,^{\alpha} \overline{\phi} \boldsymbol{\phi} {q}^{-4} q^{i}\,_{\underline{i}} q^{j}\,_{\underline{j}} \rho^{\underline{i}}\,_{\alpha} \rho^{\underline{j}}\,_{\beta} - \frac{1}{2}c \epsilon^{a}\,_{b c d} (\sigma_{a})^{\alpha}{}_{\dot{\beta}} \psi^{b}\,_{i \alpha} \overline{\psi}^{c j \dot{\beta}} \overline{\psi}^{d}\,_{j \dot{\alpha}} \phi \overline{\boldsymbol{\phi}} {q}^{-2} q^{i \underline{i}} \overline{\rho}_{\underline{i}}\,^{\dot{\alpha}} - \frac{1}{2}c \epsilon^{a}\,_{b c d} (\sigma_{a})^{\alpha}{}_{\dot{\beta}} \psi^{b}\,_{i \alpha} \overline{\psi}^{c j \dot{\beta}} \overline{\psi}^{d}\,_{j \dot{\alpha}} \overline{\phi} \boldsymbol{\phi} {q}^{-2} q^{i \underline{i}} \overline{\rho}_{\underline{i}}\,^{\dot{\alpha}}+\frac{1}{2}c \epsilon^{a}\,_{b c d} (\sigma_{a})^{\beta}{}_{\dot{\alpha}} \psi^{b}\,_{j \beta} \psi^{c j \alpha} \overline{\psi}^{d}{}_{i}^{\dot{\alpha}} \phi \overline{\boldsymbol{\phi}} {q}^{-2} q^{i}\,_{\underline{i}} \rho^{\underline{i}}\,_{\alpha}+\frac{1}{2}c \epsilon^{a}\,_{b c d} (\sigma_{a})^{\beta}{}_{\dot{\alpha}} \psi^{b}\,_{j \beta} \psi^{c j \alpha} \overline{\psi}^{d}{}_{i}^{\dot{\alpha}} \overline{\phi} \boldsymbol{\phi} {q}^{-2} q^{i}\,_{\underline{i}} \rho^{\underline{i}}\,_{\alpha}+\frac{1}{2}c \epsilon^{a}\,_{b c d} (\sigma_{a})^{\beta}{}_{\dot{\alpha}} \psi^{b}\,_{i \beta} \psi^{c j \alpha} \overline{\psi}^{d}{}_{j}^{\dot{\alpha}} \phi \overline{\boldsymbol{\phi}} {q}^{-2} q^{i}\,_{\underline{i}} \rho^{\underline{i}}\,_{\alpha}+\frac{1}{2}c \epsilon^{a}\,_{b c d} (\sigma_{a})^{\beta}{}_{\dot{\alpha}} \psi^{b}\,_{i \beta} \psi^{c j \alpha} \overline{\psi}^{d}{}_{j}^{\dot{\alpha}} \overline{\phi} \boldsymbol{\phi} {q}^{-2} q^{i}\,_{\underline{i}} \rho^{\underline{i}}\,_{\alpha} - \frac{1}{2}{\rm i} c (\sigma_{a})^{\alpha}{}_{\dot{\alpha}} \psi^{a}\,_{i}\,^{\beta} \overline{\boldsymbol{\phi}} \lambda_{j \beta} {q}^{-4} q^{i}\,_{\underline{i}} q^{j \underline{j}} \rho^{\underline{i}}\,_{\alpha} \overline{\rho}_{\underline{j}}\,^{\dot{\alpha}} - \frac{1}{2}{\rm i} c (\sigma_{a})^{\alpha}{}_{\dot{\alpha}} \psi^{a}\,_{i}\,^{\beta} \overline{\boldsymbol{\phi}} \lambda_{j \beta} {q}^{-4} q^{i \underline{j}} q^{j}\,_{\underline{i}} \rho^{\underline{i}}\,_{\alpha} \overline{\rho}_{\underline{j}}\,^{\dot{\alpha}} - \frac{1}{2}{\rm i} c (\sigma_{a})^{\alpha}{}_{\dot{\alpha}} \psi^{a}\,_{i}\,^{\beta} \overline{\phi} \boldsymbol{\lambda}_{j \beta} {q}^{-4} q^{i}\,_{\underline{i}} q^{j \underline{j}} \rho^{\underline{i}}\,_{\alpha} \overline{\rho}_{\underline{j}}\,^{\dot{\alpha}}%
 - \frac{1}{2}{\rm i} c (\sigma_{a})^{\alpha}{}_{\dot{\alpha}} \psi^{a}\,_{i}\,^{\beta} \overline{\phi} \boldsymbol{\lambda}_{j \beta} {q}^{-4} q^{i \underline{j}} q^{j}\,_{\underline{i}} \rho^{\underline{i}}\,_{\alpha} \overline{\rho}_{\underline{j}}\,^{\dot{\alpha}} - \frac{1}{2}{\rm i} c (\sigma_{b})^{\alpha}{}_{\dot{\beta}} \psi_{a i \alpha} \overline{\psi}^{a j}\,_{\dot{\alpha}} \overline{\psi}^{b}{}_{j}^{\dot{\beta}} \phi \overline{\boldsymbol{\phi}} {q}^{-2} q^{i \underline{i}} \overline{\rho}_{\underline{i}}\,^{\dot{\alpha}} - \frac{1}{2}{\rm i} c (\sigma_{b})^{\alpha}{}_{\dot{\beta}} \psi_{a i \alpha} \overline{\psi}^{a j}\,_{\dot{\alpha}} \overline{\psi}^{b}{}_{j}^{\dot{\beta}} \overline{\phi} \boldsymbol{\phi} {q}^{-2} q^{i \underline{i}} \overline{\rho}_{\underline{i}}\,^{\dot{\alpha}}+\frac{1}{2}{\rm i} c (\sigma_{a})^{\alpha}{}_{\dot{\beta}} \psi^{a}\,_{i \alpha} \overline{\psi}_{b}\,^{j \dot{\beta}} \overline{\psi}^{b}\,_{j \dot{\alpha}} \phi \overline{\boldsymbol{\phi}} {q}^{-2} q^{i \underline{i}} \overline{\rho}_{\underline{i}}\,^{\dot{\alpha}}+\frac{1}{2}{\rm i} c (\sigma_{a})^{\alpha}{}_{\dot{\beta}} \psi^{a}\,_{i \alpha} \overline{\psi}_{b}\,^{j \dot{\beta}} \overline{\psi}^{b}\,_{j \dot{\alpha}} \overline{\phi} \boldsymbol{\phi} {q}^{-2} q^{i \underline{i}} \overline{\rho}_{\underline{i}}\,^{\dot{\alpha}}+\frac{1}{2}{\rm i} c (\sigma_{b})^{\beta}{}_{\dot{\alpha}} \psi_{a j \beta} \psi^{a j \alpha} \overline{\psi}^{b}{}_{i}^{\dot{\alpha}} \phi \overline{\boldsymbol{\phi}} {q}^{-2} q^{i}\,_{\underline{i}} \rho^{\underline{i}}\,_{\alpha}+\frac{1}{2}{\rm i} c (\sigma_{b})^{\beta}{}_{\dot{\alpha}} \psi_{a j \beta} \psi^{a j \alpha} \overline{\psi}^{b}{}_{i}^{\dot{\alpha}} \overline{\phi} \boldsymbol{\phi} {q}^{-2} q^{i}\,_{\underline{i}} \rho^{\underline{i}}\,_{\alpha} - \frac{1}{2}{\rm i} c (\sigma_{b})^{\beta}{}_{\dot{\alpha}} \psi_{a i \beta} \psi^{a j \alpha} \overline{\psi}^{b}{}_{j}^{\dot{\alpha}} \phi \overline{\boldsymbol{\phi}} {q}^{-2} q^{i}\,_{\underline{i}} \rho^{\underline{i}}\,_{\alpha} - \frac{1}{2}{\rm i} c (\sigma_{b})^{\beta}{}_{\dot{\alpha}} \psi_{a i \beta} \psi^{a j \alpha} \overline{\psi}^{b}{}_{j}^{\dot{\alpha}} \overline{\phi} \boldsymbol{\phi} {q}^{-2} q^{i}\,_{\underline{i}} \rho^{\underline{i}}\,_{\alpha} - \frac{1}{2}{\rm i} c (\sigma_{b})^{\beta}{}_{\dot{\alpha}} \psi_{a j}\,^{\alpha} \psi^{b}{}^{j}_{\beta} \overline{\psi}^{a}{}_{i}^{\dot{\alpha}} \phi \overline{\boldsymbol{\phi}} {q}^{-2} q^{i}\,_{\underline{i}} \rho^{\underline{i}}\,_{\alpha} - \frac{1}{2}{\rm i} c (\sigma_{b})^{\beta}{}_{\dot{\alpha}} \psi_{a j}\,^{\alpha} \psi^{b}{}^{j}_{\beta} \overline{\psi}^{a}{}_{i}^{\dot{\alpha}} \overline{\phi} \boldsymbol{\phi} {q}^{-2} q^{i}\,_{\underline{i}} \rho^{\underline{i}}\,_{\alpha} - \frac{1}{2}{\rm i} c (\sigma_{b})^{\beta}{}_{\dot{\alpha}} \psi_{a}\,^{j \alpha} \psi^{b}\,_{i \beta} \overline{\psi}^{a}{}_{j}^{\dot{\alpha}} \phi \overline{\boldsymbol{\phi}} {q}^{-2} q^{i}\,_{\underline{i}} \rho^{\underline{i}}\,_{\alpha} - \frac{1}{2}{\rm i} c (\sigma_{b})^{\beta}{}_{\dot{\alpha}} \psi_{a}\,^{j \alpha} \psi^{b}\,_{i \beta} \overline{\psi}^{a}{}_{j}^{\dot{\alpha}} \overline{\phi} \boldsymbol{\phi} {q}^{-2} q^{i}\,_{\underline{i}} \rho^{\underline{i}}\,_{\alpha}$\end{adjustwidth}

\section{On-Shell SU(1,1)/U(1) sigma model}

The bosonic and some of the fermionic terms of the on-shell SU(1,1)/U(1) sigma model are given in eqs.~(6.18) and (6.19) of the Subsection 6.3 in the main paper. The on-shell action is derived by substituting the equations of motion for all auxiliary fields eqs.~(6.4) and (6.5) followed by the imposition of gauge-fixing conditions as seen in eqs.~(6.7) into the off-shell action of eq.~(6.3). These terms along with the remaining higher fermionic terms of the SU(1,1)/U(1) on-shell component action are given by
\bea
\cL^{{\rm on-shell,\,SU(1,1)/U(1)}}
=\cL^{{\rm on,\,SU(1,1)/U(1)}}_{{\rm bosons}}
+\cL^{{\rm on,\,SU(1,1)/U(1)}}_{2{\rm ferms}}
+\cL^{{\rm on,\,SU(1,1)/U(1)}}_{{4\rm ferms}}
~,~~~
\eea
where once again the subscript of the Lagrangian denotes the order in explicit fermions. In this section, there are no implicit fermions as the conversion has been performed for $H^{a \underline{i} \underline{j}}$, $F^{a b}$, and $\textbf{F}^{a b}$ to $\tilde{h}^{a \underline{i} \underline{j}}$, $f^{a b}$, and $\textbf{f}^{a b}$, respectively, by eqs. (2.67), (2.68), and (2.44) in the main paper and $\textbf{D}_{a}$ has been replaced by $\cD'_{a}$ plus gravitini by using eq.~(2.14) for the spin connection with $b_a=0$.

\subsection{$\cL^{{\rm on,\,SU(1,1)/U(1)}}_{{\rm bosons}}$}

\begin{adjustwidth}{1em}{4cm}${}16{e}^{4U} \tilde{h}_{a i j} \tilde{h}^{a i j} - \frac{1}{2}R+\frac{1}{2}c {y}^{-1} \overline{\boldsymbol{\phi}} f_{a b} f^{a b}+\frac{1}{4}{\rm i} c {y}^{-1} \epsilon_{a b c d} \overline{\boldsymbol{\phi}} f^{a b} f^{c d}+\frac{1}{2}c y {\overline{\boldsymbol{\phi}}}^{-1} \textbf{f}_{a b} \textbf{f}^{a b}+\frac{1}{4}{\rm i} c y \epsilon_{a b c d} {\overline{\boldsymbol{\phi}}}^{-1} \textbf{f}^{a b} \textbf{f}^{c d}+\frac{1}{2}c {y}^{-1} \boldsymbol{\phi} f_{a b} f^{a b} - \frac{1}{4}{\rm i} c {y}^{-1} \epsilon_{a b c d} \boldsymbol{\phi} f^{a b} f^{c d}+\frac{1}{2}c y {\boldsymbol{\phi}}^{-1} \textbf{f}_{a b} \textbf{f}^{a b} - \frac{1}{4}{\rm i} c y \epsilon_{a b c d} {\boldsymbol{\phi}}^{-1} \textbf{f}^{a b} \textbf{f}^{c d}+c y \mathcal{D}'_{a}{\mathcal{D}'^{a}{\overline{\boldsymbol{\phi}}}}+{c}^{2} {y}^{2} \mathcal{D}'_{a}{\overline{\boldsymbol{\phi}}} \mathcal{D}'^{a}{\overline{\boldsymbol{\phi}}}-2{c}^{2} {y}^{2} \mathcal{D}'_{a}{\boldsymbol{\phi}} \mathcal{D}'^{a}{\overline{\boldsymbol{\phi}}}+{c}^{2} {y}^{2} \mathcal{D}'_{a}{\boldsymbol{\phi}} \mathcal{D}'^{a}{\boldsymbol{\phi}}+c y \mathcal{D}'_{a}{\mathcal{D}'^{a}{\boldsymbol{\phi}}}-2\mathcal{D}'_{a}{U} \mathcal{D}'^{a}{U}+2\mathcal{D}'_{a}{\mathcal{D}'^{a}{U}} +\frac{4}{3} \xi_{i j} \e^{m n p q} h_{m np}{}^{i j} v_{q}$\end{adjustwidth}

\subsection{$\cL^{{\rm on,\,SU(1,1)/U(1)}}_{2{\rm ferms}}$}
\begin{adjustwidth}{1em}{4cm}${}  \frac{1}{4} c \overline{\bM}_{ij} {e}^{U}  \lambda^{j \alpha} \rho^{i}\,_{\alpha}-\frac{1}{16} c y e^{2U} \overline{\bM}_{i j} \rho^{i \alpha} \rho^{j}\,_{\alpha} - \frac{1}{4}{\rm  i}c  \bM_{i j} (\sigma_{a})^{\alpha}{}_{\dot{\alpha}} \psi^{a}{}^{j}_{\alpha} \overline{\lambda}^{i \dot{\alpha}}-\frac{1}{4} {\rm i} c  \overline{\bM}_{i j} (\sigma_{a})^{\alpha}{}_{\dot{\alpha}} \overline{\psi}^{a j \dot{\alpha}} \lambda^{i}\,_{\alpha}+\frac{1}{4} c {e}^{U} \bM^{i j} \overline{\lambda}_{j \dot{\alpha}} \overline{\rho}_{i}\,^{\dot{\alpha}} - \frac{1}{16} c y e^{2U} \bM^{i j} \overline{\rho}_{i \dot{\alpha}} \overline{\rho}_{j}\,^{\dot{\alpha}}+\frac{1}{4} {\rm i} c y {e}^{U} \bM_{i j} (\sigma_{a})^{\alpha}{}_{\dot{\alpha}} \psi^{a}{}^{j}_{\alpha} \overline{\rho}^{i \dot{\alpha}} + \frac{1}{4}{\rm i}c y {e}^{U} \overline{\bM}_{i j} (\sigma_{a})^{\alpha}{}_{\dot{\alpha}} \overline{\psi}^{a j \dot{\alpha}} \rho^{i}\,_{\alpha} - \frac{1}{2} c y  \bM_{i j} (\sigma_{ab})^{\alpha\beta} \psi^{a}{}^{i}_{\alpha} \psi^{b}{}^{j}_{\beta}-\frac{1}{2} c y  \overline{\bM}^{i j} (\tilde{\sigma}_{ab})_{\dot{\alpha}\dot{\beta}} \overline{\psi}^{a}{}_{i}^{\dot{\alpha}} \overline{\psi}^{b}{}_{j}^{\dot{\beta}}+\frac{1}{4}{e}^{U} \rho_{i}\,^{\alpha} \mathcal{D}'_{a}{\psi^{a}{}^{i}_{\alpha}} - \frac{1}{4}{e}^{U} \psi_{a i}\,^{\alpha} \mathcal{D}'^{a}{\rho^{i}\,_{\alpha}}+\frac{1}{4}{e}^{U} \overline{\rho}^{i}\,_{\dot{\alpha}} \mathcal{D}'_{a}{\overline{\psi}^{a}{}_{i}^{\dot{\alpha}}} - \frac{1}{4}{e}^{U} \overline{\psi}_{a}\,^{i}\,_{\dot{\alpha}} \mathcal{D}'^{a}{\overline{\rho}_{i}\,^{\dot{\alpha}}}+2{e}^{3U} \overline{\psi}_{a j \dot{\alpha}} \overline{\rho}_{i}\,^{\dot{\alpha}} \tilde{h}^{a j i}+c \psi_{a i}\,^{\alpha} \lambda^{i}\,_{\alpha} \mathcal{D}'^{a}{\overline{\boldsymbol{\phi}}}+c \overline{\psi}_{a}\,^{i}\,_{\dot{\alpha}} \overline{\lambda}_{i}\,^{\dot{\alpha}} \mathcal{D}'^{a}{\boldsymbol{\phi}}+\frac{1}{2}\epsilon^{a}\,_{b c d} (\sigma_{a})^{\alpha}{}_{\dot{\alpha}} \psi^{b}{}^{i}_{\alpha} \mathcal{D}'^{c}{\overline{\psi}^{d}{}_{i}^{\dot{\alpha}}} - \frac{1}{2}\epsilon^{a}\,_{b c d} (\sigma_{a})^{\alpha}{}_{\dot{\alpha}} \overline{\psi}^{b}{}_{i}^{\dot{\alpha}} \mathcal{D}'^{c}{\psi^{d}{}^{i}_{\alpha}}%
+\frac{1}{2}{e}^{U} (\sigma_{ab})^{\alpha\beta} \rho_{i \alpha} \mathcal{D}'^{a}{\psi^{b}{}^{i}_{\beta}} - \frac{1}{2}{e}^{U} (\sigma_{ba})^{\beta\alpha} \psi^{b}\,_{i \beta} \mathcal{D}'^{a}{\rho^{i}\,_{\alpha}}+\frac{1}{2}{e}^{U} (\tilde{\sigma}_{ab})_{\dot{\alpha}\dot{\beta}} \overline{\rho}^{i \dot{\alpha}} \mathcal{D}'^{a}{\overline{\psi}^{b}{}_{i}^{\dot{\beta}}} - \frac{1}{2}{e}^{U} (\tilde{\sigma}_{ba})_{\dot{\beta}\dot{\alpha}} \overline{\psi}^{b i \dot{\beta}} \mathcal{D}'^{a}{\overline{\rho}_{i}\,^{\dot{\alpha}}}+\frac{1}{4}{e}^{U} \psi_{a i}\,^{\alpha} \rho^{i}\,_{\alpha} \mathcal{D}'^{a}{U}+\frac{1}{4}{e}^{U} \overline{\psi}_{a}\,^{i}\,_{\dot{\alpha}} \overline{\rho}_{i}\,^{\dot{\alpha}} \mathcal{D}'^{a}{U}+4{e}^{3U} (\tilde{\sigma}_{ba})_{\dot{\beta}\dot{\alpha}} \overline{\psi}^{b}{}_{j}^{\dot{\beta}} \overline{\rho}_{i}\,^{\dot{\alpha}} \tilde{h}^{a j i}+\frac{1}{8}{\rm i} {e}^{2U} (\sigma_{a})^{\alpha}{}_{\dot{\alpha}} \rho^{i}\,_{\alpha} \mathcal{D}'^{a}{\overline{\rho}_{i}\,^{\dot{\alpha}}}+\frac{1}{8}{\rm i} {e}^{2U} (\sigma_{a})^{\alpha}{}_{\dot{\alpha}} \overline{\rho}_{i}\,^{\dot{\alpha}} \mathcal{D}'^{a}{\rho^{i}\,_{\alpha}}+2c (\sigma_{ba})^{\beta\alpha} \psi^{b}\,_{i \beta} \lambda^{i}\,_{\alpha} \mathcal{D}'^{a}{\overline{\boldsymbol{\phi}}}+2c (\tilde{\sigma}_{ba})_{\dot{\beta}\dot{\alpha}} \overline{\psi}^{b i \dot{\beta}} \overline{\lambda}_{i}\,^{\dot{\alpha}} \mathcal{D}'^{a}{\boldsymbol{\phi}}+\frac{1}{2}{e}^{U} (\sigma_{ba})^{\beta\alpha} \psi^{b}\,_{i \beta} \rho^{i}\,_{\alpha} \mathcal{D}'^{a}{U}+\frac{1}{2}{e}^{U} (\tilde{\sigma}_{ba})_{\dot{\beta}\dot{\alpha}} \overline{\psi}^{b i \dot{\beta}} \overline{\rho}_{i}\,^{\dot{\alpha}} \mathcal{D}'^{a}{U}-2{e}^{3U} \psi_{a j}\,^{\alpha} \rho_{i \alpha} \tilde{h}^{a j i} - \frac{1}{4}{y}^{-1} \overline{\psi}_{a}\,^{i}\,_{\dot{\alpha}} \overline{\psi}_{b i}\,^{\dot{\alpha}} f^{a b}+{\rm i} (\sigma_{b})^{\alpha}{}_{\dot{\alpha}} \psi_{a}\,^{i}\,_{\alpha} \overline{\psi}^{b}{}_{i}^{\dot{\alpha}} \mathcal{D}'^{a}{U}-{\rm i} (\sigma_{b})^{\alpha}{}_{\dot{\alpha}} \psi^{b}{}^{i}_{\alpha} \overline{\psi}_{a i}\,^{\dot{\alpha}} \mathcal{D}'^{a}{U} - \frac{1}{8}{\rm i} {y}^{-1} \epsilon_{a b c d} \overline{\psi}^{a i}\,_{\dot{\alpha}} \overline{\psi}^{b}{}_{i}^{\dot{\alpha}} f^{c d} - \frac{1}{2}c \psi_{a i}\,^{\alpha} \psi_{b}\,^{i}\,_{\alpha} \boldsymbol{\phi} f^{a b} - \frac{1}{2}c \psi_{a i}\,^{\alpha} \psi_{b}\,^{i}\,_{\alpha} \overline{\boldsymbol{\phi}} f^{a b}%
+\frac{1}{4}c \epsilon^{c}\,_{d a b} (\sigma_{c})^{\alpha}{}_{\dot{\alpha}} \psi^{d}{}^{i}_{\alpha} \overline{\lambda}_{i}\,^{\dot{\alpha}} \textbf{f}^{a b}+\frac{1}{4}c \epsilon^{c}\,_{d a b} (\sigma_{c})^{\alpha}{}_{\dot{\alpha}} \overline{\psi}^{d}{}_{i}^{\dot{\alpha}} \lambda^{i}\,_{\alpha} \textbf{f}^{a b} - \frac{1}{2}c y \psi_{a i}\,^{\alpha} \psi_{b}\,^{i}\,_{\alpha} \textbf{f}^{a b}+\frac{1}{2}c y \overline{\psi}_{a}\,^{i}\,_{\dot{\alpha}} \overline{\psi}_{b i}\,^{\dot{\alpha}} \textbf{f}^{a b} - \frac{1}{16}{e}^{2U} (\sigma_{ab})^{\alpha\beta} {\overline{\boldsymbol{\phi}}}^{-1} \textbf{f}^{a b} \rho_{i \alpha} \rho^{i}\,_{\beta}+\frac{1}{16}{e}^{2U} (\tilde{\sigma}_{ab})_{\dot{\alpha}\dot{\beta}} {\boldsymbol{\phi}}^{-1} \textbf{f}^{a b} \overline{\rho}^{i \dot{\alpha}} \overline{\rho}_{i}\,^{\dot{\beta}}-4{e}^{3U} (\sigma_{ba})^{\beta\alpha} \psi^{b}\,_{j \beta} \rho_{i \alpha} \tilde{h}^{a j i} - \frac{1}{16}{y}^{-1} {e}^{2U} (\sigma_{ab})^{\alpha\beta} f^{a b} \rho_{i \alpha} \rho^{i}\,_{\beta}+\frac{1}{16}{y}^{-1} {e}^{2U} (\tilde{\sigma}_{ab})_{\dot{\alpha}\dot{\beta}} f^{a b} \overline{\rho}^{i \dot{\alpha}} \overline{\rho}_{i}\,^{\dot{\beta}}+\frac{1}{4}{\rm i} c \epsilon_{a b c d} \psi^{a}\,_{i}\,^{\alpha} \psi^{b}{}^{i}_{\alpha} \boldsymbol{\phi} f^{c d}+\frac{1}{4}{\rm i} c \epsilon_{a b c d} \psi^{a}\,_{i}\,^{\alpha} \psi^{b}{}^{i}_{\alpha} \overline{\boldsymbol{\phi}} f^{c d}+\frac{1}{2}{\rm i} c (\sigma_{b})^{\alpha}{}_{\dot{\alpha}} \psi_{a}\,^{i}\,_{\alpha} \overline{\lambda}_{i}\,^{\dot{\alpha}} \textbf{f}^{a b} - \frac{1}{2}{\rm i} c (\sigma_{b})^{\alpha}{}_{\dot{\alpha}} \overline{\psi}_{a i}\,^{\dot{\alpha}} \lambda^{i}\,_{\alpha} \textbf{f}^{a b} - \frac{1}{2}{\rm i} c {y}^{-1} (\sigma_{a})^{\alpha}{}_{\dot{\alpha}} \boldsymbol{\phi} \lambda^{i}\,_{\alpha} \mathcal{D}'^{a}{\overline{\lambda}_{i}\,^{\dot{\alpha}}} - \frac{1}{2}{\rm i} c {y}^{-1} (\sigma_{a})^{\alpha}{}_{\dot{\alpha}} \boldsymbol{\phi} \overline{\lambda}_{i}\,^{\dot{\alpha}} \mathcal{D}'^{a}{\lambda^{i}\,_{\alpha}} - \frac{1}{2}{\rm i} c {y}^{-1} (\sigma_{a})^{\alpha}{}_{\dot{\alpha}} \overline{\boldsymbol{\phi}} \lambda^{i}\,_{\alpha} \mathcal{D}'^{a}{\overline{\lambda}_{i}\,^{\dot{\alpha}}} - \frac{1}{2}{\rm i} c {y}^{-1} (\sigma_{a})^{\alpha}{}_{\dot{\alpha}} \overline{\boldsymbol{\phi}} \overline{\lambda}_{i}\,^{\dot{\alpha}} \mathcal{D}'^{a}{\lambda^{i}\,_{\alpha}} - \frac{1}{2}{\rm i} c {y}^{-1} (\sigma_{a})^{\alpha}{}_{\dot{\alpha}} \lambda^{i}\,_{\alpha} \overline{\lambda}_{i}\,^{\dot{\alpha}} \mathcal{D}'^{a}{\boldsymbol{\phi}}+\frac{1}{2}{\rm i} c {y}^{-1} (\sigma_{a})^{\alpha}{}_{\dot{\alpha}} \lambda^{i}\,_{\alpha} \overline{\lambda}_{i}\,^{\dot{\alpha}} \mathcal{D}'^{a}{\overline{\boldsymbol{\phi}}}+\frac{1}{4}{\rm i} c y \epsilon_{c d a b} \psi^{c}\,_{i}\,^{\alpha} \psi^{d}{}^{i}_{\alpha} \textbf{f}^{a b}%
+\frac{1}{4}{\rm i} c y \epsilon_{c d a b} \overline{\psi}^{c i}\,_{\dot{\alpha}} \overline{\psi}^{d}{}_{i}^{\dot{\alpha}} \textbf{f}^{a b}+\frac{1}{2}{\rm i} {c}^{2} (\sigma_{a})^{\alpha}{}_{\dot{\alpha}} \boldsymbol{\phi} \lambda^{i}\,_{\alpha} \overline{\lambda}_{i}\,^{\dot{\alpha}} \mathcal{D}'^{a}{\boldsymbol{\phi}} - \frac{1}{2}{\rm i} {c}^{2} (\sigma_{a})^{\alpha}{}_{\dot{\alpha}} \boldsymbol{\phi} \lambda^{i}\,_{\alpha} \overline{\lambda}_{i}\,^{\dot{\alpha}} \mathcal{D}'^{a}{\overline{\boldsymbol{\phi}}}+\frac{1}{2}{\rm i} {c}^{2} (\sigma_{a})^{\alpha}{}_{\dot{\alpha}} \overline{\boldsymbol{\phi}} \lambda^{i}\,_{\alpha} \overline{\lambda}_{i}\,^{\dot{\alpha}} \mathcal{D}'^{a}{\boldsymbol{\phi}} - \frac{1}{2}{\rm i} {c}^{2} (\sigma_{a})^{\alpha}{}_{\dot{\alpha}} \overline{\boldsymbol{\phi}} \lambda^{i}\,_{\alpha} \overline{\lambda}_{i}\,^{\dot{\alpha}} \mathcal{D}'^{a}{\overline{\boldsymbol{\phi}}}+{\rm i} {e}^{4U} (\sigma_{a})^{\alpha}{}_{\dot{\alpha}} \rho_{i \alpha} \overline{\rho}_{j}\,^{\dot{\alpha}} \tilde{h}^{a i j}+\frac{1}{2}c y \epsilon^{a}\,_{b c d} (\sigma_{a})^{\alpha}{}_{\dot{\alpha}} \psi^{b}{}^{i}_{\alpha} \overline{\psi}^{c}{}_{i}^{\dot{\alpha}} \mathcal{D}'^{d}{\boldsymbol{\phi}} - \frac{1}{2}c y \epsilon^{a}\,_{b c d} (\sigma_{a})^{\alpha}{}_{\dot{\alpha}} \psi^{b}{}^{i}_{\alpha} \overline{\psi}^{c}{}_{i}^{\dot{\alpha}} \mathcal{D}'^{d}{\overline{\boldsymbol{\phi}}}+4{e}^{2U} \epsilon^{a}\,_{b c d} (\sigma_{a})^{\alpha}{}_{\dot{\alpha}} \psi^{b}\,_{i \alpha} \overline{\psi}^{c}{}_{j}^{\dot{\alpha}} \tilde{h}^{d i j}+\frac{1}{2}{\rm i} c y (\sigma_{b})^{\alpha}{}_{\dot{\alpha}} \psi_{a}\,^{i}\,_{\alpha} \overline{\psi}^{b}{}_{i}^{\dot{\alpha}} \mathcal{D}'^{a}{\boldsymbol{\phi}}+\frac{1}{2}{\rm i} c y (\sigma_{b})^{\alpha}{}_{\dot{\alpha}} \psi_{a}\,^{i}\,_{\alpha} \overline{\psi}^{b}{}_{i}^{\dot{\alpha}} \mathcal{D}'^{a}{\overline{\boldsymbol{\phi}}} - \frac{1}{2}{\rm i} c y (\sigma_{b})^{\alpha}{}_{\dot{\alpha}} \psi^{b}{}^{i}_{\alpha} \overline{\psi}_{a i}\,^{\dot{\alpha}} \mathcal{D}'^{a}{\boldsymbol{\phi}} - \frac{1}{2}{\rm i} c y (\sigma_{b})^{\alpha}{}_{\dot{\alpha}} \psi^{b}{}^{i}_{\alpha} \overline{\psi}_{a i}\,^{\dot{\alpha}} \mathcal{D}'^{a}{\overline{\boldsymbol{\phi}}}+\frac{1}{8}{\rm i} c y {e}^{2U} (\sigma_{a})^{\alpha}{}_{\dot{\alpha}} \rho^{i}\,_{\alpha} \overline{\rho}_{i}\,^{\dot{\alpha}} \mathcal{D}'^{a}{\boldsymbol{\phi}} - \frac{1}{8}{\rm i} c y {e}^{2U} (\sigma_{a})^{\alpha}{}_{\dot{\alpha}} \rho^{i}\,_{\alpha} \overline{\rho}_{i}\,^{\dot{\alpha}} \mathcal{D}'^{a}{\overline{\boldsymbol{\phi}}}+\frac{1}{4}c \epsilon^{c}\,_{d a b} (\sigma_{c})^{\alpha}{}_{\dot{\alpha}} \psi^{d}{}^{i}_{\alpha} \boldsymbol{\phi} {\overline{\boldsymbol{\phi}}}^{-1} \overline{\lambda}_{i}\,^{\dot{\alpha}} \textbf{f}^{a b}+\frac{1}{4}c \epsilon^{c}\,_{d a b} (\sigma_{c})^{\alpha}{}_{\dot{\alpha}} \overline{\psi}^{d}{}_{i}^{\dot{\alpha}} {\boldsymbol{\phi}}^{-1} \overline{\boldsymbol{\phi}} \lambda^{i}\,_{\alpha} \textbf{f}^{a b} - \frac{1}{4}c {y}^{-1} \epsilon^{a}\,_{b c d} (\sigma_{a})^{\alpha}{}_{\dot{\alpha}} \psi^{b}{}^{i}_{\alpha} \boldsymbol{\phi} \overline{\lambda}_{i}\,^{\dot{\alpha}} f^{c d} - \frac{1}{4}c {y}^{-1} \epsilon^{a}\,_{b c d} (\sigma_{a})^{\alpha}{}_{\dot{\alpha}} \psi^{b}{}^{i}_{\alpha} \overline{\boldsymbol{\phi}} \overline{\lambda}_{i}\,^{\dot{\alpha}} f^{c d} - \frac{1}{4}c {y}^{-1} \epsilon^{a}\,_{b c d} (\sigma_{a})^{\alpha}{}_{\dot{\alpha}} \overline{\psi}^{b}{}_{i}^{\dot{\alpha}} \boldsymbol{\phi} \lambda^{i}\,_{\alpha} f^{c d}%
 - \frac{1}{4}c {y}^{-1} \epsilon^{a}\,_{b c d} (\sigma_{a})^{\alpha}{}_{\dot{\alpha}} \overline{\psi}^{b}{}_{i}^{\dot{\alpha}} \overline{\boldsymbol{\phi}} \lambda^{i}\,_{\alpha} f^{c d} - \frac{1}{2}c y \psi_{a i}\,^{\alpha} \psi_{b}\,^{i}\,_{\alpha} {\boldsymbol{\phi}}^{-1} \overline{\boldsymbol{\phi}} \textbf{f}^{a b}+\frac{1}{2}c y \overline{\psi}_{a}\,^{i}\,_{\dot{\alpha}} \overline{\psi}_{b i}\,^{\dot{\alpha}} \boldsymbol{\phi} {\overline{\boldsymbol{\phi}}}^{-1} \textbf{f}^{a b}+\frac{1}{2}{\rm i} c (\sigma_{b})^{\alpha}{}_{\dot{\alpha}} \psi_{a}\,^{i}\,_{\alpha} \boldsymbol{\phi} {\overline{\boldsymbol{\phi}}}^{-1} \overline{\lambda}_{i}\,^{\dot{\alpha}} \textbf{f}^{a b} - \frac{1}{2}{\rm i} c (\sigma_{b})^{\alpha}{}_{\dot{\alpha}} \overline{\psi}_{a i}\,^{\dot{\alpha}} {\boldsymbol{\phi}}^{-1} \overline{\boldsymbol{\phi}} \lambda^{i}\,_{\alpha} \textbf{f}^{a b} - \frac{1}{2}{\rm i} c {y}^{-1} (\sigma_{b})^{\alpha}{}_{\dot{\alpha}} \psi_{a}\,^{i}\,_{\alpha} \boldsymbol{\phi} \overline{\lambda}_{i}\,^{\dot{\alpha}} f^{a b} - \frac{1}{2}{\rm i} c {y}^{-1} (\sigma_{b})^{\alpha}{}_{\dot{\alpha}} \psi_{a}\,^{i}\,_{\alpha} \overline{\boldsymbol{\phi}} \overline{\lambda}_{i}\,^{\dot{\alpha}} f^{a b}+\frac{1}{2}{\rm i} c {y}^{-1} (\sigma_{b})^{\alpha}{}_{\dot{\alpha}} \overline{\psi}_{a i}\,^{\dot{\alpha}} \boldsymbol{\phi} \lambda^{i}\,_{\alpha} f^{a b}+\frac{1}{2}{\rm i} c {y}^{-1} (\sigma_{b})^{\alpha}{}_{\dot{\alpha}} \overline{\psi}_{a i}\,^{\dot{\alpha}} \overline{\boldsymbol{\phi}} \lambda^{i}\,_{\alpha} f^{a b}+\frac{1}{4}{\rm i} c y \epsilon_{c d a b} \psi^{c}\,_{i}\,^{\alpha} \psi^{d}{}^{i}_{\alpha} {\boldsymbol{\phi}}^{-1} \overline{\boldsymbol{\phi}} \textbf{f}^{a b}+\frac{1}{4}{\rm i} c y \epsilon_{c d a b} \overline{\psi}^{c i}\,_{\dot{\alpha}} \overline{\psi}^{d}{}_{i}^{\dot{\alpha}} \boldsymbol{\phi} {\overline{\boldsymbol{\phi}}}^{-1} \textbf{f}^{a b}+4{\rm i} c {y}^{-1} {e}^{2U} (\sigma_{a})^{\alpha}{}_{\dot{\alpha}} \boldsymbol{\phi} \lambda_{i \alpha} \overline{\lambda}_{j}\,^{\dot{\alpha}} \tilde{h}^{a i j}+4{\rm i} c {y}^{-1} {e}^{2U} (\sigma_{a})^{\alpha}{}_{\dot{\alpha}} \overline{\boldsymbol{\phi}} \lambda_{i \alpha} \overline{\lambda}_{j}\,^{\dot{\alpha}} \tilde{h}^{a i j}$\end{adjustwidth}

\subsection{$\cL^{{\rm on,\,SU(1,1)/U(1)}}_{4{\rm ferms}}$}

\begin{adjustwidth}{1em}{5cm}${} - \frac{1}{8}{e}^{2U} \sigma^{c}\,_{a}\,^{\rho \alpha} (\sigma_{cb})^{\lambda\beta} \psi^{a}\,_{j \rho} \psi^{b}{}^{j}_{\lambda} \rho_{i \alpha} \rho^{i}\,_{\beta} - \frac{1}{8}{e}^{2U} \sigma^{c}\,_{a}\,^{\rho \beta} (\sigma_{cb})^{\lambda\alpha} \psi^{a}\,_{i \rho} \psi^{b}\,_{j \lambda} \rho^{i}\,_{\alpha} \rho^{j}\,_{\beta} - \frac{1}{32}{e}^{2U} (\sigma_{a})^{\beta}{}_{\dot{\beta}} (\sigma_{b})^{\alpha}{}_{\dot{\alpha}} \psi^{a}{}^{j}_{\beta} \overline{\psi}^{b}{}_{j}^{\dot{\beta}} \rho^{i}\,_{\alpha} \overline{\rho}_{i}\,^{\dot{\alpha}} - \frac{1}{32}{e}^{2U} (\sigma_{a})^{\alpha}{}_{\dot{\beta}} (\sigma_{b})^{\beta}{}_{\dot{\alpha}} \psi^{a}{}^{j}_{\beta} \overline{\psi}^{b}{}_{j}^{\dot{\beta}} \rho^{i}\,_{\alpha} \overline{\rho}_{i}\,^{\dot{\alpha}} - \frac{1}{32}{e}^{2U} (\sigma_{a})^{\beta}{}_{\dot{\alpha}} (\sigma_{b})^{\alpha}{}_{\dot{\beta}} \psi^{a}{}^{j}_{\beta} \overline{\psi}^{b}{}_{j}^{\dot{\beta}} \rho^{i}\,_{\alpha} \overline{\rho}_{i}\,^{\dot{\alpha}} - \frac{1}{32}{e}^{2U} (\sigma_{a})^{\alpha}{}_{\dot{\alpha}} (\sigma_{b})^{\beta}{}_{\dot{\beta}} \psi^{a}{}^{j}_{\beta} \overline{\psi}^{b}{}_{j}^{\dot{\beta}} \rho^{i}\,_{\alpha} \overline{\rho}_{i}\,^{\dot{\alpha}} - \frac{1}{32}{e}^{2U} (\sigma_{a})^{\beta}{}_{\dot{\beta}} (\sigma_{b})^{\alpha}{}_{\dot{\alpha}} \psi^{a}{}^{j}_{\beta} \overline{\psi}^{b}{}_{i}^{\dot{\beta}} \rho^{i}\,_{\alpha} \overline{\rho}_{j}\,^{\dot{\alpha}} - \frac{1}{32}{e}^{2U} (\sigma_{a})^{\alpha}{}_{\dot{\beta}} (\sigma_{b})^{\beta}{}_{\dot{\alpha}} \psi^{a}{}^{j}_{\beta} \overline{\psi}^{b}{}_{i}^{\dot{\beta}} \rho^{i}\,_{\alpha} \overline{\rho}_{j}\,^{\dot{\alpha}} - \frac{1}{32}{e}^{2U} (\sigma_{a})^{\beta}{}_{\dot{\alpha}} (\sigma_{b})^{\alpha}{}_{\dot{\beta}} \psi^{a}{}^{j}_{\beta} \overline{\psi}^{b}{}_{i}^{\dot{\beta}} \rho^{i}\,_{\alpha} \overline{\rho}_{j}\,^{\dot{\alpha}} - \frac{1}{32}{e}^{2U} (\sigma_{a})^{\alpha}{}_{\dot{\alpha}} (\sigma_{b})^{\beta}{}_{\dot{\beta}} \psi^{a}{}^{j}_{\beta} \overline{\psi}^{b}{}_{i}^{\dot{\beta}} \rho^{i}\,_{\alpha} \overline{\rho}_{j}\,^{\dot{\alpha}}+\frac{1}{8}{e}^{U} \epsilon^{a b}\,_{c d} (\sigma_{ae})^{\rho\alpha} (\sigma_{b})^{\beta}{}_{\dot{\alpha}} \psi^{c}\,_{j \beta} \psi^{e}{}^{j}_{\rho} \overline{\psi}^{d}{}_{i}^{\dot{\alpha}} \rho^{i}\,_{\alpha}+\frac{1}{8}{e}^{U} \epsilon^{a b}\,_{c d} (\sigma_{ae})^{\rho\alpha} (\sigma_{b})^{\beta}{}_{\dot{\alpha}} \psi^{c}\,_{i \beta} \psi^{e}{}^{j}_{\rho} \overline{\psi}^{d}{}_{j}^{\dot{\alpha}} \rho^{i}\,_{\alpha} - \frac{1}{8}{e}^{2U} \tilde{\sigma}^{c}\,_{a \dot{\rho} \dot{\alpha}} (\tilde{\sigma}_{cb})_{\dot{\lambda}\dot{\beta}} \overline{\psi}^{a j \dot{\rho}} \overline{\psi}^{b}{}_{j}^{\dot{\lambda}} \overline{\rho}^{i \dot{\alpha}} \overline{\rho}_{i}\,^{\dot{\beta}} - \frac{1}{8}{e}^{2U} \tilde{\sigma}^{c}\,_{a \dot{\rho} \dot{\beta}} (\tilde{\sigma}_{cb})_{\dot{\lambda}\dot{\alpha}} \overline{\psi}^{a i \dot{\rho}} \overline{\psi}^{b j \dot{\lambda}} \overline{\rho}_{i}\,^{\dot{\alpha}} \overline{\rho}_{j}\,^{\dot{\beta}} - \frac{1}{4}{e}^{U} \epsilon^{a b}\,_{c d} (\sigma_{a})^{\alpha}{}_{\dot{\beta}} (\tilde{\sigma}_{be})_{\dot{\rho}\dot{\alpha}} \psi^{c}{}^{j}_{\alpha} \overline{\psi}^{d i \dot{\beta}} \overline{\psi}^{e}{}_{j}^{\dot{\rho}} \overline{\rho}_{i}\,^{\dot{\alpha}} - \frac{1}{4}{e}^{U} \epsilon^{a b}\,_{c d} (\sigma_{a})^{\alpha}{}_{\dot{\beta}} (\tilde{\sigma}_{be})_{\dot{\rho}\dot{\alpha}} \psi^{c}{}^{i}_{\alpha} \overline{\psi}^{d j \dot{\beta}} \overline{\psi}^{e}{}_{j}^{\dot{\rho}} \overline{\rho}_{i}\,^{\dot{\alpha}} - \frac{1}{8}{e}^{U} \epsilon^{a b}\,_{c d} (\sigma_{a})^{\beta}{}_{\dot{\alpha}} (\sigma_{be})^{\rho\alpha} \psi^{c}\,_{j \beta} \psi^{e}{}^{j}_{\rho} \overline{\psi}^{d}{}_{i}^{\dot{\alpha}} \rho^{i}\,_{\alpha} - \frac{1}{8}{e}^{U} \epsilon^{a b}\,_{c d} (\sigma_{a})^{\beta}{}_{\dot{\alpha}} (\sigma_{be})^{\rho\alpha} \psi^{c}\,_{i \beta} \psi^{e}{}^{j}_{\rho} \overline{\psi}^{d}{}_{j}^{\dot{\alpha}} \rho^{i}\,_{\alpha} - \frac{1}{8}\psi_{a i}\,^{\alpha} \psi_{b}\,^{i}\,_{\alpha} \overline{\psi}^{a j}\,_{\dot{\alpha}} \overline{\psi}^{b}{}_{j}^{\dot{\alpha}}%
+\frac{1}{8}(\sigma_{a})^{\beta}{}_{\dot{\alpha}} (\sigma_{b})^{\alpha}{}_{\dot{\beta}} \psi^{a}\,_{i \alpha} \psi^{b}{}^{i}_{\beta} \overline{\psi}_{c}\,^{j \dot{\alpha}} \overline{\psi}^{c}{}_{j}^{\dot{\beta}} - \frac{1}{4}(\sigma_{b})^{\alpha}{}_{\dot{\beta}} (\sigma_{c})^{\beta}{}_{\dot{\alpha}} \psi_{a i \alpha} \psi^{b}{}^{i}_{\beta} \overline{\psi}^{a j \dot{\alpha}} \overline{\psi}^{c}{}_{j}^{\dot{\beta}}+\frac{1}{8}(\sigma_{b})^{\alpha}{}_{\dot{\beta}} (\sigma_{c})^{\beta}{}_{\dot{\alpha}} \psi_{a i \alpha} \psi^{a}{}^{i}_{\beta} \overline{\psi}^{b j \dot{\alpha}} \overline{\psi}^{c}{}_{j}^{\dot{\beta}} - \frac{1}{2}\psi_{a}\,^{i \alpha} \psi^{a}{}^{j}_{\alpha} \overline{\psi}_{b i \dot{\alpha}} \overline{\psi}^{b}{}_{j}^{\dot{\alpha}}+\frac{1}{4}\psi_{a}\,^{i \alpha} \psi_{b}\,^{j}\,_{\alpha} \overline{\psi}^{a}\,_{i \dot{\alpha}} \overline{\psi}^{b}{}_{j}^{\dot{\alpha}} - \frac{1}{8}(\sigma_{a})^{\beta}{}_{\dot{\alpha}} (\sigma_{b})^{\alpha}{}_{\dot{\beta}} \psi^{a}{}^{i}_{\alpha} \psi^{b}{}^{j}_{\beta} \overline{\psi}_{c i}\,^{\dot{\alpha}} \overline{\psi}^{c}{}_{j}^{\dot{\beta}}+\frac{1}{4}(\sigma_{b})^{\alpha}{}_{\dot{\beta}} (\sigma_{c})^{\beta}{}_{\dot{\alpha}} \psi_{a}\,^{i}\,_{\alpha} \psi^{b}{}^{j}_{\beta} \overline{\psi}^{a}{}_{i}^{\dot{\alpha}} \overline{\psi}^{c}{}_{j}^{\dot{\beta}} - \frac{1}{8}(\sigma_{b})^{\alpha}{}_{\dot{\beta}} (\sigma_{c})^{\beta}{}_{\dot{\alpha}} \psi_{a}\,^{i}\,_{\alpha} \psi^{a}{}^{j}_{\beta} \overline{\psi}^{b}{}_{i}^{\dot{\alpha}} \overline{\psi}^{c}{}_{j}^{\dot{\beta}}+\frac{1}{8}{e}^{2U} (\sigma_{ab})^{\beta\alpha} \psi^{a}{}^{j}_{\beta} \overline{\psi}^{b}\,_{i \dot{\alpha}} \rho^{i}\,_{\alpha} \overline{\rho}_{j}\,^{\dot{\alpha}} - \frac{1}{8}{e}^{2U} (\tilde{\sigma}_{ab})_{\dot{\rho}\dot{\beta}} \overline{\psi}^{a i \dot{\rho}} \overline{\psi}^{b j}\,_{\dot{\alpha}} \overline{\rho}_{i}\,^{\dot{\alpha}} \overline{\rho}_{j}\,^{\dot{\beta}}+\frac{5}{32}{e}^{U} \epsilon^{a}\,_{b c d} (\sigma_{a})^{\alpha}{}_{\dot{\beta}} \psi^{b}{}^{j}_{\alpha} \overline{\psi}^{c i \dot{\beta}} \overline{\psi}^{d}\,_{j \dot{\alpha}} \overline{\rho}_{i}\,^{\dot{\alpha}}+\frac{1}{4}{e}^{U} \epsilon^{a}\,_{b c d} (\sigma_{a})^{\alpha}{}_{\dot{\beta}} \psi^{b}{}^{i}_{\alpha} \overline{\psi}^{c j \dot{\beta}} \overline{\psi}^{d}\,_{j \dot{\alpha}} \overline{\rho}_{i}\,^{\dot{\alpha}} - \frac{3}{128}{\rm i} {e}^{3U} (\sigma_{a})^{\beta}{}_{\dot{\alpha}} \psi^{a}\,_{i}\,^{\alpha} \rho^{i}\,_{\alpha} \rho^{j}\,_{\beta} \overline{\rho}_{j}\,^{\dot{\alpha}}+\frac{5}{128}{\rm i} {e}^{3U} (\sigma_{a})^{\beta}{}_{\dot{\alpha}} \psi^{a}\,_{i \beta} \rho^{i \alpha} \rho^{j}\,_{\alpha} \overline{\rho}_{j}\,^{\dot{\alpha}}+\frac{1}{16}{\rm i} {e}^{3U} (\sigma_{a})^{\alpha}{}_{\dot{\alpha}} \psi^{a j \beta} \rho_{i \alpha} \rho^{i}\,_{\beta} \overline{\rho}_{j}\,^{\dot{\alpha}}+\frac{1}{16}{\rm i} {e}^{3U} (\sigma_{a})^{\alpha}{}_{\dot{\alpha}} \overline{\psi}^{a}\,_{i \dot{\beta}} \rho^{i}\,_{\alpha} \overline{\rho}^{j \dot{\alpha}} \overline{\rho}_{j}\,^{\dot{\beta}}+\frac{5}{128}{\rm i} {e}^{3U} (\sigma_{a})^{\alpha}{}_{\dot{\beta}} \overline{\psi}^{a j \dot{\beta}} \rho^{i}\,_{\alpha} \overline{\rho}_{j \dot{\alpha}} \overline{\rho}_{i}\,^{\dot{\alpha}} - \frac{3}{128}{\rm i} {e}^{3U} (\sigma_{a})^{\alpha}{}_{\dot{\beta}} \overline{\psi}^{a j}\,_{\dot{\alpha}} \rho^{i}\,_{\alpha} \overline{\rho}_{j}\,^{\dot{\alpha}} \overline{\rho}_{i}\,^{\dot{\beta}} - \frac{1}{16}{\rm i} {e}^{2U} \epsilon^{a b}\,_{c d} (\sigma_{a})^{\beta}{}_{\dot{\beta}} (\sigma_{b})^{\alpha}{}_{\dot{\alpha}} \psi^{c}\,_{i \beta} \overline{\psi}^{d j \dot{\beta}} \rho^{i}\,_{\alpha} \overline{\rho}_{j}\,^{\dot{\alpha}} - \frac{1}{16}{\rm i} {e}^{2U} \epsilon^{a b}\,_{c d} (\sigma_{a})^{\beta}{}_{\dot{\beta}} (\sigma_{b})^{\alpha}{}_{\dot{\alpha}} \psi^{c}{}^{j}_{\beta} \overline{\psi}^{d}{}_{i}^{\dot{\beta}} \rho^{i}\,_{\alpha} \overline{\rho}_{j}\,^{\dot{\alpha}}%
 - \frac{1}{8}{e}^{2U} (\sigma_{ab})^{\rho\beta} \psi^{a}\,_{i \rho} \psi^{b}\,_{j}\,^{\alpha} \rho^{i}\,_{\alpha} \rho^{j}\,_{\beta} - \frac{1}{8}{e}^{2U} (\tilde{\sigma}_{ab})_{\dot{\beta}\dot{\alpha}} \psi^{a j \alpha} \overline{\psi}^{b}{}_{i}^{\dot{\beta}} \rho^{i}\,_{\alpha} \overline{\rho}_{j}\,^{\dot{\alpha}} - \frac{1}{4}{e}^{U} \epsilon^{a}\,_{b c d} (\sigma_{a})^{\beta}{}_{\dot{\alpha}} \psi^{b}\,_{j \beta} \psi^{c j \alpha} \overline{\psi}^{d}{}_{i}^{\dot{\alpha}} \rho^{i}\,_{\alpha} - \frac{5}{32}{e}^{U} \epsilon^{a}\,_{b c d} (\sigma_{a})^{\beta}{}_{\dot{\alpha}} \psi^{b}\,_{i \beta} \psi^{c j \alpha} \overline{\psi}^{d}{}_{j}^{\dot{\alpha}} \rho^{i}\,_{\alpha}+\frac{1}{8}{\rm i} c {y}^{-1} {e}^{U} (\sigma_{a})^{\alpha}{}_{\dot{\alpha}} \psi^{a}\,_{j}\,^{\beta} \boldsymbol{\phi} \lambda^{j}\,_{\beta} \overline{\lambda}_{i}\,^{\dot{\alpha}} \rho^{i}\,_{\alpha} - \frac{1}{8}{\rm i} c {y}^{-1} {e}^{U} (\sigma_{a})^{\beta}{}_{\dot{\alpha}} \psi^{a}\,_{j \beta} \boldsymbol{\phi} \lambda^{j \alpha} \overline{\lambda}_{i}\,^{\dot{\alpha}} \rho^{i}\,_{\alpha} - \frac{1}{8}{\rm i} c {y}^{-1} {e}^{U} (\sigma_{a})^{\alpha}{}_{\dot{\alpha}} \psi^{a j \beta} \boldsymbol{\phi} \lambda_{i \beta} \overline{\lambda}_{j}\,^{\dot{\alpha}} \rho^{i}\,_{\alpha}+\frac{1}{8}{\rm i} c {y}^{-1} {e}^{U} (\sigma_{a})^{\beta}{}_{\dot{\alpha}} \psi^{a}{}^{j}_{\beta} \boldsymbol{\phi} \lambda_{i}\,^{\alpha} \overline{\lambda}_{j}\,^{\dot{\alpha}} \rho^{i}\,_{\alpha}+\frac{1}{8}{\rm i} c {y}^{-1} {e}^{U} (\sigma_{a})^{\alpha}{}_{\dot{\alpha}} \overline{\psi}^{a}\,_{j \dot{\beta}} \boldsymbol{\phi} \lambda^{j}\,_{\alpha} \overline{\lambda}^{i \dot{\beta}} \overline{\rho}_{i}\,^{\dot{\alpha}} - \frac{1}{8}{\rm i} c {y}^{-1} {e}^{U} (\sigma_{a})^{\alpha}{}_{\dot{\beta}} \overline{\psi}^{a}{}_{j}^{\dot{\beta}} \boldsymbol{\phi} \lambda^{j}\,_{\alpha} \overline{\lambda}^{i}\,_{\dot{\alpha}} \overline{\rho}_{i}\,^{\dot{\alpha}} - \frac{1}{8}{\rm i} c {y}^{-1} {e}^{U} (\sigma_{a})^{\alpha}{}_{\dot{\alpha}} \overline{\psi}^{a j}\,_{\dot{\beta}} \boldsymbol{\phi} \lambda^{i}\,_{\alpha} \overline{\lambda}_{j}\,^{\dot{\beta}} \overline{\rho}_{i}\,^{\dot{\alpha}}+\frac{1}{8}{\rm i} c {y}^{-1} {e}^{U} (\sigma_{a})^{\alpha}{}_{\dot{\beta}} \overline{\psi}^{a j \dot{\beta}} \boldsymbol{\phi} \lambda^{i}\,_{\alpha} \overline{\lambda}_{j \dot{\alpha}} \overline{\rho}_{i}\,^{\dot{\alpha}} - \frac{1}{4}{\rm i} c {y}^{-1} \epsilon^{a b}\,_{c d} (\sigma_{a})^{\beta}{}_{\dot{\beta}} (\sigma_{b})^{\alpha}{}_{\dot{\alpha}} \psi^{c}\,_{i \beta} \overline{\psi}^{d j \dot{\beta}} \boldsymbol{\phi} \lambda^{i}\,_{\alpha} \overline{\lambda}_{j}\,^{\dot{\alpha}} - \frac{1}{4}{\rm i} c {y}^{-1} \epsilon^{a b}\,_{c d} (\sigma_{a})^{\beta}{}_{\dot{\beta}} (\sigma_{b})^{\alpha}{}_{\dot{\alpha}} \psi^{c}{}^{j}_{\beta} \overline{\psi}^{d}{}_{i}^{\dot{\beta}} \boldsymbol{\phi} \lambda^{i}\,_{\alpha} \overline{\lambda}_{j}\,^{\dot{\alpha}}+\frac{1}{8}{\rm i} c {y}^{-1} {e}^{U} (\sigma_{a})^{\alpha}{}_{\dot{\alpha}} \psi^{a}\,_{j}\,^{\beta} \overline{\boldsymbol{\phi}} \lambda^{j}\,_{\beta} \overline{\lambda}_{i}\,^{\dot{\alpha}} \rho^{i}\,_{\alpha} - \frac{1}{8}{\rm i} c {y}^{-1} {e}^{U} (\sigma_{a})^{\beta}{}_{\dot{\alpha}} \psi^{a}\,_{j \beta} \overline{\boldsymbol{\phi}} \lambda^{j \alpha} \overline{\lambda}_{i}\,^{\dot{\alpha}} \rho^{i}\,_{\alpha} - \frac{1}{8}{\rm i} c {y}^{-1} {e}^{U} (\sigma_{a})^{\alpha}{}_{\dot{\alpha}} \psi^{a j \beta} \overline{\boldsymbol{\phi}} \lambda_{i \beta} \overline{\lambda}_{j}\,^{\dot{\alpha}} \rho^{i}\,_{\alpha}+\frac{1}{8}{\rm i} c {y}^{-1} {e}^{U} (\sigma_{a})^{\beta}{}_{\dot{\alpha}} \psi^{a}{}^{j}_{\beta} \overline{\boldsymbol{\phi}} \lambda_{i}\,^{\alpha} \overline{\lambda}_{j}\,^{\dot{\alpha}} \rho^{i}\,_{\alpha}+\frac{1}{8}{\rm i} c {y}^{-1} {e}^{U} (\sigma_{a})^{\alpha}{}_{\dot{\alpha}} \overline{\psi}^{a}\,_{j \dot{\beta}} \overline{\boldsymbol{\phi}} \lambda^{j}\,_{\alpha} \overline{\lambda}^{i \dot{\beta}} \overline{\rho}_{i}\,^{\dot{\alpha}} - \frac{1}{8}{\rm i} c {y}^{-1} {e}^{U} (\sigma_{a})^{\alpha}{}_{\dot{\beta}} \overline{\psi}^{a}{}_{j}^{\dot{\beta}} \overline{\boldsymbol{\phi}} \lambda^{j}\,_{\alpha} \overline{\lambda}^{i}\,_{\dot{\alpha}} \overline{\rho}_{i}\,^{\dot{\alpha}}%
 - \frac{1}{8}{\rm i} c {y}^{-1} {e}^{U} (\sigma_{a})^{\alpha}{}_{\dot{\alpha}} \overline{\psi}^{a j}\,_{\dot{\beta}} \overline{\boldsymbol{\phi}} \lambda^{i}\,_{\alpha} \overline{\lambda}_{j}\,^{\dot{\beta}} \overline{\rho}_{i}\,^{\dot{\alpha}}+\frac{1}{8}{\rm i} c {y}^{-1} {e}^{U} (\sigma_{a})^{\alpha}{}_{\dot{\beta}} \overline{\psi}^{a j \dot{\beta}} \overline{\boldsymbol{\phi}} \lambda^{i}\,_{\alpha} \overline{\lambda}_{j \dot{\alpha}} \overline{\rho}_{i}\,^{\dot{\alpha}} - \frac{1}{4}{\rm i} c {y}^{-1} \epsilon^{a b}\,_{c d} (\sigma_{a})^{\beta}{}_{\dot{\beta}} (\sigma_{b})^{\alpha}{}_{\dot{\alpha}} \psi^{c}\,_{i \beta} \overline{\psi}^{d j \dot{\beta}} \overline{\boldsymbol{\phi}} \lambda^{i}\,_{\alpha} \overline{\lambda}_{j}\,^{\dot{\alpha}} - \frac{1}{4}{\rm i} c {y}^{-1} \epsilon^{a b}\,_{c d} (\sigma_{a})^{\beta}{}_{\dot{\beta}} (\sigma_{b})^{\alpha}{}_{\dot{\alpha}} \psi^{c}{}^{j}_{\beta} \overline{\psi}^{d}{}_{i}^{\dot{\beta}} \overline{\boldsymbol{\phi}} \lambda^{i}\,_{\alpha} \overline{\lambda}_{j}\,^{\dot{\alpha}} - \frac{1}{8}c {y}^{-3} \boldsymbol{\phi} \lambda^{i \alpha} \lambda^{j}\,_{\alpha} \overline{\lambda}_{i \dot{\alpha}} \overline{\lambda}_{j}\,^{\dot{\alpha}} - \frac{1}{16}c {y}^{-1} {e}^{2U} \boldsymbol{\phi} \lambda^{j \alpha} \overline{\lambda}_{j \dot{\alpha}} \rho^{i}\,_{\alpha} \overline{\rho}_{i}\,^{\dot{\alpha}} - \frac{1}{8}c {y}^{-3} \overline{\boldsymbol{\phi}} \lambda^{i \alpha} \lambda^{j}\,_{\alpha} \overline{\lambda}_{i \dot{\alpha}} \overline{\lambda}_{j}\,^{\dot{\alpha}} - \frac{1}{16}c {y}^{-1} {e}^{2U} \overline{\boldsymbol{\phi}} \lambda^{j \alpha} \overline{\lambda}_{j \dot{\alpha}} \rho^{i}\,_{\alpha} \overline{\rho}_{i}\,^{\dot{\alpha}}+\frac{1}{8}c {y}^{-1} \psi_{a}\,^{i \alpha} \psi^{a}{}^{j}_{\alpha} \overline{\boldsymbol{\phi}} \overline{\lambda}_{i \dot{\alpha}} \overline{\lambda}_{j}\,^{\dot{\alpha}} - \frac{1}{16}c {y}^{-1} (\sigma_{a})^{\beta}{}_{\dot{\beta}} (\sigma_{b})^{\alpha}{}_{\dot{\alpha}} \psi^{a}{}^{i}_{\alpha} \psi^{b}{}^{j}_{\beta} \overline{\boldsymbol{\phi}} \overline{\lambda}_{i}\,^{\dot{\alpha}} \overline{\lambda}_{j}\,^{\dot{\beta}} - \frac{1}{2}c {y}^{-1} \psi_{a}\,^{j \alpha} \overline{\psi}^{a}\,_{i \dot{\alpha}} \overline{\boldsymbol{\phi}} \lambda^{i}\,_{\alpha} \overline{\lambda}_{j}\,^{\dot{\alpha}} - \frac{1}{4}c {y}^{-1} (\sigma_{a})^{\alpha}{}_{\dot{\beta}} (\sigma_{b})^{\beta}{}_{\dot{\alpha}} \psi^{a}{}^{j}_{\beta} \overline{\psi}^{b}{}_{i}^{\dot{\beta}} \overline{\boldsymbol{\phi}} \lambda^{i}\,_{\alpha} \overline{\lambda}_{j}\,^{\dot{\alpha}}+\frac{1}{8}{\rm i} {y}^{-1} (\sigma_{b})^{\alpha}{}_{\dot{\alpha}} \psi_{a}\,^{i}\,_{\alpha} \overline{\psi}^{a j}\,_{\dot{\beta}} \overline{\psi}^{b}{}_{j}^{\dot{\beta}} \overline{\lambda}_{i}\,^{\dot{\alpha}}+\frac{1}{4}c {y}^{-1} \overline{\psi}_{a i \dot{\alpha}} \overline{\psi}^{a}{}_{j}^{\dot{\alpha}} \overline{\boldsymbol{\phi}} \lambda^{i \alpha} \lambda^{j}\,_{\alpha} - \frac{1}{8}c {y}^{-1} (\sigma_{a})^{\beta}{}_{\dot{\beta}} (\sigma_{b})^{\alpha}{}_{\dot{\alpha}} \overline{\psi}^{a}{}_{i}^{\dot{\alpha}} \overline{\psi}^{b}{}_{j}^{\dot{\beta}} \overline{\boldsymbol{\phi}} \lambda^{i}\,_{\alpha} \lambda^{j}\,_{\beta}+\frac{1}{8}{\rm i} {y}^{-1} (\sigma_{b})^{\alpha}{}_{\dot{\beta}} \overline{\psi}_{a}\,^{j}\,_{\dot{\alpha}} \overline{\psi}^{a}{}_{i}^{\dot{\beta}} \overline{\psi}^{b}{}_{j}^{\dot{\alpha}} \lambda^{i}\,_{\alpha} - \frac{1}{4}c y \psi_{a i}\,^{\alpha} \psi^{a}\,_{j}\,^{\beta} \psi_{b}\,^{i}\,_{\alpha} \psi^{b}{}^{j}_{\beta} \overline{\boldsymbol{\phi}} - \frac{1}{4}c y \psi_{a i}\,^{\alpha} \psi_{b}\,^{i}\,_{\alpha} \overline{\psi}^{a j}\,_{\dot{\alpha}} \overline{\psi}^{b}{}_{j}^{\dot{\alpha}} \boldsymbol{\phi} - \frac{1}{8}c y \overline{\psi}_{a}\,^{i}\,_{\dot{\alpha}} \overline{\psi}^{a j}\,_{\dot{\beta}} \overline{\psi}_{b i}\,^{\dot{\alpha}} \overline{\psi}^{b}{}_{j}^{\dot{\beta}} \overline{\boldsymbol{\phi}} - \frac{1}{16}{\rm i} c {y}^{-1} \epsilon^{a b}\,_{c d} (\sigma_{a})^{\alpha}{}_{\dot{\alpha}} (\sigma_{b})^{\beta}{}_{\dot{\beta}} \psi^{c}{}^{i}_{\alpha} \psi^{d}{}^{j}_{\beta} \overline{\boldsymbol{\phi}} \overline{\lambda}_{i}\,^{\dot{\alpha}} \overline{\lambda}_{j}\,^{\dot{\beta}}%
+\frac{1}{16}{y}^{-1} \epsilon^{a}\,_{b c d} (\sigma_{a})^{\alpha}{}_{\dot{\alpha}} \psi^{b}{}^{i}_{\alpha} \overline{\psi}^{c j}\,_{\dot{\beta}} \overline{\psi}^{d}{}_{j}^{\dot{\beta}} \overline{\lambda}_{i}\,^{\dot{\alpha}}+\frac{3}{8}{\rm i} c {y}^{-1} \epsilon^{a b}\,_{c d} (\sigma_{a})^{\alpha}{}_{\dot{\alpha}} (\sigma_{b})^{\beta}{}_{\dot{\beta}} \overline{\psi}^{c}{}_{i}^{\dot{\alpha}} \overline{\psi}^{d}{}_{j}^{\dot{\beta}} \overline{\boldsymbol{\phi}} \lambda^{i}\,_{\alpha} \lambda^{j}\,_{\beta}+\frac{1}{16}{y}^{-1} \epsilon^{a}\,_{b c d} (\sigma_{a})^{\alpha}{}_{\dot{\beta}} \overline{\psi}^{b j}\,_{\dot{\alpha}} \overline{\psi}^{c}{}_{j}^{\dot{\alpha}} \overline{\psi}^{d}{}_{i}^{\dot{\beta}} \lambda^{i}\,_{\alpha} - \frac{1}{8}{\rm i} c y \epsilon_{a b c d} \psi^{a}\,_{i}\,^{\alpha} \psi^{b}{}^{i}_{\alpha} \psi^{c}\,_{j}\,^{\beta} \psi^{d}{}^{j}_{\beta} \overline{\boldsymbol{\phi}} - \frac{1}{16}{\rm i} \epsilon_{a b c d} \psi^{a}\,_{i}\,^{\alpha} \psi^{b}{}^{i}_{\alpha} \overline{\psi}^{c j}\,_{\dot{\alpha}} \overline{\psi}^{d}{}_{j}^{\dot{\alpha}} - \frac{1}{8}{\rm i} c y \epsilon_{a b c d} \psi^{a}\,_{i}\,^{\alpha} \psi^{b}{}^{i}_{\alpha} \overline{\psi}^{c j}\,_{\dot{\alpha}} \overline{\psi}^{d}{}_{j}^{\dot{\alpha}} \boldsymbol{\phi}+\frac{1}{16}{\rm i} c y \epsilon_{a b c d} \overline{\psi}^{a i}\,_{\dot{\alpha}} \overline{\psi}^{b}{}_{i}^{\dot{\alpha}} \overline{\psi}^{c j}\,_{\dot{\beta}} \overline{\psi}^{d}{}_{j}^{\dot{\beta}} \overline{\boldsymbol{\phi}} - \frac{1}{32}{\rm i} {y}^{-1} {e}^{2U} (\sigma_{a})^{\alpha}{}_{\dot{\alpha}} \psi^{a j \beta} \overline{\lambda}_{j}\,^{\dot{\alpha}} \rho_{i \alpha} \rho^{i}\,_{\beta} - \frac{1}{32}{e}^{2U} (\sigma_{ab})^{\alpha\beta} \psi^{a}\,_{j}\,^{\rho} \psi^{b}{}^{j}_{\rho} \rho_{i \alpha} \rho^{i}\,_{\beta} - \frac{1}{32}{e}^{2U} (\sigma_{ab})^{\alpha\beta} \overline{\psi}^{a j}\,_{\dot{\alpha}} \overline{\psi}^{b}{}_{j}^{\dot{\alpha}} \rho_{i \alpha} \rho^{i}\,_{\beta}+\frac{1}{8}c {y}^{-1} \psi_{a}\,^{i \alpha} \psi^{a}{}^{j}_{\alpha} {\boldsymbol{\phi}}^{2} {\overline{\boldsymbol{\phi}}}^{-1} \overline{\lambda}_{i \dot{\alpha}} \overline{\lambda}_{j}\,^{\dot{\alpha}} - \frac{1}{16}c {y}^{-1} (\sigma_{a})^{\beta}{}_{\dot{\beta}} (\sigma_{b})^{\alpha}{}_{\dot{\alpha}} \psi^{a}{}^{i}_{\alpha} \psi^{b}{}^{j}_{\beta} {\boldsymbol{\phi}}^{2} {\overline{\boldsymbol{\phi}}}^{-1} \overline{\lambda}_{i}\,^{\dot{\alpha}} \overline{\lambda}_{j}\,^{\dot{\beta}} - \frac{1}{2}c {y}^{-1} \psi_{a}\,^{j \alpha} \overline{\psi}^{a}\,_{i \dot{\alpha}} \boldsymbol{\phi} \lambda^{i}\,_{\alpha} \overline{\lambda}_{j}\,^{\dot{\alpha}} - \frac{1}{4}c {y}^{-1} (\sigma_{a})^{\alpha}{}_{\dot{\beta}} (\sigma_{b})^{\beta}{}_{\dot{\alpha}} \psi^{a}{}^{j}_{\beta} \overline{\psi}^{b}{}_{i}^{\dot{\beta}} \boldsymbol{\phi} \lambda^{i}\,_{\alpha} \overline{\lambda}_{j}\,^{\dot{\alpha}}+\frac{1}{4}{\rm i} c (\sigma_{b})^{\alpha}{}_{\dot{\alpha}} \psi_{a}\,^{i}\,_{\alpha} \overline{\psi}^{a j}\,_{\dot{\beta}} \overline{\psi}^{b}{}_{j}^{\dot{\beta}} {\boldsymbol{\phi}}^{2} {\overline{\boldsymbol{\phi}}}^{-1} \overline{\lambda}_{i}\,^{\dot{\alpha}}+\frac{1}{4}{\rm i} c (\sigma_{b})^{\alpha}{}_{\dot{\beta}} \overline{\psi}_{a}\,^{j}\,_{\dot{\alpha}} \overline{\psi}^{a}{}_{i}^{\dot{\beta}} \overline{\psi}^{b}{}_{j}^{\dot{\alpha}} \boldsymbol{\phi} \lambda^{i}\,_{\alpha} - \frac{1}{8}c y \overline{\psi}_{a}\,^{i}\,_{\dot{\alpha}} \overline{\psi}^{a j}\,_{\dot{\beta}} \overline{\psi}_{b i}\,^{\dot{\alpha}} \overline{\psi}^{b}{}_{j}^{\dot{\beta}} {\boldsymbol{\phi}}^{2} {\overline{\boldsymbol{\phi}}}^{-1} - \frac{1}{16}{\rm i} c {y}^{-1} \epsilon^{a b}\,_{c d} (\sigma_{a})^{\alpha}{}_{\dot{\alpha}} (\sigma_{b})^{\beta}{}_{\dot{\beta}} \psi^{c}{}^{i}_{\alpha} \psi^{d}{}^{j}_{\beta} {\boldsymbol{\phi}}^{2} {\overline{\boldsymbol{\phi}}}^{-1} \overline{\lambda}_{i}\,^{\dot{\alpha}} \overline{\lambda}_{j}\,^{\dot{\beta}}+\frac{1}{8}c \epsilon^{a}\,_{b c d} (\sigma_{a})^{\alpha}{}_{\dot{\alpha}} \psi^{b}{}^{i}_{\alpha} \overline{\psi}^{c j}\,_{\dot{\beta}} \overline{\psi}^{d}{}_{j}^{\dot{\beta}} {\boldsymbol{\phi}}^{2} {\overline{\boldsymbol{\phi}}}^{-1} \overline{\lambda}_{i}\,^{\dot{\alpha}}+\frac{1}{8}c \epsilon^{a}\,_{b c d} (\sigma_{a})^{\alpha}{}_{\dot{\beta}} \overline{\psi}^{b j}\,_{\dot{\alpha}} \overline{\psi}^{c}{}_{j}^{\dot{\alpha}} \overline{\psi}^{d}{}_{i}^{\dot{\beta}} \boldsymbol{\phi} \lambda^{i}\,_{\alpha}%
+\frac{1}{16}{\rm i} c y \epsilon_{a b c d} \overline{\psi}^{a i}\,_{\dot{\alpha}} \overline{\psi}^{b}{}_{i}^{\dot{\alpha}} \overline{\psi}^{c j}\,_{\dot{\beta}} \overline{\psi}^{d}{}_{j}^{\dot{\beta}} {\boldsymbol{\phi}}^{2} {\overline{\boldsymbol{\phi}}}^{-1}+\frac{1}{32}{\rm i} {y}^{-1} {e}^{2U} (\sigma_{a})^{\alpha}{}_{\dot{\alpha}} \psi^{a j \beta} \boldsymbol{\phi} {\overline{\boldsymbol{\phi}}}^{-1} \overline{\lambda}_{j}\,^{\dot{\alpha}} \rho_{i \alpha} \rho^{i}\,_{\beta} - \frac{1}{32}{e}^{2U} (\sigma_{ab})^{\alpha\beta} \overline{\psi}^{a j}\,_{\dot{\alpha}} \overline{\psi}^{b}{}_{j}^{\dot{\alpha}} \boldsymbol{\phi} {\overline{\boldsymbol{\phi}}}^{-1} \rho_{i \alpha} \rho^{i}\,_{\beta}+\frac{1}{4}c {y}^{-1} \psi_{a}\,^{i \alpha} \psi^{a}{}^{j}_{\alpha} \boldsymbol{\phi} \overline{\lambda}_{i \dot{\alpha}} \overline{\lambda}_{j}\,^{\dot{\alpha}} - \frac{1}{8}c {y}^{-1} (\sigma_{a})^{\beta}{}_{\dot{\beta}} (\sigma_{b})^{\alpha}{}_{\dot{\alpha}} \psi^{a}{}^{i}_{\alpha} \psi^{b}{}^{j}_{\beta} \boldsymbol{\phi} \overline{\lambda}_{i}\,^{\dot{\alpha}} \overline{\lambda}_{j}\,^{\dot{\beta}}+\frac{1}{8}c {y}^{-1} \overline{\psi}_{a i \dot{\alpha}} \overline{\psi}^{a}{}_{j}^{\dot{\alpha}} \boldsymbol{\phi} \lambda^{i \alpha} \lambda^{j}\,_{\alpha} - \frac{1}{16}c {y}^{-1} (\sigma_{a})^{\beta}{}_{\dot{\beta}} (\sigma_{b})^{\alpha}{}_{\dot{\alpha}} \overline{\psi}^{a}{}_{i}^{\dot{\alpha}} \overline{\psi}^{b}{}_{j}^{\dot{\beta}} \boldsymbol{\phi} \lambda^{i}\,_{\alpha} \lambda^{j}\,_{\beta} - \frac{1}{4}{\rm i} c (\sigma_{b})^{\alpha}{}_{\dot{\alpha}} \psi_{a j}\,^{\beta} \psi^{b}{}^{j}_{\beta} \overline{\psi}^{a}{}_{i}^{\dot{\alpha}} \boldsymbol{\phi} \lambda^{i}\,_{\alpha} - \frac{1}{8}c y \psi_{a i}\,^{\alpha} \psi^{a}\,_{j}\,^{\beta} \psi_{b}\,^{i}\,_{\alpha} \psi^{b}{}^{j}_{\beta} \boldsymbol{\phi} - \frac{1}{4}c y \overline{\psi}_{a}\,^{i}\,_{\dot{\alpha}} \overline{\psi}^{a j}\,_{\dot{\beta}} \overline{\psi}_{b i}\,^{\dot{\alpha}} \overline{\psi}^{b}{}_{j}^{\dot{\beta}} \boldsymbol{\phi} - \frac{3}{8}{\rm i} c {y}^{-1} \epsilon^{a b}\,_{c d} (\sigma_{a})^{\alpha}{}_{\dot{\alpha}} (\sigma_{b})^{\beta}{}_{\dot{\beta}} \psi^{c}{}^{i}_{\alpha} \psi^{d}{}^{j}_{\beta} \boldsymbol{\phi} \overline{\lambda}_{i}\,^{\dot{\alpha}} \overline{\lambda}_{j}\,^{\dot{\beta}}+\frac{1}{16}{\rm i} c {y}^{-1} \epsilon^{a b}\,_{c d} (\sigma_{a})^{\alpha}{}_{\dot{\alpha}} (\sigma_{b})^{\beta}{}_{\dot{\beta}} \overline{\psi}^{c}{}_{i}^{\dot{\alpha}} \overline{\psi}^{d}{}_{j}^{\dot{\beta}} \boldsymbol{\phi} \lambda^{i}\,_{\alpha} \lambda^{j}\,_{\beta}+\frac{1}{8}c \epsilon^{a}\,_{b c d} (\sigma_{a})^{\alpha}{}_{\dot{\alpha}} \psi^{b}\,_{j}\,^{\beta} \psi^{c}{}^{j}_{\beta} \overline{\psi}^{d}{}_{i}^{\dot{\alpha}} \boldsymbol{\phi} \lambda^{i}\,_{\alpha} - \frac{1}{16}{\rm i} c y \epsilon_{a b c d} \psi^{a}\,_{i}\,^{\alpha} \psi^{b}{}^{i}_{\alpha} \psi^{c}\,_{j}\,^{\beta} \psi^{d}{}^{j}_{\beta} \boldsymbol{\phi}+\frac{1}{8}{\rm i} c y \epsilon_{a b c d} \overline{\psi}^{a i}\,_{\dot{\alpha}} \overline{\psi}^{b}{}_{i}^{\dot{\alpha}} \overline{\psi}^{c j}\,_{\dot{\beta}} \overline{\psi}^{d}{}_{j}^{\dot{\beta}} \boldsymbol{\phi} - \frac{1}{32}{\rm i} {y}^{-1} {e}^{2U} (\sigma_{a})^{\alpha}{}_{\dot{\alpha}} \overline{\psi}^{a}\,_{j \dot{\beta}} \lambda^{j}\,_{\alpha} \overline{\rho}^{i \dot{\alpha}} \overline{\rho}_{i}\,^{\dot{\beta}} - \frac{1}{32}{e}^{2U} (\tilde{\sigma}_{ab})_{\dot{\alpha}\dot{\beta}} \psi^{a}\,_{j}\,^{\alpha} \psi^{b}{}^{j}_{\alpha} \overline{\rho}^{i \dot{\alpha}} \overline{\rho}_{i}\,^{\dot{\beta}} - \frac{1}{32}{e}^{2U} (\tilde{\sigma}_{ab})_{\dot{\alpha}\dot{\beta}} \overline{\psi}^{a j}\,_{\dot{\rho}} \overline{\psi}^{b}{}_{j}^{\dot{\rho}} \overline{\rho}^{i \dot{\alpha}} \overline{\rho}_{i}\,^{\dot{\beta}}+\frac{1}{4}{\rm i} c (\sigma_{b})^{\alpha}{}_{\dot{\alpha}} \psi_{a}\,^{i}\,_{\alpha} \overline{\psi}^{a j}\,_{\dot{\beta}} \overline{\psi}^{b}{}_{j}^{\dot{\beta}} \boldsymbol{\phi} \overline{\lambda}_{i}\,^{\dot{\alpha}}+\frac{1}{8}c {y}^{-1} \overline{\psi}_{a i \dot{\alpha}} \overline{\psi}^{a}{}_{j}^{\dot{\alpha}} {\boldsymbol{\phi}}^{-1} {\overline{\boldsymbol{\phi}}}^{2} \lambda^{i \alpha} \lambda^{j}\,_{\alpha}%
 - \frac{1}{16}c {y}^{-1} (\sigma_{a})^{\beta}{}_{\dot{\beta}} (\sigma_{b})^{\alpha}{}_{\dot{\alpha}} \overline{\psi}^{a}{}_{i}^{\dot{\alpha}} \overline{\psi}^{b}{}_{j}^{\dot{\beta}} {\boldsymbol{\phi}}^{-1} {\overline{\boldsymbol{\phi}}}^{2} \lambda^{i}\,_{\alpha} \lambda^{j}\,_{\beta}+\frac{1}{4}{\rm i} c (\sigma_{b})^{\alpha}{}_{\dot{\alpha}} \psi_{a j}\,^{\beta} \psi^{b}{}^{j}_{\beta} \overline{\psi}^{a}{}_{i}^{\dot{\alpha}} {\boldsymbol{\phi}}^{-1} {\overline{\boldsymbol{\phi}}}^{2} \lambda^{i}\,_{\alpha}+\frac{1}{4}{\rm i} c (\sigma_{b})^{\alpha}{}_{\dot{\beta}} \overline{\psi}_{a}\,^{j}\,_{\dot{\alpha}} \overline{\psi}^{a}{}_{i}^{\dot{\beta}} \overline{\psi}^{b}{}_{j}^{\dot{\alpha}} \overline{\boldsymbol{\phi}} \lambda^{i}\,_{\alpha} - \frac{1}{8}c y \psi_{a i}\,^{\alpha} \psi^{a}\,_{j}\,^{\beta} \psi_{b}\,^{i}\,_{\alpha} \psi^{b}{}^{j}_{\beta} {\boldsymbol{\phi}}^{-1} {\overline{\boldsymbol{\phi}}}^{2} - \frac{1}{4}c y \psi_{a i}\,^{\alpha} \psi_{b}\,^{i}\,_{\alpha} \overline{\psi}^{a j}\,_{\dot{\alpha}} \overline{\psi}^{b}{}_{j}^{\dot{\alpha}} \overline{\boldsymbol{\phi}}+\frac{1}{8}c \epsilon^{a}\,_{b c d} (\sigma_{a})^{\alpha}{}_{\dot{\alpha}} \psi^{b}{}^{i}_{\alpha} \overline{\psi}^{c j}\,_{\dot{\beta}} \overline{\psi}^{d}{}_{j}^{\dot{\beta}} \boldsymbol{\phi} \overline{\lambda}_{i}\,^{\dot{\alpha}}+\frac{1}{16}{\rm i} c {y}^{-1} \epsilon^{a b}\,_{c d} (\sigma_{a})^{\alpha}{}_{\dot{\alpha}} (\sigma_{b})^{\beta}{}_{\dot{\beta}} \overline{\psi}^{c}{}_{i}^{\dot{\alpha}} \overline{\psi}^{d}{}_{j}^{\dot{\beta}} {\boldsymbol{\phi}}^{-1} {\overline{\boldsymbol{\phi}}}^{2} \lambda^{i}\,_{\alpha} \lambda^{j}\,_{\beta} - \frac{1}{8}c \epsilon^{a}\,_{b c d} (\sigma_{a})^{\alpha}{}_{\dot{\alpha}} \psi^{b}\,_{j}\,^{\beta} \psi^{c}{}^{j}_{\beta} \overline{\psi}^{d}{}_{i}^{\dot{\alpha}} {\boldsymbol{\phi}}^{-1} {\overline{\boldsymbol{\phi}}}^{2} \lambda^{i}\,_{\alpha}+\frac{1}{8}c \epsilon^{a}\,_{b c d} (\sigma_{a})^{\alpha}{}_{\dot{\beta}} \overline{\psi}^{b j}\,_{\dot{\alpha}} \overline{\psi}^{c}{}_{j}^{\dot{\alpha}} \overline{\psi}^{d}{}_{i}^{\dot{\beta}} \overline{\boldsymbol{\phi}} \lambda^{i}\,_{\alpha} - \frac{1}{16}{\rm i} c y \epsilon_{a b c d} \psi^{a}\,_{i}\,^{\alpha} \psi^{b}{}^{i}_{\alpha} \psi^{c}\,_{j}\,^{\beta} \psi^{d}{}^{j}_{\beta} {\boldsymbol{\phi}}^{-1} {\overline{\boldsymbol{\phi}}}^{2} - \frac{1}{8}{\rm i} c y \epsilon_{a b c d} \psi^{a}\,_{i}\,^{\alpha} \psi^{b}{}^{i}_{\alpha} \overline{\psi}^{c j}\,_{\dot{\alpha}} \overline{\psi}^{d}{}_{j}^{\dot{\alpha}} \overline{\boldsymbol{\phi}}+\frac{1}{32}{\rm i} {y}^{-1} {e}^{2U} (\sigma_{a})^{\alpha}{}_{\dot{\alpha}} \overline{\psi}^{a}\,_{j \dot{\beta}} {\boldsymbol{\phi}}^{-1} \overline{\boldsymbol{\phi}} \lambda^{j}\,_{\alpha} \overline{\rho}^{i \dot{\alpha}} \overline{\rho}_{i}\,^{\dot{\beta}} - \frac{1}{32}{e}^{2U} (\tilde{\sigma}_{ab})_{\dot{\alpha}\dot{\beta}} \psi^{a}\,_{j}\,^{\alpha} \psi^{b}{}^{j}_{\alpha} {\boldsymbol{\phi}}^{-1} \overline{\boldsymbol{\phi}} \overline{\rho}^{i \dot{\alpha}} \overline{\rho}_{i}\,^{\dot{\beta}} - \frac{1}{32}{y}^{-4} \lambda^{i \alpha} \lambda^{j}\,_{\alpha} \overline{\lambda}_{i \dot{\alpha}} \overline{\lambda}_{j}\,^{\dot{\alpha}} - \frac{1}{128}{e}^{4U} \rho^{i \alpha} \rho^{j}\,_{\alpha} \overline{\rho}_{i \dot{\alpha}} \overline{\rho}_{j}\,^{\dot{\alpha}} - \frac{3}{512}{c}^{-1} {y}^{-1} {e}^{4U} {\overline{\boldsymbol{\phi}}}^{-1} \rho_{i}\,^{\alpha} \rho^{i \beta} \rho_{j \alpha} \rho^{j}\,_{\beta} - \frac{3}{512}{c}^{-1} {y}^{-1} {e}^{4U} {\boldsymbol{\phi}}^{-1} \overline{\rho}^{i}\,_{\dot{\alpha}} \overline{\rho}_{i \dot{\beta}} \overline{\rho}^{j \dot{\alpha}} \overline{\rho}_{j}\,^{\dot{\beta}}+\frac{1}{8}{\rm i} c {y}^{-1} \epsilon^{a b}\,_{c d} (\sigma_{a})^{\beta}{}_{\dot{\beta}} (\sigma_{b})^{\alpha}{}_{\dot{\alpha}} \psi^{c}{}^{j}_{\beta} \overline{\psi}^{d}{}_{j}^{\dot{\beta}} \boldsymbol{\phi} \lambda^{i}\,_{\alpha} \overline{\lambda}_{i}\,^{\dot{\alpha}}+\frac{1}{8}{\rm i} c {y}^{-1} \epsilon^{a b}\,_{c d} (\sigma_{a})^{\beta}{}_{\dot{\beta}} (\sigma_{b})^{\alpha}{}_{\dot{\alpha}} \psi^{c}{}^{j}_{\beta} \overline{\psi}^{d}{}_{j}^{\dot{\beta}} \overline{\boldsymbol{\phi}} \lambda^{i}\,_{\alpha} \overline{\lambda}_{i}\,^{\dot{\alpha}} - \frac{1}{2}c {y}^{-1} (\sigma_{ab})^{\alpha\beta} \psi^{a}{}^{i}_{\alpha} \psi^{b}{}^{j}_{\beta} \boldsymbol{\phi} \overline{\lambda}_{i \dot{\alpha}} \overline{\lambda}_{j}\,^{\dot{\alpha}}%
 - \frac{1}{2}c {y}^{-1} (\tilde{\sigma}_{ab})_{\dot{\alpha}\dot{\beta}} \overline{\psi}^{a}{}_{i}^{\dot{\alpha}} \overline{\psi}^{b}{}_{j}^{\dot{\beta}} \overline{\boldsymbol{\phi}} \lambda^{i \alpha} \lambda^{j}\,_{\alpha} - \frac{1}{8}{y}^{-2} (\sigma_{ab})^{\beta\alpha} \psi^{a}{}^{j}_{\beta} \overline{\psi}^{b}\,_{j \dot{\alpha}} \lambda^{i}\,_{\alpha} \overline{\lambda}_{i}\,^{\dot{\alpha}}+\frac{1}{16}{e}^{2U} (\sigma_{ab})^{\beta\alpha} \psi^{a}{}^{j}_{\beta} \overline{\psi}^{b}\,_{j \dot{\alpha}} \rho^{i}\,_{\alpha} \overline{\rho}_{i}\,^{\dot{\alpha}}+\frac{1}{8}{y}^{-2} (\tilde{\sigma}_{ab})_{\dot{\beta}\dot{\alpha}} \psi^{a j \alpha} \overline{\psi}^{b}{}_{j}^{\dot{\beta}} \lambda^{i}\,_{\alpha} \overline{\lambda}_{i}\,^{\dot{\alpha}} - \frac{1}{16}{e}^{2U} (\tilde{\sigma}_{ab})_{\dot{\beta}\dot{\alpha}} \psi^{a j \alpha} \overline{\psi}^{b}{}_{j}^{\dot{\beta}} \rho^{i}\,_{\alpha} \overline{\rho}_{i}\,^{\dot{\alpha}} - \frac{1}{2}c {y}^{-1} \psi_{a i}\,^{\alpha} \overline{\psi}^{a j}\,_{\dot{\alpha}} \boldsymbol{\phi} \lambda^{i}\,_{\alpha} \overline{\lambda}_{j}\,^{\dot{\alpha}} - \frac{1}{2}c {y}^{-1} \psi_{a i}\,^{\alpha} \overline{\psi}^{a j}\,_{\dot{\alpha}} \overline{\boldsymbol{\phi}} \lambda^{i}\,_{\alpha} \overline{\lambda}_{j}\,^{\dot{\alpha}} - \frac{1}{2}c {y}^{-1} (\sigma_{ab})^{\beta\alpha} \psi^{a}\,_{i \beta} \overline{\psi}^{b j}\,_{\dot{\alpha}} \boldsymbol{\phi} \lambda^{i}\,_{\alpha} \overline{\lambda}_{j}\,^{\dot{\alpha}} - \frac{1}{4}c {y}^{-1} (\tilde{\sigma}_{ab})_{\dot{\alpha}\dot{\beta}} \psi^{a}\,_{j}\,^{\alpha} \psi^{b}{}^{j}_{\alpha} \boldsymbol{\phi} \overline{\lambda}^{i \dot{\alpha}} \overline{\lambda}_{i}\,^{\dot{\beta}} - \frac{5}{128}{c}^{-1} {y}^{-1} {e}^{2U} (\tilde{\sigma}_{ab})_{\dot{\alpha}\dot{\beta}} \psi^{a}\,_{j}\,^{\alpha} \psi^{b}{}^{j}_{\alpha} {\boldsymbol{\phi}}^{-1} \overline{\rho}^{i \dot{\alpha}} \overline{\rho}_{i}\,^{\dot{\beta}}+\frac{1}{2}c {y}^{-1} (\tilde{\sigma}_{ab})_{\dot{\beta}\dot{\alpha}} \psi^{a}\,_{i}\,^{\alpha} \overline{\psi}^{b j \dot{\beta}} \boldsymbol{\phi} \lambda^{i}\,_{\alpha} \overline{\lambda}_{j}\,^{\dot{\alpha}} - \frac{3}{16}{e}^{2U} (\sigma_{ab})^{\rho\alpha} \psi^{a}\,_{j \rho} \psi^{b j \beta} \rho_{i \alpha} \rho^{i}\,_{\beta} - \frac{1}{2}c {y}^{-1} (\sigma_{ab})^{\beta\alpha} \psi^{a}\,_{i \beta} \overline{\psi}^{b j}\,_{\dot{\alpha}} \overline{\boldsymbol{\phi}} \lambda^{i}\,_{\alpha} \overline{\lambda}_{j}\,^{\dot{\alpha}} - \frac{1}{4}c {y}^{-1} (\sigma_{ab})^{\alpha\beta} \overline{\psi}^{a j}\,_{\dot{\alpha}} \overline{\psi}^{b}{}_{j}^{\dot{\alpha}} \overline{\boldsymbol{\phi}} \lambda_{i \alpha} \lambda^{i}\,_{\beta} - \frac{1}{32}{c}^{-1} {y}^{-1} {e}^{2U} (\sigma_{ab})^{\alpha\beta} \overline{\psi}^{a j}\,_{\dot{\alpha}} \overline{\psi}^{b}{}_{j}^{\dot{\alpha}} {\overline{\boldsymbol{\phi}}}^{-1} \rho_{i \alpha} \rho^{i}\,_{\beta}+\frac{1}{2}c {y}^{-1} (\tilde{\sigma}_{ab})_{\dot{\beta}\dot{\alpha}} \psi^{a}\,_{i}\,^{\alpha} \overline{\psi}^{b j \dot{\beta}} \overline{\boldsymbol{\phi}} \lambda^{i}\,_{\alpha} \overline{\lambda}_{j}\,^{\dot{\alpha}} - \frac{3}{16}{e}^{2U} (\tilde{\sigma}_{ab})_{\dot{\rho}\dot{\alpha}} \overline{\psi}^{a j \dot{\rho}} \overline{\psi}^{b}\,_{j \dot{\beta}} \overline{\rho}^{i \dot{\alpha}} \overline{\rho}_{i}\,^{\dot{\beta}} - \frac{1}{16}{e}^{2U} \psi_{a}\,^{j \alpha} \overline{\psi}^{a}\,_{j \dot{\alpha}} \rho^{i}\,_{\alpha} \overline{\rho}_{i}\,^{\dot{\alpha}}+\frac{1}{8}{\rm i} {e}^{U} (\sigma_{a})^{\alpha}{}_{\dot{\beta}} \psi^{a}{}^{j}_{\alpha} \overline{\psi}_{b}\,^{i}\,_{\dot{\alpha}} \overline{\psi}^{b}{}_{j}^{\dot{\beta}} \overline{\rho}_{i}\,^{\dot{\alpha}} - \frac{1}{8}{\rm i} {e}^{U} (\sigma_{b})^{\alpha}{}_{\dot{\beta}} \psi_{a}\,^{j}\,_{\alpha} \overline{\psi}^{a i}\,_{\dot{\alpha}} \overline{\psi}^{b}{}_{j}^{\dot{\beta}} \overline{\rho}_{i}\,^{\dot{\alpha}}%
+\frac{1}{8}c {y}^{-1} {e}^{2U} \boldsymbol{\phi} \lambda_{i}\,^{\alpha} \overline{\lambda}^{j}\,_{\dot{\alpha}} \rho^{i}\,_{\alpha} \overline{\rho}_{j}\,^{\dot{\alpha}}+\frac{1}{8}c {y}^{-1} {e}^{2U} \boldsymbol{\phi} \lambda^{j \alpha} \overline{\lambda}_{i \dot{\alpha}} \rho^{i}\,_{\alpha} \overline{\rho}_{j}\,^{\dot{\alpha}}+\frac{1}{8}c {y}^{-1} {e}^{2U} \overline{\boldsymbol{\phi}} \lambda_{i}\,^{\alpha} \overline{\lambda}^{j}\,_{\dot{\alpha}} \rho^{i}\,_{\alpha} \overline{\rho}_{j}\,^{\dot{\alpha}}+\frac{1}{8}c {y}^{-1} {e}^{2U} \overline{\boldsymbol{\phi}} \lambda^{j \alpha} \overline{\lambda}_{i \dot{\alpha}} \rho^{i}\,_{\alpha} \overline{\rho}_{j}\,^{\dot{\alpha}}+\frac{1}{16}{\rm i} {e}^{U} (\sigma_{b})^{\alpha}{}_{\dot{\alpha}} (\tilde{\sigma}_{ac})_{\dot{\beta}\dot{\rho}} \psi^{a}{}^{j}_{\alpha} \overline{\psi}^{b}{}_{j}^{\dot{\beta}} \overline{\psi}^{c i \dot{\rho}} \overline{\rho}_{i}\,^{\dot{\alpha}} - \frac{1}{16}{\rm i} {e}^{U} (\sigma_{a})^{\alpha}{}_{\dot{\alpha}} (\tilde{\sigma}_{bc})_{\dot{\beta}\dot{\rho}} \psi^{a}{}^{j}_{\alpha} \overline{\psi}^{b i \dot{\beta}} \overline{\psi}^{c}{}_{j}^{\dot{\rho}} \overline{\rho}_{i}\,^{\dot{\alpha}}+\frac{1}{16}{\rm i} {e}^{U} (\sigma_{b})^{\alpha}{}_{\dot{\beta}} (\tilde{\sigma}_{ac})_{\dot{\rho}\dot{\alpha}} \psi^{a}{}^{j}_{\alpha} \overline{\psi}^{b i \dot{\beta}} \overline{\psi}^{c}{}_{j}^{\dot{\rho}} \overline{\rho}_{i}\,^{\dot{\alpha}}+\frac{3}{32}{e}^{U} \epsilon^{a}\,_{b c d} (\sigma_{a})^{\alpha}{}_{\dot{\alpha}} \psi^{b}{}^{j}_{\alpha} \overline{\psi}^{c i}\,_{\dot{\beta}} \overline{\psi}^{d}{}_{j}^{\dot{\beta}} \overline{\rho}_{i}\,^{\dot{\alpha}}+\frac{1}{8}{\rm i} c {y}^{-1} {e}^{U} (\sigma_{a})^{\alpha}{}_{\dot{\beta}} \overline{\psi}^{a}\,_{j \dot{\alpha}} \boldsymbol{\phi} \lambda^{j}\,_{\alpha} \overline{\lambda}^{i \dot{\beta}} \overline{\rho}_{i}\,^{\dot{\alpha}}+\frac{1}{8}{\rm i} c {y}^{-1} {e}^{U} (\sigma_{a})^{\alpha}{}_{\dot{\beta}} \overline{\psi}^{a}\,_{j \dot{\alpha}} \overline{\boldsymbol{\phi}} \lambda^{j}\,_{\alpha} \overline{\lambda}^{i \dot{\beta}} \overline{\rho}_{i}\,^{\dot{\alpha}} - \frac{1}{64}{c}^{-1} {y}^{-1} {e}^{2U} (\sigma_{a})^{\alpha}{}_{\dot{\alpha}} (\sigma_{b})^{\beta}{}_{\dot{\beta}} \psi^{a}\,_{j \alpha} \psi^{b}{}^{j}_{\beta} {\boldsymbol{\phi}}^{-1} \overline{\rho}^{i \dot{\alpha}} \overline{\rho}_{i}\,^{\dot{\beta}}+\frac{1}{64}{c}^{-1} {y}^{-1} {e}^{2U} (\sigma_{a})^{\beta}{}_{\dot{\alpha}} (\sigma_{b})^{\alpha}{}_{\dot{\beta}} \psi^{a}\,_{j \alpha} \psi^{b}{}^{j}_{\beta} {\boldsymbol{\phi}}^{-1} \overline{\rho}^{i \dot{\alpha}} \overline{\rho}_{i}\,^{\dot{\beta}} - \frac{1}{64}{c}^{-1} {y}^{-1} {e}^{2U} (\sigma_{a})^{\alpha}{}_{\dot{\alpha}} (\sigma_{b})^{\beta}{}_{\dot{\beta}} \overline{\psi}^{a j \dot{\alpha}} \overline{\psi}^{b}{}_{j}^{\dot{\beta}} {\overline{\boldsymbol{\phi}}}^{-1} \rho_{i \alpha} \rho^{i}\,_{\beta}+\frac{1}{64}{c}^{-1} {y}^{-1} {e}^{2U} (\sigma_{a})^{\alpha}{}_{\dot{\beta}} (\sigma_{b})^{\beta}{}_{\dot{\alpha}} \overline{\psi}^{a j \dot{\alpha}} \overline{\psi}^{b}{}_{j}^{\dot{\beta}} {\overline{\boldsymbol{\phi}}}^{-1} \rho_{i \alpha} \rho^{i}\,_{\beta}+\frac{1}{32}{e}^{2U} \psi_{a j}\,^{\alpha} \psi^{a j \beta} \rho_{i \alpha} \rho^{i}\,_{\beta}+\frac{1}{32}{e}^{2U} \overline{\psi}_{a}\,^{j}\,_{\dot{\alpha}} \overline{\psi}^{a}\,_{j \dot{\beta}} \overline{\rho}^{i \dot{\alpha}} \overline{\rho}_{i}\,^{\dot{\beta}} - \frac{1}{32}{e}^{2U} \overline{\psi}_{a}\,^{i}\,_{\dot{\beta}} \overline{\psi}^{a j}\,_{\dot{\alpha}} \overline{\rho}_{i}\,^{\dot{\alpha}} \overline{\rho}_{j}\,^{\dot{\beta}}+\frac{1}{64}{\rm i} {y}^{-2} {e}^{U} (\sigma_{a})^{\alpha}{}_{\dot{\beta}} \overline{\psi}^{a i}\,_{\dot{\alpha}} \lambda^{j}\,_{\alpha} \overline{\lambda}_{j}\,^{\dot{\beta}} \overline{\rho}_{i}\,^{\dot{\alpha}} - \frac{1}{64}{\rm i} {y}^{-2} {e}^{U} (\sigma_{a})^{\alpha}{}_{\dot{\beta}} \overline{\psi}^{a i \dot{\beta}} \lambda^{j}\,_{\alpha} \overline{\lambda}_{j \dot{\alpha}} \overline{\rho}_{i}\,^{\dot{\alpha}} - \frac{1}{16}{\rm i} {e}^{U} (\sigma_{ab})^{\beta\rho} (\sigma_{c})^{\alpha}{}_{\dot{\alpha}} \psi^{a}\,_{i \beta} \psi^{b}{}^{j}_{\rho} \overline{\psi}^{c}{}_{j}^{\dot{\alpha}} \rho^{i}\,_{\alpha}%
+\frac{1}{16}{\rm i} {e}^{U} (\sigma_{ac})^{\beta\rho} (\sigma_{b})^{\alpha}{}_{\dot{\alpha}} \psi^{a}\,_{i \beta} \psi^{b}{}^{j}_{\rho} \overline{\psi}^{c}{}_{j}^{\dot{\alpha}} \rho^{i}\,_{\alpha}+\frac{1}{16}{\rm i} {e}^{U} (\sigma_{ac})^{\beta\alpha} (\sigma_{b})^{\rho}{}_{\dot{\alpha}} \psi^{a}{}^{j}_{\beta} \psi^{b}\,_{i \rho} \overline{\psi}^{c}{}_{j}^{\dot{\alpha}} \rho^{i}\,_{\alpha} - \frac{3}{32}{e}^{U} \epsilon^{a}\,_{b c d} (\sigma_{a})^{\alpha}{}_{\dot{\alpha}} \psi^{b}\,_{i}\,^{\beta} \psi^{c}{}^{j}_{\beta} \overline{\psi}^{d}{}_{j}^{\dot{\alpha}} \rho^{i}\,_{\alpha} - \frac{1}{8}{\rm i} {e}^{U} (\sigma_{b})^{\beta}{}_{\dot{\alpha}} \psi_{a i}\,^{\alpha} \psi^{b}{}^{j}_{\beta} \overline{\psi}^{a}{}_{j}^{\dot{\alpha}} \rho^{i}\,_{\alpha}+\frac{1}{8}{\rm i} {e}^{U} (\sigma_{b})^{\beta}{}_{\dot{\alpha}} \psi_{a i}\,^{\alpha} \psi^{a}{}^{j}_{\beta} \overline{\psi}^{b}{}_{j}^{\dot{\alpha}} \rho^{i}\,_{\alpha}+\frac{1}{8}{\rm i} c {y}^{-1} {e}^{U} (\sigma_{a})^{\beta}{}_{\dot{\alpha}} \psi^{a}\,_{j}\,^{\alpha} \boldsymbol{\phi} \lambda^{j}\,_{\beta} \overline{\lambda}_{i}\,^{\dot{\alpha}} \rho^{i}\,_{\alpha} - \frac{1}{8}{\rm i} c {y}^{-1} {e}^{U} (\sigma_{a})^{\beta}{}_{\dot{\alpha}} \psi^{a j \alpha} \boldsymbol{\phi} \lambda_{i \beta} \overline{\lambda}_{j}\,^{\dot{\alpha}} \rho^{i}\,_{\alpha} - \frac{1}{8}{\rm i} c {y}^{-1} {e}^{U} (\sigma_{a})^{\alpha}{}_{\dot{\beta}} \overline{\psi}^{a j}\,_{\dot{\alpha}} \boldsymbol{\phi} \lambda^{i}\,_{\alpha} \overline{\lambda}_{j}\,^{\dot{\beta}} \overline{\rho}_{i}\,^{\dot{\alpha}} - \frac{1}{8}{\rm i} c {y}^{-1} {e}^{U} (\sigma_{a})^{\alpha}{}_{\dot{\beta}} \overline{\psi}^{a j}\,_{\dot{\alpha}} \overline{\boldsymbol{\phi}} \lambda^{i}\,_{\alpha} \overline{\lambda}_{j}\,^{\dot{\beta}} \overline{\rho}_{i}\,^{\dot{\alpha}} - \frac{5}{128}{\rm i} {e}^{3U} (\sigma_{a})^{\alpha}{}_{\dot{\alpha}} \overline{\psi}^{a j}\,_{\dot{\beta}} \rho^{i}\,_{\alpha} \overline{\rho}_{j}\,^{\dot{\alpha}} \overline{\rho}_{i}\,^{\dot{\beta}}+\frac{1}{8}{\rm i} c {y}^{-1} {e}^{U} (\sigma_{a})^{\beta}{}_{\dot{\alpha}} \psi^{a}\,_{j}\,^{\alpha} \overline{\boldsymbol{\phi}} \lambda^{j}\,_{\beta} \overline{\lambda}_{i}\,^{\dot{\alpha}} \rho^{i}\,_{\alpha} - \frac{1}{8}{\rm i} c {y}^{-1} {e}^{U} (\sigma_{a})^{\beta}{}_{\dot{\alpha}} \psi^{a j \alpha} \overline{\boldsymbol{\phi}} \lambda_{i \beta} \overline{\lambda}_{j}\,^{\dot{\alpha}} \rho^{i}\,_{\alpha}+\frac{1}{16}{e}^{2U} \psi_{a}\,^{j \alpha} \overline{\psi}^{a}\,_{i \dot{\alpha}} \rho^{i}\,_{\alpha} \overline{\rho}_{j}\,^{\dot{\alpha}}+\frac{1}{8}{\rm i} \epsilon^{a}\,_{b c d} (\sigma_{a})^{\alpha}{}_{\dot{\beta}} (\sigma_{e})^{\beta}{}_{\dot{\alpha}} \psi^{b}{}^{i}_{\alpha} \psi^{c}{}^{j}_{\beta} \overline{\psi}^{d}{}_{j}^{\dot{\alpha}} \overline{\psi}^{e}{}_{i}^{\dot{\beta}}+\frac{1}{4}(\sigma_{a})^{\alpha}{}_{\dot{\alpha}} (\sigma_{b})^{\beta}{}_{\dot{\beta}} \psi^{a}{}^{i}_{\alpha} \psi^{b}{}^{j}_{\beta} \overline{\psi}_{c i}\,^{\dot{\alpha}} \overline{\psi}^{c}{}_{j}^{\dot{\beta}}+\frac{1}{2}(\sigma_{c})^{\alpha}{}_{\dot{\beta}} (\sigma_{b})^{\beta}{}_{\dot{\alpha}} \psi_{a}\,^{i}\,_{\alpha} \psi^{b}{}^{j}_{\beta} \overline{\psi}^{a}{}_{j}^{\dot{\alpha}} \overline{\psi}^{c}{}_{i}^{\dot{\beta}} - \frac{1}{8}(\sigma_{a})^{\beta}{}_{\dot{\beta}} (\sigma_{b})^{\alpha}{}_{\dot{\alpha}} \psi^{a}{}^{i}_{\alpha} \psi^{b}{}^{j}_{\beta} \overline{\psi}_{c i}\,^{\dot{\alpha}} \overline{\psi}^{c}{}_{j}^{\dot{\beta}} - \frac{1}{4}(\sigma_{b})^{\alpha}{}_{\dot{\beta}} (\sigma_{c})^{\beta}{}_{\dot{\alpha}} \psi_{a}\,^{i}\,_{\alpha} \psi^{b}{}^{j}_{\beta} \overline{\psi}^{a}{}_{j}^{\dot{\alpha}} \overline{\psi}^{c}{}_{i}^{\dot{\beta}}+\frac{1}{4}(\sigma_{b})^{\alpha}{}_{\dot{\alpha}} (\sigma_{c})^{\beta}{}_{\dot{\beta}} \psi_{a}\,^{i}\,_{\alpha} \psi^{a}{}^{j}_{\beta} \overline{\psi}^{b}{}_{i}^{\dot{\alpha}} \overline{\psi}^{c}{}_{j}^{\dot{\beta}}+\frac{1}{8}(\sigma_{b})^{\alpha}{}_{\dot{\beta}} (\sigma_{c})^{\beta}{}_{\dot{\alpha}} \psi_{a}\,^{i}\,_{\alpha} \psi^{a}{}^{j}_{\beta} \overline{\psi}^{b}{}_{j}^{\dot{\alpha}} \overline{\psi}^{c}{}_{i}^{\dot{\beta}}%
+\frac{1}{4}\psi_{a}\,^{i \alpha} \psi_{b}\,^{j}\,_{\alpha} \overline{\psi}^{a}\,_{j \dot{\alpha}} \overline{\psi}^{b}{}_{i}^{\dot{\alpha}} - \frac{1}{8}{\rm i} \epsilon^{a}\,_{b c d} (\sigma_{a})^{\beta}{}_{\dot{\beta}} (\sigma_{e})^{\alpha}{}_{\dot{\alpha}} \psi^{b}{}^{i}_{\alpha} \psi^{e}{}^{j}_{\beta} \overline{\psi}^{c}{}_{i}^{\dot{\alpha}} \overline{\psi}^{d}{}_{j}^{\dot{\beta}}+\frac{1}{64}{\rm i} {y}^{-2} {e}^{U} (\sigma_{a})^{\beta}{}_{\dot{\alpha}} \psi^{a}\,_{i \beta} \lambda^{j \alpha} \overline{\lambda}_{j}\,^{\dot{\alpha}} \rho^{i}\,_{\alpha} - \frac{1}{64}{\rm i} {y}^{-2} {e}^{U} (\sigma_{a})^{\beta}{}_{\dot{\alpha}} \psi^{a}\,_{i}\,^{\alpha} \lambda^{j}\,_{\beta} \overline{\lambda}_{j}\,^{\dot{\alpha}} \rho^{i}\,_{\alpha}+\frac{1}{8}{\rm i} {e}^{U} (\sigma_{b})^{\alpha}{}_{\dot{\beta}} \psi_{a}\,^{j}\,_{\alpha} \overline{\psi}^{a}\,_{j \dot{\alpha}} \overline{\psi}^{b i \dot{\beta}} \overline{\rho}_{i}\,^{\dot{\alpha}}+\frac{1}{8}{\rm i} {e}^{U} (\sigma_{a})^{\alpha}{}_{\dot{\beta}} \psi^{a}{}^{j}_{\alpha} \overline{\psi}_{b}\,^{i \dot{\beta}} \overline{\psi}^{b}\,_{j \dot{\alpha}} \overline{\rho}_{i}\,^{\dot{\alpha}} - \frac{5}{128}{\rm i} {e}^{3U} (\sigma_{a})^{\alpha}{}_{\dot{\alpha}} \psi^{a}\,_{i}\,^{\beta} \rho^{i}\,_{\alpha} \rho^{j}\,_{\beta} \overline{\rho}_{j}\,^{\dot{\alpha}} - \frac{1}{16}{\rm i} {y}^{-2} \epsilon^{a b}\,_{c d} (\sigma_{a})^{\beta}{}_{\dot{\beta}} (\sigma_{b})^{\alpha}{}_{\dot{\alpha}} \psi^{c}{}^{j}_{\beta} \overline{\psi}^{d}{}_{j}^{\dot{\beta}} \lambda^{i}\,_{\alpha} \overline{\lambda}_{i}\,^{\dot{\alpha}} - \frac{1}{32}{e}^{2U} \psi_{a i}\,^{\beta} \psi^{a}\,_{j}\,^{\alpha} \rho^{i}\,_{\alpha} \rho^{j}\,_{\beta}+\frac{1}{8}{\rm i} {e}^{U} (\sigma_{b})^{\alpha}{}_{\dot{\beta}} \psi_{a}\,^{i}\,_{\alpha} \overline{\psi}^{a j}\,_{\dot{\alpha}} \overline{\psi}^{b}{}_{j}^{\dot{\beta}} \overline{\rho}_{i}\,^{\dot{\alpha}} - \frac{1}{8}{\rm i} {e}^{U} (\sigma_{a})^{\alpha}{}_{\dot{\beta}} \psi^{a}{}^{i}_{\alpha} \overline{\psi}_{b}\,^{j \dot{\beta}} \overline{\psi}^{b}\,_{j \dot{\alpha}} \overline{\rho}_{i}\,^{\dot{\alpha}} - \frac{1}{8}{\rm i} {e}^{U} (\sigma_{b})^{\beta}{}_{\dot{\alpha}} \psi_{a j \beta} \psi^{a j \alpha} \overline{\psi}^{b}{}_{i}^{\dot{\alpha}} \rho^{i}\,_{\alpha}+\frac{1}{8}{\rm i} {e}^{U} (\sigma_{b})^{\beta}{}_{\dot{\alpha}} \psi_{a i \beta} \psi^{a j \alpha} \overline{\psi}^{b}{}_{j}^{\dot{\alpha}} \rho^{i}\,_{\alpha}+\frac{1}{8}{\rm i} {e}^{U} (\sigma_{b})^{\beta}{}_{\dot{\alpha}} \psi_{a j}\,^{\alpha} \psi^{b}{}^{j}_{\beta} \overline{\psi}^{a}{}_{i}^{\dot{\alpha}} \rho^{i}\,_{\alpha}+\frac{1}{8}{\rm i} {e}^{U} (\sigma_{b})^{\beta}{}_{\dot{\alpha}} \psi_{a}\,^{j \alpha} \psi^{b}\,_{i \beta} \overline{\psi}^{a}{}_{j}^{\dot{\alpha}} \rho^{i}\,_{\alpha}+\frac{1}{256}{\rm i} {c}^{-1} {y}^{-1} {e}^{2U} \epsilon^{a b}\,_{c d} (\sigma_{a})^{\alpha}{}_{\dot{\alpha}} (\sigma_{b})^{\beta}{}_{\dot{\beta}} \psi^{c}\,_{j \alpha} \psi^{d}{}^{j}_{\beta} {\boldsymbol{\phi}}^{-1} \overline{\rho}^{i \dot{\alpha}} \overline{\rho}_{i}\,^{\dot{\beta}} - \frac{1}{64}{\rm i} {y}^{-2} {e}^{U} (\sigma_{a})^{\alpha}{}_{\dot{\alpha}} \overline{\psi}^{a i}\,_{\dot{\beta}} \lambda^{j}\,_{\alpha} \overline{\lambda}_{j}\,^{\dot{\beta}} \overline{\rho}_{i}\,^{\dot{\alpha}}+\frac{1}{64}{\rm i} {y}^{-2} {e}^{U} (\sigma_{a})^{\alpha}{}_{\dot{\alpha}} \psi^{a}\,_{i}\,^{\beta} \lambda^{j}\,_{\beta} \overline{\lambda}_{j}\,^{\dot{\alpha}} \rho^{i}\,_{\alpha}+\frac{1}{16}{\rm i} {e}^{U} (\sigma_{b})^{\alpha}{}_{\dot{\rho}} (\tilde{\sigma}_{ac})_{\dot{\beta}\dot{\alpha}} \psi^{a}{}^{j}_{\alpha} \overline{\psi}^{b}{}_{j}^{\dot{\beta}} \overline{\psi}^{c i \dot{\rho}} \overline{\rho}_{i}\,^{\dot{\alpha}} - \frac{1}{16}{\rm i} {e}^{U} (\sigma_{a})^{\alpha}{}_{\dot{\beta}} (\tilde{\sigma}_{bc})_{\dot{\rho}\dot{\alpha}} \psi^{a}{}^{j}_{\alpha} \overline{\psi}^{b i \dot{\beta}} \overline{\psi}^{c}{}_{j}^{\dot{\rho}} \overline{\rho}_{i}\,^{\dot{\alpha}}%
+\frac{1}{16}{\rm i} {e}^{U} (\sigma_{b})^{\alpha}{}_{\dot{\alpha}} (\tilde{\sigma}_{ac})_{\dot{\beta}\dot{\rho}} \psi^{a}{}^{j}_{\alpha} \overline{\psi}^{b i \dot{\beta}} \overline{\psi}^{c}{}_{j}^{\dot{\rho}} \overline{\rho}_{i}\,^{\dot{\alpha}} - \frac{1}{16}{\rm i} {e}^{U} (\sigma_{ab})^{\rho\alpha} (\sigma_{c})^{\beta}{}_{\dot{\alpha}} \psi^{a}\,_{i \beta} \psi^{b}{}^{j}_{\rho} \overline{\psi}^{c}{}_{j}^{\dot{\alpha}} \rho^{i}\,_{\alpha}+\frac{1}{16}{\rm i} {e}^{U} (\sigma_{ac})^{\rho\alpha} (\sigma_{b})^{\beta}{}_{\dot{\alpha}} \psi^{a}\,_{i \beta} \psi^{b}{}^{j}_{\rho} \overline{\psi}^{c}{}_{j}^{\dot{\alpha}} \rho^{i}\,_{\alpha}+\frac{1}{16}{\rm i} {e}^{U} (\sigma_{ac})^{\beta\rho} (\sigma_{b})^{\alpha}{}_{\dot{\alpha}} \psi^{a}{}^{j}_{\beta} \psi^{b}\,_{i \rho} \overline{\psi}^{c}{}_{j}^{\dot{\alpha}} \rho^{i}\,_{\alpha}$\end{adjustwidth}

\begin{footnotesize}

\end{footnotesize}